\documentclass[aps,twocolumn,showpacs,pra,superscriptaddress,floatfix,10pt]{revtex4-1}

\usepackage{braket,hhline}
\usepackage{colortbl}

\usepackage{subfiles,soul}
\usepackage{amsmath, amsthm, amssymb, amsfonts, amsthm}
\usepackage[usenames,dvipsnames]{xcolor}
\usepackage{empheq}
\usepackage{graphicx}
\definecolor{textcolour}{rgb}{0.43, 0.43, 0.43}
\usepackage[caption=false]{subfig}
\usepackage{url}
\usepackage{multirow}
\usepackage{dsfont} %for identity operator
\usepackage{mathrsfs}
\usepackage{bm}
\usepackage{float}
\usepackage{tcolorbox}
\usepackage[retainorgcmds]{IEEEtrantools}
\usepackage{scalerel}
\usepackage{hhline}

\usepackage{marginnote}
\newcommand{\ken}{\hspace{0.03cm}}
\definecolor{myblue}{RGB}{100,190,180}
\definecolor{myblue2}{rgb}{0.47, 0.62, 0.8}
\definecolor{mygreen}{RGB}{154,205,50}

\definecolor{mygreen2}{rgb}{0.0, 0.5, 0.0}
\definecolor{myred}{rgb}{0.92, 0.3, 0.26}
\definecolor{mygray}{RGB}{160,160,160}
\definecolor{myorange}{RGB}{255,215,0}
\definecolor{myyellow}{rgb}{0.99, 0.93, 0.0}

\newcommand{\ketbra}[2]{|#1\rangle \! \langle #2|}

\makeatletter
\DeclareFontFamily{OMX}{MnSymbolE}{}
\DeclareSymbolFont{MnLargeSymbols}{OMX}{MnSymbolE}{m}{n}
\SetSymbolFont{MnLargeSymbols}{bold}{OMX}{MnSymbolE}{b}{n}
\DeclareFontShape{OMX}{MnSymbolE}{m}{n}{
    <-6>  MnSymbolE5
   <6-7>  MnSymbolE6
   <7-8>  MnSymbolE7
   <8-9>  MnSymbolE8
   <9-10> MnSymbolE9
  <10-12> MnSymbolE10
  <12->   MnSymbolE12
}{}
\DeclareFontShape{OMX}{MnSymbolE}{b}{n}{
    <-6>  MnSymbolE-Bold5
   <6-7>  MnSymbolE-Bold6
   <7-8>  MnSymbolE-Bold7
   <8-9>  MnSymbolE-Bold8
   <9-10> MnSymbolE-Bold9
  <10-12> MnSymbolE-Bold10
  <12->   MnSymbolE-Bold12
}{}

\let\llangle\@undefined
\let\rrangle\@undefined
\DeclareMathDelimiter{\llangle}{\mathopen}%
                     {MnLargeSymbols}{'164}{MnLargeSymbols}{'164}
\DeclareMathDelimiter{\rrangle}{\mathclose}%
                     {MnLargeSymbols}{'171}{MnLargeSymbols}{'171}
\makeatother
\begin{document}

\title{Dissipative quantum state preparation and metastability in two-photon micromasers}

\author{Andreas Kouzelis} %[a,b]
%[a]
\affiliation{School of Physics and Astronomy, University of Nottingham, Nottingham, NG7 2RD, United Kingdom}
%[b]
\affiliation{Centre for the Mathematics and Theoretical Physics of Quantum Non-Equilibrium Systems, School of Physics and Astronomy, University of Nottingham, Nottingham, NG7 2RD, United Kingdom}

\author{Katarzyna Macieszczak} %[c,a,b]
%[c]
\affiliation{TCM Group, Cavendish Laboratory, University of Cambridge, Cambridge, CB3 0HE, United Kingdom}
%[a]
\affiliation{School of Physics and Astronomy, University of Nottingham, Nottingham, NG7 2RD, United Kingdom}
%[b]
\affiliation{Centre for the Mathematics and Theoretical Physics of Quantum Non-Equilibrium Systems, School of Physics and Astronomy, University of Nottingham, Nottingham, NG7 2RD, United Kingdom}

\author{Ji\v{r}\'{i} Min\'{a}\v{r}} %[d,e,a,b]
%[d]
\affiliation{Institute for Theoretical Physics, University of Amsterdam, Science Park 904, 1098 XH Amsterdam, Netherlands}
%[e]
\affiliation{Department of Physics, Lancaster University, Lancaster, LA1 4YB, United Kingdom}
%[a]
\affiliation{School of Physics and Astronomy, University of Nottingham, Nottingham, NG7 2RD, United Kingdom}
%[b]
\affiliation{Centre for the Mathematics and Theoretical Physics of Quantum Non-Equilibrium Systems, School of Physics and Astronomy, University of Nottingham, Nottingham, NG7 2RD, United Kingdom}

\author{Igor Lesanovsky} %[a,b]
%[a]
\affiliation{School of Physics and Astronomy, University of Nottingham, Nottingham, NG7 2RD, United Kingdom}
%[b]
\affiliation{Centre for the Mathematics and Theoretical Physics of Quantum Non-Equilibrium Systems, School of Physics and Astronomy, University of Nottingham, Nottingham, NG7 2RD, United Kingdom}

\date{\today}

% -----------------------------------------------------------------------------
% -----------------------------------------------------------------------------
% -----------------------------------------------------------------------------

\begin{abstract}
We study the preparation of coherent quantum states in a two-photon micromaser for applications in quantum metrology. While this setting can be in principle realized in a host of physical systems,
we consider atoms interacting with the field of a cavity. We focus on the case of the interaction described by the Jaynes-Cummings Hamiltonian, which cannot be achieved by the conventional approach with three-level atoms coupled  to the cavity field at two-photon resonance. We find that additional levels are required in order to cancel Stark shifts emerging in the leading order. Once this is accomplished, the dynamics of the cavity features a degenerate stationary state manifold of pure states. We derive the analytic form of these states and show that they include Schr{\"o}dinger cat states with a tunable mean photon number. %as well as states reminiscent of grid states. 
We also confirm these states can be useful in phase estimation protocols with their quantum Fisher information exceeding the standard limit. To account for realistic imperfections, we consider single-photon losses from the cavity, finite lifetime of atom levels, and higher order corrections in the far-detuned limit, which result in metastability of formerly stationary cavity states and long-time dynamics with a unique mixed stationary state. Despite being mixed, this stationary state can still feature quantum Fisher information above the standard limit.
Our work delivers a comprehensive overview of the two-photon micromaser dynamics with particular focus on application in phase estimation and, while we consider the setup with atoms coupled to a cavity, the results can be directly translated to optomechanical systems.
\end{abstract}

\maketitle

\section{Introduction}

There is currently an intense effort to engineer quantum states in a number of platforms ranging from atomic ensembles to nanomechanical, cavity and circuit QED systems. The impressive experimental progress is documented by the creation of Schr{\"o}dinger cat states with more than 100 photons, together with the so-called compass states~\cite{zurek2001sub}, in circuit QED~\cite{vlastakis2013deterministically}, generation of squeezed coherent states in mechanical oscillators~\cite{wollman2015quantum,Rashid_2016,Pirkkalainen:15,Lecocq:15} and squeezed cat states using light at optical wavelengths~\cite{ourjoumtsev2007generation, Etesse_2015, Huang_2015}, traveling (itinerant) squeezed coherent states in the microwave domain~\cite{Flurin_2012, Mallet_2011, nakamura2013breakthroughs}, and spin-squeezed states in atomic ensembles~\cite{pezze2016non}. There are also experimental developments and theoretical proposals for interfacing different platforms in hybrid setups such as coupling a mechanical oscillator with passing Rydberg atoms via electric charge~\cite{stevenson2016prospects} or with Nitrogen-Vacancy center via magnetic field~\cite{Kepesidis_2016}.

Nowadays, the generation of quantum states goes beyond the well-established paradigm of squeezed coherent and cat states. A general paradigm of dissipative quantum state preparation was developed in Refs.~\cite{diehl_quantum_2008,kraus_preparation_2008} and encompasses the so-called grid states~\cite{Gottesman_2001,Duivenvoorden_2017,Terhal_2016} as well as squeezed and displaced superpositions of a finite number of phonons~\cite{Brunelli2018a,Brunelli2018b}.
The produced quantum states find applications to quantum information processing and quantum enhanced sensing~\cite{Munro_2002,gilchrist2004schrodinger, Knott_2016}, ranging from ultra sensitive force measurements in optomechanical systems~\cite{anetsberger2009near,gavartin2012hybrid} to probes of macroscopic-scale decoherence~\cite{bose1999scheme,arndt2014testing} or dark matter detection~\cite{Bateman:15}.

Among possible approaches to the robust quantum state engineering are those based on two-photon processes. In the seminal work on two-photon micromasers by Haroche and co-workers~\cite{davidovich1987two,Brune_1987}, a stream of three-level atoms passed through a microwave cavity, allowing for photon exchange between the cavity field and the atoms.  For the energy gap between the ground and the excited (top) atom levels equal to double the frequency of the cavity  and the  middle level being far detuned, the resulting dynamics corresponded to a simultaneous exchange of two photons between the atom and the cavity~\cite{Brune_1987rydberg_two_phot, davidovich1987two,Ashraf1990_2ph_micromaser_statistics,toor1996theory}. Following this work,  the two-photon resonance is now exploited in stabilization of Schr\"{o}dinger cat states~\cite{Sarlette_2011}, in ultrasensitive electro-measurements based on Rydberg atoms interacting with a microwave cavity~\cite{facon2016sensitive}, in two-photon lasing by a superconducting qubit~\cite{Neilinger_2015}, and in dynamical protection and reservoir engineering in circuit QED~\cite{mirrahimi2014dynamically, roy2015continuous, leghtas2015confining}.  Despite the importance of the two-photon interactions in generation, manipulation, and exploitation of quantum information, it has been shown that the two-photon micromasers based on three-level systems feature only squeezed vacuum (squeezed single photon) or a Fock state as their stationary states \cite{orszag1993generation}.

\begin{figure*}[t]
	\includegraphics[width=\textwidth]{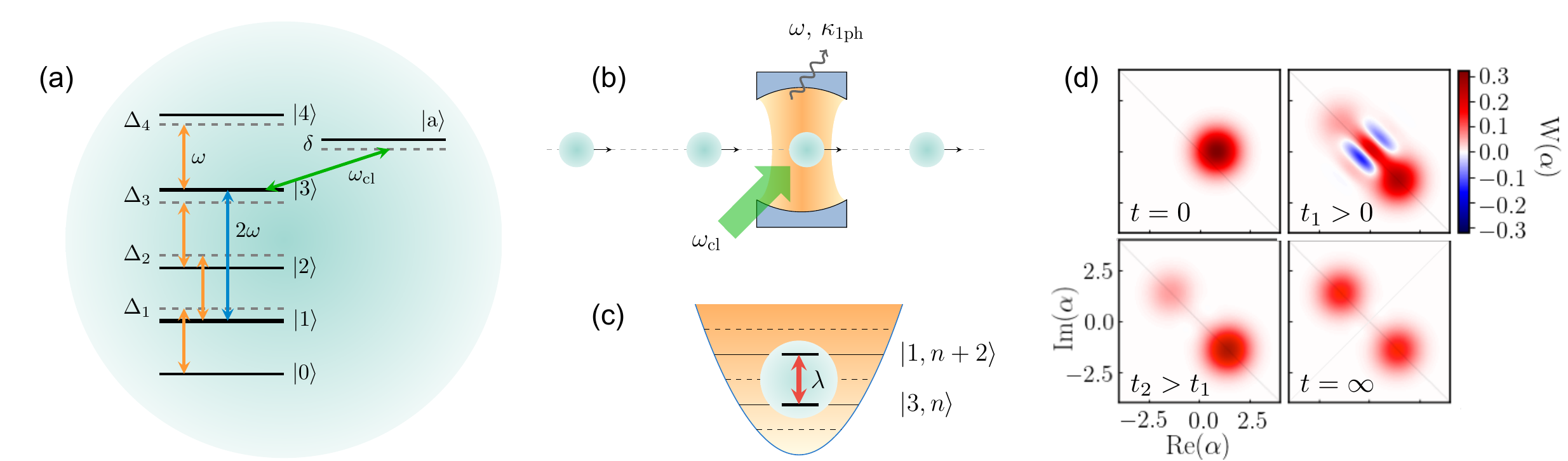}
	\caption{ {\bf (a)} \textbf{Atomic level structure}: the transitions $\ket{j-1} \leftrightarrow \ket{j}$, $j=1,..,4$ are coupled to the cavity field with the strengths $g_j$ and detunings $\Delta_j$. The transition $\ket{3} \leftrightarrow \ket{{\rm a}}$ is driven by a classical field with Rabi frequency $G$ and detuning $\delta$ (see Sec.~\ref{sec:ACinteraction}).  {\bf(b)} \textbf{Micromaser}: atoms are passing through a lossy cavity one at a time, interacting with a single-mode quantized cavity field of frequency $\omega$ (orange) and a classical Rabi field $G$ of frequency $\omega_{\rm cl}$ (green).
		{\bf (c)} \textbf{Effective dynamics}: at the two-photon resonance $\Delta_2=-\Delta_3$, the (5+1)-level model reduces to an effective two-photon Jaynes-Cummings interaction with the coupling strength $\lambda$ between the cavity field (depicted as a quantum harmonic oscillator) and the effective two-level atom with ground and exited states $|1\rangle$ and $|3\rangle$ (see Sec.~\ref{sec:2photon_int}).	
		{\bf (d)} \textbf{Micromaser dynamics in weak-coupling regime}: the Wigner function~\eqref{eq:Wigner} for the cavity state is shown. The initial coherent state $\ket{\alpha}$ with $\alpha=0.6$ evolves first into a DFS spanned by the odd and even cat states (time $t_1$), which would be stationary if not for single-photon losses from the cavity that renders it metastable. After the first metastable regime, the macroscopic coherence dephases (time $t_2$), leading to metastable mixture of coherent states. This mixture then finally relaxes into a unique stationary state~(time $t=\infty$) via mixing dynamics. In the second metastable regime ($t\geq t_2$), the system state features a single reflection symmetry, while the final parity-symmetric stationary state features two reflection symmetries (see Sec.~\ref{sec:dynamics}). The parameters are as in Fig.~\ref{fig:metastability}(b); see Sec.~\ref{sec:losses} for discussion.
	}
	\label{fig:model}\vspace*{-3mm}
	
\end{figure*}

In this work, we demonstrate that the limited set of two-photon micromaser stationary states is due to the Stark shifts present in the effective two-photon dynamics~\cite{Brune_1987rydberg_two_phot, davidovich1987two, Ashraf1990_2ph_micromaser_statistics,toor1996theory}. We show that the Stark shifts can be removed by considering a scheme with $(5+1)$-level atoms, where four single-photon transitions are driven by the cavity field and one transition is driven by a classical Rabi field (see Fig.~\ref{fig:model}). This leads to the atom-cavity interaction given by a two-photon Jaynes-Cummings Hamiltonian~\cite{gerry1988two-photon} without the spurious Stark shifts and opens doors to the dissipative generation of novel pure quantum states.

 For a pure state of incoming atoms, we derive the resulting pure stationary states, which depend both on the initial atomic state and the time integral of the atom-cavity coupling strength, in contrast with the three-level setup where the stationary states depend only on the atomic state~\cite{Orszag1992_squeezed,orszag1993generation,roa1995phase}. We investigate the usefulness of the generated state in phase estimation by means of the quantum Fisher information (QFI)~\cite{Helstrom1967,Helstrom1968,Braunstein1994} and find that a number of states yield the QFI exceeding not only the standard quantum limit, but also the performance of the squeezed coherent, cat, and squeezed cat states generated by the micromaser in the weak-coupling limit. Some of the generated states with a high QFI display a delocalized Wigner function~\cite{cahill1969density} and bear resemblance to the so-called grid states~\cite{Gottesman_2001,Duivenvoorden_2017,Terhal_2016}.

To account for cavity imperfections and finite detuning of the cavity fields from the atomic transitions, we consider single-photon losses from the cavity and higher order corrections to the effective two-photon atom-cavity interaction. In the limit of a small loss rate and large detunings, we discuss the resulting metastability of the pure states and their long-time dynamics leading to a unique mixed stationary state of the cavity field \cite{macieszczak2016towards}. In the weak-coupling regime, our results are consistent with the recent findings for the harmonic oscillator with two-photon driving and two-photon losses, which features Schr\"{o}dinger cat states as pure stationary states~\cite{mirrahimi2014dynamically,azouit2016well},  but in the presence of single-photon losses, displays mixing dynamics and a unique stationary state~\cite{minganti2016exact,Bartolo_2016,azouit2015convergence}.  Importantly, we find that, although the stationary states of the cavity  are no longer pure, their QFI can still feature enhancement beyond the standard quantum limit.

The article is structured as follows. In Sec.~\ref{sec:model}, we discuss the dynamics of micromaser with $(5+1)$-level atoms  leading to the effective two-photon dynamics in the far-detuned regime. In Sec.~\ref{sec:steadystates}, we investigate the resulting pure stationary states of the cavity field, while in Sec. \ref{sec:metastability} we include the effects of higher order corrections, single-photon losses, atom decay and distribution of atom velocities.  Motivated by the application in quantum metrology,  in  Sec.~\ref{sec:metrology} we characterize the dissipatively generated states by the QFI. Finally,  in Sec. \ref{sec:Implementations}  we discuss possible experimental platforms, and we conclude in Sec. \ref{sec:Conclusions}.

%%% 2 - MODEL %%%
%*******************************************************************************************************************************************************
%*******************************************************************************************************************************************************

\section{Two-photon micromaser with (5+1)-level atoms}
\label{sec:model}

In this section we introduce the (5+1)-level model of the atom-cavity interaction and, in the far-detuned limit, derive the effective two-photon dynamics with tunable Stark shifts. We further focus on the case when the Stark shifts are canceled,  and discuss the corresponding micromaser dynamics. This condition will prove to be crucial for dissipative generation of novel pure quantum states of the cavity presented in Sec. \ref{sec:steadystates}. 

\subsection{Atom-cavity interaction}
\label{sec:ACinteraction}

We consider (5+1)-level atoms with the levels $\ket{j}$ and the energies $E_j$, $j = 0, 1, ..., 4, {\rm a}$, and the cavity field with the frequency $\omega$. The transitions $\ket{j-1} \leftrightarrow \ket{j}$ are coupled to the cavity field with the strengths $g_j$, $j=1,..,4$, and the transition  $\ket{3} \leftrightarrow \ket{\rm a}$ to the auxiliary level $\ket{\rm a}$ is driven by a classical field of frequency $\omega_{\rm cl}$ and Rabi frequency $G$ [see Fig.~\ref{fig:model}(a)].

We assume that the detunings $\Delta_j$, $j=1,..,4$, and $\delta$, defined as
\begin{subequations}
	\label{eq:energygap}
	\begin{align}
		(E_j-E_0) &= j\omega + \sum_{i=1}^j \Delta_i ,\quad j=1,..,4, \\
		(E_{\rm a}-E_0) &= 3\omega +\sum_{i=1}^3 \Delta_i+ \omega_{\rm cl}+\delta,
	\end{align}
\end{subequations}
are much smaller than the corresponding energy gaps, $|\Delta_j|\ll  \omega$ for $j=1,..,4$, and $|\delta|\ll  \omega_\text{cl}$,  cf.~Fig.~\ref{fig:model}(a), which leads to Jaynes-Cummings Hamiltonian  via the rotating wave approximation,
\begin{subequations}
	\label{eq:H_5plus1_levels}
	\begin{align}
	H_0 &= \sum_{j=1}^4 \sigma_{jj} \sum_{i=1}^j \Delta_i +\sigma_{\rm aa} \Big(\delta+\sum_{j=1}^3 \Delta_j\Big), \label{eq:H0}\\ 
	H_{\rm int}  &= a\sum_{j=1}^{4} g_j \,\sigma_{j (j-1)} + G \sigma_{\rm a3} + {\rm H.c.},\label{eq:Hint}
	\end{align}
\end{subequations}
where $\sigma_{ij}=\ketbra{i}{j}$,   $a$ and $a^{\dagger}$ denote the cavity annihilation and creation operators, $\hbar=1$, and we consider the frame rotating with $\omega_{\rm cl}\sigma_{\rm aa}+\omega N$, where $N = a^\dag a + \sum_{j=1}^4 j\sigma_{jj}+3\sigma_{\rm aa}$ (see Appendix~\ref{app:Transformations}). Since the total number of excitations $N$ is conserved by $H=H_0 + H_{\rm int}$, the dynamics can in principle be solved by diagonalizing $H$ in six-dimensional eigenspaces of $N$.

%===============================================================================================================================================

\subsection{Effective two-photon interaction}
\label{sec:2photon_int}

In order to obtain two-photon dynamics of the atom and the cavity, 
we assume \emph{two-photon resonance}
\begin{equation}\label{eq:resonance}
\Delta_2=-\Delta_3\equiv\Delta, 
\end{equation}
which leads to degeneracy of $|1\rangle$ and $|3\rangle$ in $H_0$, and consider the levels $\ket{0}$, $\ket{2}$, $\ket{4}$, and $\ket{\rm a}$ to be far detuned from the one-photon transitions, i.e., $|g_j/\Delta_j| \ll 1$, $j=1,..,4$ and $|G/\delta| \ll 1$.  In this case, $H_{\rm int}$ in Eq.~\eqref{eq:Hint} can be treated as a perturbation of $H_0$ in Eq.~\eqref{eq:H0} by means of adiabatic elimination~\cite{Tannoudji_1998, Alexanian1995unitary,Klimov2002}. In Appendix~\ref{app:adiabatic}, we show that up to the second order the dynamics couples only the levels $\ket{1}$ and $\ket{3}$ via the effective Hamiltonian 
\begin{eqnarray}  \label{eq:Heff0}
	H_{\rm eff} &= & - \frac{g_2 g_3}{\Delta}\,a^{2}  \ken \sigma_{31} -    \frac{g_2^* g_3^*}{\Delta}\,a^{\dagger 2}  \ken \sigma_{13}\\\nonumber
	\\\nonumber
	&& + \left[\frac{|g_1|^2}{\Delta_1} -a^{\dagger}a \left(\frac{|g_2|^2}{\Delta} - \frac{|g_1|^2}{\Delta_1}\right)\right]\ken \sigma_{11}
	 \\\nonumber
		&&- \left[ \left(\frac{|G|^2}{\delta} + \frac{|g_3|^2}{\Delta}\right)+a^{\dagger}a\left(\frac{|g_4|^2}{\Delta_4} + \frac{|g_3|^2}{\Delta}\right)\right]\ken\sigma_{33}  ,
\end{eqnarray} 
where we omitted $\Delta_1\left(\sigma_{11} + \sigma_{33}\right)$ (constant in the subspace of $\ket{1}$ and $\ket{3}$). As $H_{\rm eff}$ conserves the number of excitations $N_\text{eff}=a^\dagger a+\sigma_{11}+3\sigma_{33}$, the corresponding atom-cavity dynamics can be solved exactly by diagonalizing $H_{\rm eff}$ restricted to two-dimensional $N_\text{eff}$ eigenspaces (see Appendix~\ref{app:2photon}).

The second and third lines in Eq.~\eqref{eq:Heff0} correspond to the Stark shifts, which crucially influence the dynamics of cavity coherences in the Fock basis (cf.~\cite{Toor1992validity,Boone1989effective_hamiltonians, Ashraf1990_2ph_micromaser_statistics, Dung1994}). In particular, the Stark shifts are canceled when
\begin{subequations}
	\label{eq:conditions}
	\begin{align}
		\frac{|g_1|^2}{\Delta_1} &= \frac{|g_2|^2}{\Delta}, \label{eq:conditions_a}\\
		\frac{|g_4|^2}{\Delta_4} &= -\frac{|g_3|^2}{\Delta}, \label{eq:conditions_b}\\
		\frac{|G|^2}{\delta} &= -\frac{|g_3|^2+|g_2|^2}{\Delta},\label{eq:conditions_c}
	\end{align}
\end{subequations}
in which case the Hamiltonian \eqref{eq:Heff0} reduces to the two-photon Jaynes-Cummings Hamiltonian~\cite{gerry1988two-photon} 
\begin{equation}\label{eq:Heff}
	H_{\rm eff} =\lambda\, a^2 \ken \sigma_{31}+\lambda^* a^{\dagger  2} \ken \sigma_{13} ,
\end{equation}
where $\lambda= - g_2 g_3/\Delta$ is the effective two-photon coupling strength [see Fig.~\ref{fig:model}(c)] and we omitted  $\frac{|g_2|^2}{\Delta}\left(\sigma_{11} + \sigma_{33}\right)$ (constant in the considered subspace of $\ket{1}$ and $\ket{3}$). 

We emphasize that (5+1)-level scheme in Fig.~\ref{fig:model} is a \emph{minimal} model to cancel the Stark shifts [cf.~Eq.~\eqref{eq:conditions}]. This is the case on which we focus on in this work, motivated by the dissipative generation of a plethora of distinct pure quantum states in Sec.~\ref{sec:steadystates}. Actually, in Appendix~\ref{app:2photon} we show that only in this case does the adiabatic two-photon dynamics between the cavity and the atoms generate stationary states of the cavity which are pure and dependent on both the atom state and the atom-cavity coupling. For any other setup, including the three-level scheme~\cite{Ashraf1990_2ph_micromaser_statistics,toor1996theory,Orszag1992_squeezed,orszag1993generation,Alexanian1998_trapped}
\begin{eqnarray}  \label{eq:Heff_3level}
H_{\rm eff}^{\rm 3-level} &= & - \frac{g_2 g_3}{\Delta}\,a^{2}  \ken \sigma_{31} -    \frac{g_2^* g_3^*}{\Delta}\,a^{\dagger 2}  \ken \sigma_{13}\\\nonumber
\\\nonumber
&& -\frac{|g_2|^2}{\Delta} \,a^{\dagger}a  \, \sigma_{11}- \frac{|g_3|^2}{\Delta} \left(a^{\dagger}a+1\right)  \sigma_{33}.  
\end{eqnarray} 
(obtained with $|\Delta_1|,|\Delta_4|,|\delta|\rightarrow\infty$ or equivalently $g_1=g_4=G=0$), pure stationary states, if generated, always correspond to the squeezed vacuum state and squeezed single-photon state, independently from the atom state. 
Furthermore, this means that our study, together with the earlier work~\cite{Orszag1992_squeezed,orszag1993generation,Alexanian1998_trapped}, provides the \emph{complete analysis} of dissipative  generation of pure states in two-photon micromasers based on single-photon Jaynes-Cummings interaction~\cite{jaynes1963comparison}.

%Can you extend this paragraph slightly? I think it is important to emphasize that (13) can indeed not be obtained within a "conventional" two-level on 2-photon resonance. We need to say that the 5+1 model permits to access nontrivial cavity states.

\subsection{Two-photon micromaser}
\label{sec:dynamics}

The \emph{micromaser} is a setup in which atoms pass through the cavity, one at a time, and interact with its field [see Fig.~\ref{fig:model}(b) and Appendix~\ref{app:micromaser}]. We consider atoms of the same velocity (a monochromatic beam) and  initially in a \emph{pure state} ($|c_{\rm g}|^2 + |c_{\rm e}|^2 = 1$)
\begin{equation}\label{eq:psi}
\ket{\psi_{\mathrm{at}}}=c_{\rm g} \ket{1}+c_{\rm e}\ket{3},
\end{equation}
where the amplitudes $c_e$ and $c_g$ will enable us to control the coherence of the generated cavity states. Since the effective dynamics couples only $\ket{1}$ and $\ket{3}$ levels, they can be viewed as the ground state and the excited state of the effective two-level atom interacting with the cavity.

\subsubsection{Micromaser dynamics}

In the frame rotating with the free Hamiltonian, the cavity state changes only when an atom is passing through. For an atom in a pure superposition, Eq.~\eqref{eq:psi}, interacting with the cavity for time $\tau$, the state of the cavity after passage of $k$ atoms is 
\begin{equation}
\label{eq:discrete_2}
\rho^{(k)} =\sum_{j={\rm g,e}} M_{j}\, \rho^{(k-1)} M_{j}^{\dagger }\equiv\mathcal{M}_0\,[\rho^{(k-1)}],
\end{equation}
where the Kraus operators [cf.~Eq.~\eqref{eq:Heff}] 
\begin{subequations}
	\label{eq:Kraus}
	\begin{align}
	M_{\rm g} &=\langle 1|  \mathcal{T} e^{- i\int_{0}^{\tau} d t \ken H_\text{eff}(t)}|\psi_\text{at}\rangle \nonumber \\
	&= c_{\mathrm{g}}  \ken \cos\left(\phi \sqrt{a^{\dagger  2}\ken a^2}\right)- ic_{\mathrm{e}}\ken a^{\dagger  2}\ken\frac{\sin\left(\phi \sqrt{a^2\ken a^{\dagger  2}}\right)}{\sqrt{ a^2\ken a^{\dagger  2}}},\label{eq:Kraus1}\\
	M_{\rm e} &=\langle 3| \mathcal{T} e^{- i\int_{0}^{\tau} d t \ken H_\text{eff}(t)}|\psi_\text{at}\rangle \nonumber \\
	&=- ic_{\mathrm{g}} \ken a^2\ken\frac{\sin\left(\phi \sqrt{a^{\dagger 2}\ken a^2}\right)}{\sqrt{a^{\dagger  2}\ken a^2}}+c_{\mathrm{e}}\cos\left(\phi\sqrt{a^{2}\ken a^{\dagger  2}}\right)\!\!,\label{eq:Kraus2}
	\end{align}
\end{subequations}
with $\mathcal{T}$ denoting time ordering and the integrated coupling strength $\phi = \int_{0}^{\tau} d t\ken \lambda(t)$~\footnote{Here we assumed $\lambda(t)$ real. If the coupling $\lambda(t)$ is complex,  its phase instead adds to the relative phase between the atom state amplitudes $c_{\rm g}$ and $c_{\rm e}$.}; see Appendix~\ref{app:micromaser} for derivation.

For atoms arriving to the cavity at the rate $\nu$~\cite{davidovich1987two,englert2002elements,guerra1991role}, the \emph{average} micromaser dynamics is governed by the master equation~\cite{Lindblad1976,Gorini1976} 
\begin{equation}
\frac{ d }{ d t} \rho(t)  = \nu\,\mathcal{M}_0 [\rho(t)] - \nu\,\rho(t)\equiv \mathcal{L}_0\, [\rho(t)].
\label{eq:L0}
\end{equation}
In this work, we mostly consider the continuous dynamics (\ref{eq:L0}). The comparison of the results to the case of discrete dynamics~(\ref{eq:discrete_2}) can be found in Appendix~\ref{app:numerics}.

The subscript $0$ in Eqs.~\eqref{eq:discrete_2} and~\eqref{eq:L0} indicates the far-detuned limit in which two-photon dynamics in Eq.~\eqref{eq:Heff} is achieved. We consider the effect of the higher order corrections to this limit, as well as single-photon losses, later in Sec.~\ref{sec:metastability}, while the influence of approximately fulfilled conditions of~Eqs.~\eqref{eq:resonance} and~\eqref{eq:conditions}, a mixed atom state, and a nonmonochromatic atom beam, is discussed in Appendix~\ref{app:Leff}.

\subsubsection{Properties}

The micromaser dynamics generated by~\eqref{eq:Kraus} features only two-photon transitions, so that the parity   
\begin{equation}
	P= (-1)^{a^\dagger a}
	\label{eq:parity_op}
\end{equation}	
commutes with the Kraus operators,
\begin{equation} \label{eq:parity} 
	[M_{\rm g, e},\, P] = 0,
\end{equation}
and is \emph{conserved} during the evolution, 
\begin{equation} \label{eq:parity_cons} 
\frac{ d }{ d t} \mathrm{Tr}\left[P\ken\rho(t)\right]=  \mathrm{Tr}\left\{P \mathcal{L}_0\left[\rho(t)\right]\right\}= \mathrm{Tr}\left[ \mathcal{L}_0^\dagger( P) \ken\rho(t)\right]= 0,
\end{equation}
as we have $\mathcal{M}_0^\dagger(P)=P$ and thus $\mathcal{L}_0^\dagger(P)=0$. In particular, a cavity state initially supported in the even (odd) subspace, remains there at all times, which implies the existence of \emph{even and odd stationary states}. % and the conservation of the projections on the odd and even subspace, $\mathds{1}_{\pm}=(\mathds{1} \pm P)/2$. 
We will show in Sec.~\ref{sec:steadystates} these stationary states are generally pure.

The Kraus operators (\ref{eq:Kraus}) become real upon the  transformation $a\mapsto e^{- i(\varphi/2-\pi/4)}a$ where $c_{\rm g}/c_{\rm e}={\rm e}^{i\varphi} |c_{\rm g}/c_{\rm e}|$.  Therefore, an initial state of the cavity with real-valued coefficients in the transformed basis, remains real at all times, and so the odd and even stationary states must be \emph{realvalued} in this basis.  

Conservation of the parity and real-valued dynamics are reflected in the \emph{reflection symmetries} of the Wigner function~\cite{cahill1969density,royer1977wigner} for the stationary states,
\begin{equation}
	W(\alpha)=\frac{2}{\pi}\text{Tr}\left[\rho \ken D(\alpha) \ken P \ken D(-\alpha) \right],
	\label{eq:Wigner}
\end{equation}
where $D(\alpha) = \exp \left(\alpha a^{\dagger}-\alpha^* a\right)$ is the displacement operator  [see Figs.~\ref{fig:wig_gall}(a) and~\ref{fig:hard_walls}].  First, for an even or odd $\rho$, we have $P\rho P=\rho$, while $P^2=\mathds{1}$ and $P D(\alpha) P=D(-\alpha)$, and thus  $W(\alpha)=\frac{2}{\pi}\text{Tr}\{\rho [ P \ken D(\alpha) \ken P] P[ P\ken D(-\alpha) P] \}  =W(-\alpha)$, which is the inversion symmetry. Second, for a real-valued cavity state in the transformed basis $a\mapsto e^{- i(\varphi/2-\pi/4)}a$, we have $W(\alpha)=W^*(\alpha)=\frac{2}{\pi}\mathrm{Tr}[\rho^* D(\alpha^*)P D(-\alpha^*) ]=W(\alpha^*)$, which is the reflection symmetry with respect to the real axis [cf. the system state for $t\geq t_2$ in Fig.~\ref{fig:model}(d)]. Therefore, together with the inversion symmetry, we also obtain the reflection symmetry with respect to the imaginary axis.

% *****************************************************************************************************************************************************
% *****************************************************************************************************************************************************
\section{Pure stationary states of two-photon micromaser and relaxation timescales}
\label{sec:steadystates}
% *****************************************************************************************************************************************************

We now show that the two-photon micromaser introduced in Sec.~\ref{sec:model} features pure stationary states of odd and even parities. The coherences between the states are also stationary, forming a decoherence-free subspace~\cite{Zanardi1997d,Zanardi1997e,Lidar1998c}. In particular, in the weak-coupling limit, the stationary states become odd and even Schr\"{o}dinger cat states~\cite{Dodonov1974, gerry1993non} with a tunable mean photon number. We also discuss the possibility of trapping states~\cite{orszag1993generation}, which, in turn, provides an insight into emergent slow timescales during the relaxation toward the pure stationary states.

% *****************************************************************************************************************************************************
% *****************************************************************************************************************************************************

 \subsection{Pure stationary states}
 \label{sec:recurrence}

 \begin{figure*}[t]
 	\begin{minipage}[t]{0.575\textwidth}
 		% Wig gallery
 		%\centering
 		\begin{flushleft}
 			\text{\hspace*{1mm} \textbf{(a)}}
 		\end{flushleft}
 		\vspace*{-5mm}
 		\begin{flushleft}
 			\hspace*{-9mm}
 			\includegraphics[width=0.95\textwidth]{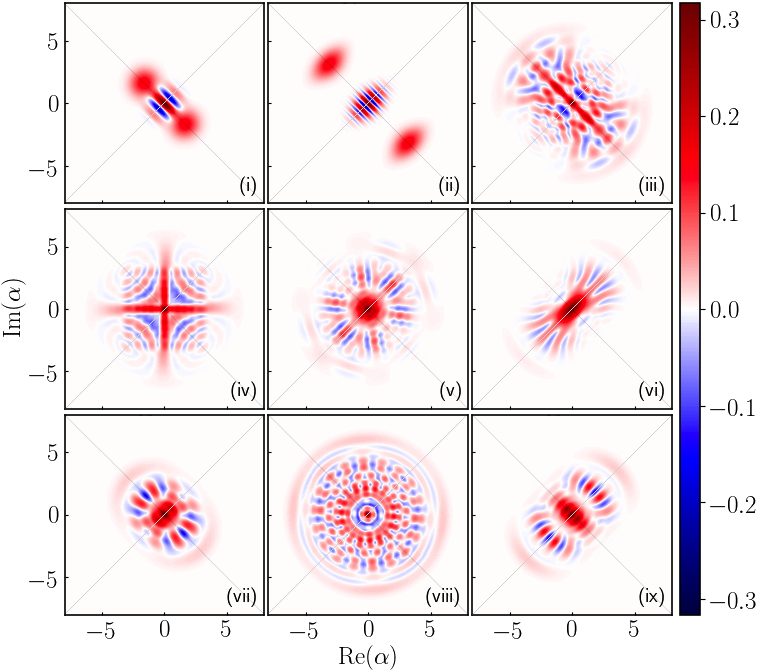}
 		\end{flushleft}
 	\end{minipage}%
 	\begin{minipage}[t]{0.425\textwidth}
 		% Table with gallery states
 		%\centering
 		
 		\begin{flushleft}\text{\hspace*{-10mm}\textbf{(b)}}\end{flushleft}
 		\vspace*{-9mm}
 		\hspace*{-13mm}
 		%\hspace*{-5mm}
 		\subfloat{\includegraphics[width=1.25\columnwidth]{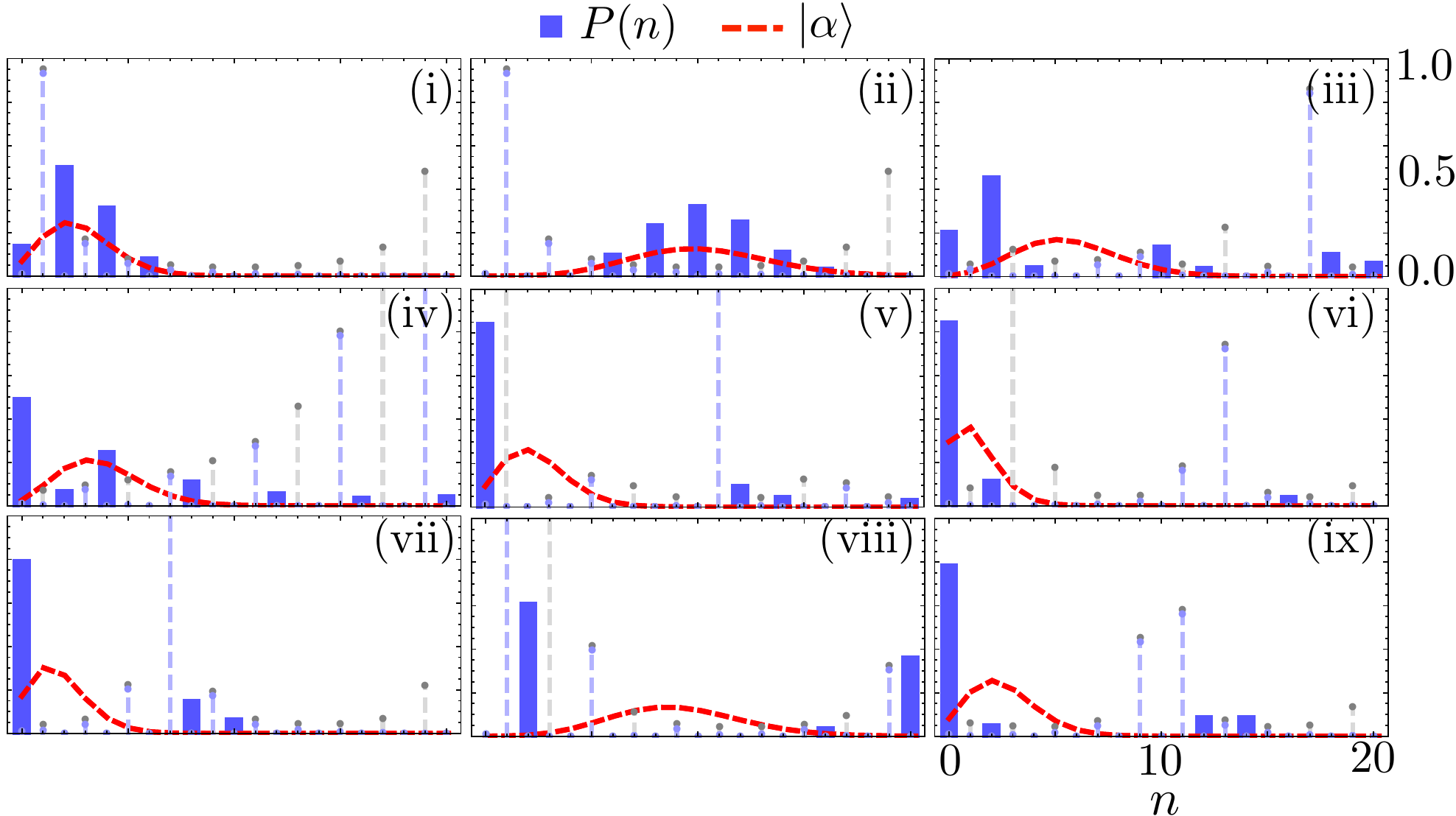}}
 		\vspace*{-5.5mm}
 		\begin{flushleft}\text{\hspace*{-10mm}\textbf{(c)}}\end{flushleft}
 		\vspace*{-2mm}
 		\hspace*{-12mm}
 		\begin{tabular}{|c||c|c|c|c|c|c|}
 			\hline			
 			State & $(K, c_{\rm e})$ & $\braket{n}$ & $\Delta n^2$ & $\!F_Q/4\!\braket{n}\!$ & $ \max\sin_{2n}^{-2}(\phi)$ & $k_{\rm ss}$\\
 			\hhline{|=||=|=|=|=|=|=|}	 	 
 			(i) &\!(1, 0.20)\!&  2.69 & 2.70 & 1.00 & 24& 100\\
 			\hline 
 			(ii) &\!(1, 0.70)\!& 10.13 & 6.36 &  0.63 & 24& 50\\
 			\hline
 			(iii) &\!(5, 0.65)\!& 5.58 & 41.20 & 7.39 & 22& $10^3$\\
 			\hline
 			(iv) &\!(11, 0.60)\!& 3.66 & 25.24 & 6.90 & 179& $10^3$\\
 			\hline
 			(v) &\!(15, 0.65)\!& 2.35 & 29.10 & 12.38 &593& $1.5\times 10^4$\\
 			\hline
 			(vi) &\!(19, 0.65)\!& 1.22 & 15.53 & 12.72 & 27& $3\times 10^4$\\
 			\hline
 			(vii) &\!(23, 0.15)\!& 1.77 & 12.08 & 6.84 & 1767& $10^4$\\
 			\hline
 			(viii) &\!(31, 0.80)\!& 8.94 & 73.98 & 8.28&6670 & $5\times 10^4$\\
 			\hline
 			(ix) &\!(41, 0.40)\!& 2.51 & 26.26 & 10.46& 15 & $2\times 10^3$\\
 			\hline
 		\end{tabular}

 	\end{minipage}
 	%\vspace*{-1mm}
 	\caption{\textbf{Pure stationary states of cavity dynamics}: \textbf{(a)}~Wigner function [Eq.~\eqref{eq:Wigner}] for even cavity stationary states corresponding to the parameters in panel (c) [and indicated in Fig.~\ref{fig:qfi}(d)].  The two reflection symmetries (along diagonal gray lines) are due to the stationary states being parity symmetric and real valued (after adding the phase $\pi/4$) (see Sec.~\ref{sec:dynamics}). \textbf{(b)}~The photon-number distribution of the states (blue bars, only even photon numbers) is compared to that of the coherent states with the same average photon number $\braket{n}$ (red dashed lines). Blue dashed lines show  $\cot_{2n}^2(\phi/2)/10$,  which diverges as $4/\sin_{2n}^2(\phi)$ [gray dashed lines] for soft walls concurring with the boundary condition for stationary states (see Sec.~\ref{sec:walls_soft}). \textbf{(c)}~Properties of stationary states (i)-(ix): the parameters $(K,\, c_{\rm e})$ [which determine $\phi$ by Eq.~\eqref{eq:phi_wall}, where the hard wall is at $m=20$; for $\phi$ see also the last panel in Fig.~\ref{fig:qfi}, while $c_g=\sqrt{1-c_e^2}$], the mean photon number $\braket{n}$, the variance $\Delta n^2$, the enhancement~\eqref{eq:qfi_enh} in phase estimation, the maximal rate related to even soft wall $\max_{0\leq 2n\leq m} 1/\sin_{2n}(\phi)$, 	and the estimated number of atoms $k_{\rm ss}$ for which the stationary states are reached, as characterized by the fidelity $F[\rho_\text{ss};\rho(k)] = {\rm Tr} \ \sqrt{\sqrt{\rho_\text{ss}}\, \rho(k) \sqrt{\rho_\text{ss}}} \geq 0.99$, for the cavity initially in the vacuum state $|0\rangle$.}
 	\label{fig:wig_gall}
 	\vspace*{-3mm}
 \end{figure*}
 
 The stationary states of the cavity satisfy $\frac{ d }{ d t} \rho_\text{ss}=\mathcal{L}_0\left(\rho_\text{ss}\right)=0$, which is equivalent to $ \mathcal{M}_0\left(\rho_\text{ss}\right)=\rho_\text{ss}$. When the stationary state is pure,  $\rho_\text{ss}=\ket{\Psi_\text{ss}}\!\!\bra{\Psi_\text{ss}}$, it is necessarily an eigenstate of all Kraus operators, 
\begin{subequations}
	\label{eq:pure_cond}
	\begin{align}
	M_{\rm g}|\Psi_\text{ss}\rangle &= \alpha|\Psi_\text{ss}\rangle, \\
	M_{\rm e}|\Psi_\text{ss}\rangle &= \beta|\Psi_\text{ss}\rangle. 
	\end{align}
\end{subequations}
Indeed, in order to maintain its purity, the cavity state must be uncorrelated from the outgoing atom state, $e^{- i\int_{0}^{\tau} d t \ken H_\text{eff}(t)}(\ket{\psi_\text{at}}\otimes\ket{\Psi_\text{ss}})=(\alpha\ket{1}+\beta\ket{3})\otimes \ket{\Psi_\text{ss}}$ [cf.~Eq.~\eqref{eq:Kraus}] and we have $|\alpha|^2 + |\beta|^2 = 1$ from the state normalization.

 \subsubsection{Recurrence relation}
For the pure stationary state $|\Psi_\text{ss}\rangle=\sum_{n=0}^\infty c_n|n\rangle$, Eq.~\eqref{eq:pure_cond} corresponds to
\begin{subequations}
	\label{eq:pure_eq}
	\begin{align}
	\alpha\, c_{n+2} &= c_{\rm g} \cos_n(\phi) \, c_{n+2} - i c_{\rm e} \sin_n(\phi) \, c_n, \label{eq:pure_eq1} \\
	\beta\, c_n &= - i c_{\rm g} \sin_n(\phi) \,c_{n+2}+c_{\rm e} \cos_n(\phi)\,c_n, \label{eq:pure_eq2}	
	\end{align}
\end{subequations}
where we defined 
		$\cos_n(\phi) =\cos[\phi\sqrt{(n+1)(n+2)}]$ and
		$\sin_n(\phi) =\sin[\phi\sqrt{(n+1)(n+2)}]$.
The solutions exist when the determinant of Eq.~\eqref{eq:pure_eq}, $\alpha\beta+c_{\rm e}c_{\rm g}-\cos_n(\phi)(\alpha c_{\rm e}+\beta c_{\rm g})$ is $0$, and thus  
\begin{equation}
\alpha=\pm c_{\rm g}, \quad \beta= \mp\,c_{\rm e}, \label{eq:atom_out}
\end{equation}
leading to recurrence relation for coefficients of the stationary states,
\begin{equation}
c_{n+2} =\mp  i\,\frac{c_{\rm e}}{c_{\rm g}}\frac{\sin_n(\phi)}{1 \mp \cos_n(\phi)}\, c_n=\mp  i\,\frac{c_{\rm e}}{c_{\rm g}}\left[\cot_n\!\left(\frac{\phi}{2}\right)\right]^{\pm} c_n.
\label{eq:psi_ss}
\end{equation}
 We note that the odd and even stationary states are determined independently by Eq.~\eqref{eq:psi_ss}, which a consequence of the parity conservation (cf.~Sec.~\ref{sec:dynamics}). Here we assumed that $c_{\rm g} \neq 0$ and $1 \mp \cos_n(\phi) \neq 0$; we revisit these assumptions in~Sec.~\ref{sec:walls_hard}.

\subsubsection{Boundary conditions}
Since $a^2|0\rangle=0=a^2|1\rangle$, from Eqs.~\eqref{eq:Kraus} and~\eqref{eq:pure_cond} we also obtain the boundary conditions
\begin{equation}
\alpha\, c_{0}= c_{\rm g}\,c_{0}, \quad \alpha\, c_{1}= c_{\rm g}\,c_{1},  \label{eq:psi_boundary}
\end{equation}
which determine the outgoing atom state as
\begin{equation}
 \alpha=c_{\rm g}\quad \text{and}\quad \beta=-c_{\rm e}, \label{eq:atom_boundary},
 \end{equation}
  independently of $\phi$ \cite{orszag1993generation,Orszag1992_squeezed}. Therefore, the recurrence relation (\ref{eq:psi_ss}) leads to the existence of odd and even pure stationary states,   
\begin{subequations}
	\label{eq:even_odd_ss}
	\begin{align}
	\ket{\Psi_{+}} &= c_{0} |0\rangle+ c_{0} \sum_{n=1}^{\infty}\left(- i\,\frac{c_{\rm e}}{c_{\rm g}}\right)^n \, \prod_{k=0}^{n-1} \cot_{2k}\!\left(\frac{\phi}{2}\right) \ket{2n}, \\
	\ket{\Psi_{-}} &= c_{1} |1\rangle + c_{1} \sum_{n=1}^{\infty}\left(- i\,\frac{c_{\rm e}}{c_{\rm g}}\right)^n \, \prod_{k=0}^{n-1} \cot_{2k+1}\!\left(\frac{\phi}{2}\right) \ket{2n+1} ,
	\end{align}
\end{subequations}
 where $c_0$ and $c_1$ are determined, up to a phase, by the state normalization. In contrast to the case of the three-level micromaser~\cite{orszag1993generation,Orszag1992_squeezed}, here the stationary states are dependent not only on the incoming atom state,~\eqref{eq:psi}, but also on the integrated coupling $\phi$, which allows for dissipative generation of plethora of distinct stationary states; in Fig.~\ref{fig:wig_gall} we show a few examples. %We discuss their properties in the context for applications for quantum metrology in Sec.~\ref{sec:metrology}.

\subsection{Stationary decoherence-free subspace}
\label{sec:DFS}

Since the eigenvalues $\alpha$ and $\beta$ of the Kraus operators $M_g$ and $M_e$ [cf.~Eqs.~\eqref{eq:pure_cond} and~\eqref{eq:atom_out}], are the same for the odd and even pure stationary states, the even-odd coherences, $\ketbra{\Psi_{+}}{\Psi_{-}}$ and $\ketbra{\Psi_{-}}{\Psi_{+}}$, are also stationary, i.e.,  
\begin{equation} \label{eq:L0coh}
\mathcal{L}_0\left(\ketbra{\Psi_{+}}{\Psi_{-}}\right)=\nu\left(\alpha_+ \alpha_-^* + \beta_+ \beta_-^*-1\right)\!\ketbra{\Psi_{+}}{\Psi_{-}}=0.
\end{equation}
Therefore, any superposition of $\ket{\Psi_{+}}$ and $\ket{\Psi_{-}} $ is stationary, forming a decoherence-free subspace (DFS)~of a qubit~\cite{Zanardi1997d,Zanardi1997e,Lidar1998c}. 
The existence of the DFS can be made apparent by choosing the shifted Kraus operators
\begin{subequations}
	\label{eq:Kraus_shifted}
	\begin{align}
	\widetilde{M}_{\rm g}&= M_{\rm g} - c_{\rm g}\mathds{1}, \\
	\widetilde{M}_{\rm e}&= M_{\rm e} + c_{\rm e}\mathds{1},
	\end{align}
\end{subequations}
as jump operators in the master equation~\eqref{eq:L0}, in which case from ${c_{\rm g}^* M_{\rm g}-c_{\rm g}M_{\rm g}^\dagger }-c_{\rm e}^* M_{\rm e}+c_{\rm e}M_{\rm e}^\dagger =0$ [cf.~Eq.~\eqref{eq:Kraus}]
\begin{equation}
\frac{ d }{ d t} \rho(t)
=
\frac{\nu}{2}\sum_{j=\rm{g,e}} \left[2\widetilde{M}_j\rho(t)\widetilde{M}_j^\dagger- \widetilde{M}_j^\dagger \widetilde{M}_j\, \rho(t) - \rho(t)\,\widetilde{M}_j^\dagger \widetilde{M}_j\right],
\label{eq:master_noloss2}
\end{equation} 
The pure stationary states $\ket{\Psi_{+}}$ and $\ket{\Psi_{-}}$ are both \emph{dark}, i.e., $\widetilde{M}_{\rm g,e}|\Psi_\pm\rangle=0$, and thus their coherences do not decay.

In general, the asymptotic state of the cavity is  
	\begin{eqnarray}\label{eq:P00}
	\lim_{t\rightarrow\infty} e^{t\mathcal{L}_0}\rho&\equiv&\Pi_0\, (\rho) 
	 \\	\nonumber
	&=& \ketbra{\Psi_+}{\Psi_+}\,\mathrm{Tr}(\mathds{1}_+ \rho)+\ketbra{\Psi_-}{\Psi_-}\,\mathrm{Tr}(\mathds{1}_- \rho)
	 \\\nonumber
	& +&\ketbra{\Psi_+}{\Psi_-}\,\mathrm{Tr}(L_{+-} \rho)+\ketbra{\Psi_-}{\Psi_+}\,\mathrm{Tr}(L_{-+} \rho), 
	\end{eqnarray}
	where the superoperator $\Pi_0$ projects the initial cavity state $\rho$ on the stationary DFS with $\mathds{1}_+=(\mathds{1}+P)/2$ and $\mathds{1}_-=(\mathds{1}-P)/2$ being the projections on the odd and even subspaces, and $L_{+-}=L_{-+}^\dagger$ with $\mathrm{Tr}(L_{+-}\ketbra{\Psi_+}{\Psi_-})=1$ supported in the even-odd coherences, in which the structure reflects the parity conservation~\cite{Buca2012,mirrahimi2014dynamically,albert2014symmetries}. Furthermore, dynamics conserves $\mathds{1}_+$, $\mathds{1}_-$, $L_{+-}$, $L_{-+}$, and thus $L_{+-}$, $L_{-+}$  can be obtained numerically as
	\begin{equation} \label{eq:Lpm}
	L_{+-}=\lim_{t\rightarrow\infty} e^{t\mathcal{L}_0^\dagger}\left(|\Psi_{-}\rangle\!\langle\Psi_{+}|\right).
	\end{equation}
	% where the choice of the initial operator in the space of odd-even coherences as $|\Psi_{-}\rangle\!\langle\Psi_{+}|$, ensures the normalization of $L_{+-}$.
	
	%The disjoint support of the conserved quantities reflects the strong parity symmetry,~\eqref{eq:parity}, which implies that the master operator $\mathcal{L}_0$ is block-diagonal with respect to: the odd subspace, the even subspace, the even-odd coherences and the odd-even coherences, so that the corresponding parts of a density matrix evolve independently~\cite{Buca2012,mirrahimi2014dynamically,albert2014symmetries}. 

%===================================================================

\subsection{Schr\"{o}dinger cat states in weak-coupling limit}
\label{sec:schrodinger_cats}

We show that in the limit of the weak coupling, Schr\"{o}dinger cat states are recovered as stationary states of the cavity and its dynamics corresponds to two-photon drive and two-photon losses~\cite{hach1994generation,gilles1994generation,mirrahimi2014dynamically,azouit2015convergence,Bartolo_2016,minganti2016exact} [see Fig.~\ref{fig:model}(d) and state (i) in Fig.~\ref{fig:wig_gall}].

\subsubsection{Steady states}
 In the limit of the weak coupling, $|\phi|\ll 1$, the recurrence relation~\eqref{eq:psi_ss} with the boundary condition~\eqref{eq:atom_boundary} can be approximated as 
\begin{equation}
\frac{c_{n+2}}{c_n} = - i \frac{c_{\rm e}}{c_{\rm g}}\frac{2 }{ \phi \sqrt{(n+1)(n+2)}}+O\left[\frac{c_{\rm e}}{c_{\rm g}}\phi\sqrt{(n+1)(n+2)}\right]\!\!,\label{eq:app_lin}
\end{equation}
identifying the stationary states as the odd and even Schr\"{o}dinger cat states~\cite{Dodonov1974, gerry1993non} [see the state (i) in Fig.~\ref{fig:wig_gall}]
\begin{equation}
|\Psi_\pm\rangle\approx\frac{|\alpha\rangle\pm|-\alpha\rangle}{\sqrt{2\pm 2 e^{-2|\alpha|^2}}},\qquad \alpha= e^{- i\frac{\pi}{4}}\sqrt{\frac{2 c_{\rm e}}{c_{\rm g} \phi}}, \label{eq:psi_ss_lin}
\end{equation}  
with the coherent state $|\alpha\rangle\equiv e^{-|\alpha|^2/2} \sum_{n=0}^\infty \alpha^n/\sqrt{n!}\, |n\rangle$.
For validity of the approximation~\eqref{eq:app_lin} we require that the neglected terms are small, e.g. the first-order corrections to the fidelity, 
\begin{equation}
\sum_{n=0}^\infty \frac{1}{3} |c_n |^2 \left|\frac{c_{\rm e}} {c_{\rm g}}\right|^2=\frac{1}{3}\left|\frac{c_{\rm e}} {c_{\rm g}}\right|^2  \ll 1. 
\end{equation}
%This is satisfied for $c_{\rm e}/c_{\rm g}=O(\phi)$. 
Therefore, the conditions for obtaining Schr\"{o}dinger cat states are
\begin{equation}
|\phi|\ll 1 \qquad \text{and} \qquad |c_{\rm e}|\ll 1. \label{eq:cond_lin}
\end{equation}
We emphasize that the conditions on the parameters $c_e$ and $\phi$ are independent, and thus their ratio, as well as the value of $\alpha$, do \emph{not} need to be small [cf. the state (i) in Fig.~\ref{fig:wig_gall}]. Indeed, for large $|\alpha|$ we have that the photon distribution is centered around $2 |c_{\rm e}/(c_{\rm g} \phi)|$, since   $\langle n\rangle\approx |\alpha|^2\approx\Delta^2 n $, and thus the approximation in Eq.~\eqref{eq:app_lin} is still valid when $|c_{\rm e}|\ll 1$ [cf.~Eq.~\eqref{eq:cond_lin}].

\subsubsection{Dynamics}
 The Kraus operators in Eq.~\eqref{eq:Kraus_shifted} can be expanded in $\phi$ up to the quadratic terms in $\phi$ and $c_e$ [cf. Eq.~\eqref{eq:cond_lin}] as 
\begin{subequations}
	\label{eq:Kraus_lin}
	\begin{align}
\widetilde{M}_g&\approx - i c_{\rm e} \phi\, a^{\dagger  2} - c_{\rm g} \frac{\phi^2}{2}\,a^{\dagger  2}a^2
\approx 0,\\
\widetilde{M}_e&\approx 2c_{\rm e}\mathds{1} -i c_{\rm g} \phi\, a^2,%- c_{\rm e}\frac{\phi^2}{2}\,a^{2}a^{\dagger 2} 
%\approx 2c_{\rm e}\mathds{1} -i c_{\rm g} \phi\, a^2,
\end{align}
\end{subequations}
where in the first line we further neglected the terms which will contribute only in the fourth order to Eq.~\eqref{eq:master_noloss2}. Therefore, we arrive at the cavity dynamics
\begin{eqnarray}\label{eq:master_lin}
\frac{ d }{ d t} \rho&\approx& -i [g_\text{2ph}^* a^2+g_\text{2ph} a^{\dagger 2},\rho] \\\nonumber
&&+ \kappa_\text{2ph}\, a^2 \rho a^{\dagger 2} -\frac{\kappa_\text{2ph}}{2}\left(a^{2\dagger }a^2\,\rho+\rho \,a^{2\dagger }a^2\right) 
\end{eqnarray}
with
\begin{equation}
g_\text{2ph}=\nu c_{\rm g}^* c_{\rm e}\, \phi\qquad \text{and}\qquad \kappa_\text{2ph}=\nu |c_{\rm g}|^2 \phi^2,
\end{equation}
which are of the second order [cf.~Eq.~\eqref{eq:cond_lin}]. Equation~\eqref{eq:master_lin} describes an extensively  studied model of two-photon drive and two-photon losses~\cite{hach1994generation,gilles1994generation,Guerra_1997_PRA,Xie_2013_ProcRomAcademy,mirrahimi2014dynamically,azouit2015convergence,Bartolo_2016,minganti2016exact} leading to $\alpha=e^{- i\pi/4}\sqrt{2 g_\text{2ph}/\kappa_\text{2ph}} $ in Eq.~\eqref{eq:psi_ss_lin}. In particular, the conserved quantities $L_{+-}$ and $L_{-+}$ in Eq.~\eqref{eq:Lpm} are known exactly~\cite{mirrahimi2014dynamically} and thus so are the asymptotic states in Eq.~\eqref{eq:P00}. In Appendixes~\ref{app:Leff:conserved:atom_mixed}-\ref{app:Leff:conserved:atom_beam}, we  show that the two-photon cavity dynamics in Eq.~\eqref{eq:master_lin} is robust to both nonmonochromaticity of the atom beam and decay of the atom state toward levels uncoupled from the cavity field, but it is modified by two-photon injections when the atom state entering the cavity is mixed rather than pure [cf.~Eq.~\eqref{eq:psi}].

% ***************************************************************************************************************************************************

\subsection{Trapping states}
\label{sec:walls_hard}

Here we characterize the atom-cavity coupling strengths leading to the disconnected cavity dynamics.  Among others, this situation allows us to prepare the cavity in fixed photon number states, so-called trapping states~\cite{orszag1993generation}. 
We also discuss the purity of the resulting coherent stationary states.

\subsubsection{Hard walls}
 The terms in the Kraus operators, Eq.~(\ref{eq:Kraus}), that connect the cavity states $\ket{m}$ and $\ket{m+2}$ are proportional to $\sin_m(\phi)$. Therefore,  when the integrated interaction strength $\phi$ gives $\sin_m(\phi)=0$ for some $m$, that is
\begin{equation}
\phi=\frac{K\pi}{\sqrt{(m+1)(m+2)}} \quad\text{with}\quad K=\pm 1, \pm 2, ...,
\label{eq:phi_wall}
\end{equation}
so that $\cos_m(\phi)=(-1)^K$, the Kraus operators become \emph{block diagonal} in the Fock space, with the dynamics on the left (photon numbers $n \leq m$) and on the right ($n > m$) being independent. As the initial cavity state supported below $m$ and of the same parity as $m$, remains supported below $m$ at all times, we refer to this case as a \emph{hard wall} at $m$. This provides natural truncation points for the cavity Hilbert space in numerical simulations, which we exploit in Figs.~\ref{fig:wig_gall}-\ref{fig:phi_spread}.

\subsubsection{Trapping states}
 For the cavity pumped by the excited atoms ($|c_{\rm e}|=1$, $c_{\rm g}=0$), a hard wall at at $m$ corresponds to a trapping state $|m\rangle$, and thus can be used to obtain a fixed photon number.  Indeed, for a first hard wall at $m_1$, when the initial cavity state is of the same parity as $m_1$ and is supported below $m_1$, the asymptotic state is the pure trapping state $|\Psi_\text{ss}\rangle=|m_1\rangle$. A general initial state of the cavity evolves into a mixed state supported on all trapping states $|m_n\rangle$, and the asymptotic distribution, $p_{n}=\langle m_n|\rho_\text{ss}|m_n\rangle$, is given by the initial supports between subsequent walls of the same parity. It is also possible for coherences between the trapping states $|m_n\rangle$ to be stationary, which takes place when $\cos_{m_n}(\phi)=(-1)^{K_n}$ [cf. Eq.~\eqref{eq:phi_wall}] are of the same sign [cf.~Eq.~\eqref{eq:L0coh}]. In contrast, in the absence of even (odd) hard walls of a given parity, the cavity energy increases without a bound and there is no even (odd) stationary state.

We now show that the cavity dynamics features either no trapping states, or infinitely many (see Appendix~\ref{app:hard_walls}). This is due to the fact that, for a given coupling strength $\phi$ and the parameters $m_1$ and $K_1$ of the first wall,  Eq.~\eqref{eq:phi_wall} for $m_n$ and $K_n$ of another wall corresponds to the \emph{Pell equation}~\cite{Barbeau, Lenstra08solvingthe},
\begin{equation}
\label{eq:pell}
x^2 - D y^2 = 1,
\end{equation}
where the arguments $x= 2m_n+3$ and $y = 2K_n/K_1$ and the parameter $D = (m_1+1)(m_1+2)$.
As $D$ is not a square of an integer, the hyperbolic equation~\eqref{eq:pell} is known to feature infinitely many integer solutions~\cite{Copley1959}, which translate into the \emph{recurrence relation}
\begin{subequations}\label{eq:mK}
	\begin{align}
	m_n&=m_{n-1}(2m_1+3)+3(m_1+1)\\\nonumber
	&\quad +2 (m_1+1)(m_1+2)\, K_{n-1}/K_1,\\
	K_n&=K_{n-1}(2m_1+3) +K_1(2 m_{n-1}+3).	
	\end{align}
\end{subequations}
%where $K_{n-1}/K_1$ is necessarily an integer (cf.~Appendix~\ref{app:hard_walls}). 
Therefore, the position $m_n$ of hard walls \emph{grows exponentially} with $n$, 
\begin{eqnarray}\label{eq:m}
m_n= \frac{\left(\!2 m_1+3+2\sqrt{D} \right)^n\!\!+\!\left(\!2 m_1+3-2\sqrt{D} \right)^n\!	}{4}-3, \qquad
\end{eqnarray}
while the parity of $m_n$ and $K_n$ is determined as in Table~\ref{tab:mK}. For $m_1$ odd only odd trapping states exists, with all coherences stationary for $K_1$ even, while for $K_1$ odd only coherences between every second trapping state do not decay.  For $m_2$ even, the coherences between all (even and odd) trapping states are stationary when $K_1$ is even, while for $K_1$ odd, only the coherences between the trapping states of the same parity remain.

\subsubsection{Coherent stationary states between hard walls}

 For atoms prepared in the superposition~\eqref{eq:psi}, a hard wall at $m$ implies \emph{boundary conditions} for pure stationary states. Namely, for  $\sin_m(\phi)=0$ and $\cos_m(\phi)=(-1)^K$, Eq.~\eqref{eq:pure_eq}  gives
\begin{equation}
\beta \ken c_m =(-1)^K \ken c_{\rm e} \ken c_m, \label{eq:psi_boundary2a}
\end{equation}
for the coefficient $c_m$ of the pure stationary state before the wall, and 
\begin{equation}
\alpha \; c_{m+2} =(-1)^K \; c_{\rm g} \; c_{m+2}, \label{eq:psi_boundary2b}
\end{equation}
for the coefficient $c_{m+2}$  of the pure stationary state after the wall. Therefore, for a pure stationary state to exist between subsequent walls of the same parity, at $m_{n}$ and $m_{n'}$, %the boundary conditions after $m_n$ and before $m_{n'}$ must be simultaneously fulfilled, 
$\cos_{m_n}(\phi)=-\cos_{m_{n'}}(\phi)$, and, thus, odd $K_{n'}-K_{n}$ is required [see Eq.~\eqref{eq:phi_wall}]; otherwise a stationary state between $m_{n}$ and $m_{n'}$ is mixed, but still coherent in the photon-number basis [cf. Appendix~\ref{app:Leff:conserved:atom_mixed}].

In general, from  Table~\ref{tab:mK}, the stationary states between the walls are pure only when both $m_1$ and $K_1$ are odd; i.e., there are only odd hard walls. Otherwise, the stationary states must be mixed, except for the stationary state before the first wall for odd $K_1$ [cf.~Eq.~\eqref{eq:psi_boundary}]. They can be approximately pure if the support of the state vanishes at one of the hard walls~\footnote{This corresponds to the presence of a soft wall with {$\cos_{m'}(\phi)=-1$} between the hard walls (see Appendix~\ref{app:soft_walls}).}; see~Fig.~\ref{fig:hard_walls}(b). 

All coherences between pure and mixed stationary states decay~\footnote{Coherence to a pure stationary state {$|\Psi\rangle$} would be of the form {$|\Psi\rangle\!\langle \Phi|$}, where {$|\Phi\rangle$} is within the disjoint support of a mixed stationary state $\rho$. Since {$|\Psi\rangle$} can be considered a dark state [cf.~Eq.~\eqref{eq:Kraus_shifted}], any coherence decays with the corresponding effective Hamiltonian, unless it is a coherence to another dark  state, i.e., pure state with the same boundary conditions as {$|\Psi\rangle$}. {$|\Phi\rangle$} however corresponds to the mixed boundary conditions, since $\rho$ is mixed.}, and only the coherences between the pure stationary states with the same boundary condition are stationary (every second state for $m_1$ and $K_1$ odd), as the boundary conditions in Eqs.~\eqref{eq:psi_boundary2a} and~\eqref{eq:psi_boundary2b} determine the eigenvalues of the Kraus operators [cf.~Eqs.~\eqref{eq:L0coh}]. The latter is a consequence of the hard wall imprinting, with every passing atom, the opposite phases on the two stationary states before and after the wall, so that on average the coherence undergoes \emph{dephasing} and decays at the rate $2\nu$~\footnote{The coherence can be maintained using a feedback mechanism of counting the total number of atoms passing through the cavity, and then applying a phase flip, if necessary, to the final state (cf.~\cite{minganti2016exact}).} [see~Fig.~\ref{fig:hard_walls}(a)].
\begin{table}[t]
	\begin{tabular}{|c||c |c|}	 
		\hline
		& \multicolumn{1}{c|}{\multirow{2}{*}{$\cos_{m_1}(\phi)=1$}} & \multicolumn{1}{c|}{\multirow{2}{*}{$\cos_{m_1}(\phi)=-1$}}\\
		& \multicolumn{1}{c|}{} & \multicolumn{1}{c|}{} \\		
		\hline
		\hline
		\multirow{2}{*}{$m_1$ even} &  $m_{2n}$ odd, $m_{2n+1}$ even & $m_{2n}$ odd, $m_{2n+1}$ even  \\
		& $\cos_{m_n}(\phi)=1$ & $\cos_{m_n}(\phi)=(-1)^n$ \\
		\hline
		\multirow{2}{*}{$m_1$ odd}  & $m_n$ odd & \cellcolor{blue!10} $m_n$ odd \\
		&  $\cos_{m_n}(\phi)=1$ & \cellcolor{blue!10} $\cos_{m_n}(\phi)=(-1)^n$ \\		
		\hline
	\end{tabular}
	\caption{ \small \textbf{Parity of hard walls} located at $m_n$ from Eq.~\eqref{eq:mK} [cf.~Eq.~\eqref{eq:phi_wall}]. The blue shaded case is the only situation leading to pure coherent states between the hard walls [cf.~Eqs.~\eqref{eq:psi_boundary2a} and~\eqref{eq:psi_boundary2b}].
	}
	\label{tab:mK}
\end{table}
\begin{figure}[t]
	\begin{flushleft}
		\hspace*{-6mm}
		\includegraphics[width=1.2\columnwidth]{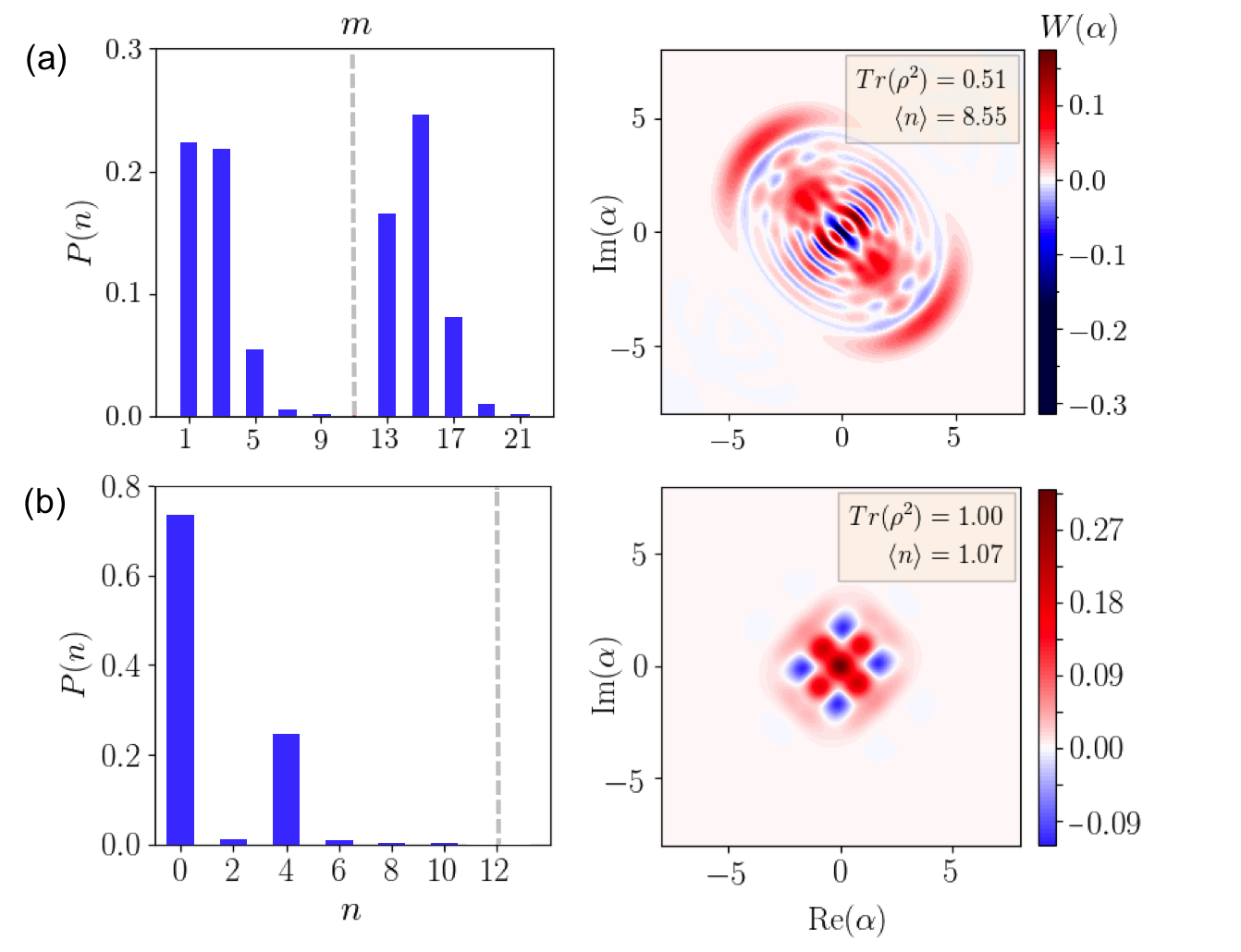}
	\end{flushleft}
	\vspace*{-6mm}
	\caption{
		\textbf{Steady states in the presence of hard walls}. The photon-number distribution $P(n)$ and the Wigner function [Eq.~\eqref{eq:Wigner}] for: \textbf{(a)} the equal mixture of the odd pure stationary states obtained from the initial superposition of odd Fock states $ (\ket{1} + \ket{15})/\sqrt{2}$ for $c_{\rm e}=0.3$  and the hard wall (dashed gray) at $m = 11$ with $K=1$ ($\phi\approx 0.252$) and \textbf{(b)} the approximately pure stationary state obtained from the initial vacuum state $|0\rangle$ for $c_{\rm e}=0.4$ and the hard wall at $m=12$ with $K=8$ ($\phi\approx 0.593$). 
	}\vspace*{-3mm}
	\label{fig:hard_walls}
\end{figure}

\subsection{Relaxation timescales}
\label{sec:walls_soft}
We now discuss how slow timescales arise in the relaxation toward pure stationary states as a result of approximately disconnected cavity dynamics. Furthermore, such structure of the dynamics facilitates multimodal photon number distributions in the stationary states. Derivations and further discussion can be found in Appendix~\ref{app:soft_walls}.

\subsubsection{Soft walls}

 We now consider the case when the terms in the Kraus operators of Eq.~(\ref{eq:Kraus}) that connect the cavity states $\ket{m}$ and $\ket{m+2}$ are close to $0$, i.e., $\sin_m (\phi) \approx 0$, but not equal $0$, so that the Kraus operators are only approximately block diagonal. We refer to this situation as a \emph{soft wall} at $m$. %with the height proportional to  $\sin_m^{-2}(\phi)$.  

\begin{figure}[t]
	\centering
	\hspace*{-5.5mm}
	\includegraphics[width=1.1\columnwidth]{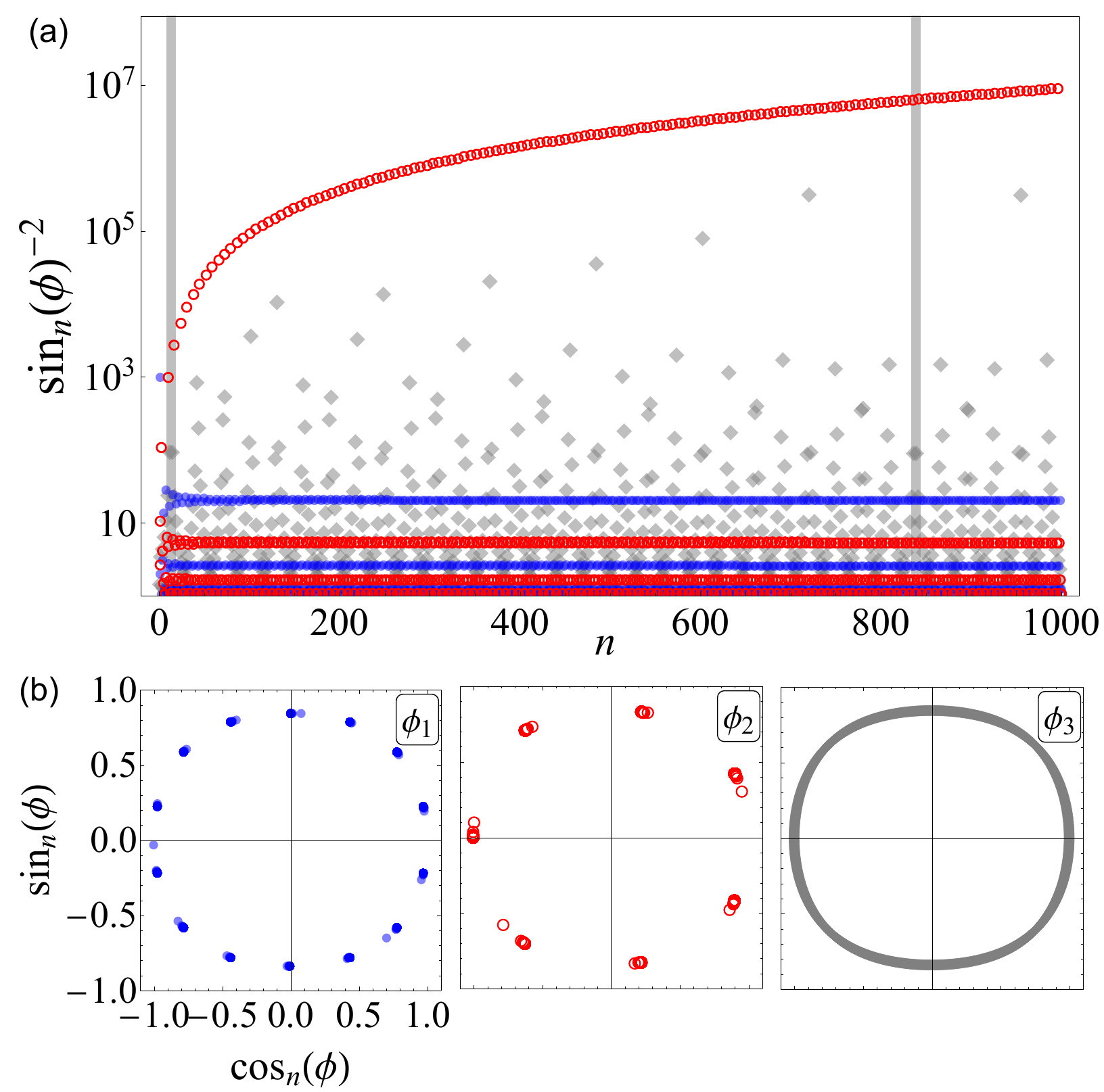}
	\caption{\textbf{Soft walls}. {\bf (a)} The function $\sin_n^{-2}(\phi)$ for:  rational $\phi_1/\pi=5/7$ (blue dots), $\phi_2/\pi=6/7$ (red circles), and irrational $\phi_{3}/\pi= 7/\sqrt{210}$ (gray diamonds), with the  hard walls (gray lines) at $m_1=13$ and $m_2=839$.  For $\phi_1$ the walls remain finite, in contrast to $\phi_2$, where $\sin_n^{-2}(\phi)$ diverges as $n^{-2}$ [cf.~Eq.~\eqref{eq:phi_taylor}] and $\phi_3$, where soft walls appear due to recurrence of the irrational rotation.  \textbf{(b)} The orbits for both  $\phi_1$ and $\phi_2$ are approximately periodic (with periods 14 and 7), while for $\phi_3$ the orbit is dense.   
	}				
	\label{fig:soft_walls}\vspace*{-3mm}
\end{figure}

 We confirm that the effects from soft walls indeed play an important role in the cavity dynamics, as for any integrated coupling strength $\phi$ such that $\phi/\pi$ is irrational or $\phi/\pi=p/q$ is rational with the even irreducible numerator $p$, there exists infinitely many soft walls (see Fig.~\ref{fig:soft_walls}).
 % [for a hard wall with $\phi$ given by Eq.~\eqref{eq:phi_wall}, $\phi/\pi$ is irrational, but with a rational square]. 
 %
 Indeed, from Taylor series 
 \begin{equation}\label{eq:phi_taylor}
 \phi\sqrt{(n+1)(n+2)} =\phi \left(\!n+\frac{3}{2}\right)+O\left(\frac{\phi}{n}\right),
 \end{equation}
 so that for $n$ large,  $\sin_n (\phi)\approx \sin[\phi (n +3/2) ]$ corresponds to $n$ rotations of a unit circle by $\phi$ with the initial phase $ 3 \phi/2$. 
 For an irrational $\phi/\pi$, values $e^{i\phi n}$ for all $n$ are dense in the circle, so that they pass within an arbitrary proximity by any point on the circle, and this takes place infinitely many times by Poincar\'e recurrence theorem. Therefore,  the cavity dynamics features infinitely many soft walls with $\sin_n(\phi)$ arbitrarily close to $0$, for both parities~\footnote{The odd or even parity correspond to $\sin_{2n} (\phi)$ or $\sin_{2n+1} (\phi)$, respectively, which are approximated as $n$ rotations by $2\phi$ with the initial phase shift $3\phi/2$ or $5\phi/2$, respectively. Thus, as for the irrational $\phi/\pi$, $2\phi/\pi$ and $3\phi/2$ is also irrational, there exist infinitely many soft walls of both parities.} (see $\phi_3$ in  Fig.~\ref{fig:soft_walls}). % As these properties of the rotation are not changed by the initial phase,
 In contrast, for a rational $\phi/\pi=p/q$, values of $e^{i\phi n}$ are periodic with the period $q$ for even $p$, and $2q$ for odd $p$. Therefore, from Eq.~\eqref{eq:phi_taylor}, the values of $\sin_n (\phi)$ become approximately periodic for large $n$, but with a shift in phase by $3 \phi/2$ [see $\phi_1$ and $\phi_2$ in Fig.~\ref{fig:soft_walls}(b)]. Nevertheless, soft walls  appear periodically when $\sin [\phi (m +3/2) ]=0=\sin (k\pi)$, which requires $ (2m+ 3) p = 2 k q$, i.e., $p$ to be even. In this case, soft walls appear at $m$ of both parities ($q$ is odd) with $\cos_m(\phi)\approx \cos[\phi (m +3/2) ]=(-1)^{p/2}$ and $\sin_m^{-2}(\phi)\approx (8m+12)^2/\phi^2$ [see $\phi_2$ in Fig.~\ref{fig:soft_walls}(a)]. 
 %no metastable pure states for periodic m, since the wall alternates.

\subsubsection{Slow relaxation}
The cavity dynamics with soft walls, $\sin_m (\phi) \approx 0$, can be considered as a local perturbation of the dynamics where the soft walls are replaced by hard walls, $\sin_m (\phi)= 0$. As discussed in Sec.~\ref{sec:walls_hard}, this auxiliary dynamics features stationary states supported between the introduced hard walls. As in reality the walls are soft, those states are not stationary, but become \emph{metastable}~\cite{macieszczak2016towards,Rose2016} and at long times undergo the effective dynamics at rates proportional to the perturbation size, i.e., $\nu \sin_m^2(\phi)$. Since the perturbation is local, the effective dynamics  connects only states across a single soft wall or introduces coherences between states separated by two walls. % (see Appendix~\ref{app:soft_walls_dynamics}). 
Furthermore, the dynamics rates are proportional to the state amplitude directly next to the soft wall, so that for the small amplitude the timescales of the dynamics are further extended [cf. the last two columns in~Fig.~\ref{fig:wig_gall}(c)].

\subsubsection{Multimodal pure stationary states}
 It follows from Eq.~\eqref{eq:even_odd_ss} that the stationary state of long-time dynamics across soft walls of a given parity must be pure. It is, however, approximately composed only from the metastable states supported between the walls, which can be pure or mixed depending on the wall boundary conditions (see Sec.~\ref{sec:walls_hard}). Therefore, the stationary state is approximately supported only on the pure metastable states, which, furthermore, obey the same boundary conditions as in Eq.~\eqref{eq:psi_boundary} to ensure metastable coherences between them [in Fig.~\ref{fig:wig_gall}(b) the states after blue and before gray soft walls]. As a result, the photon number distribution in the stationary state is \emph{multimodal}, as the pure states with the same boundary conditions are separated at least by two walls [cf. Eqs.~\eqref{eq:psi_boundary2a} and~\eqref{eq:psi_boundary2b}, and see~Fig.~\ref{fig:wig_gall}(b)]. Although, in general long experimental timescales are needed to prepare such multimodal pure states, they can be highly useful for quantum metrology applications, which we will discuss in Sec.~\ref{sec:metrology}.  

%distinct modes...

%
%
%
%

% *****************************************************************************************************************************************************
% *****************************************************************************************************************************************************

% *****************************************************************************************************************************************************
% *****************************************************************************************************************************************************

\section{Noise and higher order effects in micromaser dynamics}
\label{sec:metastability}

%From the answer to the refereee...

%In contrast, in our work we consider generation of pure steady states both for the weak and finite coupling. We do consider single-photon losses, as Ref. [77,78], but in place of numerical results, we use metastability theory to obtain analytic insight into the timescales and structure of the long-time dynamics both for all coupling strengths. Even in the already studied weak-coupling limit, we uncover the emergence of second classical metastable regime for finitely occupied cat states, (see Sec. IV B; earlier Sec. V B). Furthermore, these analytic results enable us to quantitatively consider the resulting change in the usefulness of generated states for potential applications, e.g., for phase estimation, as discussed in Sec. V B (earlier Sec. VI B). Finally, we also explicitly study the corrections to the two-photon dynamics using the (5+1) model obtained from first principles, which provides us with a direct check of validity of the used approximations. 

In Secs.~\ref{sec:model} and~\ref{sec:steadystates}, we considered the cavity dynamics in the far-detuned limit, where the interaction with  atoms was given by the two-photon Jaynes-Cummings Hamiltonian in Eq.~\eqref{eq:Heff}. The parity of photon number in the cavity was conserved leading to existence of even and odd stationary states, which were in general pure and with coherences between them also stationary.   

\begin{figure*}[ht]
	\includegraphics[width=0.9\textwidth]{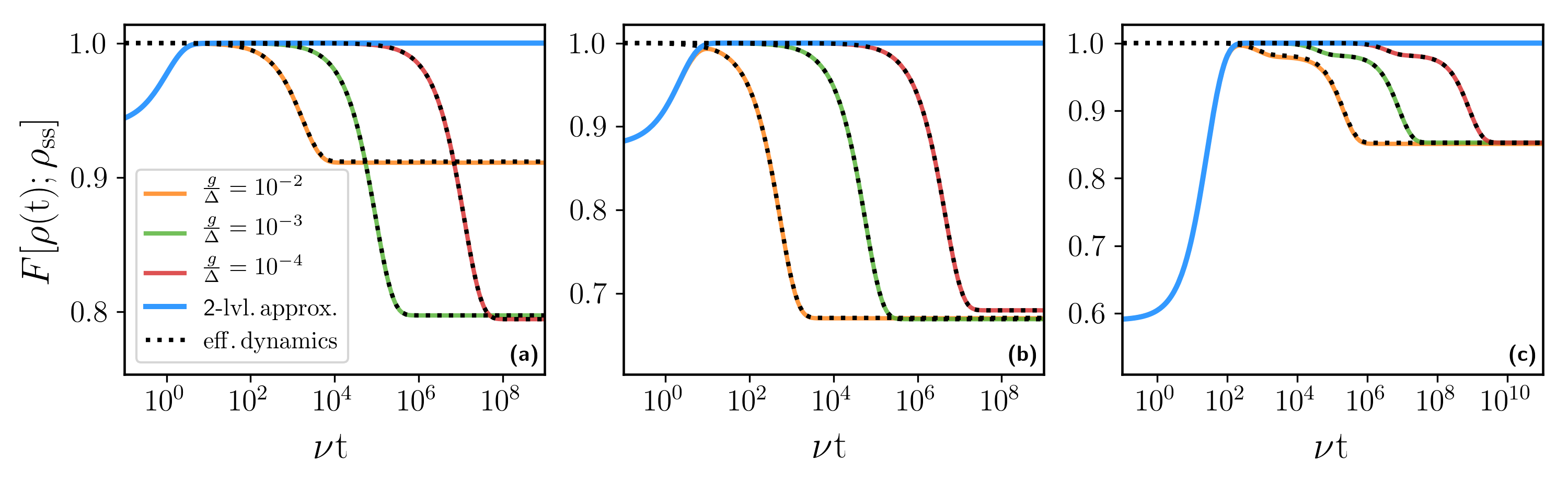}
	\caption{\textbf{Dynamics of (5+1)-level micromaser  vs effective 2-photon micromaser}. The fidelities $F[\rho_\text{ss};\rho(t)] = {\rm Tr} \ \sqrt{\sqrt{\rho_\text{ss}}\, \rho(t) \sqrt{\rho_\text{ss}}}$ of the stationary state $\rho_{\rm ss}$ in the two-photon micromaser with respect to its evolving state $\rho(t)$ (blue solid line), Eq. (\ref{eq:L0}), and to the evolving state $\rho(t)$ of (5+1)-micromaser, Eq.~\eqref{eq:master}, for increasing values of detuning (orange, green, red solid lines), while keeping the integrated coupling $\phi$ constant. Excellent agreement is observed during the metastable regime, whose length increases with the square of the detuning and coupling strength ratio, and is followed by the long-time dynamics well approximated by the effective dynamics in the DFS  (black dotted lines), Eq.~\eqref{eq:Leff_corr}. These results are observed for different atom states, coupling strengths, and initial cavity states: 
		\textbf{(a)} $c_{\rm e} = 0.3$, $\phi = 1.0$, $\ket{\psi_{\rm in}}=\ket{0}$ (the vacuum), \textbf{(b)} $c_{\mathrm e} = 0.2$, $\phi = 0.3$, $\ket{\psi_{\rm in}} = \ket{1}$ (the single-photon state), and \textbf{(c)} $c_{\mathrm e} = 0.1$, $\phi = 0.1$, $\ket{\psi_{\rm in}} = \ket{\alpha}$, $\alpha=1$ (a coherent state). The coupling strengths and the detunings in the (5+1)-level model are chosen uniformly as $g_1=g_2=g_3=g_4=g$ and $\Delta_1=\Delta_2=-\Delta_3=-\Delta_4=\Delta$, together with $G=2g$ and $\delta=2\Delta$, and thus satisfy Eq.~\eqref{eq:conditions}. 
		%\\\textbf{KM: Please check whether $\delta=\pm2\Delta$, which is necessary for the conditions to be satisfied. }
	}
	\label{fig:comparison}
	\vspace*{-4mm}
\end{figure*}

Here we discuss how the dynamics and the stationary states of the cavity are modified because of imperfections of the two-photon setup and the presence of noise affecting atoms or the cavity. That is, we consider higher order corrections to the two-photon approximation of Eq.~\eqref{eq:Heff} and approximately fulfilled conditions in~\eqref{eq:conditions} (Sec.~\ref{sec:higher-order}), nonmonochromaticity of atom beam and decay of atom levels (Sec.~\ref{sec:meta_mixed}), and single-photon losses from the cavity (Sec.~\ref{sec:losses}). 

%decide what order!!!

In order to understand how robust the results of Sec.~\ref{sec:steadystates} are, we consider weak noise and small higher order effects. The distinct parameter scales lead to a clear separation of timescales in the dynamics, known as \emph{metastability}~\cite{macieszczak2016towards}. This, together with weak or strong parity symmetries~\cite{Buca2012,albert2014symmetries}, enables us to obtain the analytic insight both into long-time dynamics and stationary states in a realistic micromaser. Furthermore, we revisit this assumption in Sec.~\ref{sec:beyond}, where we consider the noise faster than the longest timescales of the dynamics set by the relaxation across soft walls (cf. Sec.~\ref{sec:walls_soft}). The derivations can be found in Appendix~\ref{app:Leff}, while a short review of metastability theory for open quantum systems is provided in Appendix~\ref{app:meta_theory}.

%===================================================================================================================================================
%===================================================================================================================================================

\subsection{Higher order corrections in the far-detuned limit}
\label{sec:higher-order}

The two-photon micromaser investigated in Secs. \ref{sec:2photon_int} and \ref{sec:steadystates} relies on the assumption of the far-detuned limit, i.e., $|g_j/\Delta_j|, |G/\delta| \ll 1$, $j=0,...,4$ (cf.~Fig.~\ref{fig:model}). Now, we discuss how the micromaser dynamics is changed by the higher order corrections to the atom-cavity interaction.

\subsubsection{Weak parity symmetry and breaking of parity conservation}
 Recall that beyond the far-detuned limit,~\eqref{eq:Heff}, the atom-cavity interaction,~\eqref{eq:Hint}, couples all atom levels. This corresponds to six, rather than only two, Kraus operators [cf.~Eqs.~\eqref{eq:KrausFULL} and~\eqref{eq:Kraus}]
\begin{equation}
M_j=\langle j|U(\tau)|\psi_\text{at}\rangle,\quad j=0,...,4,\text{a}, \label{eq:KrausFull}
\end{equation}
where  $U(\tau)$, describes the atom-cavity interaction during time $\tau$ when the atom, initially in $|\psi_\text{at}\rangle$, passes through the cavity. These Kraus operators either \emph{conserve or swap the cavity parity} $P$ [cf.~Eq.~\eqref{eq:parity_op}] depending on $j$,
\begin{eqnarray}
M_j\,P &=& \langle j|  U(\tau)  P  |\psi_\text{at}\rangle =  - \langle j|  U(\tau)  (-1)^{N} |\psi_\text{at}\rangle  \nonumber \\ 
&=& -\langle j|  (-1)^N U(\tau)   |\psi_\text{at}\rangle =  (-1)^{j+1} \,P\, M_j,\label{eq:Kraus_parity}
\end{eqnarray}
where we used the fact that the dynamics conserves the total number of excitations $N=a^\dagger  a+ \sum_{j=1}^4 j\,\sigma_{jj}+3\sigma_\text{aa} $, i.e., $[U(\tau),N]=0$, while $(-1)^N |j\rangle= (-1)^j P |j\rangle$ and thus $(-1)^N |\psi_\text{at}\rangle=- P  |\psi_\text{at}\rangle$ for the initial atom state as in Eq.~\eqref{eq:psi}. For $j=0,2,4$, the Kraus operator swaps the parity, $M_j\,P+P\,M_j=0$, while for $j=1,3,\text{a}$, the Kraus operator conserves the parity, $M_j\,P-P\,M_j=0$. Therefore, beyond the far-detuned limit, although the cavity dynamics in Eqs.~\eqref{eq:discrete} and~\eqref{eq:master} does no longer conserve the parity,~\eqref{eq:parity_cons},  it still features \emph{weak parity symmetry}~\cite{Buca2012,albert2014symmetries},
\begin{equation} 
\label{eq:weak_symmetry}
[\mathcal{P}, \mathcal{L}]=0=[\mathcal{P}, \mathcal{M}],
\end{equation}
where the parity superoperator $\mathcal{P}(\rho)=P \rho P$ (cf.~Sec.~\ref{sec:dynamics}). From the weak parity symmetry, it follows that $\mathcal{L}$ is block diagonal in the eigenspaces of $\mathcal{P}$; i.e., odd-even and even-odd coherences evolve independently from the mixtures of even and odd states. In particular, if $\mathcal{L}$ features a unique stationary state, it must be a mixture of odd and even states without coherences between them.

\subsubsection{Higher order corrections to cavity dynamics}
 The approximation of far-detuned regime yields two-photon interaction of the cavity with only two atomic levels $|1\rangle$ and $|3\rangle$, Eq.~\eqref{eq:Heff}, and thus two parity-conserving Kraus operators $M_1$ and $M_3$ [denoted as $M_g$ and $M_e$ in Eq.~\eqref{eq:Kraus}]. Beyond this approximation, the remaining Kraus operators, $M_0,M_2,M_4, M_\text{a}$, also contribute to the cavity dynamics and enter as the first-order corrections in  $|g_j/\Delta_j|,|G/\delta| \ll 1$, $j=1,..,4$, while $M_1$ and $M_3$ are altered only in the second order as a consequence of the parity conservation (see Appendix~\ref{app:Kraus}).

\subsubsection{Metastability and perturbation theory}
 In Fig.~\ref{fig:comparison}, we compare the dynamics of the (5+1) micromaser, Eq.~(\ref{eq:master}), with the two-photon dynamics, Eq.~(\ref{eq:L0}), obtained in the far-detuned limit. We observe that the (5+1) micromaser features the \emph{initial relaxation} to the DFS of even and odd pure stationary states of the two-photon dynamics [Eq.~\eqref{eq:even_odd_ss}]. This is followed by the regime of apparent stationarity, i.e., the \emph{metastable regime}, before the \emph{final relaxation} toward the true stationary state at much longer times. Furthermore, the metastable regime becomes more pronounced with the increasing detuning, as the far-detuned limit is approached, but the asymptotic stationary state remains manifestly different from the metastable one. This indicates that higher order corrections to the atom-cavity dynamics affect the micromaser dynamics in a perturbative way, and, due to parity breaking, lift the degeneracy of the (formerly) stationary states. We therefore adapt it as the working assumption, which will enable us to analytically derive and investigate the long-time dynamics of the micromaser. We note, however, that the numerical simulations in this work are performed for \emph{truncated} cavity space, which is infinite (see also Sec.~\ref{sec:walls_hard}). Although for finitely dimensional systems the perturbative approach we utilize here is known to be convergent~\cite{Kato1995}, the cavity is an infinitely dimensional system and its unperturbed dynamics in principle features infinitely many timescales. Therefore, in principle, a formal analysis as in Ref.~\cite{azouit2015convergence} should be performed.

~\\ 
The DFS of pure stationary states of the cavity (see Sec.~\ref{sec:DFS}) correspond to the eigenmodes with eigenvalue $0$ of the master dynamics $\mathcal{L}_0$ in Eq.~\eqref{eq:L0}. To investigate the full dynamics $\mathcal{L}$ of the cavity in Eq. \eqref{eq:master}, we consider it as the perturbation of $\mathcal{L}_0$. In this case, the higher order corrections in the far-detuned limit of the cavity and atom interactions lift the degeneracy of zero eigenmodes, thus introducing their long-time dynamics (see~Appendix~\ref{app:Leff:weak} for derivation):
\begin{widetext}
	\begin{equation}
	\frac{ d }{ d t}\,\rho(t)= \nu
	\left[\begin{array}{cccc}
	-  \langle X\rangle_{+} &     \phantom{+}\langle X\rangle_{-}&0&0\\
	\phantom{+}\langle X\rangle_{+} &    -\langle X\rangle_{-} &0&0\\
	0&0&-i\Omega -\frac{1}{2}\left(\langle X\rangle_{+}+\langle X\rangle_{-}\right)& \eta \sqrt{\langle X\rangle_{+} \langle X\rangle_{-}} \\ 
	0&0& \eta^* \sqrt{\langle X\rangle_{+} \langle X\rangle_{-}} &i\Omega -\frac{1}{2}\left(\langle X\rangle_{+}+\langle X\rangle_{-}\right)\\
	\end{array}\right] \rho(t), \label{eq:Leff_corr}
	\end{equation}
\end{widetext}
where $\rho(t)$ belongs to the DFS spanned by $|\Psi_+\rangle$ and $|\Psi_-\rangle$ (we assumed there is a unique stationary state of even and odd parities, i.e., there are no hard walls of $\mathcal{L}_0$). The long-time dynamics is expressed in the DFS basis $ \ketbra{\Psi_+}{\Psi_+}, \ketbra{\Psi_-}{\Psi_-}, \ketbra{\Psi_+}{\Psi_-}, \ketbra{\Psi_-}{\Psi_+}$. The nontrivial long-time dynamics of the pure states of the cavity means that they are \emph{metastable} and at long times relax to a \emph{unique stationary state} approximated by the stationary state of Eq.~\eqref{eq:Leff_corr} (cf.~Fig.~\ref{fig:comparison}):
\begin{equation}
\rho_\text{ss}\approx\frac{\langle X\rangle_{-}}{\langle X\rangle_{-}+\langle X\rangle_{+}}\ketbra{\Psi_+}{\Psi_+}+\frac{\langle X\rangle_{+}}{\langle X\rangle_{-}+\langle X\rangle_{+}}\ketbra{\Psi_-}{\Psi_-}. \label{eq:rhoss_corr}
\end{equation}

The \emph{block-diagonal} structure of the effective dynamics generator in Eq.~\eqref{eq:Leff_corr}, with the coherences $\ketbra{\Psi_+}{\Psi_-}, \ketbra{\Psi_-}{\Psi_+}$, evolving independently from $ \ketbra{\Psi_+}{\Psi_+}, \ketbra{\Psi_-}{\Psi_-}$,  reflects the weak parity symmetry of dynamics, Eq.~\eqref{eq:weak_symmetry}, which further manifests in diagonal structure of the stationary state in Eq.~\eqref{eq:rhoss_corr}.  The dynamics features the Hamiltonian part~\cite{zanardi_coherent_2014,zanardi_geometry_2015} from the second-order corrections in the parity-conserving Kraus operators $M_1$ and $M_3$, with the frequency 
\begin{eqnarray}\label{eq:Omega}
\Omega &\equiv & \mathrm{Im}\langle c_{\rm g}(M_1-M_{\rm g})^\dagger-c_{\rm e}(M_3-M_{\rm e})^\dagger\rangle_+ \nonumber \\
&& - \mathrm{Im}   \langle c_{\rm g}( M_1-M_{\rm g})^\dagger-c_{\rm e}(M_3-M_{\rm e})^\dagger\rangle_-, 
\end{eqnarray} 
and the dissipative counterpart~\cite{zanardi_dissipative_2016} induced by the (first-order) corrections in the parity swapping operators, where
\begin{equation}\label{eq:X}
X\equiv M_0^\dagger M_0+ M_2^\dagger M_2 + M_4^\dagger M_4,
\end{equation}
so that $\langle X\rangle_\pm$ is positive and of the second order,
\begin{widetext}
	\begin{eqnarray}\nonumber
	&&\langle X\rangle_\pm = 2 |c_{\rm g}|^2 \frac{|g_2|^2}{\Delta\Delta_1} \left \langle\! (n+1)-(n+1)\cos\left[\tau \Delta_1 +\tau\frac{|g_2|^2}{\Delta}\left ( n+2  \right)  \right] \right\rangle_\pm\!\!\!
	+2|c_{\rm g}|^2 \frac{|g_2|^2}{\Delta^2} \left \langle\! n-n\cos\left[\tau \Delta +\tau\frac{|g_2|^2+|g_3|^2}{\Delta}\left ( n-1  \right)  \right] \right\rangle_\pm \!\!
	\\\nonumber
	&&
	+2|c_{\rm e}|^2 \frac{|g_3|^2}{\Delta^2} \left \langle\!(n+1)+(n+1)\cos\left[\tau \Delta +\tau\frac{|g_2|^2+|g_3|^2}{\Delta}\left ( n+1  \right)  \right] \right\rangle_\pm\!\!\!
	- 2 |c_{\rm e}|^2 \frac{|g_3|^2}{\Delta\Delta_4} \left \langle\! n+n\cos\left[\tau \left(\Delta_4 -\frac{|g_2|^2}{\Delta}-\frac{|g_3|^2}{\Delta} n  \right)  \right] \right\rangle_\pm\!\!
	\\\label{eq:Xav}
	&& +2\left \langle\! -  i \frac{g_2^* g_3}{\Delta^2} (a^\dagger)^2  \sin\left[\tau\Delta + \tau \left(n+1\right)  \frac{|g_2|^2 +|g_3|^2 }{\Delta }   \right]   +  i \frac{g_2 g_3^*}{\Delta^2}\sin\left[\tau\Delta + \tau \left(n+1\right)  \frac{|g_2|^2 +|g_3|^2 }{\Delta }   \right] a^2 \right\rangle_\pm\!\!, 
	\end{eqnarray}
\end{widetext}
with  $\langle\,\cdot\,\rangle_\pm= \langle \Psi_{\pm}|\,\cdot\,|\Psi_{\pm}\rangle$ and $n=a^\dagger a$.
We note that the parity-conserving $M_\text{a}$ does not contribute to the second-order dynamics [cf.~Eqs.~\eqref{eq:X} and~\eqref{eq:eta_corr}], as the pure stationary states are eigenstates of $M_\text{a}$ in the first order (see~Appendix~\ref{app:Leff:conserved}). % i.e $M_\text{a}|\Psi_{\pm}\rangle = - \frac{ G}{\delta} [  1   -     e^{- i\tau ( \delta- \frac{2|g_2|^2 +|g_3|^2 }{\Delta })} ] c_{\rm e}|\Psi_{\pm}\rangle$.  
Furthermore, the dynamics of coherences depends on 
\begin{equation}
\eta = \frac{\mathrm{Tr}\left(L_{+-} \sum_{j=0,2,4} M_j\ketbra{\Psi_-}{\Psi_+}M_j^\dagger\right)}{\sqrt{\langle X\rangle_{+}\langle X\rangle_{-}}} ,
\label{eq:eta_corr}
\end{equation}
where $ L_{+-} $ is a conserved quantity in the far-detuned limit corresponding to the coherence  $\ketbra{\Psi_+}{\Psi_-}$, and $\eta$ can be obtained numerically without diagonalizing $\mathcal{L}_0$ from Eq.~\eqref{eq:Lpm}. From the complete positivity  of the perturbative long-time dynamics~\cite{macieszczak2016towards}, we have $|\eta|\leq 1$. We note that the effective dynamics in Eq.~\eqref{eq:Leff_corr} depends via $\Omega$ and $\langle X\rangle_{\pm}$ on the second order of the corrections to the far-detuned limit, $|g_j/\Delta_j|,|G/\delta| \ll 1$, $j=1,..,4$, as well as the interaction time $\tau$, rather than only the integrated coupling $\phi$.

In Fig.~\ref{fig:comparison}, we compare the dynamics of the cavity in (5+1) model (solid lines), Eq.~\eqref{eq:master}, to the effective long-time dynamics within the DFS (dotted lines), Eq.~\eqref{eq:Leff_corr}, and observe a very good agreement in the relaxation  after the metastable regime toward the stationary state, Eq.~\eqref{eq:rhoss_corr}. %Since the effective dynamics depends via $\Omega$ and $\langle X\rangle_{\pm}$ both on the second order of the corrections to the far-detuned limit, $|g_j/\Delta_j|,|G/\delta| \ll 1$, $j=1,..,4$, as well as the interaction time $\tau$, the structure of the final stationary state can be changed without altering the initial relaxation toward the metastable DFS, which depends only on the integrated coupling $\phi$ [see Fig.~\ref{fig:comparison}(a)].
Therefore,  Eq.~\eqref{eq:Leff_corr} determines the final relaxation timescales toward the unique stationary state. These timescales are inversely proportional to the second order of the corrections to the far-detuned limit [cf. Eqs.~\eqref{eq:Omega},~\eqref{eq:X}, and~\eqref{eq:eta_corr}], and thus the free parameters $g_2$, $g_3$, $\Delta$, $\Delta_1/\Delta>0$, $\Delta_4/\Delta<0$, and   $\delta/\Delta<0$ in Eq.~\eqref{eq:conditions} can be further optimized in order to extend the length of metastability regime, while keeping $\phi$ constant.

\subsubsection{Approximate cancellation of Stark shifts}
  Finally, we note that relaxing of the conditions of Eq.~\eqref{eq:conditions}, which we have chosen to obtain the two-photon Jaynes-Cummings Hamiltonian in Eq.~\eqref{eq:Heff}, will lead to a perturbation of this Hamiltonian [cf.~Eq.~\eqref{eq:Heff0}] and thus corrections to parity-conserving Kraus operators $M_g$ and $M_e$. Therefore, analogously to Eq.~\eqref{eq:Omega}, in the lowest order only a unitary dynamics will be induced in DFS, with dephasing possibly entering in higher orders (see Sec.~\ref{sec:meta_mixed} and Appendix~\ref{app:Leff:conserved:cond}  for further discussion). We can conclude that the (5+1) design is \emph{stable}, which is necessary to achieve any experimental implementation of the desired two-photon dynamics.

%===================================================================================================================================================
%===================================================================================================================================================

%===================================================================================================================================================
%===================================================================================================================================================
\subsection{Single-photon losses}
\label{sec:losses}

We now turn to investigate a realistic cavity undergoing \emph{single-photon losses}~\cite{englert2002elements}, typically due to imperfect mirrors,
\begin{eqnarray}
\label{eq:L1ph}
\mathcal{L}_{1{\rm ph}} \left[\rho(t)\right] &=& \kappa \, a \,\rho(t)\, a^\dagger- \frac{\kappa}{2} \left [a^\dagger a\, \rho(t) +\rho(t)\,a^\dagger a\right],\quad
\end{eqnarray}
where $\kappa$ is the single-photon loss rate. Provided that losses of photons can be assumed to take place when no atom is found within the cavity, i.e., the atom passage time $\tau$ is such that  $\kappa \ken \tau \ll 1$, the single-photon losses can be considered independent of the atom-cavity dynamics \cite{englert2002elements,davidovich1987two}, so that the cavity state evolves as
\begin{equation}
\frac{\rm d}{{\rm d}t} \rho(t) = \left(\mathcal{L}_0 + \mathcal{L}_{\rm 1ph} \right) \left[ \rho(t)\right].
\label{eq:master_loss}
\end{equation}
In~Eq.~\eqref{eq:master_loss}, we  assumed the far-detuned limit of Eq.~\eqref{eq:L0}.

\subsubsection{Weak parity symmetry}
 The single-photon loss swaps the parity, similarly to the case of higher order corrections in the far-detuned limit  [cf. Eq.~\eqref{eq:Kraus_parity}],
\begin{equation}
\label{eq:loss_parity}
a\, P=-P\,a.
\end{equation}
This leads to the weak parity symmetry of the dynamics [cf.~Eq.~\eqref{eq:weak_symmetry}],
\begin{equation} 
\label{eq:weak_symmetry_loss}
[\mathcal{P}, \mathcal{L}_0 +\mathcal{L}_{1{\rm ph}}]=0.
\end{equation}

\subsubsection{Metastability and perturbation theory}
\begin{figure}[b]
	\centering
	\vspace*{3mm}
	\includegraphics[width=0.5\textwidth]{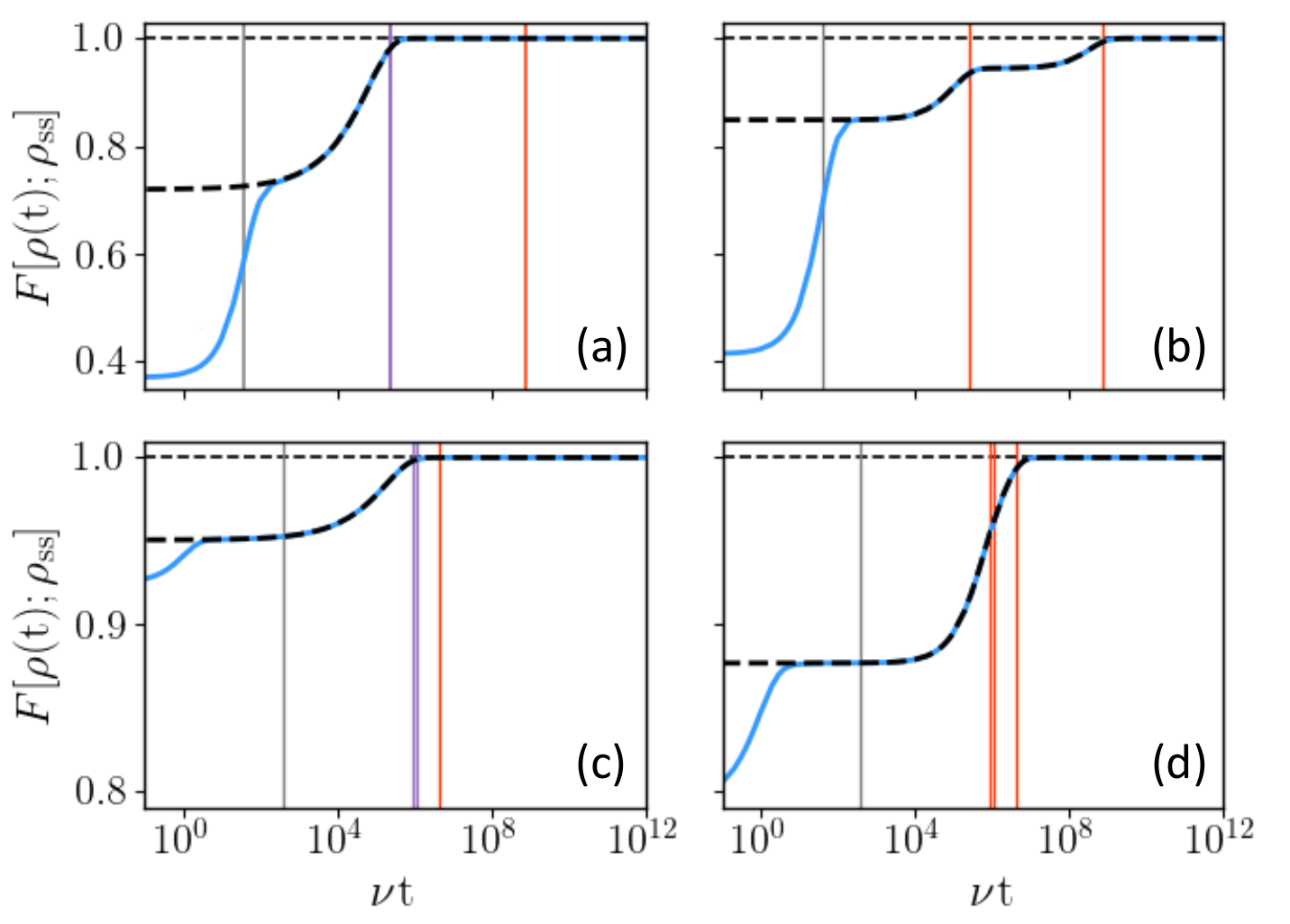}
	\caption{\textbf{Dynamics of micromaser with single-photon losses}. 
		The fidelity [cf.~Fig.~\ref{fig:comparison}] between the cavity state $\rho(t)$ and the stationary state~\eqref{eq:rhoss_loss} is compared for the dynamics of two-photon micromaser with single-photon losses (blue solid line),~\eqref{eq:master_loss}, and the effective dynamics in the DFS (black dashed line),~\eqref{eq:Leff_loss}. The effective dynamics approximates well the long-time dynamics of the cavity for the initial states $\ket{0}$ [(a),(c)] and $\ket{\alpha}$, $\alpha = 0.6$ [(b),(d)], both in the weak-coupling limit [$c_{\mathrm e} = 0.1$, $\phi = 0.1$ in panels (a) and (b)], where additional metastable regime (second plateau) is observed (b), and at the finite coupling [$c_{\mathrm e} = 0.2$, $\phi = 1.0$ in panels (c) and (d)]. The loss rate was chosen as $\kappa/\nu= 10^{-6}$, and the vertical lines indicate the timescales of the dynamics determined by the eigenvalues of Eq.~\eqref{eq:master_loss}, $(-\mathrm{Re} \lambda_k)^{-1}$%, where $\lambda_k$ is the $k$th eigenvalue of $\mathcal{L}$ (decreasing order) 
		for $k=5,4,2$ (black, purple, red), which are ordered in decreasing real value [see also Eqs.~\eqref{eq:Leff_modes} and~\eqref{eq:Leff_modes2} and cf. Appendix~\ref{app:meta_theory}].
	}	
	\label{fig:metastability}
	\vspace*{-3mm}
\end{figure}

 In Fig.~\ref{fig:metastability}, we show the cavity dynamics in the presence of small losses (blue solid lines),  Eq.~\eqref{eq:master_loss} with $\kappa\ll \nu$, and observe a plateau in the relaxation toward the unique stationary state of the dynamics. This manifests a \emph{metastable regime} in the dynamics when cavity states appear stationary for different initial conditions, although the true stationary state has not been achieved [see also Fig~\ref{fig:model}(d)].

If the losses are treated as a perturbation of the cavity dynamics $\mathcal{L}_0$, the formerly stationary states in the DFS of $|\Psi_+\rangle$ and $|\Psi_-\rangle$, Eq.~\eqref{eq:even_odd_ss}, undergo the following dynamics [cf.~Eq.~\eqref{eq:Leff_corr}]
\begin{widetext}
	\begin{eqnarray} \label{eq:Leff_loss}
	\frac{ d }{ d t} \rho(t)&=&
	\kappa\left[\begin{array}{cccc}
	-\langle n\rangle_{+}& \phantom{+}\langle n\rangle_{-}&0&0\\
	\phantom{+}\langle n\rangle_{+}& -\langle n\rangle_{-}&0&0\\
	0&0&-\frac{1}{2}(\langle n\rangle_{+}+\langle n\rangle_{-})&\eta_{\rm loss}\sqrt{\langle n\rangle_{+}\langle n\rangle_{-}}\\ 
	0&0&\eta_{\rm loss}\sqrt{\langle n\rangle_{+}\langle n\rangle_{-}}&-\frac{1}{2}(\langle n\rangle_{+}+\langle n\rangle_{-})\\
	\end{array}\right] \rho(t),
	\end{eqnarray}
\end{widetext}
where we expressed the dynamics in the basis $\{\ketbra{\Psi_+}{\Psi_+}, \ketbra{\Psi_-}{\Psi_-}, \ketbra{\Psi_+}{\Psi_-}, \ketbra{\Psi_-}{\Psi_+}\}$, and denoted the average loss rate as $\kappa\langle n\rangle_{\pm}=\kappa\langle\Psi_\pm|a^\dagger  a|\Psi_\pm\rangle$. The dynamics is \emph{block-diagonal}, with the densities and the coherences evolving independently, due to the weak parity symmetry, Eq.~\eqref{eq:weak_symmetry_loss}. The dynamics of coherences further depends on the real coefficient [cf. Eq.~\eqref{eq:eta_corr}]
\begin{eqnarray}\label{eq:eta_loss}
\eta_\text{loss} = \frac{\mathrm{Tr}\left(L_{+-} \,a\ketbra{\Psi_-}{\Psi_+}a^\dagger\right)}{\sqrt{\langle n\rangle_{+}\langle n\rangle_{-}}} ,&&\qquad |\eta_\text{loss}|\leq 1,
\end{eqnarray}
that can be determined numerically from Eq.~\eqref{eq:Lpm}. In particular, in the weak-coupling regime, where the DFS corresponds to Schr\"{o}dinger-cat states, we have $\eta_\text{loss} =1$, as the photon loss preserves the DFS (see Sec.~\ref{sec:schrodinger_cats} and Ref.~\cite{mirrahimi2014dynamically}).  

In  Fig.~\ref{fig:metastability}, the effective dynamics of Eq.~\eqref{eq:Leff_loss} (black dashed line) indeed approximates well the long-time dynamics of the cavity. This confirms that the initial relaxation of the cavity state takes the system into the DFS spanned by $|\Psi_+\rangle$ and $|\Psi_-\rangle$ [cf.~Eq.~\eqref{eq:P00}]. The DFS then remains metastable until timescales inversely proportional to the average loss rates. Then, the final relaxation takes place  into a \emph{unique stationary state}, well approximated by the stationary state of Eq.~\eqref{eq:Leff_loss},
\begin{equation}
\label{eq:rhoss_loss}
\rho_{\rm ss} \approx \frac{\langle n\rangle_{-}}{\langle n\rangle_{-}+\langle n\rangle_{+}}\ketbra{\Psi_+}{\Psi_+} + \frac{\langle n\rangle_{+}}{\langle n\rangle_{-}+\langle n\rangle_{+}}\ketbra{\Psi_-}{\Psi_-},
\end{equation}
cf. Refs.~\cite{minganti2016exact,Bartolo_2016}. The stationary state does not feature odd-even coherences because of the weak parity symmetry in Eq.~\eqref{eq:weak_symmetry_loss} (see~Fig.~\ref{fig:bloch}). Finally, we note that the rates of the effective dynamics are proportional to the average photon number, so that, as expected, the states with more photons are more sensitive to losses. In particular, in the stationary state~\eqref{eq:rhoss_loss} the state with the lower average photon number has larger weight.   

An analogous result to Eq.~\eqref{eq:Leff_loss} can be obtained for a cavity in a \emph{thermal environment}. In this case, photons are lost from the cavity at the rate $\kappa (n_{\text{th}}+1)$, but they are also injected to the cavity [which process is described as by replacing  $a$  by $a^\dagger$ in~Eq.~\eqref{eq:L1ph}] at the rate $\kappa \,n_{\text{th}}$, and $n_{\text{th}}$ is a average photon number of the environment.

\subsubsection{Emergent classical metastability in weak-coupling limit}
 The timescales of the long-time dynamics are determined by the eigenvalues of Eq.~\eqref{eq:Leff_loss} (see also Appendix~\ref{app:meta_theory}). The stationary state in Eq.~\eqref{eq:rhoss_loss} necessarily corresponds to the eigenvalue $\lambda_1=0$, while
\begin{subequations}
	\label{eq:Leff_modes}
	\begin{align}
	\lambda_2 &= -\frac{\kappa}{2}\left(\langle n\rangle_{+}+\langle n\rangle_{-} - 2|\eta_\text{loss}|\sqrt{\langle n\rangle_{+}\langle n\rangle_{-}}\right),\\
	\lambda_3 &= -\frac{\kappa}{2}\left(\langle n\rangle_{+}+\langle n\rangle_{-} + 2|\eta_\text{loss}|\sqrt{\langle n\rangle_{+}\langle n\rangle_{-}}\right),\\
	\lambda_4 &= -\kappa\left(\langle n\rangle_{+}+\langle n\rangle_{-}\right),
	\end{align}
\end{subequations}
ordered in decreasing real part.

\begin{figure}[t]
	\begin{flushleft}
		\subfloat{\includegraphics[width=1.1\columnwidth]{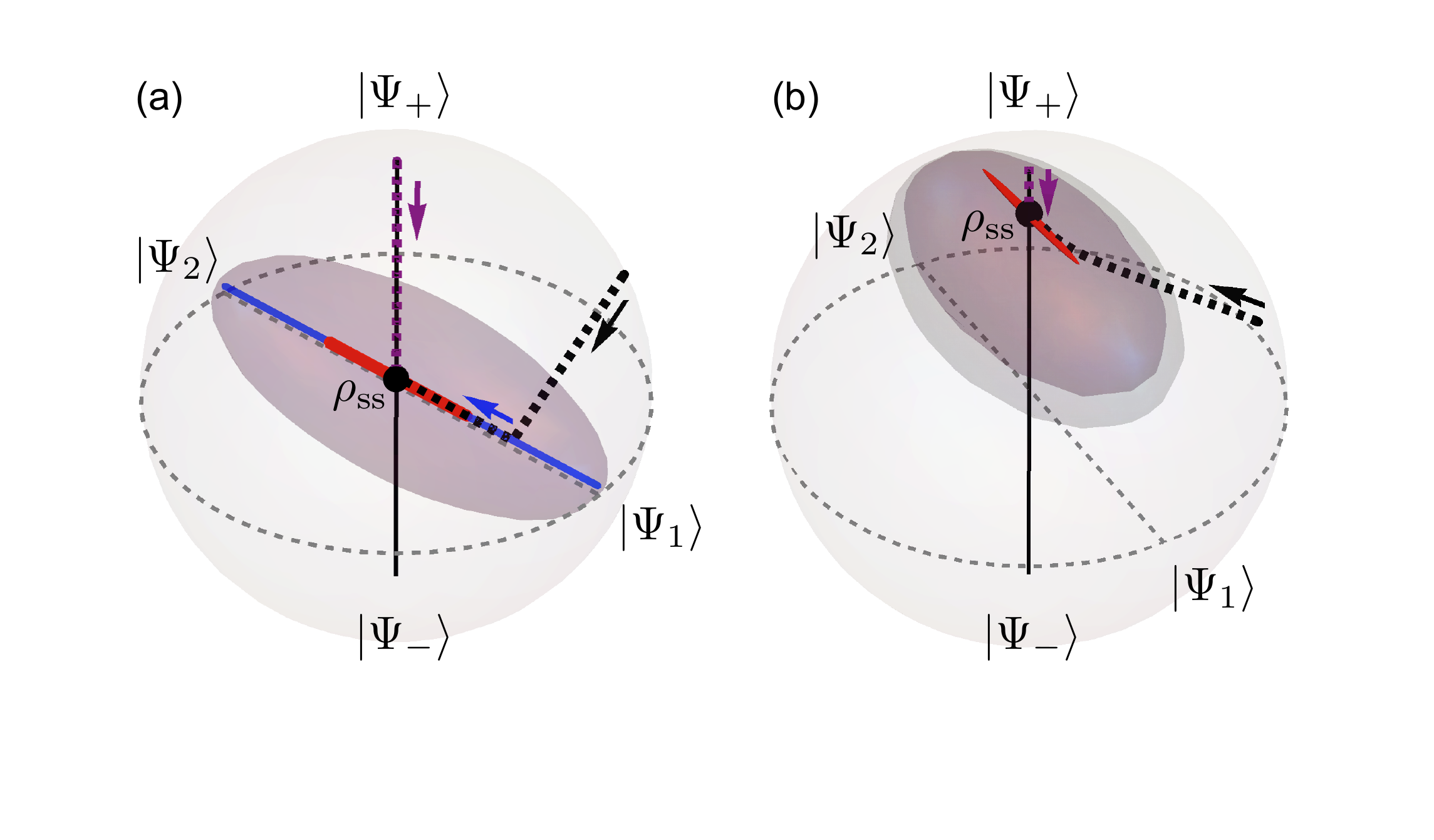}}
	\end{flushleft}
	\vspace*{-10mm}
	\caption{\textbf{Effective long-time dynamics due to single-photon losses}. The DFS of the odd and even states~\eqref{eq:even_odd_ss} (the Bloch sphere in light gray) is shown under the effective dynamics in Eq.~\eqref{eq:Leff_loss}, for times $t=(-\lambda_4)^{-1},\, (-\lambda_3)^{-1}$, and $(-\lambda_2)^{-1}$ (gray, purple, red) [see Eq.~\eqref{eq:Leff_modes} and vertical lines in Fig.~\ref{fig:metastability}]. Because of the weak parity symmetry, the stationary state (black dot), Eq.~\eqref{eq:rhoss_loss}, is found on the vertical axis (black line) representing mixtures of even and odd states, while when the initial state is odd or even, its dynamics remains confined to the vertical axis at all times (cf.~purple dashed trajectory). As the effective dynamics is also real, the coherence eigenmodes correspond to the axis between the states in Eq.~\eqref{eq:ems} (dashed gray) and the axis perpendicular to it that also crosses the equator. The trajectories for two initial states are also shown: $|\Psi_+\rangle$ (dashed purple) and $\cos(\pi/6)|\Psi_+\rangle+e^{i\pi/4}\sin(\pi/6)|\Psi_-\rangle$ (dashed black).  In panel (a), due to separation of the characteristic timescales of the dynamics as given by Eq.~\eqref{eq:Leff_modes2}, classical metastable manifold emerges [blue; the image of DFS under the dynamics of at $t=(-\lambda_2)^{-1}/100$], well approximated by mixtures of the states $|\Psi_1\rangle$ and $|\Psi_2\rangle$ in Eq.~\eqref{eq:ems} (dashed gray axis). Here an initial state first relaxes onto the manifold (black arrow along black dashed trajectory), and only at later times relaxes toward the stationary state (blue arrow) [see also Fig.~\ref{fig:metastability}(b)]. Parameters: \textbf{(a)} as in Figs.~\ref{fig:metastability}(a) and ~\ref{fig:metastability}(b) leading to $\eta_\text{loss}\approx 1.00$, $\langle n\rangle_+\approx1.92 $ and $\langle n\rangle_-\approx 2.07$; \textbf{(b)} as in Figs.~\ref{fig:metastability}(c) and ~\ref{fig:metastability}(d) leading to $\eta_\text{loss}\approx 0.99$, $\langle n\rangle_+\approx 0.11$, and $\langle n\rangle_-\approx 1.01$.
		\vspace*{-3mm}
	}
	\label{fig:bloch}
\end{figure}

For small interactions, $|\phi|\ll 1$, where the stationary states of the lossless cavity are approximated by Schr\"{o}dinger-cat states~\cite{gilles1994generation,Guerra_1997_PRA} the dynamics in Fig.~\ref{fig:metastability}(b) features two plateaus corresponding to \emph{two metastability regimes} [see also Fig.~\ref{fig:model}(d)]. Indeed, in this case, $\eta_{\rm loss} = 1$ in~Eq.~\eqref{eq:eta_loss}, so that $\lambda_{2,3}= -\kappa \,(\sqrt{\langle n\rangle_{+}}\mp\sqrt{\langle n\rangle_{-}})^2/2$. Therefore, when the average photon numbers in the even and odd Schr\"{o}dinger-cat states are similar [$\langle n \rangle_+= |\alpha|^2\tanh(|\alpha|^2)$, $\langle n \rangle_-=|\alpha|^2\coth(|\alpha|^2)$ with $|\alpha|^2= 2 |c_{\rm e}/c_{\rm g} \phi|\gg 1$; cf.~Eq.~\eqref{eq:cond_lin}], a \emph{separation in the spectrum} of the long time-dynamics emerges:
\begin{equation}\label{eq:Leff_modes2}
-\lambda_2\approx  \frac{\kappa}{4}\frac{\left(\langle n\rangle_{+}-\langle n\rangle_{-}\right)^2}{\langle n\rangle_{+}+\langle n\rangle_{-}} \ll-\lambda_3\approx \kappa\left(\langle n\rangle_{+}+\langle n\rangle_{-}\right)=\lambda_4.
\end{equation}
This separation is responsible for the second plateau in Fig.~\ref{fig:metastability}(b), as it leads to metastability regime for times $(-\lambda_3)^{-1}\ll t\ll(-\lambda_2)^{-1}$ when the faster eigenmodes of the long-time dynamics  corresponding to $\lambda_3$ and  $\lambda_4$ have decayed, while the decay of the slow mode corresponding to $\lambda_2$ is negligible (see~Appendix~\ref{app:meta_theory}). Only the stationary state and the slow eigenmode then contribute to the cavity state~\cite{macieszczak2016towards,Rose2016,Minganti2018}, 
\begin{equation}\label{eq:rho_meta1}
\rho(t)\approx \rho_\text{ss} + c_\text{Re} \left(\ket{\Psi_+}\!\bra{\Psi_-} + \ket{\Psi_-}\!\bra{\Psi_+} \right),
\end{equation}
where $c_\text{Re}= \mathrm{Re}[\mathrm{Tr}(L_{+-} \rho )]$ [cf. Eq.~\eqref{eq:P00}] and $\rho_\text{ss}\approx\left(\ket{\Psi_+}\!\bra{\Psi_+} + \ket{\Psi_-}\!\bra{\Psi_-} \right)/2$ [cf. Eq.~\eqref{eq:rhoss_loss}]. Therefore, the second metastable regime is observed only for initial states with feature odd-even coherences [cf. Figs.~\ref{fig:metastability}(a) and~\ref{fig:metastability}(b)]. Furthermore, during the metastable regime the cavity state can be also be regarded as a \emph{classical mixture}~\cite{Rose2016}, with the probability $p=1/2+c_\text{Re}$, 
\begin{equation}\label{eq:rho_meta2}
\rho(t) \approx p \,\ketbra{\Psi_1}{\Psi_1}+(1-p) \ketbra{\Psi_2}{\Psi_2}
\end{equation}
of the \emph{coherent states} [cf.~Fig.~\ref{fig:model}(d) and see  Fig.~\ref{fig:bloch}(a)]:
\begin{equation}
\label{eq:ems}
\ket{\Psi_{1,2}} =\frac{1}{\sqrt{2}}(\ket{\Psi_+} \pm \ket{\Psi_-})\approx|\pm\alpha\rangle. 
\end{equation}
Note that classical metastability can occur also beyond weak-coupling limit if both $|\eta_{\text{loss}}| \approx 1$ and  $\langle n \rangle_+\approx\langle n \rangle_-$.
\\

%Finally, after, the metastability regime, $t\gg (-\lambda_2)^{-1}$, the cavity state relaxes into the stationary state due to the decay of the slow mode, cf.~Eqs.~\eqref{eq:rho_meta1},
%\begin{equation}\label{eq:rho_meta3}
%\rho(t)\approx \rho_\text{ss} +  e^{t\lambda_2} \,c\left(\ket{\Psi_-}\!\bra{\Psi_+} + \ket{\Psi_+}\!\bra{\Psi_-}\right).
%\end{equation}

%
%

The origin of the classical metastability can be understood by representing Eq.~(\ref{eq:Leff_loss}) in terms of the master equation within the DFS~\cite{Lindblad1976,Gorini1976} (here $\eta_\text{loss}=1$; for a general case, see Appendix~\ref{app:Leff:weak}), 
\begin{equation}
\frac{ d }{ d t}\rho(t)=   \gamma_\text{loss} \, J\rho(t) J^\dagger -\frac{ \gamma_\text{loss}}{2}\left[J^\dagger  J\, \rho(t) +\rho(t)\,J^\dagger  J\right],
\label{eq:master_flip main}
\end{equation}
where the dissipation rate is given by the average photon loss
\begin{equation}
\label{eq:rate_spin_flip}
\gamma_\text{loss}= \kappa \,\frac{\langle n \rangle_++\langle n \rangle_- } {2},
\end{equation}
and the jump operator $J$ describes the effect of a single-photon loss on the DFS by \emph{flipping the parity} (cf. Refs.~\cite{minganti2016exact,azouit2015convergence})
\begin{align}
J &=  \frac{1}{N}\left[\left(\langle n \rangle_+ +  \sqrt{\langle n \rangle_+ \langle n \rangle_-}\right) \right.  \ketbra{\Psi_+}{\Psi_-} \nonumber\\
& \phantom{=} + \left.\left(\langle n \rangle_-+ \sqrt{\langle n \rangle_+ \langle n \rangle_-}\right)  \ketbra{\Psi_-}{\Psi_+} \right],
\label{eq:op_spin_flip}
\end{align}
with the normalization factor $N=\sqrt{\langle n \rangle_++\langle n \rangle_-}\,( \sqrt{\langle n \rangle_+}+\sqrt {\langle n \rangle_-}) $. When the average photon number in the even and odd states is similar, $\langle n \rangle_+\approx\langle n \rangle_-$, as takes place for large enough $|\alpha|$ of Schr\"{o}dinger cat states in Eq.~\eqref{eq:psi_ss_lin}, the jump operator in Eq.~\eqref{eq:op_spin_flip} can be approximated as the spin flip
\begin{align}\label{eq:op_spin_flip2}
J \approx& \,\,\frac{1}{\sqrt{2}}\left(\ketbra{\Psi_+}{\Psi_-} + \ketbra{\Psi_-}{\Psi_+} \right)
\\\nonumber
&=\frac{1}{\sqrt{2}}\left(\ketbra{\Psi_1}{\Psi_1} - \ketbra{\Psi_2}{\Psi_2} \right),
\end{align}
which causes \emph{dephasing} of coherences between the states in Eq.~\eqref{eq:ems} [see Figs.~\ref{fig:model}(d) and~\ref{fig:bloch}] at the rate $\gamma_\text{loss}$. The states $\ket{\Psi_1}$ and $\ket{\Psi_2}$ are metastable, as they are unchanged by the dephasing, and only the higher corrections in Eq.~\eqref{eq:op_spin_flip2} ultimately lead to their mixing toward the stationary state in Eq.~\eqref{eq:rhoss_loss} approximated by $\left(\ket{\Psi_+}\!\bra{\Psi_+} + \ket{\Psi_-}\!\bra{\Psi_-} \right)/2$.

%===================================================================================================================================================

\subsection{Decay of atom levels and nonmonochromatic atom beam}
\label{sec:meta_mixed}

In Sec.~\ref{sec:steadystates}, we considered micromaser dynamics arising from interaction of the cavity with atoms prepared in a pure state in Eq.~\eqref{eq:psi}, lasting time $\tau$ leading to integrated coupling $\phi$ [cf. Eq.~\eqref{eq:Kraus}], which  results in pure stationary states dependent both on $\phi$ and atom amplitudes [cf. Eq.~\eqref{eq:even_odd_ss}]. 

In a realistic setup, the lifetime of atom levels is finite, leading to dissipative \emph{decay of atom state}, which modifies the dynamics in two ways. First, due to the decay during time $T$ between the atom preparation and entering the cavity, atoms arrive at the cavity in a mixed state (see Appendix~\ref{app:Leff:conserved:atom_mixed}), and at a reduced rate if decay takes place toward levels not coupled to the cavity field.  Second, possible atom decay during time $\tau$ of the interaction with the cavity introduces modified Kraus operators determined by times of decay events (see Appendix~\ref{app:Leff:conserved:atom_decay}). On the other hand, the velocity of the atoms is usually described by a distribution rather than a single value, which we refer to as \emph{nonmonochromatic} atom beam.  leading to fluctuating interaction time $\tau$ and thus the fluctuating integrated coupling $\phi$ in Eq.~\eqref{eq:Kraus}  (see Appendix~\ref{app:Leff:conserved:atom_beam}). Nevertheless, in the far-detuned limit, the parity of photon number is conserved in the presence of such noise, so that there still exist (at least) two stationary states of odd and even parities (see Sec.~\ref{sec:dynamics}).

\subsubsection{Parity conservation and metastability}
 We now consider a limit of weak atom decay with respect to time $T$ and $\tau$, and a narrow distribution of velocities. From the perturbation theory, the  first-order dynamics in the DFS basis $\ketbra{\Psi_+}{\Psi_+}$, $\ketbra{\Psi_-}{\Psi_-}$, $\ketbra{\Psi_+}{\Psi_-}$, $\ketbra{\Psi_-}{\Psi_+}$ is
\emph{diagonal} as a consequence of parity conservation  [cf. Eq.~\eqref{eq:parity_cons}],
\begin{equation}
\frac{ d }{ d t}\rho(t)= 
\left[\begin{array}{cccc}
0&0&0&0\\
0& 0&0&0\\
0&0&- i\Omega-\gamma_\text{deph}&0\\ 
0&0&0& i\Omega-\gamma_\text{deph}\\
\end{array}\right] \rho(t). \label{eq:Leff_deph_0}
\end{equation}
For the discrete dynamics described by $\mathcal{M}$ rather than $\mathcal{M}_0$ in Eq.~\eqref{eq:discrete_2}, we have $- i\Omega-\gamma_\text{deph}=\mathrm{Tr}[L_{+-}\delta\mathcal{M}(\ketbra{\Psi_+}{\Psi_-})]$, where the perturbation $\delta \mathcal{M}\equiv\mathcal{M}-\mathcal{M}_0$, which value can be found numerically from  Eq.~\eqref{eq:Lpm}. The effective dynamics in Eq.~\eqref{eq:Leff_deph_0} describes  \emph{dephasing} between $|\Psi_{+}\rangle$ and $|\Psi_{-}\rangle$ at the rate  $\gamma_\text{deph}$ and unitary dynamics at the frequency $\Omega$,
\begin{eqnarray}
\rho(t)&=& p\,\ketbra{\Psi_{+}}{\Psi_{+}}+(1-p)\,\ketbra{\Psi_{-}}{\Psi_{-}}, 
\\\nonumber
&&+ e^{-i\Omega t -\gamma_\text{deph}t} c\,\ketbra{\Psi_+}{\Psi_-} + e^{i\Omega t -\gamma_\text{deph} t} c^*\ketbra{\Psi_+}{\Psi_-}, 
\end{eqnarray} 
where $p= \mathrm{Tr}(\mathds{1}_{+} \rho)$, $c= \mathrm{Tr}(L_{+-} \rho ) $ [cf. Eq.~\eqref{eq:P00}]. 

The asymptotic state of the dynamics in Eq.~\eqref{eq:Leff_deph_0} is not unique. Therefore, when $\gamma_{\rm deph}> 0$, the asymptotic state of the micromaser dynamics is approximated by a mixture of the odd and even pure states 
\begin{equation}\label{eq:rhoss_deph}
\rho_\text{ss}\approx p\,\ketbra{\Psi_{+}}{\Psi_{+}}+(1-p)\,\ketbra{\Psi_{-}}{\Psi_{-}}.
\end{equation}
with $p$ determined by the initial support in the even subspace, which reflects the conservation of the odd and even subspaces of the parity by the dynamics. Furthermore, dephasing in Eq.~\eqref{eq:Leff_deph_0}   manifests the fact that the odd and even stationary states of the cavity are actually no longer pure, but \emph{mixed} (for further discussion, see Appendixes~\ref{app:Leff:conserved:atom_mixed}-\ref{app:Leff:conserved:atom_beam}).

\subsubsection{Atom decay}
 The rate of dephasing dynamics caused by atom decay can be bounded by (cf.~Appendix~\ref{app:Leff:conserved:atom_decay})
\begin{eqnarray}\nonumber
\frac{\gamma_\text{deph}}{\nu}&\leq& 2\!\left[\left(\Gamma_1-\gamma_{1}\right) |c_g|^2\!\!+\!2\!\left(\Gamma_3-\gamma_{3} -\gamma_{13}|c_g|^2 \right)\!  |c_e|^2\right]T \\
&&+2\,\max\left(\Gamma_1,\Gamma_3\right) \tau,
\label{eq:deph_bound1_ad}
\end{eqnarray}
while the frequency
\begin{eqnarray}
\frac{|\Omega|}{\nu}&\leq&  \left[\gamma_{01} |c_g|^2+\left(\gamma_{03}+\gamma_{23}\right) |c_e|^2\right] T \nonumber\\
&& +\left(\gamma_{01}+\gamma_{03}+\gamma_{23}\right) \tau,
\label{eq:deph_bound2_ad}
\end{eqnarray} 
where $\gamma_{jk}$ denotes the decay rate from the atom level $|k\rangle$ to $|j\rangle$ (where $E_j<E_k$), $\gamma_{k}$ is the decay rate from $|k\rangle$  to the levels not coupled to the cavity, and  $\Gamma_{k}=\sum_{j: E_j<E_k}\gamma_{jk} +\gamma_{k} $ is the overall decay rate, so that the average \emph{lifetime} of the level $|k\rangle$ is $\Gamma_k^{-1}$.
The atom rate is, in turn, reduced to
\begin{equation}\label{eq:rate_deph_ad}
\frac{\overline \nu}{\nu}= 1-\left(\gamma_{1} |c_g|^2+\gamma_{3}|c_e|^2\right) T.
\end{equation}
From Eq.~\eqref{eq:deph_bound1_ad}, we observe that there is no contribution to the dephasing from time $T$ in the case of the decay to only uncoupled levels, $\Gamma_j=\gamma_j$ for $j=1,3$. Indeed, in this case, effectively, the atom arriving to the cavity is pure for all $T$, but the atom rate is still reduced [cf. Eq.~\eqref{eq:rate_deph_ad}]. Furthermore, for no decay to levels $|0\rangle$ and $|2\rangle$, we obtain $\Omega=0$, which reflects that in that case dynamics is real valued (cf. Sec.~\ref{sec:dynamics}).

\subsubsection{Nonmonochromatic atom beam}
 For the small fluctuations of the integrated coupling, $\delta \phi \, n \ll 1$ for $n$ within the support of stationary states with $\phi=\overline{\phi}+\delta \phi$, the nonmonochromatic beam effectively leads to two-photon decay and two-photon injections at the respective rates $\nu\overline{\delta\phi^2} |c_g|^2/2$ and $\nu\overline{\delta\phi^2} |c_e|^2/2$ (see~Appendix~\ref{app:Leff:conserved:atom_beam}). Here $\overline{\phi}$ is the average and  $\overline{\delta\phi^2}$ is the standard deviation of the resulting distribution of integrated coupling $\phi$. The rate of dephasing due to nonmonochromatic beam can, in turn, be bounded as
\begin{eqnarray}
\frac{\gamma_\text{deph}}{\nu}&\leq& \frac{\overline{\delta\phi^2}}{2} \bigg[|c_g|^2 \!\left(\!\sqrt{\langle a^{\dagger 2}\,a^2 \rangle_+}+\sqrt{\langle  a^{\dagger 2}\,a^2\rangle_-} \right)\!^2  \label{eq:deph_bound1_ab}\\
&&\qquad+|c_e|^2\!\left(\!\sqrt{\langle a^2 \,a^{\dagger 2}\rangle_+}+\sqrt{\langle a^2 \,a^{\dagger 2}\rangle_-} \right)\!^2\bigg],\nonumber
\end{eqnarray}
Furthermore, due to the real-valued dynamics, there is no unitary dynamics, 
\begin{eqnarray}\label{eq:deph_bound2_ab}
\frac{\Omega}{\nu}&=& 0. 
\end{eqnarray}

%===================================================================================================================================================
%===================================================================================================================================================

\subsection{Metastable dynamics of realistic micromaser}
 In Secs.~\ref{sec:higher-order}-\ref{sec:meta_mixed}, we have considered perturbative contributions from noise or higher order corrections to the long-time dynamics of the cavity.  In general, the micromaser dynamics is described by the sum of all the contributions, i.e., the sum of Eqs.~\eqref{eq:Leff_corr},~\eqref{eq:Leff_loss}, and~\eqref{eq:Leff_deph_0}, and features the unique stationary state
\begin{eqnarray}
\label{eq:rhoss_combined}
\rho_{\rm ss} &\approx& p\,\ketbra{\Psi_+}{\Psi_+} + \left(1-p\right) \ketbra{\Psi_-}{\Psi_-},\\\nonumber
p&=&\frac{\kappa\langle n\rangle_{-}+\nu \langle X\rangle_{-}}{\kappa\left( \langle n\rangle_{-}+\langle n\rangle_{+}\right)+\nu \left( \langle X\rangle_{-}+\langle X\rangle_{+}\right)}.
\end{eqnarray}  
Note that dephasing or unitary dynamics of odd-even coherences do not determine the approximation [cf. Eq.~\eqref{eq:rhoss_deph}].

\subsection{Beyond weak noise and small corrections}
\label{sec:beyond}

In Secs.~\ref{sec:higher-order}-\ref{sec:meta_mixed}, we assumed that noise and imperfections in the (5+1) micromaser setup contribute to the slowest timescales in the cavity dynamics, that is the inverse of the gap in the effective dynamics of Eqs.~\eqref{eq:Leff_corr},~\eqref{eq:Leff_loss}, and~\eqref{eq:Leff_deph_0},  [e.g., given by $-\lambda_2$ in Eq.~\eqref{eq:Leff_modes}] is much larger than the relaxation to the DFS of the pure stationary states $|\Psi_+\rangle$ and $|\Psi_-\rangle$. These results provide insight in the robustness of the dissipative preparation by indicating timescales on which the noise becomes relevant; i.e., the effective dynamics is no longer negligible. In particular, we find that the effective rate  for single-photon losses, higher order corrections, and nonmonochromatic beam depends on the average photon number, and thus the states with higher photon numbers are less robust to such noise.

However, as we discussed in  Sec.~\ref{sec:walls_soft}, the relaxation timescales may be significantly extended due to low amplitudes of connecting certain photon numbers, e.g., $|m\rangle$ and $|m+2\rangle$ due to $\sin_m\approx0$ in Eq.~\eqref{eq:Kraus}, that is as a soft wall at $m$.  Thus, when the noise is comparable with such timescales, we can no longer approximate the dynamics as taking place inside the DFS. Nevertheless, we can instead consider the effective dynamics 
among states which would be stationary in the approximation $\sin_m=0$, i.e., the states supported between soft walls. Such dynamics features two contributions: transitions across soft walls (considered in Sec.~\ref{sec:walls_soft}) and dynamics due to noise and imperfections of the micromaser, which we discuss now.

As a wall affects only the neighboring states of the same parity, any perturbations in the dynamics that swap the parity allow for circumventing walls by connecting the states of opposite parity, with rates simply proportional to the support of the perturbed state between walls of the opposite parity. This is the case for the higher order corrections in the far-detuned limit and single-photon losses from the cavity leading to the decay at the rate $\kappa \langle n\rangle_k^\pm +\nu \langle X\rangle_k^\pm$ of the $k$th state of the even(odd) parity $\rho_{k}^\pm$. Furthermore, as they change the photon number in the cavity only by one, they  enable \emph{local transitions}, that is for the state $\rho_k^\pm$ between walls at $m_k^\pm$ and $m_{k+1}^{\pm}$  only to the states $\rho_{k'}^\mp$ such that $m_k^\pm\leq  m_{k'}^\mp\leq m_{k+1}^\pm$ or $m_k^\pm\leq  m_{k'+1}^\mp\leq m_{k+1}^\pm$. Additionally, coherences between such states of the same parity, $\rho_{k'}^\mp$, can be created if they are pure and obey the same boundary conditions (cf. Secs.~\ref{sec:walls_hard} and~\ref{sec:walls_soft}). In turn, such coherences only get connected to coherences of the opposite parity.  
On the other hand, the atom decay or nonmonochromatic atom beam effectively lead to a random distribution of interaction times between atoms and the cavity, which results in fluctuations of the integrated coupling $\phi$, and thus also changing positions of walls. In turn, the states between walls again undergo local transitions to the preceding and following states of the same parity, i.e., $\rho_k^\pm$ to $\rho_{k-1}^\pm$ and $\rho_{k+1}^\pm$ at the respective rates $  \nu|c_e|^2 [\Gamma_1 \tau/2 + (m_{k}^\pm\!+\!1)(m_{k}^\pm\!+\!2)\overline{\delta\phi^2}]$ and $  \nu|c_g|^2 [(4\Gamma_3-3\gamma_3) \tau/8 + (m_{k+1}^\pm\!+\!1)(m_{k+1}^\pm\!+\!2)\overline{\delta\phi^2}]$. Here no coherences are created and, moreover, the effective dynamics obeys \emph{detailed balance}. 
Therefore, overall, we obtain a unique stationary state without even-odd coherences, which is consistent with the weak parity symmetry [cf. Eqs.~\eqref{eq:weak_symmetry} and~\eqref{eq:weak_symmetry_loss}].

In particular, in the case when the dynamics induced by noise is even faster than the timescales of relaxation across soft walls, we can neglect the latter by means of the almost degenerate perturbation theory. The resulting stationary state describes the state prepared in the cavity beyond the nondegenerate perturbative approximation of Secs.~\ref{sec:higher-order}-\ref{sec:meta_mixed}, and allows us to understand the change in the usefulness of generated states for potential applications, e.g., for phase estimation (see Sec.~\ref{sec:QFI_loss}). 
In general, the dynamics needs to be calculated individually for each value $\phi$ as positions of soft walls strongly depend on the integrated coupling, but for hard walls, with known positions given in Sec.~\ref{sec:walls_hard}, we derive the analytic description of the effective dynamics and find the resulting stationary states in Appendix~\ref{app:Leff:hard_walls}.
In particular, for the cavity being pumped by the excited atoms, $|c_{\rm e}|=1$, we find that the long-time dynamics due to noise or imperfections leads to local transitions between trapping states that always increase the photon number with the lifetime of a trapping state $|m\rangle$ given by $[\kappa\, m+ \nu \langle m|X|m\rangle + \nu\, \Gamma_1 \tau/2 + \nu(m\!+\!1)(m\!+\!2)\overline{\delta\phi^2}]^{-1}$. Therefore, in a realistic micromaser there are no trapping states.

% *****************************************************************************************************************************************************
% *****************************************************************************************************************************************************

\section{Application in phase estimation}
\label{sec:metrology}

In Secs.~\ref{sec:2photon_int} and~\ref{sec:steadystates}, we discussed the dynamics of two-photon micromaser with atom-cavity interactions described by Jaynes-Cunnings Hamiltonian, Eq.~\eqref{eq:Heff}. This dynamics lead to pure stationary state of the cavity dependent on both the initial atom state and the integrated coupling strength, Eq.~\eqref{eq:even_odd_ss}. Below we investigate the usefulness of the generated states for applications in phase estimation setups. We find that weak coupling does not yield a quantum enhancement in estimation precision, but strong coupling creates states which lead to an enhanced sensitivity. Although experimental imperfections, such as single-photon losses, lead to mixed states, we find that they can still enable enhancement in phase estimation.

\subsection{Quantum Fisher information}
 We consider a phase $\varphi$ which is to be estimated as unitarily encoded in a cavity state $\rho$  by the photon number operator $n=a^\dagger a$,
\begin{equation}
\rho_{\varphi}=e^{- i\varphi\ken n}\ken \rho \,\ken e^{ i\varphi\ken n}.
\label{eq:rho_phi}
\end{equation}
This corresponds to the situation when, after dissipatively preparing the cavity in the state $\rho$ by atom passages, the phase is subsequently encoded in the cavity state, e.g.,  by changing the cavity frequency by $\delta\omega$ to induce the phase $\varphi=\delta \omega\, t$ over time $t$~\cite{gilchrist2004schrodinger}. The errors in the unbiased estimation of $\varphi$ are then bounded, $\Delta^2 \varphi\geq F_Q(\rho)^{-1}$, by the inverse of the quantum Fisher information~\cite{Helstrom1967,Helstrom1968,Helstrom1968detection,Braunstein1994},
\begin{equation}\label{eq:qfi_formula}
F_Q(\rho)=2\sum_{j,j'}\frac{(p_j-p_{j'})^2}{p_j+p_{j'}}\left|\bra{E_{j}} n \ket{E_{j'}}\right|^2,
\end{equation}
where Eq.~\eqref{eq:qfi_formula} is expressed in the orthonormal eigenbasis of the state $\rho=\sum_j p_j \ketbra{E_j}{E_j}$. In particular, for pure states, $\rho=|\Psi\rangle\!\langle\Psi|$, the QFI is simply proportional to the the photon number variance, 
\begin{equation}\label{eq:qfi_formula2}
F_Q(|\Psi\rangle)= 4\left(\langle \Psi|n^2|\Psi\rangle-\langle \Psi|n|\Psi\rangle^2\right). 
\end{equation}
For example, for the coherent state $|\alpha\rangle$, the photon distribution is Poissonian, and thus $F_Q(|\alpha\rangle)=4\langle n\rangle=4 |\alpha|^2$, which is referred to as \emph{standard quantum limit}. Therefore, the phase estimation with $\rho$ features the \emph{quantum enhancement} over the classical strategy using the same amount of resources, i.e., the coherent state with the same average photon number, whenever~\cite{Giovannetti2004,Giovannetti2006,Giovannetti2011}
\begin{equation}\label{eq:qfi_enh}
\frac{F_Q(\rho)}{4 \langle n\rangle}>1.
\end{equation}
Considering this figure of merit is motivated by experimental limitations on the allowed energy, $\hbar\omega\langle n\rangle$, of the probe photon field. In such a case, further increase in the phase estimation precision can be achieved only by nonclassical distribution of the field, e.g.,  squeezing.

%===================================================================================================================================================
\subsection{QFI for micromaser in far-detuned limit}
\label{sec:QFI_lossless}

\begin{figure}[t!]
	\centering
	\includegraphics[width=1.05\columnwidth]{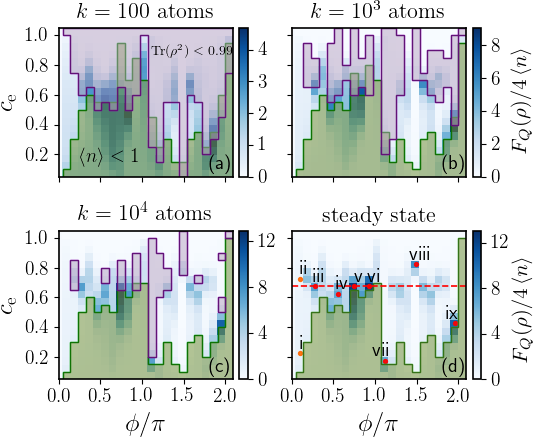}
	\vspace*{-6mm}
	\caption{\textbf{Phase estimation with dissipatively generated cavity states}. The four panels show the ratio of the QFI to the performace of the corresponding coherent state, $F_Q(\rho)/4 \braket{n}$  for the cavity initially in the vacuum $\ket{0}$ after the passage of $k = 100$, $10^3$, $10^4$ atoms and for the stationary state [Eq.~\eqref{eq:even_odd_ss}]. The enhancement i shown  as a function of the atom state [Eq.~\eqref{eq:psi}] and integrated coupling $\phi$. We sample the $\phi$ axis for $\phi_{20, K}$, Eq.~\eqref{eq:phi_wall}, with odd $K=1,3,..., 43$, which gives the hard wall at $m=20$ and allows convergence to stationary state also for $c_{\rm e}>1/\sqrt{2}$ (note that a larger $m$ would generally allow higher $\braket{n}$ and could also enable a higher enhancement in precision).
		The purple shading shows regions with reduced purity ${\rm Tr}(\rho^2)<0.99$, whereas the green shading excludes low average photon number, $\braket{n}<1$. 
		The red dots in the steady-state panel mark the stationary states (i)-(ix) analyzed in Fig.~\ref{fig:wig_gall}. The states (iii)-(ix) correspond to the states at the local maxima of the precision enhancement, while (i) and (ii) correspond to the standard and squeezed Schr\"{o}dinger cat states. A complex phase of $c_{\rm e}$ does not change the results, but the stationary states are not periodic in $\phi$, and thus here we show only a part of the parameter space.}
	\label{fig:qfi}
	\vspace*{-5mm} 
	
\end{figure}

In Fig.~\ref{fig:qfi}, we consider the QFI for an evolving cavity state and for the asymptotic stationary state. The QFI varies significantly across the parameter space of the atom state and integrated coupling strength. Importantly, multiple distinct stationary states achieve \emph{high enhancement over the classical limit}. %[red dots show states  (iii-ix) with locally highest enhancement]. 

\subsubsection{Wigner function and QFI}
The QFI,~\eqref{eq:qfi_formula}, which quantifies how sensitive is a state $\rho$ to phase rotations, is directly related to the Wigner function, Eq.~\eqref{eq:Wigner}. The QFI equals the speed of change in the overlap between the Wigner functions for $\rho$ and $\rho_\varphi$ [Eq.~\eqref{eq:rho_phi}]~\cite{Knott_2016}. Furthermore, the Wigner function for  $\rho_\varphi$ is simply the Wigner function for $\rho$ but rotated by $\varphi$. Therefore, for the states (iii)-(ix) with high values of the QFI the sign of the Wigner function highly oscillates [see Fig.~\ref{fig:wig_gall}(a)], thus ensuring a high QFI.

\subsubsection{Enhancement in precision due to soft walls}
We now argue that the enhancement above the classical limit, Eq.~\eqref{eq:qfi_enh}, is facilitated by the presence of soft walls in the dynamics. 

The stationary states, Eq.~\eqref{eq:even_odd_ss}, are dependent on the initial atom state and the integrated coupling strength, but the atom parameters alone imply the exponential decay in the photon number distribution for $|c_{\rm e}|\leq 1/\sqrt{2}$. The integrated coupling can instead facilitate a sharp revival in the occupation probability via a soft wall; for the wall at $m$, $\sin_m(\phi)\approx0$ with $\cos_m(\phi)\approx1$, we have $c_{m+2}/c_{m}\approx - i 2 \sin_m^{-1}(\phi)\, c_{\rm e}/c_{\rm g} $ (see Appendix~\ref{app:soft_walls} for further discussion). The revivals correspond directly to \emph{multimodal photon number distribution} [see Fig.~\ref{fig:wig_gall}(b)].
Since the considered stationary states are pure, their QFI is simply proportional to the photon number variance~\eqref{eq:qfi_formula2} and features the \emph{square} of distance between modes averages 
\begin{eqnarray}\label{eq:qfi_modes}
F_Q(|\Psi\rangle)&=&\sum_k p_k \,F_Q(|\Psi_k\rangle)\\\nonumber
&&+4\sum_k \sum_{k'> k}    p_k p_{k'} \left(\langle n\rangle_k-\langle n\rangle_{k'}\right)^2,
\end{eqnarray}
where $|\Psi\rangle=\sum_k \sqrt{p_k}|\Psi_k\rangle$ and $|\Psi_k\rangle$ represents the orthonormal $k$th mode. Thus, the QFI features quadratic rather than linear scaling with the average, which may lead to the precision enhancement, Eq.~\eqref{eq:qfi_enh}.  Multiple soft walls in close proximity can also lead to a unimodal distribution, but with a spread significantly wider than for the corresponding coherent states [see state (iv) in Fig.~\ref{fig:wig_gall}(b)]. The same mechanism is present for the stationary states of both parities [cf.~Fig.~\ref{fig:qfi2}].

The presence of soft walls introduces, however, long timescales of reaching pure stationary states, with cavity states being mixed at earlier times (purple shading in Fig.~\ref{fig:qfi}), even when the initial parity is fixed [see Sec.~\ref{sec:walls_soft} and Fig.~\ref{fig:wig_gall}(c)]. The mixedness of the cavity state in general lowers the estimation precision, which is captured by convexity of the QFI. Nevertheless, in Fig.~\ref{fig:qfi} we observe that the local maxima in the enhancement (iii)-(ix) are already present after passage of $100$ atoms, and their value increases with time as the corresponding pure stationary states are approached (cf.~the scale bars).

The revivals in photon probability distribution are highly sensitive to the coupling $\phi$ value, with their derivative proportional to $m$ and $\sin_m^{-1}(\phi)$. Therefore, the structure of  the cavity states varies significantly with $\phi$, allowing for preparations of distinct states (see Fig.~\ref{fig:wig_gall}) and is the reason for strong variations of the QFI in Fig.~\ref{fig:qfi}~\footnote{We note that in general steady states can vary very strongly with $\phi$. In particular, the support of the stationary state is limited by hard walls which exist at a given $m$ only for values of $\phi$ given by Eq.~\eqref{eq:phi_wall}. Nevertheless, when $\phi$ is varied, a given hard wall becomes a soft wall and the stationary state becomes metastable, so that it can be dissipatively generated in the corresponding metastable regime [cf.~Secs.~\ref{sec:walls_hard} and~\ref{sec:walls_soft}].}.

\subsubsection{Absence of enhancement in weak-coupling limit}
 In the weak-coupling limit, the cat and squeezed-cat states are generated, examples of which are marked as states (i) and(ii) in Figs.~\ref{fig:wig_gall} and~\ref{fig:qfi}. These states, although nonclassical, do not feature the enhancement in the phase estimation precision. The parity symmetry  allows for a superposition of the coherent states with the opposite phase, $\pm\alpha$, but with the same average photon number, $|\alpha|^2$. Therefore, the photon number distribution remains unimodal with the spread of the coherent state [cf.~Fig.~\ref{fig:wig_gall}(b)]. We note, however, that the enhancement proportional to $|\alpha|^2$ can be achieved via the linear operation of displacing the cat state in Eq.~\eqref{eq:psi_ss_lin} by $\pm\alpha$, which would give a bimodal photon distribution with the modes centered at $0$ and $|\alpha|^2$.    

%no isolated soft walls present in the weak-coupling limit!

\subsubsection{Coherence in DFS and QFI}
 In a general, an initial cavity state evolves into a mixed state inside the stationary DFS, but this cannot significantly reduce the enhancement present in the pure stationary states of fixed parity.  

From the conservation of the parity by the phase generator, $[n,P]=0$, we have that $\langle \Psi_+|n |\Psi_-\rangle=0$. This simplifies the QFI for any state within the DFS, 
\begin{eqnarray}\label{eq:rho}
\rho&=& p\,|\Psi_+\rangle\!\langle \Psi_+|+ (1-p)|\Psi_-\rangle\!\langle \Psi_-|\\\nonumber
&&+\, c\,|\Psi_+\rangle\!\langle \Psi_-|+c^*|\Psi_-\rangle\!\langle \Psi_+|,
\end{eqnarray}
where $|c|^2\leq p(1-p)$, to~\footnote{Let {$\rho= p_1|E_1\rangle\!\langle E_1|+p_2|E_2\rangle\!\langle E_2|$}, with {$|E_1\rangle=c_+|\Psi_+\rangle+c_-|\Psi_-\rangle$} and {$|E_2\rangle=c_-^*|\Psi_+\rangle-c_+^*|\Psi_-\rangle$}. From~\eqref{eq:qfi_formula}, we have {$F_Q(\rho)= 4 (p_1-p_2)^2 |\langle  E_1| n|E_2\rangle|^2+4 p_1 [\text{Var}(n,|E_1\rangle)-|\langle  E_1| n|E_2\rangle|^2] +4 p_2 [\text{Var}(n,|E_2\rangle)-|\langle  E_1| n|E_2\rangle|^2] $}, where {$\text{Var}(n,|E_{1,2}\rangle)$} denotes the variance of $n$ in {$|E_{1,2}\rangle$} [cf.~Eq.~\eqref{eq:qfi_formula2}]. Furthermore, {$\text{Var}(n,|E_1\rangle)= |c_+|^2 \text{Var}(n,|\Psi_+\rangle)+|c_-|^2 \text{Var}(n,|\Psi_-\rangle) + |c_+ c_- |^2(\langle n\rangle_+-\langle n\rangle_-)^2 $} [cf.~Eq.~\eqref{eq:qfi_modes}], while from the parity conservation by {$n$}, {$|\langle  E_1| n|E_2\rangle|^2= |c_+ c_-|^2(\langle n\rangle_+-\langle n\rangle_-)^2$}. Identifying {$p=p_1|c_+|^2+p_2|c_-|^2$} and {$c= (p_1-p_2) c_+ c_-^*$}  in~\eqref{eq:rho}, we arrive at~\eqref{eq:qfi_rho}.}
\begin{eqnarray}\label{eq:qfi_rho}
F_Q(\rho)&=& p \,F_Q(|\Psi_+\rangle)+(1-p)\, F_Q(|\Psi_-\rangle)
\\&&+ 4 \,|c|^2 \left(\langle n\rangle_+ -\langle n\rangle_-\right)^2.\nonumber
\end{eqnarray}
Therefore, the QFI increases with coherence $|c|$. It is  maximal for the pure state $\sqrt{p}|\Psi_+\rangle+ \sqrt{1-p}|\Psi_-\rangle$  [here $c=\sqrt{p(1-p)}$], and minimal for the mixed state $p\,|\Psi_+\rangle\!\langle \Psi_+|+ (1-p)|\Psi_-\rangle\!\langle \Psi_-|$~\footnote{In general the maximal QFI is achieved for: $p=0$ when {$F_Q(|\Psi_+\rangle)\geq F_Q(|\Psi_-\rangle)+4(\langle n\rangle_+-\langle n\rangle_-)^2$}, $p=1$  when {$F_Q(|\Psi_-\rangle)\geq F_Q(|\Psi_+\rangle)+4(\langle n\rangle_+-\langle n\rangle_-)^2$} and {$p=1/2+ [F_Q(|\Psi_+\rangle)-F_Q(|\Psi_-\rangle)]/[8(\langle n\rangle_+-\langle n\rangle_-)^2]$} otherwise. The minimal QFI is {$\min[F_Q(|\Psi_+\rangle),F_Q(|\Psi_-\rangle)]$}}. %(here,  due to parity conservation, the QFI is linear in $p$, rather than convex). 
Moreover,  the precision enhancement, Eq.~\eqref{eq:qfi_enh}, behaves as the QFI, since for all $c$ the average photon number remains constant, $\langle n\rangle =p\,\langle n\rangle_+ +(1-p)\langle n\rangle_-$.

% we could add a phase to the pure state

%what is the maximal enhancement? the derivative is proportional to third order plynomial ---> difficult to solve.

If the average photon number is similar in the odd and even states, the lack of coherence does not significantly affect the precision. More generally, if the odd and even stationary states feature the enhancement, $F_Q(|\Psi_\pm\rangle)/4\langle n\rangle_\pm \geq 1$, this is the case for any $\rho$, as
\begin{eqnarray}\label{eq:av_expanded}
&&\frac{F_Q(\rho)}{4 \braket{n}} = \bar{p} \,\frac{F_Q(\ket{\Psi_+})}{4 \braket{n}_+} + \left(1-\bar{p}\right)
\frac{F_Q(|\Psi_-\rangle)}{4 \braket{n}_-}\\\nonumber
&&\qquad\qquad+|c|^2\, \frac{ \left(\langle n\rangle_+ -\langle n\rangle_-\right)^2}{\langle n\rangle},
%&&\qquad\qquad+\frac{|c|^2}{p(1-p)}\,\bar{p}(1-\bar{p})\, \frac{ \left(\langle n\rangle_+ -\langle n\rangle_-\right)^2}{\bar{p}\,\langle n\rangle_-+(1-\bar{p})\langle n\rangle_+},
\end{eqnarray}
where
\begin{equation}
0 \leq \bar{p}= \frac{p \braket{n}_+}{p \braket{n}_+ + (1-p)\braket{n}_-}\leq 1.
\end{equation}
Furthermore, even if only the even (or the odd) stationary state features the enhancement, the precision of a mixed state in Eq.~\eqref{eq:rho} still beats the standard quantum limit provided the probability $p$ of the even [$(1-p)$ of the odd] stationary state is sufficiently large~\footnote{The minimal enhancement in the DFS is {$\min[F_Q(|\Psi_+\rangle)/\langle n\rangle_+,F_Q(|\Psi_-\rangle)/\langle n\rangle_-]$}.}; cf.~Eq.~\eqref{eq:av_expanded}.

\subsubsection{Cavity coherence from atom coherence}
 The high QFI in Fig.~\ref{fig:qfi} relies on the existence of pure coherent even and odd stationary states of the cavity.  This crucial coherence of the stationary states of fixed parity is created by the passage of pure coherent states of atoms, Eq.~\eqref{eq:psi}, which establish a phase reference for the cavity phase,~Eq.~\eqref{eq:even_odd_ss}. Indeed, whenever the atom state is mixed, but non-diagonal in the atom level basis, the even and odd stationary states of the cavity are non-diagonal in the photon-number basis, and thus feature nonzero QFI (see Appendix~\ref{app:Leff:conserved:atom_mixed}). In contrast, for diagonal states of atoms, the phase reference is absent, and the resulting cavity state is diagonal in photon-number basis (with the zero QFI), as the cavity achieves equilibrium with the effective atom temperature given by the relative population of the two atomic levels (see Appendix~\ref{app:classical}).

Mixed, but coherent atom states can be a consequence of finite lifetime of atom levels, discussed in Sec.~\ref{sec:meta_mixed}. Furthermore, this additionally lowers the purity of cavity states by possible decay events during the atom interaction with the cavity (see also Appendix~\ref{app:Leff:conserved:atom_decay}).

%===================================================================================================================================================
\subsection{QFI for micromaser with single-photon losses}
\label{sec:QFI_loss}

\begin{figure}[t!]
	\begin{flushright}
		\includegraphics[width=1\columnwidth]{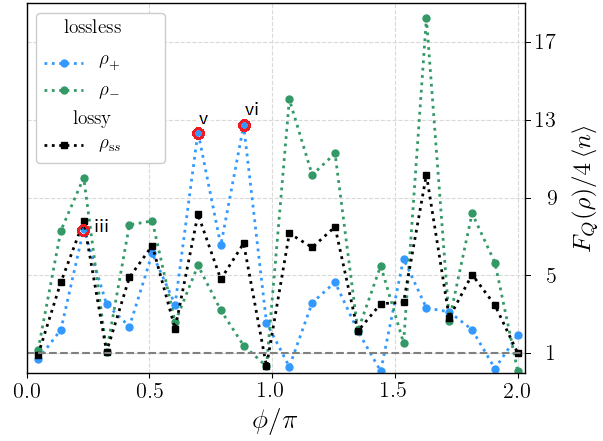}
	\end{flushright}
	\vspace*{-4mm}
	\caption{ \textbf{Effect of single-photon losses on phase estimation precision}. \textbf{(a)} The enhancement~\eqref{eq:qfi_enh} in the phase estimation is shown as a function of the integrated coupling $\phi$ [$c_{\rm e} = 0.65$ corresponding to dashed red line in Fig.~\ref{fig:qfi}]. The enhancement in the stationary state of lossy dynamics (black) [Eq.~\eqref{eq:qfi_rhoss_loss}] is shown against the enhancement in the even (blue) and odd (green) states that are stationary for lossless cavity.
		For the majority of parameter space, we observe the enhancement in phase estimation, i.e., $F_Q(\rho)/4 \braket{n}>1$ (values above the horizontal dashed gray line). Here the lossy stationary state is given by perturbative Eq.~\eqref{eq:rhoss_loss}. %was obtained after passage of $k=10^4$ atoms with $\kappa/\nu= 10^{-6}$. %Red circles denote the even states with locally highest enhancement, cf.~Fig.~\ref{fig:qfi}.  
		\textbf{(b)} Average photon number in even and odd stationary states. We observe the correlation of high photon number to when the QFI of a lossy stationary state differs from  Eq.~\eqref{eq:qfi_rhoss_loss} in panel (a), as it determines the size of the correction from the single-photon losses (together with the relaxation timescales in the lossless case [cf.~Fig.~\ref{fig:qfi}]. }
	\label{fig:qfi2}
	\vspace*{-3mm}
\end{figure}

In Secs.~\ref{sec:higher-order} and~\ref{sec:losses}, we have shown that due to the finite detunings or the presence of single-photon losses, the pure stationary states $\ket{\Psi_+}$ and $\ket{\Psi_-}$ of two-photon micromaser,  Eq.~\eqref{eq:even_odd_ss}, are rendered metastable, and the cavity dynamics leads instead to a unique stationary state approximated by their classical mixture [see Eqs.~\eqref{eq:rhoss_corr},~\eqref{eq:rhoss_loss}, and~\eqref{eq:rhoss_combined}]. Below we argue that in this limit the introduced mixedness does not significantly reduce the enhancement in the phase estimation precision. Therefore, the dissipatively generated cavity states can still be used quantum enhanced phase estimation.

% due to parity conservation by the photon number generator

\subsubsection{Enhancement in precision for lossy cavity}
 The stationary state of a lossy cavity, Eq.~\eqref{eq:rhoss_loss}, is approximated by a mixture of the even and odd states, $\rho_\text{ss}\approx\rho$ with $p= \braket{n}_-/(\braket{n}_++\braket{n}_-)$. In this case [cf.~Eq.~\eqref{eq:qfi_rho}], 
\begin{eqnarray}\label{eq:qfi_rhoss_loss}
&&\frac{F_Q(\rho)}{4 \braket{n}} = \frac{1}{2} \left[\frac{F_Q(\ket{\Psi_+})}{4 \braket{n}_+} + 
\frac{F_Q(|\Psi_-\rangle)}{4 \braket{n}_-}\right],
\end{eqnarray}
so that the enhancement higher than $2$ present in the even or the odd state implies $\frac{F_Q(\rho)}{4 \braket{n}}>1$ [cf.~Fig.~\ref{fig:qfi2}]. Note that we assume  losses to take place only during the generation of the cavity state, but not during the phase encoding [cf.~Eq.~\eqref{eq:rho_phi}].

It is important to comment here on corrections to Eq.~\eqref{eq:rhoss_loss} and thus to Eq.~\eqref{eq:qfi_rhoss_loss}. In derivation of the effective dynamics induced by single-photon losses, Eq.~\eqref{eq:rhoss_loss}, we assumed that the losses act as a perturbation of the cavity dynamics; i.e., timescales of lossy dynamics are much longer than the timescale $\tau$ of the relaxation into the pure stationary states~\eqref{eq:even_odd_ss}. In this case, the corrections to the stationary state in Eq.~\eqref{eq:rhoss_loss} are proportional to $\kappa\tau$~\cite{Kato1995,macieszczak2016towards}. Note that this perturbative approximation is \emph{limited by two factors}.

First, the influence of the single-photon losses is proportional to the average-photon number [cf.~Eq.~\eqref{eq:rhoss_loss}] as losses affects each photon independently. Therefore, states with higher photon number are more fragile to losses. This is also the reason, why losses present during the phase encoding (i.e., for \emph{fixed} strength of noise, $\kappa t$ for $\varphi=\delta\omega t$) lead to the enhancement in phase estimation limited to a constant [$(e^{\kappa t}-1)$] above the standard scaling~\cite{Kolodynski2010,Escher2011,Demkowicz2012}. 

Second, the soft walls which facilitate multimodal distribution, and thus the enhancement in precision, imply long relaxation time $\tau$. The relaxation timescales due to soft walls are, however, not directly related to the average photon number (cf.~Sec.~\ref{sec:walls_soft}).

Beyond the perturbative approximation, i.e., when losses take place at earlier timescales than $\tau$, they instead lead to the mixing dynamics of the metastable states between soft walls, as discussed in Sec.~\ref{sec:beyond}. This dynamics results in the stationary state being a mixture of pure and mixed states between soft walls, with possible coherences between pure states with the same boundary conditions, $\rho_\text{ss}\approx \sum_l p_l^\text{ss}\rho_l +\sum_k p_k^\text{ss}|\Psi_k\rangle\!\langle\Psi_k| + \sum_{k,k'} (c_{k,k'}^\text{ss} |\Psi_k\rangle\!\langle\Psi_{k'}| +\text{H.c.})$, where we explicitly distinguish between pure and mixed states and $|c_{k,k'}^\text{ss} |^2\leq  p_k^\text{ss}p_{k'}^\text{ss}$ (see also Appendix~\ref{app:Leff:hard_walls}). Thus, the QFI becomes (cf.~Eq.~\eqref{eq:qfi_modes} and see Ref.~\cite{Girolami2015})
\begin{eqnarray}\label{eq:qfi_modes_loss}
F_Q(\rho_\text{ss})&\gtrsim& \sum_l p_l^\text{ss} \,F_Q(\rho_l)+\sum_k p_k^\text{ss} \,F_Q(|\Psi_k\rangle) \\\nonumber
&&+4\sum_k \sum_{k'> k}    |c_{k,k'}^\text{ss} |^2 \left(\langle n\rangle_k-\langle n\rangle_{k'}\right)^2,
\end{eqnarray}
where the (approximate) inequality is saturated for $c_{k,k'}^\text{ss}\rightarrow0$ and in the lowest order of corrections $|c_{k,k'}^\text{ss} |^2$ is replaced by $|c_{k,k'}^\text{ss} |^2/(p_k^\text{ss}+p_{k'}^\text{ss})$.
Therefore, the precision enhancement is significantly reduced if the coherences are negligible, $|c_{k,k'}^\text{ss} |\ll  |p_k^\text{ss}-p_{k'}^\text{ss}|$, in which case it is crucial to reduce noise in an experiment to remain within the perturbative approximation of Eq.~\eqref{eq:qfi_rho}. For this, it is necessary that $\kappa\tau$ decreases inversely with the average photon number of the even and odd stationary states of the lossless cavity. Importantly, this requirement can be achieved by increasing the rate $\nu$ of atom passages,  since $\tau\propto\nu^{-1}$ [cf.~Eq.~\eqref{eq:master}].

\subsubsection{Other noise}
Similarly to single-photon losses, the higher order corrections in the far-detuned limit will lead to the mixed stationary state approximated by Eq.~\eqref{eq:rhoss_corr}, and thus Eq.~\eqref{eq:av_expanded} with $p=\langle X\rangle_-/(\langle X\rangle_++\langle X\rangle_-)$ and $c=0$. Here, however, the corrections cannot be minimized by increasing the rate $\nu$, but only by increasing atom detunings [see Fig.~\ref{fig:model}(a)].  

\begin{figure}[t]
	\centering
	\subfloat{\includegraphics[width=0.49\textwidth]{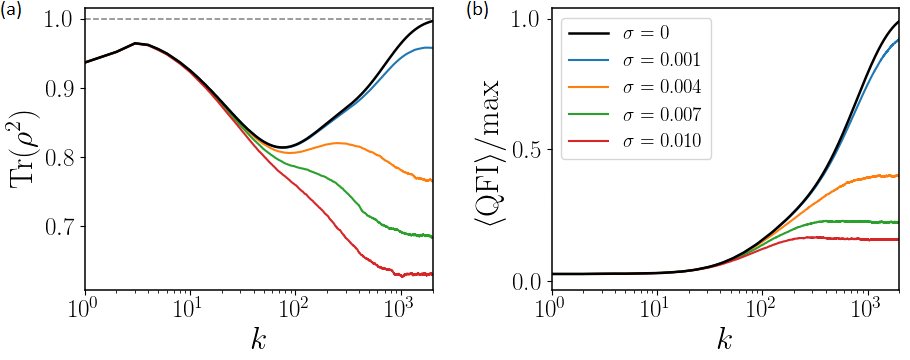}}
	\caption{\textbf{Effects of the nonmonochromaticity of atomic beam}. 
		Dynamics of the purity \textbf{(a)}, $\mathrm{Tr}(\rho^2),$   and  the QFI \textbf{(b)} [Eq.~\eqref{eq:qfi_formula}, normalized by the maximum value  $F_Q(\rho)/4\langle n\rangle=7.39$ in dynamics with the monochromatic beam], with the number of atoms $k$ passing the cavity, is shown for different widths $\sigma$ of the integrated coupling distribution, which for simplicity is assumed normal, $g(\phi) = \exp[-(\phi-\braket{\phi})^2/2 \sigma^2]/\sqrt{2 \pi \sigma^2}$. The initial state is the vacuum $\ket{0}$, the atom state is $c_{\rm e} = 0.65$, and the coupling $\overline \phi  = \phi_{20,\, 5} \approx 0.73707$ [equal to the parameters of the stationary state (iii) in Figs.~\ref{fig:wig_gall} and~\ref{fig:qfi}]. Dynamics was averaged over 100 random trajectories [cf.~Eq.~\eqref{eq:Kraus_average}]. Note the control of the order of $0.1\%$ in the velocity spread is required in order to achieve $\mathrm{Tr}(\rho^2)>0.9$  and $>90\%$ of the QFI that was obtained with a monochromatic beam.
	}	
	\label{fig:phi_spread}
	\vspace*{-2mm}
\end{figure}

The nonmonochromatic atom beam also influences the precision enhancement, as in the lowest order it leads to dephasing of odd-even coherences leading to $c=0$ in Eq.~\eqref{eq:rho} [cf. Eq.~\eqref{eq:rhoss_deph}]. Although $p=0$ is not fixed, we find that the QFI is still reduced (see Fig. \ref{fig:phi_spread}), as a result of the lowered purity of odd and even stationary states, and only small deviations in atom velocity are permitted if the purity of the produced states is to be maintained. Indeed, as discussed in Sec.~\ref{sec:beyond}, for slow relaxation across soft walls, the stationary state features no coherences, $c_{k,k'}^\text{ss}=0$, and thus the quadratic scaling is lost in Eq.~\eqref{eq:qfi_modes_loss}. Furthermore, this will also be the case for micromaser with atom levels of finite lifetime (cf. Appendix~\ref{app:Leff:hard_walls}).

%HERE THE DISTRIBUTION DOES NOT MAKE MUCH SENSE, IT SHOULD COME FROM THE DSITRIBUTION OF VELOCITIES, If they were Gaussian, i.e., Maxwell-Bolzmann distribution with a shift, then distribution of the coupling is $p(\phi)\propto \phi^{-2} \exp[-m \langle v\rangle^2/(2k_B T) (\phi-\langle \phi\rangle)^2/\phi^2 ] $.

%===================================================================================================================================================
%===================================================================================================================================================

\section{Experimental considerations}
\label{sec:Implementations}

Finally, we briefly review possible platforms to implement the Hamiltonian in Eq. (\ref{eq:Heff}).

\subsection{Rydberg atoms}
 Atoms excited to their higher principal quantum number states, so called Rydberg atoms, interacting with a microwave cavity are the setup where two-photon micromasers were originally developed~\cite{davidovich1987two, Brune_1987}. The interaction time is given by $\tau=w/v$, where $w$ is the cavity of mode waist and $v$ is the speed of atoms passing through the cavity. For $w \approx 2 \times 10^{-3}$ m and $v \approx 10^2$ ${\rm m} \, {\rm s}^{-1}$ \footnote{We have used $w \approx 2$ mm as compared with $w \approx 10$ mm used in \cite{Sarlette_2011} to make it compatible with the mode volume $V=70$ mm${}^3$ taken from \cite{Brune_1987}}, we have $\tau \approx 2\times 10^{-5} $ s. Therefore, a finite integrated coupling strength $\phi\approx 1$ requires  the coupling strength $\lambda\approx 10$ kHz, already achievable in three-level micromasers  \cite{Brune_1987}. We note, however, that currently typical single-photon loss rate $\kappa \approx 100$ Hz~\cite{Sarlette_2011}, while in order for the loss to be treated as the perturbation in the cavity dynamics the relaxation timescale must be much shorter than $\kappa^{-1}$ [see Fig.~\ref{fig:wig_gall}(c) and cf.~Sec.~\ref{sec:losses}], and thus loss rate $\kappa$  would need to be significantly lower (or $\tau$ shorter to allow for higher atom rate $\nu$).  

Nevertheless, in order to consider effective two-photon coupling $\lambda$ in a (5+1) model realized with Rydberg atoms, we aimed to identify five Rydberg levels fulfilling the conditions in Eqs.~(\ref{eq:conditions_a}) and (\ref{eq:conditions_b}) [the condition in (\ref{eq:conditions_c}) can be satisfied by appropriate choice of the Rabi frequency $G$ and the detuning $\delta$ of the classical field]. We performed a preliminary search using the ARC (Alkali.ne Rydberg Calculator) package~\cite{ARC_package,Sibalic_CompPhysComm_2017} among 30 basis states close to the levels realizing two-photon micromaser in Ref. \cite{Brune_1987}, $39{\rm S}_{\frac{1}{2}} \leftrightarrow 39{\rm P}_{\frac{3}{2}} \leftrightarrow 40{\rm S}_{\frac{1}{2}}$. We identified the transitions $37{\rm S}_{\frac{1}{2}} \leftrightarrow 37{\rm P}_{\frac{3}{2}} \leftrightarrow 38{\rm S}_{\frac{1}{2}} \leftrightarrow 38{\rm P}_{\frac{3}{2}} \leftrightarrow 39{\rm S}_{\frac{1}{2}}$ with  $\omega \approx 500$ GHz, $|\Delta_j| \approx 21$ GHz, and $g_j \approx 0.3$ MHz leading to $|g_1|^2/\Delta_1 = 0.95 |g_2|^2/\Delta$, $|g_4|^2/\Delta_4 = -1.02 |g_3|^2/\Delta$ [cf.~Eq. (\ref{eq:conditions})]. The effective coupling strength $|\lambda| \approx 5$ Hz leads only to the weak-coupling regime with $\phi\approx 10^{-4}$, where the Schr\"{o}dinger cat states could be generated (see Sec.~\ref{sec:schrodinger_cats}).  Considering larger set of basis states and an external electric field enabling tunable detunings $\Delta_j$ through the static Stark effect, could, however, yield transitions with stronger effective interaction.  See Appendix~\ref{app:experimental} for further discussion.

\subsection{Circuit QED}
Circuit QED represents a versatile platform to realize Hamiltonians with strong higher order photon processes \cite{mirrahimi2014dynamically, roy2015continuous, leghtas2015confining, Campagne-Ibarcq_PRL_2018}. In particular, a scheme studied in Ref.~\cite{Zeytinoglu_2015} realized a system with a tunable coupling between a transmon qubit and a microwave resonator with the effective single-photon Jaynes-Cummings Hamiltonian, $H = \lambda(t) \tilde{a}^\dag \tilde{\sigma}_- + \lambda(t)^* \tilde{a} \tilde{\sigma}_+$, where $\tilde{a}, \tilde{\sigma}$ are the effective photonic and atomic operators dressed by the anharmonic Jaynes-Cummings Hamiltonian of the qubit-cavity system. It remains an open question whether the two-photon Jaynes-Cummings Hamiltonian in Eq.~\eqref{eq:Heff} can also be achieved.

%===================================================================================================================================================

%===================================================================================================================================================

%===================================================================================================================================================

%===================================================================================================================================================
%===================================================================================================================================================

\section{Conclusions}
\label{sec:Conclusions}

We have proposed a scheme to realize two-photon micromasers exploiting a (5+1)-level structure of atoms passing through a cavity. We have shown that the atom parameters can be tuned to achieve an effective two- photon interaction Hamiltonian without the Stark shifts, unlike in the three-level micromasers.  We have found this enables dissipative generation of pure states with high quantum Fisher information for phase estimation. Furthermore, we have found that the pure odd- and even-parity stationary states span a decoherence-free subspace. Thus, in addition to phase estimation, the discussed scheme could be exploited in quantum information processing (cf. Ref.~\cite{mirrahimi2014dynamically}), as a quantum memory or as a quantum processor with unitary operations implemented by perturbing the micromaser dynamics~\cite{beige2000quantum,zanardi_coherent_2014,zanardi_geometry_2015}.

To account for realistic imperfections, we have considered effects of higher order corrections in the far-detuned limit, single-photon losses from the cavity, finite lifetime of atom levels, and nonmonochromatic atom beam. For small enough imperfections, there exists a pronounced metastable regime with metastable states corresponding to the formerly stationary states. After the metastable regime, the relaxation to a unique stationary state takes place. Importantly, we found, that even after the metastable regime, the generated stationary states, although mixed, can still feature a significant enhancement in phase estimation precision.  

Future research directions include identifying experimental schemes to implement the (5+1)-level model and constructing feedback schemes to counteract the mixing dynamics of metastable states due to single-photon losses.

%\bgroup\let\addcontentsline=\nocontentsline
\section*{Acknowledgments}
%\egroup

We would like to thank A. Armour, C. Davis-Tilley, M. Marcuzzi, M. Guta and J. Home for informative and useful discussions. J.M. would like to thank R. J. C. Spreeuw and H.B. van Linden van den Heuvell for useful discussion on Rydberg atoms. The  research  leading  to  these  results  has  received  funding from  the  European  Research  Council  under  the  European Union’s Seventh Framework Programme (FP/2007-2013) [ERC Grant Agreement No.  335266 (ESCQUMA)] and  from  the  European  Union’s  H2020  research  and innovation  programme  [Grant  Agreement  No.   800942 (ErBeStA)]. Funding was also received from the EPSRC [Grants  No. EP/M014266/1 and No. EP/P026133/1]. I.L.  gratefully  acknowledges  funding  through  the  Royal  Society  Wolfson  Research  Merit  Award. K.M. was supported through the Henslow Research Fellowship.

%\bibliography{bibfile}
%\bibliographystyle{apsrev4-1}


% ================================================================================================================================================
% ================================================================================================================================================

\appendix

%===================================================================================================================================================
%===================================================================================================================================================
%=====================================================================   APPENDIX   ================================================================
%===================================================================================================================================================
%===================================================================================================================================================

\onecolumngrid

\section{Atom-cavity interaction}
\label{app:model}

\subsection{(5+1) Jaynes-Cummings Hamiltonian}
\label{app:Transformations}

Here we present the details of the transformations leading to Eq.~(\ref{eq:H_5plus1_levels}).

We consider (5+1)-level atoms the cavity field of frequency $\omega$ with the free Hamiltonian [see Fig.~\ref{fig:model}(a)]
\begin{equation}
H'_0 = \omega \left(a^{\dagger}a+\frac{1}{2}\right)+ \sum_{j = 0,..., 4, {\rm a}} E_j \,\sigma_{jj},
\label{eq:H0prime}
\end{equation}
where $\sigma_{ij}=\ketbra{i}{j}$,   $a$ and $a^{\dagger}$ denote the cavity annihilation and creation operators, and $\hbar=1$.
The atom is coupled to the cavity field and a classical field of frequency $\omega_{\rm cl}$ and Rabi frequency $G$~\footnote{Alternatively, the auxiliary level $\ket{\rm a}$ can be coupled to the level $\ket{1}$, instead of $\ket{3}$, in order to compensate for the Stark shifts in the effective Hamiltonian. The coupling of three-level atom to a classical Rabi field was considered in~\cite{Kazakov_JOptB_2001}.}
\begin{eqnarray} 
H'_{\rm int}(t)  &=& \left(a+  a^{\dagger }\right) \sum_{j=1}^4 g_j\, \sigma_{j(j-1)} \nonumber \\
&& + (G e^{- i \omega_{\rm cl} t} + G^* e^{ i \omega_{\rm cl} t})\ken \sigma_{\rm a3}+ {\rm H.c.}
\label{eq:Hfull}
\end{eqnarray}

 In the frame rotating with the free Hamiltonian $H_0'$, Eq.~\eqref{eq:H0prime}, the interaction Hamiltonian~\eqref{eq:Hfull} becomes
\begin{equation} 
 e^ {it H_0'} {H}'_{{\rm int}}  e^ {-it H_0'}  = \left(a+ e^{ i 2\omega t} a^{\dagger }\right) \sum_{j=1}^4 g_j\, e^{ i \Delta_j t}\,\sigma_{j(j-1)}  + (G + G^* e^{ i 2\omega_{\rm cl} t})\ken e^{ i \delta t}\sigma_{\rm a3}+ {\rm H.c.} \label{eq:HfullROT}
\end{equation}
Since the detunings are assumed much smaller than the corresponding energy gaps, $|\Delta_j|, |\delta| \ll \omega, \omega_{\rm cl}$ are assumed, we can perform the rotating-wave approximation by neglecting the counter-rotating terms in~\eqref{eq:HfullROT} (see, e.g., Ref.~\cite[ch. 5.2.2]{AtomicOpticsBook}). This leads to the atom-cavity interaction described by multilevel Jaynes-Cummings Hamiltonian~\cite{jaynes1963comparison}
\begin{eqnarray} 
 {H}''_{\rm int}(t)   &=& a \sum_{j=1}^4 g_j\, e^{ i \Delta_j t}\,\sigma_{j(j-1)} + G\ken e^{ i\delta t}\sigma_{\rm a3}+ {\rm H.c.}
\label{eq:HfullRWAROT}
\end{eqnarray}
while in the initial frame we have
\begin{eqnarray} 
{H}_{\rm JC}(t) =  e^ {-it H_0'}{H}''_{\rm int}(t)e^ {it H_0'}  &=& a\, \sum_{j=1}^4 g_j\, \sigma_{j(j-1)} 
+  G e^{- i \omega_\text{cl} t}\ken \sigma_{\rm a3}+ {\rm H.c.}. 
\label{eq:HfullRWA}
\end{eqnarray}
It is important to note that the new dynamics, $H'_0+{H}_{\rm JC}(t)  $, conserves the number of excitations $N = n + \sum_{j=1}^4 j\sigma_{jj}+3\sigma_{\rm aa}$, where $n=a^\dagger a$ is the cavity photon number operator, i.e., $[N, H'_0+{H}_{\rm JC}(t)] = 0$.  Moreover,  it is possible and relevant (see Appendix~\ref{app:micromaser}) to remove time-dependence from the dynamics~\eqref{eq:HfullRWA}, by considering the frame rotating with
$(\omega N + \omega_{\rm cl}\sigma_{\rm aa})$, which leads to the dynamics governed by Eq.~(\ref{eq:H_5plus1_levels}).

\subsection{Effective two-photon interaction}
\label{app:adiabatic}
Here we consider adiabatic elimination~\cite{Alexanian1995unitary,Klimov2002} for atom-cavity dynamics described by $H_0+H_\text{int}$ of~\eqref{eq:H0} and~\eqref{eq:Hint} at the resonance~\eqref{eq:resonance}. We derive the effective two-photon Hamiltonian of Eqs.~\eqref{eq:Heff0} and~\eqref{eq:Heff}, which arise in the second-order of couplings $g_1$, $g_2$, $g_3$, $g_4$ and $G$ [see Fig.~\ref{fig:model}(a)].

~\\
Adiabatic elimination can be viewed as formally diagonalizing $H=H_0+H_\text{int}$,~\eqref{eq:H0} and~\eqref{eq:Hint}, by perturbation theory with respect to $H_\text{int}$. The Hamiltonian $H$ is diagonalized by a unitary transformation $e^{S}$, where the anti-Hermitian operator $S$ is assumed to be expanded  in the coupling strength, $S=S_1+S_2+...$ . Therefore, 
\begin{eqnarray}
H_\text{diag}&=& e^{S}(H_0+H_\text{int})e^{-S}=H_0+H_\text{int}+[S,H_{0}+H_\text{int}]+\frac{1}{2!}[S,[S,H_{0}+H_\text{int}]+... \nonumber \\
&=&H_0+\left(H_\text{int}+[S_1,H_0]\right)+\left([S_2,H_0]+[S_1,H_\text{int}]+\frac{1}{2!}[S_1,[S_1,H_0]]\right)+..., \label{eq:expansion}
\end{eqnarray}
where we ordered the second line of~\eqref{eq:expansion} in increasing power of the interaction strength.  Note that 
$H_\text{diag}$ is assumed diagonal up to initial degeneracy in $H_0$ of the atomic levels $|1\rangle$ and $|3\rangle$,  which is due to the resonance~\eqref{eq:resonance}. Therefore, from~\eqref{eq:expansion}, $S$ is perturbatively determined \footnote{$S_k$ is only determined up to an anti-symmetric operator commuting with $H_0$ [cf.~Eq.~\eqref{eq:Scond}]. This freedom, therefore, corresponds only to unitary transformations in degenerate eigenbasis of $H_0$, so that $H_\text{diag}$ in~\eqref{eq:expansion} remains diagonal up to this degeneracy. Here we assume this transformation to be identity.} as [cf.~\cite{Klimov2002}]
\begin{equation}
-[S_1,H_0]=H_\text{int},\quad -[S_2,H_0]=\left([S_1,H_\text{int}]+\frac{1}{2!}[S_1,[S_1,H_0]]\right)', \quad..., \label{eq:Scond}
\end{equation}
where $(X)'$ denotes the off-diagonal elements of $X$ in the eigenbasis of $H_0$. The first condition simplifies Eq.~\eqref{eq:expansion} to only even-number corrections,
\begin{equation}
H_\text{diag}= H_0+\left([S_2,H_0]+\frac{1}{2}[S_1,H_\text{int}]\right)+...\label{eq:expansion2},
\end{equation}
which is a consequence of the assumed two-photon resonance in $H_0$ and single-photon interactions in $H_{\text{int}}$.
Substituting~\eqref{eq:Scond} to \eqref{eq:expansion2}, we obtain 
\begin{equation}
H_\text{diag}=H_0+ \left[
\begin{array}{cccccc}
-a^\dagger a \frac{|g_1|^2 }{\Delta_1}& 0 & 0 & 0 & 0 & 0 \\
0 & a\,a^\dagger \frac{|g_1|^2  }{\Delta_1 }-a^\dagger a\frac{g_2^2  }{ \Delta_2}  & 0 & a^{\dagger 2}\frac{ g_2^*g_3^* (\Delta_2-\Delta_3)}{2 \Delta_2 \Delta_3} & 0 & 0 \\
0 & 0 & a\, a^\dagger  \frac{|g_2|^2  }{\Delta_2 }-a^\dagger a\frac{|g_3|^2  }{ \Delta_3} & 0 & 0 & 0 \\
0 & a^2\frac{g_2g_3  (\Delta_2-\Delta_3)}{2 \Delta_2 \Delta_3} & 0 & a\,a^\dagger\frac{ |g_3|^2 }{\Delta_3}-a^\dagger a\frac{ |g_4|^2 }{\Delta_4} -\frac{ |G|^2}{\delta }& 0 & 0 \\
0 & 0 & 0 & 0 &  a\,a^\dagger \frac{|g_4|^2 }{\Delta_4} & 0 \\
0 & 0 & 0 & 0 & 0 & \frac{|G|^2}{\delta } \\
\end{array}
\right]+\,... \label{eq:Hdiag}
\end{equation}
for the operators
\begin{eqnarray}
\label{eq:S1}
S_1&=&\left[
\begin{array}{cccccc}
0 & -a^\dagger\frac{ g_1^* }{\Delta_1} & 0 & 0 & 0 & 0 \\
a\frac{ g_1 }{\Delta_1} & 0 & -a^\dagger\frac{ g_2^*}{\Delta_2} & 0 & 0 & 0 \\
0 & a\frac{ g_2}{\Delta_2} & 0 & -a^\dagger\frac{ g_3^* }{\Delta_3} & 0 & 0 \\
0 & 0 & a\frac{ g_3\ }{\Delta_3} & 0 & -a^\dagger\frac{ g_4^* }{\Delta_4} & -\frac{G^* }{\delta } \\
0 & 0 & 0 & a\frac{ g_4 }{\Delta_4} & 0 & 0 \\
0 & 0 & 0 & \frac{G }{\delta } & 0 & 0 \\
\end{array}
\right],
\\ 
\label{eq:S2}
S_2&=&\left[
\begin{array}{cccccc}
0 & 0 & -a^{\dagger 2}\frac{g_1^*g_2^*  (\Delta_1-\Delta_2)}{2 \Delta_1 \Delta_2 (\Delta_1+\Delta_2)} & 0 & 0 & 0 \\
0 & 0 & 0 & 0 & 0 & 0 \\
a^2\frac{g_1g_2 (\Delta_1-\Delta_2)}{2 \Delta_1 \Delta_2 (\Delta_1+\Delta_2)} & 0 & 0 & 0 & -a^{\dagger 2}\frac{g_3^*g_4^* (\Delta_3-\Delta_4)}{2 \Delta_3 \Delta_4 (\Delta_3+\Delta_4)} & -a^\dagger\frac{ g_3^* G^*(\Delta_3-\delta)}{2   \Delta_3\delta (\delta +\Delta_3)} \\
0 & 0 & 0 & 0 & 0 & 0 \\
0 & 0 & a^2\frac{g_3 g_4 (\Delta_3-\Delta_4)}{2 \Delta_3 \Delta_4 (\Delta_3+\Delta_4)} & 0 & 0 & -a \frac{ g_4 G^*(\Delta_4+\delta )}{2 \Delta_4\delta  ( -\Delta_4+\delta) } \\
0 & 0 & a\,\frac{g_3 G (\Delta_3-\delta)}{2   \Delta_3\delta (\delta +\Delta_3)} & 0 & a^\dagger\frac{ g_4^* G (\Delta_4+\delta)}{2 \delta \Delta_4  ( -\Delta_4+\delta) } & 0 \\
\end{array}
\right].\quad
\end{eqnarray}
It should be emphasized that atom-cavity interaction, Eq.~\eqref{eq:Hdiag}, takes place in the diagonalizing basis [cf.~Eq.~\eqref{eq:expansion}] given by   $e^{S}(|j\rangle\otimes|n\rangle)=|j\rangle\otimes|n\rangle+S_1(|j\rangle\otimes|n\rangle)+... $, where the atom levels are labeled by $j=0,..,4,a$, while $n=0,1,2...$ denotes a photon number in the cavity. In the far-detuned limit of $|g_j/\Delta_j|\ll1$ for $j=1,..,4,\mathrm{a}$ and  $|G/\delta|\ll1$, the lowest-order corrections, the diagonalizing basis corresponds to the original atomic levels $|0\rangle$, ..., $|4\rangle$ and $|\mathrm{a}\rangle$, in tensor product with the photon-number basis of the cavity states. In particular, in Sec.~\ref{sec:2photon_int}, $H_\text{diag}$ restricted to the levels  $|1\rangle$ and $|3\rangle$ is considered [cf.~Eqs.~\eqref{eq:H0} and~\eqref{eq:Hint}].  In Appendix~\ref{app:Kraus} we consider corrections to the dynamics beyond this approximation.\\

%\noindent
%\emph{Two-photon resonance and adiabatic elimination}. 
We note that the results in Eqs.~(\ref{eq:Hdiag})-(\ref{eq:S2}) do not require the resonance condition in Eq.~\eqref{eq:resonance}. When this condition is not fulfilled, $H_0$ contributes a static Stark shift $(\Delta_2+\Delta_3) \sigma_{33}$ to the effective Hamiltonian in Eq.~\eqref{eq:Heff0}. This effect can be eliminated by adjusting $\omega_\text{cl}$ (and thus $\delta$) or $G$ of the classical field [cf.~Eq.~\eqref{eq:conditions}]. \\

%\noindent
%\emph{Convergence of perturbation theory}. 
Due to conservation of the number of excitations, $N=a^\dagger  a+ \sum_{j=1}^4 j\,\sigma_{jj}+3\sigma_\text{aa} $,  although the cavity space dimension is infinite, the perturbation theory above is effectively performed on (at most) six-dimensional subspaces spanned by $|0\rangle\otimes|n\rangle$, $|1\rangle\otimes|n-1\rangle$, $|2\rangle\otimes|n-2\rangle$, $|3\rangle\otimes|n-3\rangle$, $|4\rangle\otimes|n-4\rangle$ and $|\text{a}\rangle\otimes|n-3\rangle$, for $n=0,1,...$, denoting the photon number in the cavity. For given $N$, the effective perturbation size can be approximated as $\lVert H_\text{int}\rVert\lVert (H_0-\Delta_1)^+\rVert=O [\max (N \max_j |g_j|,G)/\min(|\Delta_1|,|\Delta_2|,|\Delta_4|,\delta)  ]$, where $(H_0-\Delta_1)^+$ denotes the pseudoinverse~\cite{Kato1995,zanardi_coherent_2014,zanardi_geometry_2015,zanardi_dissipative_2016}. This defines the far-detuned limit for a given $N$. When the dynamics in the two-level approximation~\eqref{eq:Kraus} features well-defined stationary states and the initial cavity state is bounded, i.e., it has a finite support below $n_{\rm in}$, we expect the stationary state to be achieved at a finite-relaxation time $\tau_\text{relax}$ exploring effectively a finite cavity space; cf., e.g., Ref.~\cite{azouit2015convergence}. If the perturbation size is small for $N\gg \nu \tau_\text{relax}$, for full atom-cavity dynamics given by $H_0+H_\text{int}$  there exists a metastable regime where two-level approximation holds and a metastable state is given by the former stationary state. 	At longer times, the effective dynamics resulting from the higher order corrections takes place and leads to a unique stationary state (see also Sec.~\ref{sec:higher-order}). In the next section we consider these higher order corrections to dynamics.

%*******************************************************************************************************************************************************
\section{Micromaser}

Here we discuss general dynamics of a micromaser and the assumptions leading to the Markovian time-homogeneous dynamics of the cavity, the case of which  is discussed for the far-detuned limit in Sec.~\ref{sec:dynamics}. In Appendix~\ref{app:Kraus}, we derive the higher order-corrections to the two-photon dynamics described by the Kraus operators~\eqref{eq:Kraus}, which lead to metastability and mixing long-time dynamics discussed in Sec.~\ref{sec:higher-order}.	

\subsection{General dynamics}
\label{app:micromaser}
%*******************************************************************************************************************************************************
A \emph{micromaser} is a setup in which atoms pass through the cavity, one at a time, and interact with its field [see Fig.~\ref{fig:model}(b)]. 

\subsubsection{Assumptions for Markovian time-homogenous micromaser}
We now discuss three assumptions leading to \emph{Markovian} \emph{time-homogenous} dynamics of the micromaser (cavity)~\cite{englert2002elements}.\\
\begin{comment}
\begin{enumerate}
	\item  Atoms are prepared \emph{identically and independently} (in a product state)  with respect to one another and the cavity. 
	
	\item  The atomic beam is \emph{monochromatic}, i.e., velocity of all atoms is the same.  
	
	\item  The atom state is \emph{invariant} under the dynamics in Eq.~\eqref{eq:H0}, i.e., $e^{- i t H_0} \rho_\text{at} e^{ i t H_0}=\rho_\text{at}$.
\end{enumerate}
\end{comment}

 \emph{Assumption 1.} Atoms are prepared \emph{identically and in a product state} with respect to one another and the cavity.
 
 Let  $\rho^{(k)}$ be the state of  the cavity  after the interaction with $k$ atoms. In the frame rotating with the free Hamiltonian~\eqref{eq:H0prime}, the cavity state changes only when an atom is passing through. For an initial state of the cavity and the atoms given by tensor product $\rho^{(0)}\otimes (\rho_{\mathrm{at}}\otimes \cdots\otimes\rho_{\mathrm{at}}\otimes \cdots)$, the state $\rho^{(k)}$ of the cavity depends only on its state $\rho^{(k-1)}$ before the interaction with $k$th atom, %this is due only to tensor product, but we want the dynamics to be time-homogenous and that is why we assume the state of the atoms to be identical
	\begin{equation}\label{eq:discrete1}
		\rho^{(k)}=\text{Tr}_{\mathrm{at}}\left\{U(t_k,\tau_k) \ken \left[ \rho_{\mathrm{at}}\otimes \rho^{(k-1)} \right] \ken U^{\ken\dagger }(t_k,\tau_k)\right\}, 
	\end{equation}
	where $t_k$ and $\tau_k$ denote the arrival time of the $k$th atom and the duration of its interaction with the cavity field, respectively, while
	\begin{equation}
	U(t,\tau)=\mathcal{T}\exp\left\{- i\int_{t}^{t+\tau}   d  t'  H''_{\rm int}(t')  \right\}
	\end{equation}
	is the time-ordered evolution operator for the interaction~\eqref{eq:HfullRWAROT}. Equation~\eqref{eq:discrete1} represents \emph{Markovianity} of the cavity dynamics. \\

\emph{Assumption 2.}  The atomic beam is \emph{monochromatic}, i.e., velocity of all atoms is the same.
	
	 In this case, the interaction time with the cavity is the same for all atoms,  $\tau_k\equiv\tau$.  Note  that the atomic state $\rho_{\rm at}$ is typically not initialized for all atoms at $t=0$ as written formally in Assumption 1. In practice, a state in Eq.~\eqref{eq:psi} can be prepared by atoms passing on their way to the cavity through a laser resonant with the transition $\ket{1} \leftrightarrow \ket{3}$, which, for atoms with the same velocity, leads to the identical state (as the laser phase is constant in the frame rotating with  $H_0'$). For discussion of changes in micromaser dynamics due to nonmonochromatic atomic beam see Appendix~\ref{app:Leff:conserved:atom_beam}.\\

\emph{Assumption 3.}  The atom state is \emph{invariant} under the dynamics~\eqref{eq:H0}, $e^{- i t H_0} \rho_\text{at} e^{ i t H_0}=\rho_\text{at}$.
	
	With Assumption 2., the cavity dynamics~\eqref{eq:discrete1} depends on time only via the time-dependent interaction Hamiltonian~\eqref{eq:HfullRWAROT}. The interaction Hamiltonian is, however, time independent in the frame of $(\omega N + \omega_{\rm cl}\sigma_{\rm aa})$ [cf.~Eq.~\eqref{eq:Hint}], which differs from the frame of $H_0'$ by the Hamiltonian $-H_0$ [Eq. ~\eqref{eq:H0}]
	\begin{equation}\label{eq:U0}
		 U(t,\tau)  = e^{- i t H_0} \left\{\mathcal{T} e^{- i \int_{0}^\tau  d t' [ H_\text{int}(t')+ H_0]}  \right\} e^{ i (t+\tau) H_{0}}.
	\end{equation}
	Since $H_0$ acts only on the atom state, we have
	\begin{equation}\label{eq:discrete2}
	\rho^{(k)}=\text{Tr}_{\mathrm{at}}\left\{ U(\tau) \ken \left[e^{ i (t_k+\tau) H_{0}} \rho_\text{at} e^{- i (t_k+\tau) H_{0}}\otimes \rho^{(k-1)} \right] \ken U^{\ken\dagger }(\tau)\right\}, 
	\end{equation}
	where we introduced
	\begin{equation}\label{eq:U}
	U(\tau)  = \mathcal{T} e^{- i \int_{0}^\tau  d t [ H_\text{int}(t)+ H_0]} ,
	\end{equation}
	so that for the invariant atom state the cavity dynamics simplifies to
	\begin{equation}\label{eq:discrete0}
	\rho^{(k)}=\text{Tr}_{\mathrm{at}}\left\{U(\tau) \ken \left[ \rho_{\mathrm{at}}\otimes \rho^{(k-1)} \right] \ken U^{\ken\dagger }(\tau)\right\}. 
	\end{equation}	
The time dependence of the interaction on $t$ in Eq.~\eqref{eq:U} is due to the coupling strengths $g_j(t)$, $j=1,...,4$, and $G(t)$ being in general dependent on the atom position within the cavity, which changes in time $t$ [cf.~Eqs.~\eqref{eq:H0} and~\eqref{eq:Hint}].

In order for an atom state to be invariant, it cannot feature coherences between $H_0$ eigenstates of different energy; e.g., for a nondegenerate $H_0$ it must be diagonal.  In order for the \emph{pure coherent state} in Eq.~\eqref{eq:psi} to be invariant, we require the \emph{two-photon resonance} in Eq.~\eqref{eq:resonance}, which leads to degeneracy of $|1\rangle$ and $|3\rangle$ in $H_0$.

We note, however, that when the resonance condition cannot be met, time-homogeneous cavity dynamics can be, in principle, achieved by preparing the atoms in states with a phase dependent on arrival time, e.g., for the state in Eq.~\eqref{eq:psi} by preparing the state with an off-resonant Rabi driving detuned by $-(\Delta_2+\Delta_3)$ [cf.~Eq.~\eqref{eq:H0}] (cf. discussion of \emph{Assumption 2.}).

\subsubsection{Discrete dynamics of micromaser}
 For a pure invariant atom state, $\rho_\text{at}=\ket{\psi_\text{at}}\!\!\bra{\psi_\text{at}}$ [e.g., Eq.~\eqref{eq:psi}], the dynamics in Eq.~\eqref{eq:discrete0} can be expressed with the Kraus operators
\begin{equation}
M_{j} = \bra{j}U(\tau)\ket{\psi_\text{at}} \label{eq:KrausFULL}
\end{equation}
as
\begin{equation}
\label{eq:discrete}
\rho^{(k)}= \sum_{j = 0, ..., 4, {\rm a}} M_{j}\ken \rho^{(k-1)} \ken M_{j}^{\dagger}\equiv\mathcal{M}\left[\rho^{(k-1)}\right],
\end{equation}
where $\mathcal{M}$ denotes the corresponding superoperator. We have $\sum_{j= 0, ..., 4, {\rm a}}\, M_{j}^{\dagger} M_{j}=\mathds{1}$, which guarantees the trace-preserving dynamics $\mathcal{M}^\dagger(\mathds{1})=\mathds{1}$.  The general case of the dynamics with a mixed state instead of the pure state in Eq.~\eqref{eq:psi} is discussed in Appendix~\ref{app:Leff:conserved:atom_mixed}.

\subsubsection{Continuous dynamics of micromaser}
Assuming that time at which atoms arrive to the cavity is exponentially distributed at the rate $\nu$ (see below and Refs.~\cite{davidovich1987two,englert2002elements,guerra1991role}), the \emph{average} dynamics of the cavity, coarse-grained in time over intervals $\tau$, is governed by the time-homogeneous master equation~\cite{Lindblad1976,Gorini1976} 
\begin{eqnarray}
\frac{ d }{ d t} \rho(t)&=&\nu\,\mathcal{M}\left[\rho(t) \right]-\nu \,\rho(t)\equiv\mathcal{L}\left[ \rho(t)\right].
\label{eq:master}
%&&+ \,\kappa\, a\ken\rho(t)\ken a^\dagger   -\frac{\kappa}{2}\,(n \rho(t)+ \rho(t) n). \nonumber
\end{eqnarray}
The dynamics is trace preserving, $\mathcal{L}^\dagger(\mathds{1})=0$, which follows from the properties of the Kraus operators.  \\

%\emph{Exponential arrival times to the cavity}. 
Note that in the micromaser setup, it is assumed that at most one atom is found in the cavity at a time [see Fig.~\ref{fig:model}(b)].  A possible approach used to obtain this is for the levels $\ket{j}$, $j=0,1,...,4,c$ in Fig.~\ref{fig:model}(a) to be a subset of highly excited levels (e.g.,  Rydberg levels) in a multilevel atom~\cite{englert2002elements,guerra1991role}. The initial state of the atoms is then prepared by passing a stream of atoms, initially in a low-energy state, through the excitation region where the states $\ket{j}$, $j=0,1,...,4,c$, can be excited.  If the probability of excitation from the low-energy state is small, due to the law of rare events, the number of atoms that arrive to the cavity prepared in the relevant states $\ket{j}$, $j=0,1,...,4,c$,  up to times $t$ is approximated by a Poisson distribution with the average $\nu t$, while the waiting time between the arrival of the consecutive excited atoms is given by the exponential distribution with the rate $\nu$.
%

%===================================================================================================================================================
%=================================================================================================================================================

		\subsection{Higher order corrections to cavity dynamics}
		\label{app:Kraus}

The cavity dynamics generated by the effective Hamiltonian in Eq.~\eqref{eq:Heff} corresponds to the adiabatic elimination carried out to the lowest nontrivial order in $g_j/\Delta_j$, $j=1,2,3,4$ and $G/\delta$. We now discuss how the effective micromaser dynamics in Eq.~\eqref{eq:discrete2}, which is parity preserving, is modified by higher order corrections to the far-detuned limit. The analysis below is for a general setup of Fig.~\ref{fig:model}(a), with two photon resonance in Eq.~\eqref{eq:resonance} and  the effective Hamiltonian in Eq.~\eqref{eq:Heff0}. Therefore, the results apply both to three-level model~\cite{Brune_1987rydberg_two_phot, davidovich1987two,Ashraf1990_2ph_micromaser_statistics,Orszag1992_squeezed,orszag1993generation,toor1996theory,Alexanian1998_trapped} and (5+1) model in Sec.~\ref{sec:dynamics}, where the Stark shifts can be removed in the far-detuned limit [cf.~Eq.~\eqref{eq:Heff}]. We discuss the influence of the higher order corrections on the latter case in Sec.~\ref{sec:higher-order}.

       \subsubsection{Kraus operators}
        The Kraus operators, which describe the change in the cavity state due to passage of a single atom, are given by [cf.~Eq.~\eqref{eq:KrausFull}] 
        \begin{eqnarray}
        \label{eq:Kraus_corr0}
        M_j&=&\langle j|U(\tau) |\psi_\text{at}\rangle =\langle j|e^{-S} U_{\text{diag}} (\tau) e^{S}|\psi_\text{at}\rangle
        \qquad j=0,...,4,\text{a},\end{eqnarray}
        where %$U(\tau)=\mathcal{T} e^{- i\int_0^\tau dt \,[H_\text{int}(t)+H_0]} $ 
        $U(\tau)= e^{- i\tau [H_\text{int}+H_0]} $, $U_{\text{diag}}(\tau)= e^{- i\tau H_\text{diag}}$,  $e^{-S}$ diagonalizes $H_\text{int}+H_0$ and $\ket{\psi_{\rm at}}$ is the pure state of the atom entering the cavity. We have assumed for simplicity that the field-atom coupling strength is constant, $H_\text{int}(t)=H_\text{int}$. Considering $\ket{\psi_{\rm at}}$ to be given by (\ref{eq:psi}), i.e., a superposition between $|1\rangle$ and $|3\rangle$, the Kraus operators (derived below) $M_0$, $M_2$, and $M_4$ swap the parity, while  the Kraus operators $M_1$, $M_3$, and $M_\text{a}$, conserve the parity [cf.~Eq.~\eqref{eq:Kraus_parity}].

        \subsubsection{Time-independent corrections}
         As in Appendix~\ref{app:adiabatic}, we now consider the expansion of~\eqref{eq:Kraus_corr0} with respect to $S=S_1+S_2+...$, where $j$ in $S_j$ denotes the power of the coupling strength $g,G$ (the time-dependent perturbative corrections in $U_{\text{diag}}(\tau)$  will be discussed later).
        We have
				\begin{eqnarray}
				M_j
				&=&\langle j| U_{\text{diag}} (\tau) |\psi_\text{at}\rangle+ \left(-\langle j| S_1 U_{\text{diag}} (\tau) |\psi_\text{at}\rangle   + \langle j| U_{\text{diag}} (\tau)  S_1 |\psi_\text{at}\rangle\right) \nonumber \\
				&&+ \left[-\langle j| S_1 U_{\text{diag}} (\tau) S_1 |\psi_\text{at}\rangle   + \langle j| \left(\frac{S_1^2}{2}-S_2\right) U_{\text{diag}} (\tau)  |\psi_\text{at}\rangle+ \langle j|  U_{\text{diag}} (\tau)  \left(\frac{S_1^2}{2}+S_2\right) |\psi_\text{at}\rangle\right]+..., 					\label{eq:Kraus_corr1}
				\end{eqnarray}
		where the last two terms in the first line and the second line correspond to the first- and second-order corrections. The operators $S_1$ and $S_2$ are given by Eqs.~\eqref{eq:S1} and~\eqref{eq:S2}, which leads to the parity-conserving Kraus operators given by
		\begin{eqnarray}
		\label{eq:KrausM10}
		M_1&=& 		U_{\text{diag}}^{11}(\tau) c_{\rm g} +U_{\text{diag}}^{13}(\tau) c_{\rm e}+\frac{1}{2}\left(-a a^\dagger\frac{ |g_1|^2 }{\Delta_1^2}- a^\dagger a\frac{ |g_2|^2}{\Delta_2^2}\right) \left[	U_{\text{diag}}^{11}(\tau) c_{\rm g} +U_{\text{diag}}^{13}(\tau) c_{\rm e}		  \right]
			\\\nonumber &&
			+  a^{\dagger 2}\frac{ g_2^* g_3^*}{2\Delta_2\Delta_3}\left[ U_{\text{diag}}^{31}(\tau)\, c_{\rm g}  + U_{\text{diag}}^{33}(\tau)\, c_{\rm e}  \right]
			\\\nonumber &&
		 +\frac{1}{2}U_{\text{diag}}^{11}(\tau) \left[\left(-a a^\dagger\frac{ |g_1|^2 }{\Delta_1^2}- a^\dagger a\frac{ |g_2|^2}{\Delta_2^2}\right) c_{\rm g} + (a^\dagger)^2\frac{ g_2^* g_3^*}{\Delta_2\Delta_3} c_{\rm e}\right] 
		\\\nonumber && +\frac{1}{2} U_{\text{diag}}^{13}(\tau) \left[\left(- a a^\dagger\frac{ |g_3|^2}{\Delta_3^2}  -a^\dagger  a\frac{ |g_4|^2}{\Delta_4^2}-\frac{|G|^2}{\delta^2}\right) c_{\rm e} +a^2\frac{ g_2 g_3}{\Delta_2\Delta_3} c_{\rm g}\right]
		\\\nonumber &&+a \,U_{\text{diag}}^{00}(\tau) \,a^\dagger\frac{ |g_1| ^2}{\Delta_1^2} c_{\rm g}+a^\dagger\frac{ g_2^*}{\Delta_2} \,U_{\text{diag}}^{22}(\tau)\left(\! a\frac{ g_2}{\Delta_2} c_{\rm g} - a^\dagger\frac{ g_3^*}{\Delta_3} c_{\rm e}\!\right) +...,
		\end{eqnarray}
		\begin{eqnarray}
		\label{eq:KrausM30}
		M_3&=& U_{\text{diag}}^{31}(\tau) c_{\rm g} + U_{\text{diag}}^{33}(\tau) c_{\rm e} +\frac{1}{2}\left(- a a^\dagger\frac{ |g_3|^2}{\Delta_3^2}  -a^\dagger  a\frac{ |g_4|^2}{\Delta_4^2}-\frac{|G|^2}{\delta^2}\right) \left[U_{\text{diag}}^{31}(\tau) c_{\rm g} + U_{\text{diag}}^{33}(\tau) c_{\rm e} \right] 
		\\\nonumber &&
		+a^2\frac{ g_2 g_3}{2\Delta_2\Delta_3}\left[ U_{\text{diag}}^{11}(\tau)\, c_{\rm g} + U_{\text{diag}}^{13}(\tau)\, c_{\rm e} \right]
		\\\nonumber && +\frac{1}{2} U_{\text{diag}}^{31}(\tau)  \left[\left(-a a^\dagger\frac{ |g_1|^2 }{\Delta_1^2}- a^\dagger a\frac{ |g_2|^2}{\Delta_2^2}\right) c_{\rm g} + (a^\dagger)^2\frac{ g_2^* g_3^*}{\Delta_2\Delta_3} c_{\rm e}\right]
		\\\nonumber &&+\frac{1}{2}
		U_{\text{diag}}^{33}(\tau) \left[\left(- a a^\dagger\frac{ |g_3|^2}{\Delta_3^2}  -a^\dagger  a\frac{ |g_4|^2}{\Delta_4^2}-\frac{|G|^2}{\delta^2}\right) c_{\rm e} +a^2\frac{ g_2 g_3}{\Delta_2\Delta_3} c_{\rm g}\right]
		\\\nonumber &&-a \frac{ g_3}{\Delta_3}\,U_{\text{diag}}^{22}(\tau)\left(\! a\frac{ g_2}{\Delta_2} c_{\rm g} - a^\dagger\frac{ g_3^*}{\Delta_3} c_{\rm e}\!\right)  +a^\dagger U_{\text{diag}}^{44}(\tau) \,a\frac{ |g_4|^2}{\Delta_4^2}c_{\rm e}+  U_{\text{diag}}^{\text{aa}} (\tau)\,\frac{ |G|^2}{\delta^2}  c_{\rm e} +...,
		\end{eqnarray}
		and
		\begin{eqnarray}
		M_\text{a}&=&  - \frac{ G}{\delta}  \left[ U_{\text{diag}}^{31}(\tau)\, c_{\rm g}  + U_{\text{diag}}^{33}(\tau)\, c_{\rm e}   -    U_{\text{diag}}^{\text{aa}}(\tau)\,c_{\rm e} \right]+ ...,
		\end{eqnarray}
		where $U_{\text{diag}}^{jk}(\tau)\equiv\langle j|U_{\text{diag}} (\tau)|k\rangle$ for $j,k=0,...,4,\text{a}$ [cf.~Eq.~\eqref{eq:Kraus}]. Note that  $M_1$ and $M_3$ do not feature first-order corrections [the second and third terms in~\eqref{eq:Kraus_corr1}], due to their parity conservation, as $S_1$ swaps the cavity parity, except for the atom in levels $|3\rangle$ and $|\text{a}\rangle$, so that $M_\text{a}$ is of the first order. For this reason, the parity-swapping Kraus operators are of the first order,
		\begin{eqnarray}
		M_0&=&  a^\dagger\frac{ g_1^* }{\Delta_1} \left[ U_{\text{diag}}^{11}(\tau)\, c_{\rm g} + U_{\text{diag}}^{13}(\tau)\, c_{\rm e} \right]- U_{\text{diag}}^{00}(\tau) \,a^\dagger\frac{ g_1^* }{\Delta_1} c_{\rm g}+   ...,\\
		M_2&=& - a\frac{ g_2}{\Delta_2}  \left[ U_{\text{diag}}^{11}(\tau)\, c_{\rm g} + U_{\text{diag}}^{13}(\tau)\, c_{\rm e}\right]
		+ a^\dagger\frac{ g_3^*}{\Delta_3} \left[ U_{\text{diag}}^{31}(\tau)\, c_{\rm g}  + U_{\text{diag}}^{33}(\tau)\, c_{\rm e}  \right] +U_{\text{diag}}^{22}(\tau) \left(\! a\frac{ g_2}{\Delta_2} c_{\rm g} - a^\dagger\frac{ g_3^*}{\Delta_3} c_{\rm e}\!\right) +...,\qquad\,\,\,\,\\
		M_4&=& - a\frac{ g_4}{\Delta_4} \left[  U_{\text{diag}}^{31}(\tau)\, c_{\rm g} + U_{\text{diag}}^{33}(\tau)\, c_{\rm e}   \right]  + U_{\text{diag}}^{44}(\tau) \,a\frac{ g_4}{\Delta_4}c_{\rm e} +... \,.
		\end{eqnarray}

		\subsubsection{Time-dependent corrections}
		 We now discuss time-dependent corrections to $U_{\text{diag}}(\tau)= e^{- i\tau H_\text{diag}}$ from the diagonal Hamiltonian $H_\text{diag}=H_0+H_2^\text{diag}+H_4^\text{diag}+...$,~\eqref{eq:Hdiag}, where $H_k^\text{diag}$ denotes $k$th order corrections. As $H_0$ commutes by definition with $H_\text{diag}$, we have
		 \begin{equation}
		 U_{\text{diag}} (\tau)=e^{- i \tau H_0} e^{- i \tau \left(H_2^\text{diag}+H_4^\text{diag}+...\right)} = e^{- i \tau (H_0+H_2^\text{diag})}\left(1- i \int_0^{\tau} d t\, e^{ it H_2^\text{diag}} H_4^\text{diag} e^{- i t H_2^\text{diag}}+...\right), \label{eq:Dyson}
		 \end{equation} 
		 where in the last equality we used the Dyson series. The correction 
		 \begin{equation}\label{eq:Heff_delta}
		 \int_0^{\tau} d t\, e^{ it H_2^\text{diag}} H_4^\text{diag} e^{- i t H_2^\text{diag}}\equiv\tau \, \delta H_{\text{eff}}(\tau)
		 \end{equation}
		  can be considered as the contribution from the time-averaged $H_4^\text{diag}$ in the rotating frame of $H_2^\text{diag}$. For the interaction time $\tau$ chosen so that the second-order dynamics in the two-level approximation [Eqs.~\eqref{eq:Heff} and~\eqref{eq:Kraus}] is finite, the correction  $\tau\, \delta H_{\text{eff}}(\tau)$ contributes as the second-order to $U_{\text{diag}} (\tau)$. We thus have [cf.~Eq.~\eqref{eq:Dyson} and~\eqref{eq:Kraus}] 
		 \begin{subequations}
		 \begin{align}
		 &U_{\text{diag}}^{11}(\tau)\, c_{\rm g} + U_{\text{diag}}^{13}(\tau)\, c_{\rm e} \equiv e^{- i\tau\Delta_1} \left[ M_{\rm g} +  \delta M_g \right]+...\\\nonumber
		 &\qquad= e^{- i\tau\Delta_1} \,\langle 1|e^{- i t H_2^\text{diag}}  |\psi_\text{at}\rangle-  i\tau e^{- i\tau\Delta_1} \left[ \cos (\phi \sqrt{a^{\dagger 2} a^2})\, \langle 1|\delta H_\text{eff}(\tau)|\psi_\text{at}\rangle -i a^{\dagger 2} \frac{\sin \left(\phi \sqrt{ a^2 a^{\dagger 2}}\right)}{\sqrt{ a^2 a^{\dagger 2}}}\, \langle 3|\delta H_\text{eff}(\tau)|\psi_\text{at}\rangle\right]+...,\\
		 &U_{\text{diag}}^{31}(\tau)\, c_{\rm g}  + U_{\text{diag}}^{33}(\tau)\, c_{\rm e} \equiv e^{- i\tau\Delta_1} \left[ M_{\rm e} +  \delta M_e \right]+...\\\nonumber &\qquad=e^{- i\tau\Delta_1} \,  \langle 3|e^{- i t H_2^\text{diag}}  |\psi_\text{at}\rangle-  i\tau e^{- i\tau\Delta_1} \left[ -i a^{ 2} \frac{\sin \left(\phi \sqrt{ a^{\dagger 2} a^2 }\right)}{\sqrt{ a^{\dagger 2}a^2 }}\,   \langle 1|\delta H_\text{eff}(\tau) |\psi_\text{at}\rangle + \cos (\phi \sqrt{a^{\dagger 2} a^2})\,\langle 3|\delta H_\text{eff}(\tau) |\psi_\text{at}\rangle \right]+...,
		 \end{align}
		 \end{subequations} 
		 where we defined the zeroth-order Kraus operators $M_g$ and $M_e$,~cf.~Eq.~\eqref{eq:Kraus} [note that the Kraus operators in~\eqref{eq:Kraus} differ by the global phase $e^{ i\tau \frac{|g_2|^2}{\Delta} }$, which was additionally neglected in~\eqref{eq:Heff}]. For the Kraus operators in  other micromaser setups, including the three-level model (see  Appendix~\ref{app:2photon}).

		Therefore, up to the second order in the coupling strength, the cavity dynamics~\eqref{eq:master} is determined by the first-order Kraus operators,
	\begin{eqnarray}
	\label{eq:KrausM0time}
	e^{ i\tau\Delta_1} 	M_0&=&  a^\dagger\frac{ g_1^* }{\Delta_1} \,M_{\rm g} - e^{ i\tau \left(\Delta_1+ \frac{|g_1|}{\Delta_1} a^\dagger a \right)}  \,a^\dagger\frac{ g_1^* }{\Delta_1} c_{\rm g}+   ...,
	\\
	\label{eq:KrausM2time}
	e^{ i\tau\Delta_1} 	M_2&=& - a\frac{ g_2}{\Delta_2}  \, M_{\rm g} 
	+ a^\dagger\frac{ g_3^*}{\Delta_3} \, M_{\rm e}  +\,e^{- i\tau\left(\Delta_2 + a\, a^\dagger  \frac{|g_2|^2  }{\Delta_2 }-a^\dagger a\frac{|g_3|^2  }{ \Delta_3} \right)} \left(\! a\frac{ g_2}{\Delta_2} c_{\rm g} - a^\dagger\frac{ g_3^*}{\Delta_3} c_{\rm e}\!\right) +...,\qquad\,\,\,\,
	\\
	\label{eq:KrausM4time}
	e^{ i\tau\Delta_1} 	M_4&=& - a\frac{ g_4}{\Delta_4} \,M_{\rm e}  +  e^{- i\tau\left (\sum_{k=2}^4\Delta_k +  a\,a^\dagger \frac{|g_4|^2 }{\Delta_4}\right)}  \,a\frac{ g_4}{\Delta_4}c_{\rm e} +... \,,
	\\
	\label{eq:KrausMatime}
	e^{ i\tau\Delta_1} 	M_\text{a}&=&  - \frac{ G}{\delta}  \left[  M_{\rm e}   -     e^{- i\tau\left (\sum_{k=2}^3\Delta_k + \delta +\frac{|G|^2}{\delta }\right)}\,c_{\rm e} \right]+ ..., 
	\end{eqnarray}
	where fourth-order corrections to $H_\text{diag}$ in~\eqref{eq:Hdiag} are neglected. Similarly,
			\begin{eqnarray}
			\label{eq:KrausM1time}
			e^{ i\tau\Delta_1} M_1&=& 	  M_{\rm g} +\frac{1}{2}\left(-a a^\dagger\frac{ |g_1|^2 }{\Delta_1^2}- a^\dagger a\frac{ |g_2|^2}{\Delta_2^2}\right) \, M_{\rm g} +  a^{\dagger 2}\frac{ g_2^* g_3^*}{2\Delta_2\Delta_3}\,M_{\rm e} 
			\\\nonumber &&
			+\frac{1}{2}M_{gg}  \left[\left(-a a^\dagger\frac{ |g_1|^2 }{\Delta_1^2}- a^\dagger a\frac{ |g_2|^2}{\Delta_2^2}\right) c_{\rm g} + (a^\dagger)^2\frac{ g_2^* g_3^*}{\Delta_2\Delta_3} c_{\rm e}\right] 
			\\\nonumber && +\frac{1}{2}M_{ge}  \left[\left(- a a^\dagger\frac{ |g_3|^2}{\Delta_3^2}  -a^\dagger  a\frac{ |g_4|^2}{\Delta_4^2}-\frac{|G|^2}{\delta^2}\right) c_{\rm e} +a^2\frac{ g_2 g_3}{\Delta_2\Delta_3} c_{\rm g}\right]
			\\\nonumber &&+a \,e^{ i\tau \left(\Delta_1+ \frac{|g_1|^2}{\Delta_1} a^\dagger a\right) } \,a^\dagger\frac{ |g_1| ^2}{\Delta_1^2} c_{\rm g}+a^\dagger\frac{ g_2^*}{\Delta_2} \,e^{- i\tau\left(\Delta_2 + a\, a^\dagger  \frac{|g_2|^2  }{\Delta_2 }-a^\dagger a\frac{|g_3|^2  }{ \Delta_3} \right)} \left(\! a\frac{ g_2}{\Delta_2} c_{\rm g} - a^\dagger\frac{ g_3^*}{\Delta_3} c_{\rm e}\!\right) 
			\\\nonumber &&-  i\tau \langle 1|\delta H_\text{eff}(\tau) \ket{\psi_{\rm at}}
			+...,\\
			\label{eq:KrausM3time}
			e^{ i\tau\Delta_1} M_3&=&  M_{\rm e}  +\frac{1}{2}\left(- a a^\dagger\frac{ |g_3|^2}{\Delta_3^2}  -a^\dagger  a\frac{ |g_4|^2}{\Delta_4^2}-\frac{|G|^2}{\delta^2}\right)   M_{\rm e} 
			+a^2\frac{ g_2 g_3}{2\Delta_2\Delta_3} \, M_{\rm g}
			\\\nonumber && +\frac{1}{2} M_{eg}  \left[\left(-a a^\dagger\frac{ |g_1|^2 }{\Delta_1^2}- a^\dagger a\frac{ |g_2|^2}{\Delta_2^2}\right) c_{\rm g} + (a^\dagger)^2\frac{ g_2^* g_3^*}{\Delta_2\Delta_3} c_{\rm e}\right]
			\\\nonumber &&+\frac{1}{2} M_{ee} \left[\left(- a a^\dagger\frac{ |g_3|^2}{\Delta_3^2}  -a^\dagger  a\frac{ |g_4|^2}{\Delta_4^2}-\frac{|G|^2}{\delta^2}\right) c_{\rm e} +a^2\frac{ g_2 g_3}{\Delta_2\Delta_3} c_{\rm g}\right]
			\\\nonumber &&-a \frac{ g_3}{\Delta_3}\,e^{- i\tau\left(\Delta_2 + a\, a^\dagger  \frac{|g_2|^2  }{\Delta_2 }-a^\dagger a\frac{|g_3|^2  }{ \Delta_3} \right)}\left(\! a\frac{ g_2}{\Delta_2} c_{\rm g} - a^\dagger\frac{ g_3^*}{\Delta_3} c_{\rm e}\!\right)  +a^\dagger \, e^{- i\tau\left (\sum_{k=2}^4\Delta_k +  a\,a^\dagger \frac{|g_4|^2 }{\Delta_4}\right)}  \,a\frac{ |g_4|^2}{\Delta_4^2}c_{\rm e}
			\\\nonumber &&+  e^{- i\tau\left (\sum_{k=2}^3\Delta_k + \delta +\frac{|G|^2}{\delta }\right)}\,\frac{ |G|^2}{\delta^2}  c_{\rm e} 
			-  i\tau \langle 3|\delta H_\text{eff}(\tau) |\psi_\text{at}\rangle+...\,,
			\end{eqnarray}
			and we defined $M_{\mu\nu}\equiv M_\mu$ with $c_\nu=1$, where $\mu,\nu=g,e$. The global phase factor  $e^{ i\tau\Delta_1}$ in Eqs.~(\ref{eq:KrausM0time})-(\ref{eq:KrausM3time}) corresponds to a global phase neglected in~\eqref{eq:Heff0}. Furthermore, for the (5+1)-model, the conditions in Eq.~\eqref{eq:conditions} leading to cancellation of the Stark shifts, establish dependent variables: ${ g_1^* }/{\Delta_1}={ g_2 }/\sqrt{\Delta_1 \Delta}$,  $-{ g_4/}{\Delta_4}={ g_3 }/{\sqrt{\Delta_4 \Delta}}$, and  $- { G}/{\delta}=\sqrt{|g_2|^2 +|g_3|^2}/\sqrt{\Delta\delta}$.
			
	%	Although, as we discuss in Sec.~\ref{sec:higher-order}, the contribution of higher order corrections to~\eqref{eq:Hdiag} does not affect the long-time dynamics in the second-order, 
	For completeness, we now provide fourth-order corrections to~\eqref{eq:Hdiag}, which contribute to Eq.~\eqref{eq:Heff_delta},  for the case $g_1=g_2=g_3=g_4=g$, $\Delta_1=\Delta_2=-\Delta_3=-\Delta_4=\Delta$, and $G^2/\delta = -2 g^2/\Delta$,
	\begin{equation}
	H_4^\text{diag}=\left[
	\begin{array}{cc}
	\frac{4 g^4 \left[\,a^\dagger a\left(a^\dagger a-3\right)-1\right]}{3 \Delta ^3} &  a^\dagger \frac{8 g^4 \left(g^2+G^2 a^\dagger a\right)}{3 G^2 \Delta ^3} \,a^\dagger \\
	a\,\frac{8 g^4 \left(g^2+G^2 a^\dagger a \right)}{3 G^2 \Delta ^3} \,a&  -\frac{4 g^4 \left[4 g^2-G^2 \left( a^\dagger a a a^\dagger  +1\right)\right]}{3 G^2 \Delta ^3} \\
	\end{array}
	\right],
	\label{eq:Hdiag4}  
	\end{equation}
    which are expressed for $e^S(|1\rangle\otimes|n\rangle)$ and $e^S(|3\rangle\otimes|n\rangle)$, i.e., the diagonal basis of the atom-cavity Hamiltonian [cf.~Eq.~\eqref{eq:expansion}]. Here (\ref{eq:Hdiag4}) was obtained from (\ref{eq:expansion}) by considering the expansion of $S$ up to the fourth order, i.e., $S =S_1+S_2+S_3+S_4+... $.

	\subsubsection{Higher order corrections in the three-level model}
	 In the three-level model [see Eq.~\eqref{eq:Heff} with $g_1=0, g_4=0, G=0$] at resonance~\eqref{eq:resonance}, the stationary state is known to be \emph{pure} for all detunings and given by  the squeezed vacuum~\cite{orszag1993generation,Orszag1992_squeezed}. We will now recover this result by showing that this state is not affected by the parity-swapping Kraus operator $M_2$ [cf.~Eq.~\eqref{eq:KrausFull}]. Indeed, beyond adiabatic limit we have [cf.~Eq.~\eqref{eq:KrausM2time}]
	\begin{align}
	\label{eq:KrausM2_3level}
	M_2 = - a\frac{ g_2}{\Delta}  \, M_{\rm g} - a^\dagger\frac{ g_3^*}{\Delta} \, M_{\rm e} 
	  +\,e^{- i\tau\left(\Delta + a\, a^\dagger  \frac{|g_2|^2  }{\Delta }+a^\dagger a\frac{|g_3|^2  }{ \Delta} \right)} \left(\! a\frac{ g_2}{\Delta} c_{\rm g} + a^\dagger\frac{ g_3^*}{\Delta} c_{\rm e}\!\right) +...
	\end{align}
	where $M_{\rm g}$ and $M_{\rm e}$ correspond to three-level dynamics. $M_2$ operator, however, is $0$ in the first order on the squeezed vacuum state $|\Psi_+\rangle$, as 
	\begin{eqnarray}
	&& \left(- a\frac{ g_2}{\Delta}  \, M_{\rm g} - a^\dagger\frac{ g_3^*}{\Delta} \, M_{\rm e} \right)|\Psi_+\rangle  = -\left( a\frac{ g_2}{\Delta}  \, c_{\rm g} + a^\dagger\frac{ g_3^*}{\Delta} \, c_{\rm e} \right)|\Psi_+\rangle \nonumber \\
	&& = \sum_{n=0}^\infty \left( \sqrt{2n+2}\frac{ g_2}{\Delta}  \, c_{\rm g} \,c_{2n+2} + \sqrt{2n+1}\frac{ g_3^*}{\Delta} \, c_{\rm e}\,  c_{2n}\right)|2n+1\rangle  = 0,
	\end{eqnarray}
	where in the first equality we used that in three-level model we have $M_{\rm e}|\psi_\pm\rangle=c_{\rm e}$ and $M_{\rm g}|\psi_\pm\rangle=c_{\rm g}$ (up to a global phase) (see Appendix~\ref{app:2photon}). The last equality follows from the recurrence relation for the pure stationary states (cf.~Appendix~\ref{app:2photon})
	\begin{equation}
	\frac{c_{n+2}}{c_{n}}=-\frac{c_{\rm e}}{c_{\rm g}} \frac{g_3^*}{g_2} \frac{\sqrt{n+1}}{\sqrt{n+2}}.
	\end{equation}
	It is worthwhile to emphasize that for the state of the negative parity (odd $n$), the parity-swapping Kraus operator $M_2$ does not vanish on its one-photon component, thus leading to its decay and a unique stationary state of the dynamics given by the squeezed vacuum~\cite{orszag1993generation} (see also Appendix~\ref{app:Leff:weak}).

%===================================================================================================================================================

%===================================================================================================================================================

	\section{Pure stationary states of two-photon micromasers}
	\label{app:2photon}

	In Appendix~\ref{app:adiabatic}, we derived the effective two-photon Hamiltonian, Eq.~\eqref{eq:Heff0}, describing the far-detuned limit of the cavity interaction with a multilevel atom in the ladder configuration [see Fig.~\ref{fig:model}(a)]. Here we discuss pure stationary states of \emph{general two-photon dynamics}, with a Hamiltonian of the same functional form as (\ref{eq:Heff0}) but with arbitrary Stark shifts and two-photon couplings. 	We show that beyond the stationary states in Eq.~\eqref{eq:psi_ss}, the only pure states correspond to the stationary states of the three-level model~\cite{Orszag1992_squeezed,orszag1993generation,Alexanian1998_trapped}.% We also discuss conditions on generation of trapping states~\cite{orszag1993generation}.
	~\\

		\subsection{Effective Hamiltonian}
 	Within RWA, i.e., for dynamics based single-photon Jaynes-Cummings interactions, the adiabatic limit of far-detuned levels with a two-photon resonance [Eq.~\eqref{eq:resonance}] leads in the second-order to the effective Hamiltonian
	\begin{equation}  \label{eq:HeffGEN}
	H_{\rm eff} = 
	\begin{bmatrix}
	A  a^{\dagger}a + B\, \mathds{1}  && C^* a^{\dagger 2} \\
	C\,a^{2} && D\, a^{\dagger}a + E\, \mathds{1}
	\end{bmatrix},
	\end{equation}
where $A, B, D, E \in\mathbb{R}$  and $C\in\mathbb{C}$ and the basis is given by the resonant levels $\ket{1}$, $\ket{3}$ [cf.~Appendix~\ref{app:adiabatic} and~Eq.~\eqref{eq:Heff0}]. The constants $A$, $B$,  $D$, and $E$ describe the Stark shifts, while $C$ determines the effective two-photon coupling strength. 
	
	\subsection{Pure stationary states} We are interested in the case when the two Kraus operators corresponding to the Hamiltonian~\eqref{eq:HeffGEN} feature the same cavity state $|\Psi_\text{ss}\rangle=\sum_{n=0}^\infty c_n|n\rangle$ as an eigenvector. This corresponds to the following set of equations [cf.~Eqs.~\eqref{eq:Kraus} and~\eqref{eq:pure_eq}]
	\begin{subequations}
	\label{eq:pure_eqGEN}
	\begin{align}
	\alpha\, c_{n+2} &= e^{- i \varphi_n }\left\{ c_{\rm g} \left[ \cos(\phi_n) -  i \tau s_n^z \frac{\sin(\phi_n)}{\phi_n}  \right] \, c_{n+2} - i c_{\rm e}  \tau (s_n^x)^* \frac{\sin(\phi_n)}{\phi_n} \, c_n\right\}, \\
	\beta\, c_n &= e^{- i \varphi_n }\left\{- i c_{\rm g}  \tau s_n^x \frac{\sin(\phi_n)}{\phi_n} \,c_{n+2}+c_{\rm e} \left[ \cos(\phi_n) +  i \tau s_n^z \frac{\sin(\phi_n)}{\phi_n}  \right]\,c_n\right\}, 
	\end{align}
\end{subequations}
	where 
	\begin{eqnarray}
	s_n^z=\frac{A(n+2)+B-D n-E}{2}, &\qquad& s_n^x= C \sqrt{ (n+1)(n+2)}, \label{eq:s_zx} \\
	\phi_n= \tau \sqrt{ (s_n^z)^2 + |s_n^x|^2  },&\qquad& \varphi_n= \tau\frac{(A+D)n+2A+B+E}{2}.\label{eq:phi_n} 
	\end{eqnarray}
 Equations~\eqref{eq:pure_eqGEN} feature a nontrivial solution when the corresponding determinant is $0$ independently of $n$,
		\begin{equation}\label{eq:det_GEN}
	\alpha  \beta + e^{ - i 2\varphi_n } c_{\rm e} c_{\rm g} -e^{ - i \varphi_n } \cos (\phi_n )  (\alpha  c_{\rm e}+\beta  c_{\rm g}) -  i e^{ - i\varphi_n } \tau s_n^z \frac{\sin (\phi_n )}{\phi_n}  (\alpha  c_{\rm e}-\beta  c_{\rm g})=0,
		\end{equation}
		where on the left-hand size we used the fact $ \tau^2[ (s_n^z)^2 + |s_n^x|^2] =	\phi_n^2$.  
		
		Note that in the absence of coupling, $C=0$, we obtain that $s_n^x\equiv 0$, and dynamics corresponds to the dephasing of coherences, which is caused by the Stark shifts in~\eqref{eq:HeffGEN}. This leads to a stationary state of the cavity given by the diagonal of an initial state (a classical state without coherences), unless both  $\phi_n$ and $\varphi_n$ are independent of $n$ (this takes place  when $A=0=D$, in which case the Stark shift is independent from the cavity field, and instead of dephasing the passage of atoms only changes the global phase). 
		
		For the case of $C\neq 0$, the last term in  Eq.~\mbox{\eqref{eq:phi_n}} with $ s_n^z \sin (\phi_n )/\phi_n$, is an \emph{independent} function of $n$, from both $\cos (\phi_n )$ and $e^{- i \varphi_n }$, $e^{- i 2\varphi_n }$; i.e., it cannot be canceled by the other terms for all $n$. Therefore, for Eq.~\mbox{\eqref{eq:phi_n}} to hold, it is necessary for the last term to vanish for all $n$, which takes place when $s_n^z=0$ or $\alpha  c_{\rm e}-\beta  c_{\rm g}=0$, which define \emph{two complementary cases} we now discuss.

			\subsubsection{Case A} 
			 Lets first consider  $s_n^z=0$, which from~\eqref{eq:s_zx} yields the effective Hamiltonian coefficients as
			\begin{equation}
			A=D\qquad \text{and} \qquad B=E+2A,
			\end{equation}
			As $C\neq 0$, $\phi_n$ depends on $n$, and furthermore $\cos (\phi_n )$ is an independent function from $e^{-  i \varphi_n }$ and $e^{-2 i \varphi_n }$. Therefore, it is required that  $\alpha  c_{\rm e}+\beta  c_{\rm g}=0$, so that the outgoing state of atoms are given by [cf.~Eq.~\eqref{eq:atom_out}]
			\begin{equation} \label{eq:pure_eq_GEN1}
			\alpha= e^{- i\varphi} c_{\rm g} ,\qquad\beta  =- e^{- i\varphi}c_{\rm e} ,\qquad
			\end{equation}
			This in turn simplifies the first two terms in~\eqref{eq:phi_n} as $\alpha  \beta +c_{\rm e} c_{\rm g} e^{-2 i \varphi_n }=  -2  i e^{- i (\varphi_n+\varphi)} c_{\rm e} c_{\rm g}  \sin(\varphi_n-\varphi)$, which thus requires $\varphi_n=\varphi+ k\pi $, where $k\in\mathbb{Z}$, so that 
			\begin{equation}
			A=-D=0\qquad \text{and} \qquad \varphi=2B \tau+ k\pi,
			\end{equation}
		    and there are no Stark shifts (except the global phase $\varphi$): $A=D=0$, $B=E$. This is exactly the case discussed at length  in  this work, which leads to the stationary states given by the recurrence relation~\eqref{eq:psi_ss} [by choosing $k=0,1$ in $\varphi$].

		    \subsubsection{Case B} 
		     In order to remove the amplitude of the last term in Eq.~\eqref{eq:phi_n}, we now consider  $\alpha  c_{\rm e}-\beta  c_{\rm g}=0$, which determines the outgoing state of atoms as [cf.~Eq.~\eqref{eq:pure_eq_GEN1}]
		    \begin{equation} \label{eq:pure_eq_GEN2}
		    \alpha= e^{- i\varphi} c_{\rm g} ,\qquad\beta  =e^{- i\varphi} c_{\rm e} .
		    \end{equation}
		    In this case, we have for the remaining terms
		    \begin{eqnarray}
		    \alpha  \beta +e^{-2 i \varphi_n }c_{\rm e} c_{\rm g}   -e^{- i \varphi_n }\cos (\phi_n )  (\alpha  c_{\rm e}+\beta  c_{\rm g}) 
		    &=& e^{- i(\varphi_n+\varphi)} c_{\rm e} c_{\rm g}  \left[  \cos (\varphi_n - \varphi) -\cos (\phi_n )  \right] \\\nonumber
		    &=& -2 e^{- i(\varphi_n+\varphi)} c_{\rm e} c_{\rm g}  \sin \left( \frac{\varphi_n+\phi_n - \varphi}{2} \right)\sin \left( \frac{\varphi_n-\phi_n - \varphi}{2}   \right).
		    \end{eqnarray}
		    Therefore, we require $\varphi_n - \varphi + 2k \pi =\phi_n $ or $\varphi_n - \varphi + 2k \pi =-\phi_n $ with $k\in\mathbb{Z}$, which expressions squared (and divided by $\tau^2$) yield the condition
		    \begin{equation} 
		   n^2 \left(A D-|C|^2\right)+ n \left[2 A D-3 |C|^2 +A E+DB -(A+D) \omega\right]+(2 A +B) E-2 |C|^2-(2 A +B+E)\omega+\omega^2=0,
		    \end{equation}
		    where $\omega=(\varphi-2 k\pi)/\tau$. Requiring that the above expression holds for all $n$, we arrive at the following conditions on the effective Hamiltonian coefficients,
		    \begin{equation}
		   |C|^2=AD >0,\qquad
		   \omega=\frac{A(-D +E)+DB} {A+D},\qquad (B + D - E) (A + B + 2 D - E)=0
		    \end{equation}
		    where $A+D\neq 0$ follows from $A\neq -D$ as $AD >0$. We note that there are two solutions (from the last condition) with $B= -D+E$ ($\varphi=\tau B+2 k\pi$) and $B+A+D= -D + E$ [$\varphi=\tau(A+B)+2 k\pi$], but yield the same stationary state given by the recurrence relation [cf.~Eq.~\eqref{eq:psi_ss}]
		    \begin{equation} \label{eq:psi_ssGEN2}
		    \frac{c_{n+2}}{c_n}=-\frac{c_{\rm e}}{c_{\rm g}} \frac{C^*}{A}\sqrt{\frac{n+1}{n+2}}.
		    \end{equation}
		   In the even-parity subspace, this is a squeezed vacuum state, whose squeezing can be regulated by the ratio of the dynamical shifts $\frac{|C|}{A}=\sqrt{\frac{A}{D}}$. 
		    In particular, the micromaser with three-level atoms~\cite{Orszag1992_squeezed,orszag1993generation,Alexanian1998_trapped} corresponds to the former solution with  $A=-\frac{|g_2|^2}{\Delta}$, $B=0$,  $C=-\frac{g_2g_3}{\Delta}$, and $D=E=-\frac{|g_3|^2}{\Delta}$ [cf.~Eq.~\eqref{eq:Heff_3level}]; here the squeezing is regulated by the ratio $|g_3/g_2|$.

\begin{comment}		    
~\\ \emph{Trapping states}~\cite{orszag1993generation}. For completeness we also discuss possibility of obtaining cavity states of a fixed photon number in the two-photon dynamics generated by Eq.~\eqref{eq:HeffGEN} [for the case of the effective Hamiltonian in Eq.~\eqref{eq:Heff} see Sec.~\ref{sec:walls_hard}]. A cavity state of a fixed photon number can be prepared deterministically when two subsequent number states of the same parity are not connected by the dynamics [cf.~Eq.~\eqref{eq:pure_eqGEN}]  
\begin{equation} \label{eq:TRAP}
 (s_n^x)^* \frac{\sin(\phi_n)}{\phi_n}=0
\end{equation}
(for the states $|n\rangle$ and $|n+2\rangle$), and the atom is prepared in either the ground state ($|c_g|=1$ for $|n+2\rangle$) or the excited state ($|c_e|=1$ for $|n+2\rangle$) (see Sec.~\ref{sec:walls_hard}). The condition~\eqref{eq:TRAP} is fulfilled in two cases. In the first case $s_n^x=0$. This however requires $C=0$ [cf.~Eq.~\eqref{eq:s_zx}] when the micromaser dynamics corresponds to the dephasing, and a stationary state is in general a mixture of photon number states. In the second case $\phi_n=k\pi$ with $k=\pm 1, \pm 2,... $. [cf.~Eq.~\eqref{eq:phi_n}].
 \end{comment}

	%===================================================================================================================================================

	\section{Hard walls and Pell equation}
	\label{app:hard_walls}

	In Sec.~\ref{sec:walls_hard} of the main text, we have discussed hard walls in the cavity dynamics, i.e., when the integrated coupling strength $\phi$ leads to $\sin_m(\phi)=0$ for certain $m$, so that the cavity states $|m\rangle$ and $|m+2\rangle$ are no longer coupled. Here we show that the condition in Eq.~(\ref{eq:phi_wall}) corresponds for the subsequent walls to Pell equation~\cite{Barbeau, Lenstra08solvingthe}, and derive the recurrence relation for positions of these hard walls.

	 \subsection{Pell equation}
	  For a given integrated coupling strength $\phi$, let us assume that  $m$ is the position of the \emph{first} wall with the corresponding $K$. Any other wall at $m' > m$ must fulfill, from~(\ref{eq:phi_wall}),  
	\begin{equation}
	(m'+1)(m'+2) = \left(\frac{K'}{K}\right)^2 (m+1)(m+2).
	\end{equation}
	for a certain integer $K'$. By setting $D := (m+1)(m+2)$, $x := 2m'+3$ and $y := 2 K'/K$, we get the Pell equation~\cite{Barbeau, Lenstra08solvingthe}
	\begin{equation}
	\label{eq:pell1}
	x^2 - D y^2 = 1.
	\end{equation}	
	We assume $\phi>0$ and thus $K>0$ [cf.~Eq.~\eqref{eq:phi_wall}] (otherwise we equivalently consider positive integers $-K$ and $-K'$). Since $D$ is not a perfect square, Eq.~\eqref{eq:pell1} has infinitely many positive integer solutions $(x_n,\, y_n)$, $n \geq 1$. If the solutions are ordered by the magnitude of $x_n$, the $n$th solution is given by the recurrence relation \cite{Copley1959}
	\begin{subequations}
		\label{eq:pell_recurrence}
		\begin{align}
		x_{n} &= x_1 x_{n-1}  + D y_1 y_{n-1}, \label{eq:pell_recurrence_a}\\
		y_{n} &= x_1 y_{n-1} + y_1 x_{n-1}, \label{eq:pell_recurrence_b}
		\end{align}
	\end{subequations}
    or equivalently
    	\begin{equation}
    	\label{eq:pell_recurrence2}
    	x_{n} +\sqrt{D} y_{n}= \left(x_1 + \sqrt{D} y_1 \right)^n, 
    \end{equation}
	where $(x_1,\, y_1) = (2m+3,\, 2)$ is the first nonzero integer solution, called the fundamental solution.

	\subsection{Recurrence relation for hard walls}
	 From Eq.~\eqref{eq:pell_recurrence}, we note that, since $x_1$ is odd, $x_n$ is always odd, while $y_n$ is always even as $y_1$ is even [this is a consequence of $D$ being even; cf.~Eq.~\eqref{eq:pell1}]. Therefore, each solution with $x_n$ and $y_n$, corresponds directly to a hard wall in the dynamics at $m_n= (x_n-3)/2$, and with $K_n=y_n K/2$ being a multiple of $K$. Furthermore, Eq.~\eqref{eq:pell_recurrence} yields the recurrence relation
		\begin{subequations}\label{eq:mK0}
		\begin{align}
	m_n&=m_{n-1}(2m+3)+3(m+1)+2(m+1)(m+2) \,K_{n-1}/K,\label{eq:mK0a}\\
	K_n&=K_{n-1}(2m+3) +K(2 m_{n-1}+3),	\label{eq:mK0b}
	\end{align}
\end{subequations}
    and we conclude there are infinitely many hard walls in the dynamics. From Eq.~\eqref{eq:mK0} we have that for the first hard wall at even $m_1$, the parity of the $n$th wall, $m_n$, oscillates with period $2$, while for odd $m_1$, all walls are found at $m_n$ odd. Similarly, for even $K_1$, $K_n$ is always even and thus $\cos_{m_n}(\phi)=(-1)^{K_n}=1$, while for odd $K_1$, the $K_n$ parity oscillates with period $2$, and so does $\cos_{m_n}(\phi)$. These results are summarized in Table~\ref{tab:mK}.

    We note, however, that we are also interested in solutions of~\eqref{eq:pell1}, in which $x$ is an (odd) integer, while $y$ is a rational number, i.e., when $2K'$ is not a multiple of $K$. As we show below, however, the position of walls fulfils the recurrence relation~\eqref{eq:mK0} and $K$ is always a multiplicity of $K$.  \\

    \emph{Proof}. Suppose that there exists a hard wall at $m'$ with $2 K'$ not divisible by $K$, $\overline{K} := \gcd(K,\,2K')<K$. We  have
 \begin{equation}\label{eq:Dtilde}
 (2m'+3)^2-1 = D\left(\frac{\overline{K}}{K}\right)^2  \left(\frac{2K'}{\overline{K}} \right)^2.
 \end{equation}
  Since the greatest common factor of the integers $2K'/\overline{K}$ and  $\widetilde{K}:=K/\overline{K}$ is $1$ by definition, it follows, from the left-hand side of~\eqref{eq:Dtilde} being an integer,  that $D$ must be divisible by $\widetilde{K}^2$. Therefore, $\widetilde{D}:=  D/\widetilde{K}^2<D $ is an integer, and since $D$ was not a square of integer, neither is $\widetilde{D}$. We thus arrive at a new Pell equation
  \begin{equation}
  \label{eq:pell2}
  \widetilde{x}^2 - \widetilde{D} \widetilde{y}^2 = 1,
  \end{equation}	
  where the new integer variable $\widetilde{y} := \widetilde{K} y $, while $\widetilde{x}:=x$ as before. %Note that the parent equation~\eqref{eq:pell2} can be found directly from~\eqref{eq:pell1} by considering the greatest integer $\widetilde{K}$ such that,   $\widetilde{K}$ divides $K$, while $\widetilde{K}^2$ divides $D=(m+1)(m+2)$.
  We will now show that, as $\widetilde{x}=x$  remains unchanged, the recurrence relation~\eqref{eq:mK0} stays the same.
  
  The position of the first hard wall $m$, together with $K$, yield an integer solution of Eq.~\eqref{eq:pell2}:  $\widetilde{x}=2m+3$, $\widetilde{y}=2 \widetilde{K}$.  Therefore, it must appear in the recurrence relation in~Eq.~\eqref{eq:pell_recurrence} with $D$ replaced by $\widetilde{D}$. If $\widetilde{x}_1$ is odd (i.e.  when $\widetilde{D}$ is even), $\widetilde{x}_n$ is also odd, and thus corresponds to a hard wall at an integer $\widetilde{m}_n$. In particular, the fundamental solution corresponds to the first hard wall, i.e., $m=(\widetilde{x}_1-3)/2$ [where $\widetilde{x}_1\geq 3$ follows from $\widetilde{y}_1>0$ required by the assumed positive integrated coupling $\phi>0$, cf.~Eq.~\eqref{eq:phi_wall}]. Thus, we again obtain the recurrence relation in Eq.~\eqref{eq:mK0} [as in the recurrence equation for $\widetilde{x}_n\equiv x_n$ we have that  $\widetilde{D}$ simplifies with $\widetilde{y}_1 \widetilde{y}_{n-1}$ to $D y_n y_{n-1}$ in Eq.~\eqref{eq:pell_recurrence_a}, while the recurrence equation for $\widetilde{y}_n$ can be divided by $\widetilde{K}$ yielding Eq.~\eqref{eq:pell_recurrence_b}, since $\widetilde{y}_n$ is divisible by $\widetilde{K}$ as so is $\widetilde{y}_1$].

  When $\widetilde{D}$ is odd, it is possible that $x_1$ is even (and $y_1$ odd), in which case the parity of $x_n$ (and $y_n$) oscillates with period 2. In particular, the first hard wall corresponds to the second solution, $m=(\widetilde{x}_2-3)/2=(\widetilde{x}_1^2+\widetilde{D} \widetilde{y}_1^2-3)/2$, while other hard walls correspond to $x_{2n}$. Nevertheless, from~\eqref{eq:pell_recurrence2} we have 
  \begin{equation}
  \label{eq:pell_recurrence3}
  \widetilde x_{2n} +\sqrt{ \widetilde D} \,\widetilde y_{2n}= \left(\widetilde x_2 + \sqrt{\widetilde D} \,\widetilde y_2 \right)^n, 
  \end{equation}
  so that the odd solutions also obey the recurrence relation Eq.~\eqref{eq:pell_recurrence}, but with the fundamental solution chosen as $x_2$ and $y_2$, instead of $x_1$ and $y_1$. Therefore, analogously as in the case of $\widetilde{D}$ being even, the walls are again determined by Eq.~\eqref{eq:mK0}. This concludes the proof.

	\section{Pure stationary states and relaxation times with soft walls}
	\label{app:soft_walls}
	
	In Sec.~\ref{sec:walls_soft}, we introduced the notion of a soft wall. Here we discuss the structure of stationary states in the cavity in the presence of soft walls and also discuss the induced long-time dynamics leading to those stationary states.

	\subsection{Distribution of pure stationary states between soft walls}
	
	We now discuss the structure of the stationary state between soft walls and argue that they are supported only after the walls corresponding to the boundary condition Eq.~\eqref{eq:psi_boundary}. We assume coherent dynamics $c_{\rm e},c_{\rm g}\neq 0$.\\

	Dynamics with soft walls features pure states given in Eq.~\eqref{eq:even_odd_ss}. In general, the even state can be written as a sum of contributions with support between subsequent pairs of walls located at $m^+_k$ and $m_{k+1}^+$ as
	\begin{eqnarray}\label{eq:psi_soft_wall}
	|\Psi_+\rangle
	&=& \sum_{n=0}^{m_1^+/2}  c_{2n} \,|2n\rangle +\sum_{k=1}^\infty \sum_{n=1+m_{k}^+\!/2}^{m_{k+1}^+/2}  c_{2n} \,|2n\rangle \\\nonumber
	&=& \sum_{n=0}^{m_1^+/2}  c_{2n} \,|2n\rangle + \sum_{k=1}^\infty c_{m_{k}^+} (- i)\,\frac{c_{\rm e}}{c_{\rm g}}\frac{\sin_{m_{k}^+}\!(\phi)}{1 - \cos_{m_{k}^+}\!(\phi)}   \sum_{n=1+m_k^+\!/2}^{m_{k+1}^+/2}  \frac{c_{2n}}{c_{m_{k}^++2}} \,|2n\rangle
	\\\nonumber
	&=& c_0 \left\{ \,\mathcal{N}_{0}^+ |\Psi_0^+\rangle   +  \sum_{k=1}^\infty  \left[\prod_{l=1}^k \frac{c_{m_{l}^+}}{c_{m_{l-1}^++2}}\right](- i)^k\,\left(\frac{c_{\rm e}}{c_{\rm g}}\right)^k\left[\prod_{l=1}^k \frac{\sin_{m_{l}^+}\!(\phi)}{1 - \cos_{m_{l}^+}\!(\phi)} \right] \,\mathcal{N}_{k}^+ |\Psi_k^+\rangle \right\},
	\end{eqnarray}
	where $m_k^+$ labels the walls of the even parity [cf.~Eq.~\eqref{eq:psi_ss}] and we introduced normalization  $(\mathcal{N}_{k}^+)^2 = \sum_{n=1+m_{k}^+\!/2}^{m_{k+1}^+/2}  |c_{2n}/c_{m_{k}^++2}|^2$ and the state after the $k$th even wall $|\Psi_k^+\rangle = \sum_{n=1+m_{k}^+\!/2}^{m_{k+1}^+/2}  c_{2n} \,|2n\rangle / c_{m_{k}^++2}/\mathcal{N}_{k}^+ $ (where for $|\Psi_0^+\rangle$ we formally define $m_0^+=-2$). The analogous construction holds for the odd state $|\Psi_-\rangle$ in Eq.~\eqref{eq:even_odd_ss}.

	In Eq.~\eqref{eq:psi_soft_wall}, we can identify that $c_{m_{k+1}^+}/c_{m_{k}^++2}$ is the ratio between the last and the first coefficients in the state after $k$th wall, $|\Psi_k^+\rangle$ and thus we expect it to be finite (as there are no soft walls within the state). Similarly, the norm $\mathcal{N}_{k}^+$ of the $k$th state is finite. In contrast, the remaining terms in Eq.~\eqref{eq:psi_soft_wall} can lead either to the suppression or the increase of the $k$th state contribution, depending whether the boundary condition after $k$th even soft wall,  Eq.~\eqref{eq:psi_boundary2b}, coincides with the boundary condition of the state $|\Psi_+\rangle$ in Eq.~\eqref{eq:psi_boundary}, 
	\begin{subequations}\label{eq:soft_wall_height}
		\begin{align}
		\text{when} \qquad \cos_{m_{k}^+}\!(\phi)\approx 1,\qquad &\frac{\sin_{m_{k}^+}\!(\phi)}{1 - \cos_{m_{k}^+}\!(\phi)}\approx \frac{2}{\sin_{m_{k}^+}\!(\phi)}\longrightarrow  \pm\infty  \\
		\text{when} \qquad\cos_{m_{k}^+}\!(\phi)\approx -1,  \qquad & \frac{\sin_{m_{k}^+}\!(\phi)}{1 - \cos_{m_{k}^+}\!(\phi)}\approx \frac{\sin_{m_{k}^+}\!(\phi)}{2}\longrightarrow  0
		\end{align}
	\end{subequations}
	where the arrows correspond to the limit of soft wall being hard. Noticing that $c_0$ in Eq.~\eqref{eq:psi_soft_wall} also changes with the height of the walls in order to keep the norm of $|\Psi_+\rangle$ equal $1$, we arrive at the following approximation 
	\begin{eqnarray}\label{eq:psi_soft_wall_limit}
	|\Psi_+\rangle
	&\approx&  \alpha_{0}^+\,\mathcal{N}_{0}^+ |\Psi_0^+\rangle   +  \sum_{k=1}^\infty  \left[\prod_{l=1}^k \frac{c_{m_{l}^+}}{c_{m_{l-1}^++2}}\right](- i)^k\,\left(\frac{c_{\rm e}}{c_{\rm g}}\right)^k \alpha_{k}^+  \,\mathcal{N}_{k}^+ |\Psi_k^+\rangle
	\\\nonumber 
	&=:&  \beta_{0}^+\,|\Psi_0^+\rangle   +  \sum_{k=1}^\infty  \beta_{k}^+\, |\Psi_k^+\rangle,
	\end{eqnarray}
	where we defined the \emph{hard wall limit} as
	\begin{equation}\label{eq:soft_wall_height_total}
	c_0\,\prod_{l=1}^k \frac{\sin_{m_{l}^+}\!(\phi)}{1 - \cos_{m_{l}^+}\!(\phi)} \longrightarrow  \alpha_{k}^+,
	\end{equation} 
	so that we choose $\alpha_{k}^+= 0$ if $\cos_{m_{k}^+}\!(\phi)\approx -1$ [cf.~Eq.~\eqref{eq:soft_wall_height}]. 
	
	In Eq.~\eqref{eq:psi_soft_wall_limit}, only the states after the soft walls with the boundary condition $\cos_{m_{k}^+}\!(\phi)\approx 1$ can be present [cf.~Fig.~\ref{fig:wig_gall}(b) and see the example in  Table~\ref{tab:2walls}]. Therefore, the state $|\Psi_0^+\rangle$ can be present only for the first wall with $\cos_{m_{1}^+}\!(\phi)\approx -1$ [cf.~Fig.~\ref{fig:wig_gall}(b) for the states (ii) and (viii)]. We further note that several subsequent walls with $\cos_{m_{k}^+}\!(\phi)\approx 1$ may be needed to counteract the suppression due to an earlier wall, in which case only the state after the last such a wall is present;  see Fig.~\ref{fig:wig_gall}(b) for the states (iii) (vi) (vii) and (ix).  %This is determined by the corresponding convergence rates in~\eqref{eq:soft_wall_height_total}.  	
	The same results follow from considering soft walls as a perturbation away from auxiliary hard walls (see below). 
	
	Finally, we note that for finite walls, the coefficients $\beta_{k}^+$ in Eq.~\eqref{eq:psi_soft_wall_limit} depend also on the distribution of the states $|\Psi_k^+\rangle$ between the walls, e.g.,  whether the state is supported only close to one of the walls. In particular, in the case of $|c_{\rm g}|=1$, we simply have $|\Psi_+\rangle=|\Psi_0\rangle=|0\rangle$.

\begin{comment}
	\begin{table}[h]
		\normalsize{
	\begin{tabular}{|c||c|c|c|}
		
	\hline			
	 %\backslashbox{Approx.}{Condition} 
	  \multirow{2}{*}{\textbf{Case}} & \multirow{2}{*}{$\,\cos_{m_1^+}\!(\phi)\, $} &   \multirow{2}{*}{$\,\cos_{m_2^+}\!(\phi)\, $}  & \multirow{2}{*}{$|\Psi_+\rangle$} \\
	  &&&\\	
	\hhline{|=||=|=|=|}	  
     1.&$-1+\frac{\delta_1^2}{2}$& $-1+\frac{\delta_2^2}{2}$&$|\Psi_0^+\rangle+O(\delta_1)$\\
     2. &$\phantom{+}1-\frac{\delta_1^2}{2}$& $-1+\frac{\delta_2^2}{2}$& $|\Psi_0^+\rangle+O(\delta_1,\delta_2)$\\
     3.&   $-1+\frac{\delta_1^2}{2}$& $\phantom{+}1-\frac{\delta_2^2}{2}$& $\beta_0^+|\Psi_0^+\rangle+\beta_2^+|\Psi_2^+\rangle+O[\min(\delta_1,\delta_2)]$\\
     4.&   $\phantom{+}1-\frac{\delta_1^2}{2}$& $\phantom{+}1-\frac{\delta_2^2}{2}$& $|\Psi_2^+\rangle+O(\delta_2)$\\
	\hline
\end{tabular}}
\begin{caption}{\textbf{Steady state between two soft walls} from Eqs.~\eqref{eq:psi_soft_wall} and~\eqref{eq:soft_wall_height}.  \\In case 3. $\beta_2^+/\beta_0^+= -(c_{\rm e}/c_{\rm g})^2 \mathcal{N}_{2}^+/\mathcal{N}_{0}^+ [ c_{m_{2}^+}/c_{m_{1}^++2} ][c_{m_{1}^+}/c_{0}]\,\times 4 \lim \delta_1/\delta_2$.
	\label{tab:2walls}}
\vspace*{-10mm}
\end{caption}
\end{table}
\end{comment}

\begin{table}
	\normalsize{
		\begin{tabular}{|c||c|c|c||c|c|c|}
			\hline			
			%\backslashbox{Approx.}{Condition} 
			\multirow{3}{*}{\textbf{Case}} &\multicolumn{3}{c||}{\textbf{Soft walls}}& \multicolumn{3}{c|}{\textbf{\,Hard walls} ($\delta_1,\delta_2=0$)}\\
			\cline{2-7}
			& \multirow{2}{*}{$\,\cos_{m_1^+}\!(\phi)\, $} &   \multirow{2}{*}{$\,\cos_{m_2^+}\!(\phi)\, $}  & \multirow{2}{*}{$|\Psi_+\rangle$} & \multirow{2}{*}{$\rho_0^+$} & \multirow{2}{*}{$\rho_1^+$}& \multirow{2}{*}{$\rho_2^+$} \\
			&&& & &&\\	
			\hhline{|=||=|=|=||=|=|=|}	  
			1.&$-1+\frac{\delta_1^2}{2}$& $-1+\frac{\delta_2^2}{2}$&$|\Psi_0^+\rangle+O(\delta_1)$&$|\Psi_0^+\rangle$ &mixed&mixed\\
			2. &$\phantom{+}1-\frac{\delta_1^2}{2}$& $-1+\frac{\delta_2^2}{2}$& $|\Psi_1^+\rangle+O(\delta_1,\delta_2)$& \,\,mixed\,\, &$|\Psi_1^+\rangle$&mixed\\
			3.&   $-1+\frac{\delta_1^2}{2}$& $\phantom{+}1-\frac{\delta_2^2}{2}$& $\beta_0^+|\Psi_0^+\rangle+\beta_2^+|\Psi_2^+\rangle+O[\min(\delta_1,\delta_2)]$& $|\Psi_0^+\rangle$ &$|\overline\Psi_1^+\rangle$ & $|\Psi_2^+\rangle$\\
			4.&   $\phantom{+}1-\frac{\delta_1^2}{2}$& $\phantom{+}1-\frac{\delta_2^2}{2}$& $|\Psi_2^+\rangle+O(\delta_2)$& mixed & \,\,mixed\,\, & $|\Psi_2^+\rangle$\\
			\hline
	\end{tabular}}
	\begin{caption}{\textbf{Steady state between two soft walls vs. two hard walls}. The stationary state with soft walls approximately corresponds  only to the pure stationary states of hard walls that obey the same boundary conditions. \\%We discuss 4 possible cases of the structure of the stationary state $|\Psi_+\rangle$ in Tab.~\ref{tab:2walls}.
			For soft walls:	 $|\Psi_+\rangle$ from Eqs.~\eqref{eq:psi_soft_wall} and~\eqref{eq:soft_wall_height}. In case 3. $\beta_2^+/\beta_0^+= -(c_{\rm e}/c_{\rm g})^2 \mathcal{N}_{2}^+/\mathcal{N}_{0}^+ [ c_{m_{2}^+}/c_{m_{1}^++2} ][c_{m_{1}^+}/c_{0}]\,\times 4 \lim \delta_1/\delta_2$.  
			For hard walls:  $|\overline\Psi_k^+\rangle$ refers to the $k$th pure stationary state with with boundary conditions at the $(k-1)$-th and $k$-th wall which are opposite to Eq.~\eqref{eq:psi_boundary}. To consider a finite number of walls, we have assumed a third even wall to be hard, with $\cos_{m_{3}^+}\!(\phi)= -1$, so that pure stationary states before that wall exist [cf.~Eqs.~\eqref{eq:psi_boundary} and~\eqref{eq:psi_boundary2a}]. The~same results hold for the odd stationary state.
			\label{tab:2walls}}
	\end{caption}
\end{table}

	\subsection{Dynamics with soft walls}
	\label{app:soft_walls_dynamics}
	Here we discuss timescales of achieving pure stationary states, Eq.~\eqref{eq:psi_soft_wall_limit}, by considering dynamics in the presence of soft walls as a perturbation of auxiliary dynamics with hard walls.\\

	%Why QFI high already earlier if such timescales required to get a coherence?
   
	\subsubsection{Dynamics of soft walls as perturbation of hard walls} 
	The dynamics of the cavity with soft walls can be formally considered as a \emph{perturbation of}  an auxiliary dynamics $M_{\rm g}^{(0)}$, $M_{\rm e}^{(0)}$ with \emph{hard walls} replacing soft walls, 
	\begin{subequations}
		\begin{align}
		M_{\rm g}-M_{\rm g}^{(0)}&\equiv\delta M_{\rm g}=\sum_{k=1}^\infty \left(- i c_{\rm e}  \sin_{m_k}(\phi) \ketbra{m_k\!+\!2}{m_k}+ c_{\rm g} [\cos_{m_k}(\phi)\mp 1] \ketbra{m_k\!+\!2}{m_k\!+\!2} \right) , \\
		M_{\rm e}-M_{\rm e}^{(0)}&\equiv\delta M_{\rm e}=\sum_{k=1}^\infty \left( c_{\rm e} [\cos_{m_k}(\phi)\mp 1] \ketbra{m_k}{m_k}-  i c_{\rm g}  \sin_{m_k}(\phi) \ketbra{m_k}{m_k\!+\!2}\right),
		\end{align}
		\label{eq:Kraus_soft_wall}
	\end{subequations}
	where we consider  $\cos_{m_k}(\phi)\approx \pm 1$, so that $\cos_{m_k}(\phi)\mp 1\approx 0$.  In this section, we discuss the order of the perturbation in the powers of a small parameter $\delta_k$ of  the $k$th wall, where 
	\begin{equation}\label{eq:soft_wall_order}
	\cos_{m_k} (\phi) \approx \pm \left(\!1 - \frac{\delta_k^2}{2}\right), \; \sin_{m_k}(\phi) \approx \pm \delta_k
	\end{equation}
	[see Eq.~(\ref{eq:Kraus_soft_wall}) and Table \ref{tab:2walls}].

	\subsubsection{Steady state with soft walls vs. stationary states of hard walls} 
	
	The stationary state in Eq.~\eqref{eq:even_odd_ss} is pure and fulfills the boundary condition~\eqref{eq:psi_boundary}.  In contrast, each soft wall present in the dynamics can be approximated by a hard wall that determines boundary conditions for a state before and after that wall [Eqs.~\eqref{eq:psi_boundary2a} and~\eqref{eq:psi_boundary2b}]. 
	
	~\\
	\emph{Steady states of hard walls}. First, the $k$th stationary state $\rho_k^\pm$, between subsequent walls of the same parity at $m_{k}^\pm$ and $m_{k+1}^\pm$, is pure only if $\cos_{m_k^\pm}\!(\phi)=- \cos_{m_{k+1}^\pm}\!(\phi)$. Otherwise, that stationary state is mixed. Second, even if the stationary state is pure, when its boundary condition differs from~\eqref{eq:psi_boundary}, it does not correspond to the stationary state with soft walls $|\Psi_\pm\rangle$; i.e., it differs from its projection  $|\Psi_k^{\pm}\rangle$ between the $k$th and $(k\!+\!1)$th walls, as $|\langle \overline\Psi_k^{\pm}|\Psi_k^{\pm}\rangle|^2<1$, unless $m_{k+1}^+-m_{k}^+ =2$ and it is a fixed photon state, $|\overline\Psi_k^{\pm}\rangle=|m_{k+1}^+\rangle$. Indeed,  from Eq.~\eqref{eq:psi_ss}, when $c_{\rm e},c_{\rm g}\neq 0$,  $|\langle \overline\Psi_k^{\pm}|\Psi_k^{\pm}\rangle|^2=1$ requires $\cot_{k}(\phi/2)=-\tan_{k}(\phi/2)$ [for all $m_{k}^\pm+2 < k\leq m_{k+1}^\pm$ such that $(-1)^k=\pm 1$], which is never true. Furthermore, the coherences between pure stationary states corresponding to opposite boundary condition [i.e., opposite eigenvalues of Kraus operators, see Eq.~\eqref{eq:atom_out}] and between the pure and mixed stationary states are not stationary (cf.~Sec.~\ref{sec:walls_hard}).  
	%\begin{equation}
	%|\langle \Psi_k^{0,\pm}|\Psi_k^{\pm}\rangle|^2= \frac{1- %\left(-\left|\frac{c_{\rm e}}{c_{\rm g}}\right|^{2}\right)^{\frac{m_{k+1}^\pm-m_{k}^\pm}{2}}}{\left(1+\left|\frac{c_{\rm e}}{c_{\rm g}}\right|^{2}\right)\mathcal{N}_k^{0,\pm}\,\mathcal{N}_k^\pm}.   
	%\end{equation}

	~\\
	\emph{Consequences for stationary state with soft walls}.  The perturbative dynamics defined in Eq.~\eqref{eq:Kraus_soft_wall} should recover the true stationary state in Eq.~\eqref{eq:even_odd_ss}. In particular, in the zeroth order, the solution is a linear combination of the stationary states between hard walls~\cite{Kato1995, macieszczak2016towards}. Therefore, in agreement with Eq.~\eqref{eq:psi_soft_wall_limit}, the stationary states in Eq.~\eqref{eq:even_odd_ss} can be \emph{approximated} only by the \emph{pure stationary states between hard walls that are consistent with the boundary conditions~\eqref{eq:psi_boundary}}, i.e., $\cos_{m_k^\pm}\!(\phi)=1=- \cos_{m_{k+1}^\pm}\!(\phi)$. See  Table~\ref{tab:2walls} for the example of two walls.
	
	%This argument does not realy work for the trapping state. There we can argue that the eigenvalue of M_e does not agree. It would be a third approach, to consider perturbation theory for the Kraus operators themselves. I want to understand how the eigenspace [c_{\rm g},-c_{\rm e}] changes. If it is perturbative indeed, I can only use the solutions with the same boundary condition from hard-wall dynamics. 
	
	%ALthough simpler, this wouldn't indtroduce the perturbative dynamics, which we are after though.
	
	%This is also the case, when the mixed boundary conditions do not lead to a mixed state.

	\subsubsection{Perturbative dynamics} 
	
	Below we derive the long-time dynamics due to the presence of the soft walls. We prove that this dynamics is second order in $\sin_{m_n}(\phi)$. Because of locality of the perturbation in Eq.~\eqref{eq:Kraus_soft_wall} only neighboring states get connected, or coherences between states separated by two walls are created.  Furthermore,  the perturbation depends on the amplitude of the states directly next to the walls. We discuss how the closed form of the long-time dynamics generator can be found using the structure of the stationary state Eq.~\eqref{eq:psi_soft_wall_limit} .

 ~\\	
\emph{First and second-order perturbations}.
The difference $\delta \mathcal{L}$ between the dynamics generated by $M_{\rm g}$, $M_{\rm e}$~\eqref{eq:Kraus} and the modified Kraus operators with hard walls $M_{\rm g}^{(0)}$ and $M_{\rm e}^{(0)}$  feature the first and second order perturbations  in  $\delta M_{\rm g}$ and $\delta M_{\rm e}$ [cf.~~\eqref{eq:Kraus_soft_wall}]
\begin{eqnarray} \label{eq:L_soft_wall}
\nu^{-1} \mathcal{L}\,(\rho)&=& M_{\rm g} \rho M_{\rm g}^\dagger +M_{\rm g} \rho M_{\rm g}^\dagger -\rho  = M_{\rm g}^{(0)} \rho \,[M_{\rm g}^{(0)}]^\dagger +M_{\rm e}^{(0)} \rho\, [M_{\rm e}^{(0)}]^\dagger - \rho +\\
&&+ \left\{ \delta M_{\rm g}\rho \,[M_{\rm g}^{(0)} ]^\dagger  + \delta M_{\rm e}\rho \,[M_{\rm e}^{(0)} ]^\dagger  + \text{H.c.}  \right \}  + \,\delta M_{\rm g}\rho \,\delta M_{\rm g}^\dagger + \delta M_{\rm e}\rho \,\delta M_{\rm e}^\dagger, 
\nonumber
%\\&\equiv& \nu^{-1} \left (\mathcal{L}_0 \,\rho +\delta  \mathcal{L}_1\,\rho+\delta \mathcal{L}_2\,\rho \right),\nonumber
\end{eqnarray}
cf.~Eq.~\eqref{eq:master}. The perturbations in the Kraus operators themselves, $\delta M_{\rm g}$ and $\delta M_{\rm e}$  in Eq.~\eqref{eq:Kraus_soft_wall}, feature first- and second-order perturbations [cf.~Eq.~\eqref{eq:soft_wall_order}]
\begin{subequations}
\begin{align}
\delta M_{\rm g}^{(1)}=- i c_{\rm e}\sum_{k=1}^\infty  \sin_{m_k}(\phi) \ketbra{m_k\!+\!2}{m_k}, &\qquad \delta M_{\rm g}^{(2)}=c_{\rm g}\sum_{k=1}^\infty  [\cos_{m_k}(\phi)\mp 1] \ketbra{m_k\!+\!2}{m_k\!+\!2} ,\\
\delta M_{\rm e}^{(1)}=-  i c_{\rm g} \sum_{k=1}^\infty \sin_{m_k}(\phi) \ketbra{m_k}{m_k\!+\!2},&\qquad \delta M_{\rm e}^{(2)}=c_{\rm e}\sum_{k=1}^\infty  [\cos_{m_k}(\phi)\mp 1] \ketbra{m_k}{m_k}.
\end{align}
\label{eq:Kraus_soft_wall_order}
\end{subequations}
Therefore, we can identify the first- and second-order perturbations to the master equation~\eqref{eq:L_soft_wall} as
\begin{eqnarray} \label{eq:L_soft_wall_order1}
\delta\mathcal{L}_1&=&  \delta M_{\rm g}^{(1)}\rho \,[M_{\rm g}^{(0)} ]^\dagger  + \delta M_{\rm e}^{(1)}\rho \,[M_{\rm e}^{(0)} ]^\dagger  + \text{H.c.}  ,\\
\delta\mathcal{L}_2&=& \delta M_{\rm g}^{(1)}\rho \,[\delta M_{\rm g}^{(1)}]^\dagger + \delta M_{\rm e}^{(1)}\rho \,[\delta M_{\rm e}^{(1)}]^\dagger+\left\{ \delta M_{\rm g}^{(2)}\rho \,[M_{\rm g}^{(0)} ]^\dagger  + \delta M_{\rm e}^{(2)}\rho \,[M_{\rm e}^{(0)} ]^\dagger  + \text{H.c.}  \right \}   .\label{eq:L_soft_wall_order2}
\end{eqnarray}
Below we focus on the second-order corrections to the dynamics, and thus we neglect the third- and-fourth order perturbations in~\eqref{eq:L_soft_wall}.

~\\	
\emph{Absence of first-order corrections}.  We show now  that dynamics feature no contribution from $\mathcal{L}_1$ in~\eqref{eq:L_soft_wall_order1}. We consider only even or odd states, but we drop the superscript $\pm$ in $|\Psi_{k}^\pm\rangle$, $\rho_{k}^\pm$ and $m_k^\pm$ for convenience.

Noting that for pure stationary state between the $k$th and $(k\!+\!1)$th walls, we have $M_g^{(0)}|\Psi_{k}\rangle=\pm c_{\rm g}$ and $M_e^{(0)}|\Psi_{k}\rangle=\mp c_{\rm e}$,
\begin{equation}\label{eq:Leff_soft_wall1a}
\nu^{-1}\delta\mathcal{L}_1 \left(\ketbra{\Psi_{k}}{\Psi_{k}} \right)= \pm  i  c_{\rm g}c_{\rm e}^*\,c_{m_{k}+2}^{(k)} \,    \sin_{m_k}(\phi)\,\ketbra{m_{k}}{\Psi_{k}} \mp  i c_{\rm e}  c_{\rm g}^* \,c_{m_{k+1}}^{(k)} \,   \sin_{m_{k+1}}(\phi)\,\ketbra{m_{k+1}\!+\!2}{\Psi_{k}} + \text{H.c.},
\end{equation}
where $c_{n}^{(k)}$ is the amplitude (coefficient) of $n$ photons in  the pure stationary state between the $k$th and $(k\!+\!1)$th walls. Analogously, for the coherences between the states with the same boundary conditions,
\begin{eqnarray}\label{eq:Leff_soft_wall1b}
\nu^{-1}\delta\mathcal{L}_1 \left(\ketbra{\Psi_{k_1}}{\Psi_{k_2}} \right)&=& \pm  i  c_{\rm g}c_{\rm e}^*\,c_{m_{k_1}+2}^{(k_1)} \,    \sin_{m_{k_1}}(\phi)\,\ketbra{m_{k_1}}{\Psi_{k_2}} \mp  i c_{\rm e}  c_{\rm g}^* \,c_{m_{k_1+1}}^{(k_1)} \,   \sin_{m_{k_1+1}}(\phi)\,\ketbra{m_{k_1+1}\!+\!2}{\Psi_{k_2}}\\\nonumber
&&\mp  i  c_{\rm g}^* c_{\rm e} \left(c_{m_{k_2}+2}^{(k_2)}\right)^* \,    \sin_{m_{k_2}}(\phi)\,\ketbra{\Psi_{k_1}}{m_{k_2}} \pm  i c_{\rm e}^*  c_{\rm g} \left(c_{m_{k_2+1}}^{(k_2)}\right)^* \,   \sin_{m_{k_2+1}}(\phi)\,\ketbra{\Psi_{k_1}}{m_{k_2+1}\!+\!2}.
\end{eqnarray}
Similarly, for the mixed state $\rho_k$ (mixed due to different boundary conditions implied by $k$th and $(k\!+\!1)$th walls) we have
\begin{eqnarray}\label{eq:Leff_soft_wall1c}
\nu^{-1}\delta\mathcal{L}_1 \left(\rho_k \right)&=& -  i  c_{\rm g}\, \sin_{m_k}(\phi)\,|m_k\rangle\,  \langle m_{k}\!+\!2|\rho_k [M_e^{(0)}]^\dagger  -  i c_{\rm e}    \,   \sin_{m_{k+1}}(\phi)\,|m_{k+1}\!+\!2\rangle\, \langle m_{k+1} |\rho_k  [M_g^{(0)}]^\dagger +\text{H.c.}.
\end{eqnarray}
As stationary coherences can only exist between pure stationary states which are separated by at least two walls [cf.~Eqs.~\eqref{eq:psi_boundary2a} and~\eqref{eq:psi_boundary2b}],  there are no first-order corrections to the dynamics (cf.~Eq.~\eqref{eq:1storder} and Refs.~\cite{Kato1995,zanardi_dissipative_2016, macieszczak2016towards}) 
\begin{eqnarray}
\Pi_0\,\delta\mathcal{L}_1 \left(\ketbra{\Psi_{k}}{\Psi_{k}} \right)&=& 0,\\
\Pi_0\,\delta\mathcal{L}_1 \left(\ketbra{\Psi_{k_1}}{\Psi_{k_2}} \right)&=& 0,\\
\Pi_0\,\delta\mathcal{L}_1 \left(\rho_k\right)&=& 0,
\end{eqnarray}
where $\Pi_0$ denotes the projection onto the stationary states of dynamics $\mathcal{L}_0$ with hard walls.

~\\
\noindent
\emph{Second-order corrections}. We now derive the effective dynamics in the second order of the corrections  in $\delta M_g$ and $\delta M_e$, Eq.~\eqref{eq:Kraus_soft_wall}. We consider both the corrections from $\mathcal{L}_2$, as well as the contribution from  $\mathcal{L}_1$ in Eqs.~\eqref{eq:L_soft_wall_order1} and~\eqref{eq:L_soft_wall_order2}, as the second-order corrections are given by~\cite{Kato1995,zanardi_dissipative_2016, macieszczak2016towards}
\begin{equation}\label{eq:Leff_2ndorder}
\Pi_0\mathcal{L}_2\Pi_0-\Pi_0\mathcal{L}_1 \mathcal{S}_0 \mathcal{L}_1 \Pi_0,
\end{equation}
where $\mathcal{S}_0$  is the resolvent for the dynamics $\mathcal{L}_0$ with hard walls (evaluated at $0$), i.e., $\mathcal{S}_0\mathcal{L}_0=\mathcal{L}_0\mathcal{S}_0=\mathcal{I}-\Pi_0$.
\\

First, we consider second-order corrections $\Pi_0\mathcal{L}_2\Pi_0$ due to the second-order perturbation $\mathcal{L}_2$ (cf.~Eq.~\eqref{eq:1storder} and Refs.~\cite{Kato1995,zanardi_dissipative_2016, macieszczak2016towards}). We have
%\begin{eqnarray}
%\nu^{-1} \,\delta \mathcal{L}_2 \left(\ketbra{\Psi_k}{\Psi_k}\right)&=& |c_{\rm e}|^2 \sin_{m_{k+1}}^2(\phi)\, |c_{m_{k+1}}^{(k)}|^2\ketbra{m_{k+1}\!+\!2}{m_{k+1}\!+\!2}+  |c_{\rm g}|^2 \sin_{m_{k}}^2(\phi)\, |c_{m_{k}+2}^{(k)}|^2\ketbra{m_{k}}{m_{k}}\\
%&& \pm\left\{|c_{\rm g}|^2\left[\cos_{m_k}(\phi)\mp 1\right] c_{m_{k}+2}^{(k)}\, \ketbra{m_{k}\!+\!2}{\Psi_k} - |c_{\rm e}|^2\left[\cos_{m_{k+1}}(\phi)\pm 1\right] c_{m_{k+1}}^{(k)}\, \ketbra{m_{k+1}}{\Psi_k} +\text{H.c}\right\}, \nonumber
%\end{eqnarray}
% Since the projection $\Pi_0$ on the states between the hard walls is given by the supports between the walls, 
\begin{eqnarray}\label{eq:Leff_soft_wall2a}
\nu^{-1} \,\Pi_0 \delta \mathcal{L}_2 \left(\ketbra{\Psi_k}{\Psi_k}\right)&=& |c_{\rm e}|^2 \sin_{m_{k+1}}^2(\phi)\, |c_{m_{k+1}}^{(k)}|^2\,\rho_{k+1}+  |c_{\rm g}|^2 \sin_{m_{k}}^2(\phi)\, |c_{m_{k}+2}^{(k)}|^2\,\rho_{k-1} \nonumber \\
&& \pm 2\left\{|c_{\rm g}|^2\left[\cos_{m_k}(\phi)\mp 1\right] |c_{m_{k}+2}^{(k)}|^2  - |c_{\rm e}|^2\left[\cos_{m_{k+1}}(\phi)\pm 1\right]   |c_{m_{k+1}}^{(k)}|^2 \right\}\ketbra{\Psi_k}{\Psi_k},
\end{eqnarray}
where $\rho_{k\mp1}$ denotes (note necessarily mixed) $(k\!\mp\!1)$th stationary state. We used the fact that the projection $\Pi_0$ on the states between the hard walls is given by the supports between the walls, so that  $\Pi_0(\ketbra{m_{k+1}}{\Psi_k})=(c_{m_{k+1}}^{(k)})^*\ketbra{\Psi_k}{\Psi_k}$ and $\Pi_0(\ketbra{m_{k}\!+\!2}{\Psi_k})=(c_{m_{k}+2}^{(k)})^*\ketbra{\Psi_k}{\Psi_k}$. We assumed the boundary conditions $\cos_{m_{k}}(\phi)\approx \pm 1\approx-\cos_{m_{k+1}}(\phi)$, so that up to the second order of perturbation,  we have $ \cos_{m_k}(\phi)\mp 1= \mp\sin_{m_{k}}^2(\phi)/2$ and $\cos_{m_{k+1}}(\phi)\pm 1=\pm\sin_{m_{k+1}}^2(\phi)/2$ [cf.~Eq.~\eqref{eq:soft_wall_order}].  %effective trace preservation!

Similarly, for the coherences between states $\ket{\Psi_{k_1}}$ and $\ket{\Psi_{k_2}}$ with the same boundary conditions,
%\begin{eqnarray}
%\nu^{-1} \,\delta \mathcal{L}_2 \left(\ketbra{\Psi_{k_1}}{\Psi_{k_2}}\right)&=& |c_{\rm e}|^2 \sin_{m_{k_1+1}}(\phi) \sin_{m_{k_2+1}}(\phi)\, c_{m_{k_1+1}}^{(k_1)}\left[ c_{m_{k_2+1}}^{(k_2)}\right]^*\ketbra{m_{k_1+1}\!+\!2}{m_{k_2+1}\!+\!2}\\\nonumber
%&&+  |c_{\rm g}|^2 \sin_{m_{k_1}}(\phi)\sin_{m_{k_2}}(\phi)\, c_{m_{k_1}+2}^{(k_1)}\left[c_{m_{k_2}+2}^{(k_2)}\right]^*\ketbra{m_{k_1}}{m_{k_2}}\\\nonumber
%&& \pm|c_{\rm g}|^2\left\{ \left[\cos_{m_{k_1}}(\phi)\mp 1\right] c_{m_{k_1}+2}^{(k_1)}\, \ketbra{m_{k_1}\!+\!2}{\Psi_{k_2}}+ \left[\cos_{m_{k_2}}(\phi)\mp 1\right] \left[c_{m_{k_2}+2}^{(k_2)}\right]^*\, \ketbra{\Psi_{k_1}}{m_{k_2}\!+\!2}\right\} \\&&\mp |c_{\rm e}|^2\left\{\left[\cos_{m_{k_1+1}}(\phi)\pm 1\right] c_{m_{k_1+1}}^{(k_1)} \ketbra{m_{k_1+1}}{\Psi_{k_2}} +\left[\cos_{m_{k_2+1}}(\phi)\pm 1\right] \left[c_{m_{k_2+1}}^{(k_2)}\right]^* \ketbra{\Psi_{k_1}}{m_{k_2+1}}\right\}. \nonumber
%\end{eqnarray}
%In the third and the forth line we have coherences to the stationary states $\ket{\Psi_{k_1}}$ and $\ket{\Psi_{k_2}}$. These pure stationary states are necessarily dark in shifted dynamics, cf.~Eq.~\eqref{eq:Kraus_shifted} for the boundary conditions in Eq.~\eqref{eq:psi_boundary}, and thus  the coherences are orthogonally  projected by $\Pi_0$, cf.~Eq.~\eqref{eq:Pdark} and see~\cite{albert2016geometry},
\begin{eqnarray}\label{eq:Leff_soft_wall2b}
\nu^{-1} \,\Pi_0\delta \mathcal{L}_2 \left(\ketbra{\Psi_{k_1}}{\Psi_{k_2}}\right)&=& |c_{\rm e}|^2 \sin_{m_{k_1+1}}(\phi) \sin_{m_{k_2+1}}(\phi)\, c_{m_{k_1+1}}^{(k_1)}\left[ c_{m_{k_2+1}}^{(k_2)}\right]^*\eta_{k_1,k_2}^+ \ketbra{\Psi_{k_1+1}}{\Psi_{k_2+1}} \nonumber \\\nonumber
&&+  |c_{\rm g}|^2 \sin_{m_{k_1}}(\phi)\sin_{m_{k_2}}(\phi)\, c_{m_{k_1}+2}^{(k_1)}\left[c_{m_{k_2}+2}^{(k_2)}\right]^*\eta_{k_1,k_2}^- \ketbra{\Psi_{k_1-1}}{\Psi_{k_2-1}}\\\nonumber
&& \pm|c_{\rm g}|^2\left\{ \left[\cos_{m_{k_1}}(\phi)\mp 1\right] |c_{m_{k_1}+2}^{(k_1)}|^2+ \left[\cos_{m_{k_2}}(\phi)\mp 1\right] |c_{m_{k_2}+2}^{(k_2)}|^2\right\} \ketbra{\Psi_{k_1}}{\Psi_{k_2}}\\&&\mp |c_{\rm e}|^2\left\{\left[\cos_{m_{k_1+1}}(\phi)\pm 1\right] |c_{m_{k_1+1}}^{(k_1)}|^2  +\left[\cos_{m_{k_2+1}}(\phi)\pm 1\right] |c_{m_{k_2+1}}^{(k_2)}|^2 \right\}\ketbra{\Psi_{k_1}}{\Psi_{k_2}}.
\end{eqnarray}
where we introduced $\eta_{k_1,k_2}^+=\langle \Psi_{k_1+1}|\Pi_0(\ketbra{m_{k_1+1}\!+\!2}{m_{k_2+1}\!+\!2})|\Psi_{k_2+1}\rangle$ and $\eta_{k_1,k_2}^-=\langle \Psi_{k_1-1}|\Pi_0(\ketbra{m_{k_1}}{m_{k_2}})|\Psi_{k_2-1}\rangle$, which are $0$ if the pure stationary states $|\Psi_{k_1+1}\rangle$, $|\Psi_{k_2+1}\rangle$, or $|\Psi_{k_1-1}\rangle$, $|\Psi_{k_2-1}\rangle$, do not exist. In the derivation of Eq.~\eqref{eq:Leff_soft_wall2b}, we used the fact that pure stationary states are necessarily dark in shifted dynamics [cf.~Eq.~\eqref{eq:Kraus_shifted} for the boundary conditions in Eq.~\eqref{eq:psi_boundary}], and thus the coherences  to them are orthogonally  projected by $\Pi_0$ (cf.~Eq.~\eqref{eq:Pdark} and see Ref.~\cite{albert2016geometry}), e.g.,  $\Pi_0(\ketbra{m_{k_1}\!+\!2}{\Psi_{k_2}})=(c_{m_{k}+2}^{(k)})^*\ketbra{\Psi_{k_1}}{\Psi_{k_2}}$. 

Finally, for the mixed stationary state $\rho_k$ [due to mixed boundary conditions from after $k$th and before $(k\!+\!1)$th wall; cf.~Eqs.~\eqref{eq:psi_boundary2a} and~\eqref{eq:psi_boundary2b}],
%\begin{eqnarray}
%\nu^{-1} \,\delta \mathcal{L}_2 \left(\rho_k\right)&=& |c_{\rm e}|^2 \sin_{m_{k+1}}^2(\phi)\, \langle m_{k+1}|\rho_k|m_{k+1}\rangle \ketbra{m_{k+1}\!+\!2}{m_{k+1}\!+\!2}+  |c_{\rm g}|^2 \sin_{m_{k}}^2(\phi)\, \langle m_{k}\!+\!2|\rho_k|m_{k}\!+\!2\rangle\,\ketbra{m_{k}}{m_{k}}\qquad\qquad\\
%&& +\left\{c_{\rm g} \left[\cos_{m_k}(\phi)\mp 1\right] |m_{k}\!+\!2\rangle\, \langle m_{k}\!+\!2|\rho_k M_g^{(0)}  + c_{\rm e}\left[\cos_{m_{k+1}}(\phi)\pm 1\right]  \ket{m_{k+1}}\, \langle m_{k+1}|\rho_k M_e^{(0)} +\text{H.c}\right\}, \nonumber
%\end{eqnarray}
%which gives 
\begin{eqnarray}
\nu^{-1} \,\Pi_0\delta \mathcal{L}_2 \left(\rho_k\right)&=& |c_{\rm e}|^2 \sin_{m_{k+1}}^2(\phi)\, \langle m_{k+1}|\rho_k|m_{k+1}\rangle \,\rho_{k+1}+  |c_{\rm g}|^2 \sin_{m_{k}}^2(\phi)\, \langle m_{k}\!+\!2|\rho_k|m_{k}\!+\!2\rangle\,\rho_{k-1}\qquad\qquad\\
&& \pm 2\left\{|c_{\rm g}|^2 \left[\cos_{m_k}(\phi)\mp 1\right]  \langle m_{k}\!+\!2|\rho_k |m_{k}\!+\!2 \rangle - |c_{\rm e}|^2\left[\cos_{m_{k+1}}(\phi)\pm 1\right]   \langle m_{k+1}|\rho_k |m_{k+1}\rangle \right\}\rho_k, \nonumber
\end{eqnarray}
where we again used  the fact that the projection $\Pi_0$ on the states between the hard walls is given by the support between the walls, and
from Eq.~\eqref{eq:Kraus_soft_wall} $\langle m_{k}\!+\!2|\rho_k [M_g^{(0)}]^\dagger|m_{k}\!+\!2\rangle=\pm c_{\rm g} \langle m_{k}\!+\!2|\rho_k |m_{k}\!+\!2\rangle $ and $\langle m_{k+1}|\rho_k [M_e^{(0)}]^\dagger | m_{k+1}\rangle = \mp c_{\rm e}^*\langle m_{k+1}|\rho_k  | m_{k+1}\rangle $. \\

Second, we consider the second-order corrections from the first-order  perturbation $\mathcal{L}_1$  in Eq.~\eqref{eq:L_soft_wall_order1}, which contributes as $-\Pi_0\mathcal{L}_1 \mathcal{S}_0 \mathcal{L}_1 \Pi_0$~\cite{Kato1995,zanardi_dissipative_2016, macieszczak2016towards} [cf.~Eq.~\eqref{eq:Leff_2ndorder}].
 
From Eqs.~\eqref{eq:Leff_soft_wall1a} and~\eqref{eq:Leff_soft_wall1b} for pure stationary states and coherences between them the first-order perturbation creates coherences to pure stationary states. As a pure stationary state corresponds to the dark state of  shifted dynamics, the coherences to such state decay with the corresponding effective Hamiltonian~\cite{albert2016geometry}
\begin{equation}\label{eq:Heff_hard_wall}
H_\pm\equiv - i\nu[\mathds{1}\pm(-c_{\rm g}^* M_g -c_{\rm g} M_g^\dagger +c_{\rm e}^* M_e +c_{\rm e} M_e^\dagger)/2],
\end{equation}
where we assumed the state with boundary condition  the same or opposite to Eq.~\eqref{eq:psi_boundary} [see Eq.~\eqref{eq:Kraus_shifted}]. In particular, the coherence $|\psi\rangle\!\langle \Psi_k|$ between the dark state and any state between hard walls with different boundary conditions to $|\Psi_k\rangle$  decays to $0$, i.e., $\Pi_0 |\psi\rangle\!\langle \Psi_k| = 0$. Furthermore,  as $\mathcal{S}_0=-\int_{0}^\infty  d t (e^{t\mathcal{L}_0}-\Pi_0)$, we have that the resolvent $\mathcal{S}_0$ simplifies to the \emph{pseudoinverse} of the effective Hamiltonian
\begin{eqnarray}
\mathcal{S}_0 (|\psi\rangle\!\langle \Psi_k| ) &=&-\int_{0}^\infty  d t \,e^{- i H_\pm }|\psi\rangle\!\langle \Psi_k|     = \left(- i H_\pm\right)^{-1}|\psi\rangle\!\langle \Psi_k|\\\nonumber
&=& -\nu^{-1}\left[\mathds{1}\pm\frac{-c_{\rm g}^* M_g -c_{\rm g} M_g^\dagger +c_{\rm e}^* M_e +c_{\rm e} M_e^\dagger}{2}\right]^{-1}|\psi\rangle\!\langle \Psi_k|  
\end{eqnarray}
As the effective Hamiltonian~\eqref{eq:Heff_hard_wall} does not change the support of the state between the hard walls, we have
\begin{eqnarray}\label{eq:Leff_soft_wall2d}
&&\nu^{-1}\Pi_0\delta\mathcal{L}_1 \mathcal{S}_0 \delta\mathcal{L}_1 \left(\ketbra{\Psi_{k}}{\Psi_{k}} \right)\\\nonumber
&&\quad=    \left[|c_{\rm g}c_{\rm e}^*|^2\,|c_{m_{k}+2}^{(k)}|^2 \,    \sin_{m_k}^2\!(\phi)\, \bra{m_{k}}\!\left(- i H_\pm\right)^{-1}\!\ket{m_{k}}  +|c_{\rm g}c_{\rm e}^*|^2\,|c_{m_{k+1}}^{(k)}|^2 \,    \sin_{m_{k+1}}^2\!(\phi)\, \bra{m_{k+1}\!+\!2}\!\left(- i H_\pm\right)^{-1}\!\ket{m_{k+1}\!+\!2}  \right]\ketbra{\Psi_{k}}{\Psi_{k}}\\\nonumber
&&\quad\phantom{=} -  (c_{\rm g}c_{\rm e}^*)^2|\,\left[c_{m_{k-1}}^{(n-2)} \right]^* c_{m_{k}+2}^{(k)} \,  \sin_{m_{k-1}}(\phi)  \sin_{m_k}(\phi)\, \bra{m_{k-1}\!+\!2}\!\left(- i H_\pm\right)^{-1}\!\ket{m_{k}} \ketbra{\Psi_{k-2}}{\Psi_{k}} \\\nonumber
&&\quad\phantom{=} - (c_{\rm e}  c_{\rm g}^*)^2 \,\left[c_{m_{k+2}+2}^{(k+2)}\right]^*c_{m_{k+1}}^{(k)} \,  \sin_{m_{k+2}}(\phi) \sin_{m_{k+1}}(\phi)\, \bra{m_{k+2}}\!\left(- i H_\pm\right)^{-1}\!\ket{m_{k+1}\!+\!2}\ketbra{\Psi_{k+2}}{\Psi_{k}}
\\\nonumber
&&\quad\phantom{=} -  |c_{\rm g}c_{\rm e}^*|^2\,|c_{m_{k}+2}^{(k)}|^2 \,    \sin_{m_k}^2(\phi)\, \langle m_{k}|\left(- i H_\pm\right)^{-1}|m_{k}\rangle\,\rho_{k-1}\\\nonumber
&&\quad\phantom{=}- |c_{\rm e}c_{\rm g}^*|^2   \,|c_{m_{k+1}}^{(k)}|^2 \,   \sin_{m_{k+1}}^2(\phi)\, \langle m_{k+1}\!+\!2|\left(- i H_\pm\right)^{-1}|m_{k+1}\!+\!2\rangle\,\rho_{k+1}\\\nonumber
&&\quad\phantom{=}
%\mp  c_{\rm g}(c_{\rm e}^*)^2\,c_{m_{k}+2}^{(k)} \left[c_{m_{k+1}}^{(k)}\right]\,    \sin_{m_k}(\phi)\sin_{m_{k+1}}(\phi)\, M_g^{(0)}\left(- i H_\pm\right)^{-1}\ketbra{m_{k}}{m_{k+1}\!+\!2}\\\nonumber
+  \left(c_{\rm g}c_{\rm e}^*\right)^2\,c_{m_{k}+2}^{(k)} \left[c_{m_{k+1}}^{(k)}\right]^*\,    \sin_{m_k}(\phi)\sin_{m_{k+1}}(\phi)\,  \eta_{k}^{-+}\, \ketbra{\Psi_{k-1}}{\Psi_{k+1}}\\\nonumber
&&\quad\phantom{=}
%\pm  c_{\rm e}  (c_{\rm g}^*)^2 \,c_{m_{k+1}}^{(k)} \left[c_{m_{k}+2}^{(k)}\right]^2\,   \sin_{m_{k+1}}(\phi) \sin_{m_k}(\phi)\, M_e^{(0)} \left(- i H_\pm\right)^{-1}\ketbra{m_{k+1}\!+\!2}{m_{k}}
+  \left(c_{\rm e}  c_{\rm g}^*\right)^2 \,c_{m_{k+1}}^{(k)} \left[c_{m_{k}+2}^{(k)}\right]^*\,   \sin_{m_{k+1}}(\phi) \sin_{m_k}(\phi)\,\eta_{k}^{+-}\,\ketbra{\Psi_{k+1}}{\Psi_{k-1}}
\\\nonumber
&&\quad\phantom{=}
+ \text{H.c.}.
\end{eqnarray}
%trace-preservation!
where we introduced $\eta_{k}^{-+}=\mp c_{\rm g}^{-1}\langle \Psi_{k-1}|\Pi_0[M_g^{(0)}\left(- i H_\pm\right)^{-1}\ketbra{m_{k}}{m_{k+1}\!+\!2}]| \Psi_{k+1}\rangle$,  $\eta_{k}^{+-}=\pm c_{\rm e}^{-1}\langle \Psi_{k+1}|\Pi_0[M_e^{(0)}\left(- i H_\pm\right)^{-1}\ketbra{m_{k+1}\!+\!2}{m_{k}}]| \Psi_{k-1}\rangle$, and $\eta_{k}^{\pm\mp}=0$ if the pure stationary states $|\Psi_{k-1}\rangle$ and  $|\Psi_{k+1}\rangle$ do not exist. We also assumed that the pure states  $|\Psi_{k-2}\rangle$ and  $|\Psi_{k+2}\rangle$ with same boundary condition as $|\Psi_k\rangle$ exist; otherwise the terms with corresponding coherences are absent in Eq.~\eqref{eq:Leff_soft_wall2d}. To derive first, fourth and fifth lines, we used the fact that the projection $\Pi_0$ on the states between the hard walls is given by the supports between the walls, and in the second and third lines, that the projection $\Pi_0$ of the coherence to the dark state reduces to the orthogonal projection on dark states. 

Similarly, for the coherences between states $\ket{\Psi_{k_1}}$ and $\ket{\Psi_{k_2}}$ with the same boundary conditions [cf.~Eq.~\eqref{eq:Leff_soft_wall1b}]
\begin{eqnarray}\label{eq:Leff_soft_wall2e}
&&\nu^{-1}\Pi_0\delta\mathcal{L}_1 \mathcal{S}_0 \delta\mathcal{L}_1 \left(\ketbra{\Psi_{k_1}}{\Psi_{k_2}} \right)=\\\nonumber
&&\quad=    |c_{\rm g}c_{\rm e}^*|^2\,|c_{m_{k_1}+2}^{(k_1)}|^2 \,    \sin_{m_{k_1}}^2\!(\phi)\, \bra{m_{k_1}}\!\left(- i H_\pm\right)^{-1}\!\ket{m_{k_1}}  \ketbra{\Psi_{k_1}}{\Psi_{k_2}}\\\nonumber
&&\quad\phantom{=}  +|c_{\rm g}c_{\rm e}^*|^2\,|c_{m_{k_1+1}}^{(k_1)}|^2 \,    \sin_{m_{k_1+1}}^2\!(\phi)\, \bra{m_{k_1+1}\!+\!2}\!\left(- i H_\pm\right)^{-1}\!\ket{m_{k_1+1}\!+\!2}  \ketbra{\Psi_{k_1}}{\Psi_{k_2}}\\\nonumber
&&\quad\phantom{=}
-  (c_{\rm g}c_{\rm e}^*)^2|\,\left[c_{m_{k_1-1}}^{(k_1-2)} \right]^* c_{m_{k_1}+2}^{(k_1)} \,  \sin_{m_{k_1-1}}(\phi)  \sin_{m_{k_1}}(\phi)\, \bra{m_{k_1-1}\!+\!2}\!\left(- i H_\pm\right)^{-1}\!\ket{m_{k_1}} \ketbra{\Psi_{k_1-2}}{\Psi_{k_2}} \\\nonumber
&&\quad\phantom{=} - (c_{\rm e}  c_{\rm g}^*)^2 \,\left[c_{m_{k_1+2}+2}^{(k_1+2)}\right]^*c_{m_{k_1+1}}^{(k_1)} \,  \sin_{m_{k_1+2}}(\phi) \sin_{m_{k_1+1}}(\phi)\, \bra{m_{k_1+2}}\!\left(- i H_\pm\right)^{-1}\!\ket{m_{k_1+1}\!+\!2}\ketbra{\Psi_{k_1+2}}{\Psi_{k_2}}
\\\nonumber
&&\quad\phantom{=} -  |c_{\rm g}c_{\rm e}^*|^2\, c_{m_{k_1}+2}^{(k_1)} [c_{m_{k_2}+2}^{(k_2)}]^* \,    \sin_{m_{k_1}}(\phi)\sin_{m_{k_2}}(\phi)\,   \eta_{k_1,k_2}^{--}\,\ketbra{\Psi_{k_1-1}}{\Psi_{k_2-1}}\\\nonumber
&&\quad\phantom{=}- |c_{\rm e}c_{\rm g}^*|^2   \,c_{m_{k_1+1}}^{(k_1)} [c_{m_{k_2+1}}^{(k_2)}]^* \,   \sin_{m_{k_1+1}}(\phi)\sin_{m_{k_2+1}}(\phi)\, \,\eta_{k_1,k_2}^{++}\,\ketbra{\Psi_{k_1+1}}{\Psi_{k_2+1}}\\\nonumber
&&\quad\phantom{=}
%\mp  c_{\rm g}(c_{\rm e}^*)^2\,c_{m_{k}+2}^{(k)} \left[c_{m_{k+1}}^{(k)}\right]\,    \sin_{m_k}(\phi)\sin_{m_{k+1}}(\phi)\, M_g^{(0)}\left(- i H_\pm\right)^{-1}\ketbra{m_{k}}{m_{k+1}\!+\!2}\\\nonumber
+ \left(c_{\rm g}c_{\rm e}^*\right)^2\,c_{m_{k_1}+2}^{(k_1)} \left[c_{m_{k_2+1}}^{(k_2)}\right]^*\,    \sin_{m_{k_1}}(\phi)\sin_{m_{k_2+1}}(\phi)\,  \eta_{k_1,k_2}^{-+}\, \ketbra{\Psi_{k_1-1}}{\Psi_{k_2+1}}\\\nonumber
&&\quad\phantom{=}
%\pm  c_{\rm e}  (c_{\rm g}^*)^2 \,c_{m_{k+1}}^{(k)} \left[c_{m_{k}+2}^{(k)}\right]^2\,   \sin_{m_{k+1}}(\phi) \sin_{m_k}(\phi)\, M_e^{(0)} \left(- i H_\pm\right)^{-1}\ketbra{m_{k+1}\!+\!2}{m_{k}}
+ \left(c_{\rm e}  c_{\rm g}^*\right)^2 \,c_{m_{k_1+1}}^{(k_1)} \left[c_{m_{k_2}+2}^{(k_2)}\right]^*\,   \sin_{m_{k_1+1}}(\phi) \sin_{m_{k_2}}(\phi)\,\eta_{k_1,k_2}^{+-}\,\ketbra{\Psi_{k_1+1}}{\Psi_{k_2-1}}
\\\nonumber
&&\quad\phantom{=}
+ (\text{H.c.})^{k_1\leftrightarrow\, k_2}.
\end{eqnarray}
where we introduced $\eta_{k_1,k_2}^{++}=\pm c_{\rm e}^{-1}\langle\Psi_{k_2+1}|\Pi_0[M_e^{(0)}(- i H_\pm )^{-1}|m_{k_1+1}\!+\!2\rangle\!\langle m_{k_2+1}\!+\!2|] \Psi_{k_2+1}\rangle$, $\eta_{k_1,k_2}^{--}=\mp c_{\rm g}^{-1}\langle\Psi_{k_2-1}|\Pi_0[(- i H_\pm )^{-1}|m_{k_1}\rangle\!\langle m_{k_2}|] \Psi_{k_2-1}\rangle$,  $\eta_{k_1,k_2}^{-+}=\mp c_{\rm g}^{-1}\langle \Psi_{k_1-1}|\Pi_0[M_g^{(0)}\left(- i H_\pm\right)^{-1}\ketbra{m_{k_2}}{m_{k_1+1}\!+\!2}]| \Psi_{k_2+1}\rangle$ and $\eta_{k_1,k_2}^{+-}=\pm c_{\rm e}^{-1}\langle \Psi_{k_1+1}|\Pi_0[M_e^{(0)}\left(- i H_\pm\right)^{-1}\ketbra{m_{k_1+1}\!+\!2}{m_{k_2}}]| \Psi_{k_2-1}\rangle$, while $(\text{H.c.})^{k_1\leftrightarrow\, k_2}$ denotes the Hermitian conjugate but with swapped indices $k_1$ and $k_2$. 
 
Finally, for the mixed state $\rho_k$
\begin{eqnarray}\label{eq:Leff_soft_wall2f}
\nu^{-1}\Pi_0\delta\mathcal{L}_1 \mathcal{S}_0 \delta\mathcal{L}_1 \left(\rho_k\right)&=&\mp  |c_{\rm g}|^2\,c_{\rm e}\, \sin_{m_k}^2(\phi)\,\langle m_k| \mathcal{S}_0\left[|m_k\rangle \langle m_{k}\!+\!2|\rho_k [M_e^{(0)}]^\dagger\right] |m_k\!+\!2\rangle \,\left(\rho_k -\rho_{k-1}\right)\\\nonumber
&&\pm |c_{\rm e}|^2  c_{\rm g}  \,   \sin_{m_{k+1}}^2(\phi)\,\langle m_{k+1}\!+\!2| \mathcal{S}_0\left[|m_{k+1}\!+\!2\rangle \langle m_{k+1} |\rho_k  [M_g^{(0)}]^\dagger\right] |m_{k+1}\rangle\,\left(\rho_k -\rho_{k+1}\right)+ \text{H.c.}.
\end{eqnarray}
%trace-preservation!
%no coherences can be stationary in the neighbours, if the mixed boundary conditions! 

~\\
\emph{Additional information from stationary state}. Although in Eqs.~(\ref{eq:Leff_soft_wall2d})-(\ref{eq:Leff_soft_wall2f}) we do not give closed formulas for  the terms corresponding to the resolvent (with $H_{\pm}$ or $\mathcal{S}_0$) and the projection on the coherences, the knowledge of the stationary state in Eqs.~\eqref{eq:even_odd_ss} and~\eqref{eq:psi_soft_wall_limit} can be used to further determine the second-order corrections to the dynamics across soft walls. Namely, the condition $\mathcal{L}_\text{eff}\rho_\text{ss}=0$, gives $D$ conditions on the effective second-order dynamics $\mathcal{L}_\text{eff}$, where $D$ is the dimension of the subspace, on which the dynamics takes place.

~\\
\emph{Example of two walls.} We consider Case 3. from Table \ref{tab:2walls}, where we have three pure stationary states among the walls  
$|\Psi_0\rangle$,  $|\overline\Psi_0\rangle$, and $|\Psi_2\rangle$, and the coherences $|\Psi_0\rangle\!\langle \Psi_2|$ and $|\Psi_2\rangle\!\langle \Psi_0|$ are also stationary ($D=5$) (cf.~Sec.~\ref{sec:DFS}).
\\

We have [cf.~Eq.~\eqref{eq:Leff_soft_wall2a}] $\Pi_0 \delta \mathcal{L}_2 (\ketbra{\Psi_0}{\Psi_0},\ketbra{\Psi_1}{\Psi_1},\ketbra{\Psi_2}{\Psi_2})=$
\begin{eqnarray}
 \nu \left[\begin{array}{ccc}
 - |c_{\rm e}|^2 \,\sin_{m_{1}}^2(\phi)\,   |c_{m_{1}}^{(0)}|^2&\phantom{-} |c_{\rm g}|^2 \,\sin_{m_{1}}^2(\phi)\,   |c_{m_{1}+2}^{(1)}|^2&0\\
 \phantom{-} |c_{\rm e}|^2 \,\sin_{m_{1}}^2(\phi)\,   |c_{m_{1}}^{(0)}|^2&- |c_{\rm g}|^2 \,\sin_{m_{1}}^2(\phi)\,   |c_{m_{1}+2}^{(1)}|^2- |c_{\rm e}|^2 \,\sin_{m_{2}}^2(\phi)\,   |c_{m_{2}}^{(1)}|^2&\phantom{-}|c_{\rm g}|^2 \,\sin_{m_{2}}^2(\phi)\,   |c_{m_{2}+2}^{(2)}|^2\\0&\phantom{-} |c_{\rm e}|^2 \,\sin_{m_{2}}^2(\phi)\,   |c_{m_{2}}^{(1)}|^2& -|c_{\rm g}|^2 \,\sin_{m_{2}}^2(\phi)\,   |c_{m_{2}+2}^{(2)}|^2 \end{array}\right] \left(\begin{array}{c}\ketbra{\Psi_0}{\Psi_0}\\\ketbra{\Psi_1}{\Psi_1}\\\ketbra{\Psi_2}{\Psi_2}\end{array}\right),\qquad\quad
\end{eqnarray}
and [cf.~Eq.~\eqref{eq:Leff_soft_wall2b}]
\begin{eqnarray}
 \,\Pi_0\delta \mathcal{L}_2 \left(\ketbra{\Psi_{0}}{\Psi_{2}}\right)&=&  -\frac{\nu}{2} \left[ |c_{\rm e}|^2\,\sin_{m_{1}}^2(\phi)\,|c_{m_{1}}^{(0)}|^2 +|c_{\rm g}|^2 \,\sin_{m_{2}}^2(\phi)\, |c_{m_{2}+2}^{(2)}|^2\right] \ketbra{\Psi_{0}}{\Psi_{2}}.
\end{eqnarray}
On the other hand, [cf.~Eq.~\eqref{eq:Leff_soft_wall2d}]
\begin{eqnarray}\label{eq:Leff_soft_wall2d_0}
&&\nu^{-1}\Pi_0\delta\mathcal{L}_1 \mathcal{S}_0 \delta\mathcal{L}_1 \left(\ketbra{\Psi_{0}}{\Psi_{0}} \right)=\\nonumber
&&\quad=   2 |c_{\rm g}c_{\rm e}^*|^2\,|c_{m_{1}}^{(0)}|^2 \,    \sin_{m_{1}}^2\!(\phi)\, \bra{m_{1}\!+\!2}\!\left(- i H_+\right)^{-1}\!\ket{m_{1}\!+\!2} \left(\ketbra{\Psi_{0}}{\Psi_{0}}-\ketbra{\Psi_{1}}{\Psi_{1}}\right)\\\nonumber
&&\quad\phantom{=} -\left\{ (c_{\rm e}  c_{\rm g}^*)^2 \,\left[c_{m_{2}+2}^{(2)}\right]^*c_{m_{1}}^{(0)} \,  \sin_{m_{2}}(\phi) \sin_{m_{1}}(\phi)\, \bra{m_{2}}\!\left(- i H_+\right)^{-1}\!\ket{m_{1}\!+\!2}\ketbra{\Psi_{2}}{\Psi_{0}} +\text{H.c.}\right\},
\end{eqnarray}
\begin{eqnarray}\label{eq:Leff_soft_wall2d_1}
&&\nu^{-1}\Pi_0\delta\mathcal{L}_1 \mathcal{S}_0 \delta\mathcal{L}_1 \left(\ketbra{\Psi_{1}}{\Psi_{1}} \right)\\\nonumber
&&\quad=   2 \left[|c_{\rm g}c_{\rm e}^*|^2\,|c_{m_{1}+2}^{(1)}|^2 \,    \sin_{m_1}^2\!(\phi)\, \bra{m_{1}}\!\left(- i H_-\right)^{-1}\!\ket{m_{1}}  +|c_{\rm g}c_{\rm e}^*|^2\,|c_{m_{2}}^{(1)}|^2 \,    \sin_{m_{2}}^2\!(\phi)\, \bra{m_{2}\!+\!2}\!\left(- i H_-\right)^{-1}\!\ket{m_{2}\!+\!2}  \right]\ketbra{\Psi_{1}}{\Psi_{1}}\\\nonumber
&&\quad\phantom{=} - 2 |c_{\rm g}c_{\rm e}^*|^2\,|c_{m_{1}+2}^{(1)}|^2 \,    \sin_{m_1}^2(\phi)\, \langle m_{1}|\left(- i H_-\right)^{-1}|m_{1}\rangle\,\ketbra{\Psi_{0}}{\Psi_{0}}\\\nonumber
&&\quad\phantom{=}-2 |c_{\rm e}c_{\rm g}^*|^2   \,|c_{m_{2}}^{(1)}|^2 \,   \sin_{m_{2}}^2(\phi)\, \langle m_{2}\!+\!2|\left(- i H_-\right)^{-1}|m_{2}\!+\!2\rangle\,\ketbra{\Psi_{2}}{\Psi_{2}}\\\nonumber
&&\quad\phantom{=}
%\mp  c_{\rm g}(c_{\rm e}^*)^2\,c_{m_{k}+2}^{(k)} \left[c_{m_{k+1}}^{(k)}\right]\,    \sin_{m_k}(\phi)\sin_{m_{k+1}}(\phi)\, M_g^{(0)}\left(- i H_\pm\right)^{-1}\ketbra{m_{k}}{m_{k+1}\!+\!2}\\\nonumber
+  2\left(c_{\rm g}c_{\rm e}^*\right)^2\,c_{m_{1}+2}^{(1)} \left[c_{m_{2}}^{(1)}\right]^*\,    \sin_{m_1}(\phi)\sin_{m_{2}}(\phi)\,  \eta_{1}^{-+}\, \ketbra{\Psi_{0}}{\Psi_{2}}\\\nonumber
&&\quad\phantom{=}
%\pm  c_{\rm e}  (c_{\rm g}^*)^2 \,c_{m_{k+1}}^{(k)} \left[c_{m_{k}+2}^{(k)}\right]^2\,   \sin_{m_{k+1}}(\phi) \sin_{m_k}(\phi)\, M_e^{(0)} \left(- i H_\pm\right)^{-1}\ketbra{m_{k+1}\!+\!2}{m_{k}}
+ 2 \left(c_{\rm e}  c_{\rm g}^*\right)^2 \,c_{m_{2}}^{(1)} \left[c_{m_{1}+2}^{(1)}\right]^*\,   \sin_{m_{2}}(\phi) \sin_{m_1}(\phi)\,\eta_{1}^{+-}\,\ketbra{\Psi_{2}}{\Psi_{0}},
\end{eqnarray}
and
\begin{eqnarray}\label{eq:Leff_soft_wall2d_2}
&&\nu^{-1}\Pi_0\delta\mathcal{L}_1 \mathcal{S}_0 \delta\mathcal{L}_1 \left(\ketbra{\Psi_{2}}{\Psi_{2}} \right)\\\nonumber
&&\quad=  2  |c_{\rm g}c_{\rm e}^*|^2\,|c_{m_{2}+2}^{(2)}|^2 \,    \sin_{m_2}^2\!(\phi)\, \bra{m_{2}}\!\left(- i H_+\right)^{-1}\!\ket{m_{2}}\left( \ketbra{\Psi_{2}}{\Psi_{2}}-\ketbra{\Psi_{1}}{\Psi_{1}}\right)\\\nonumber
&&\quad\phantom{=} - \left\{ (c_{\rm g}c_{\rm e}^*)^2\,\left[c_{m_{1}}^{(0)} \right]^* c_{m_{2}+2}^{(2)} \,  \sin_{m_{1}}(\phi)  \sin_{m_2}(\phi)\, \bra{m_{1}\!+\!2}\!\left(- i H_+\right)^{-1}\!\ket{m_{2}} \ketbra{\Psi_{0}}{\Psi_{2}} +\text{H.c.}\right\},
\end{eqnarray}
while
\begin{eqnarray}\label{eq:Leff_soft_wall2d_02}
&&\nu^{-1}\Pi_0\delta\mathcal{L}_1 \mathcal{S}_0 \delta\mathcal{L}_1 \left(\ketbra{\Psi_{0}}{\Psi_{2}} \right)\\\nonumber
&&\quad=  |c_{\rm g}c_{\rm e}^*|^2 \left[ |c_{m_{2}+2}^{(2)}|^2 \,    \sin_{m_{2}}^2\!(\phi)\, \bra{m_{2}}\!\left(- i H_+\right)^{-1}\!\ket{m_{2}} +|c_{m_{1}}^{(0)}|^2 \,    \sin_{m_{1}}^2\!(\phi)\, \bra{m_{1}\!+\!2}\!\left(- i H_+\right)^{-1}\!\ket{m_{1}\!+\!2} \right] \ketbra{\Psi_{0}}{\Psi_{2}}\\\nonumber
&&\quad\phantom{=} - \left(c_{\rm e}  c_{\rm g}^*\right)^2 \,c_{m_{1}}^{(0)}\left[c_{m_{2}+2}^{(2)}\right]^* \,  \sin_{m_{1}}(\phi)\sin_{m_{2}}(\phi) \, \bra{m_{2}}\!\left(- i H_+\right)^{-1}\!\ket{m_{1}\!+\!2} \left( \ketbra{\Psi_{0}}{\Psi_{0}}+\ketbra{\Psi_{2}}{\Psi_{2}} \right)
\\\nonumber
&&\quad\phantom{=}
%\pm  c_{\rm e}  (c_{\rm g}^*)^2 \,c_{m_{k+1}}^{(k)} \left[c_{m_{k}+2}^{(k)}\right]^2\,   \sin_{m_{k+1}}(\phi) \sin_{m_k}(\phi)\, M_e^{(0)} \left(- i H_\pm\right)^{-1}\ketbra{m_{k+1}\!+\!2}{m_{k}}
+ 2 \left(c_{\rm e}  c_{\rm g}^*\right)^2 \,c_{m_{1}}^{(0)} \left[c_{m_{2}+2}^{(2)}\right]^*\,   \sin_{m_{1}}(\phi) \sin_{m_{2}}(\phi)\,\bra{m_{2}}\!\left(- i H_+\right)^{-1}\!\ket{m_{1}\!+\!2}\,\ketbra{\Psi_{1}}{\Psi_{1}},
\end{eqnarray}
cf.~Eq.~\eqref{eq:Leff_soft_wall2e}. In the above expression, we used the fact that $- i H_\pm$ is Hermitian [cf.~Eq.~\eqref{eq:Heff_hard_wall}].

The stationary state is [cf.~Table~\ref{tab:2walls}]
\begin{equation}\label{eq:rhoss_soft_wall}
\rho_\text{ss}=|\beta_0|^2 \ketbra{\Psi_{0}}{\Psi_{0}} + |\beta_2|^2 \ketbra{\Psi_{2}}{\Psi_{2}} + \beta_0\beta_2^* \ketbra{\Psi_{0}}{\Psi_{2}}+ \beta_2\beta_0^*\ketbra{\Psi_{2}}{\Psi_{0}},
\end{equation}
where $|\beta_0|^2+|\beta_2|^2=1$. Therefore, from Eq.~\eqref{eq:Leff_2ndorder}, we have $\Pi_0\delta\mathcal{L}_2(\rho_\text{ss}) = \Pi_0\delta\mathcal{L}_1 \mathcal{S}_0 \delta\mathcal{L}_1 (\rho_\text{ss} )$, which can be written as
\begin{eqnarray}
\small
Y
\left[\begin{array}{c}
\sin_{m_1}^2\!(\phi)\bra{m_{1}\!+\!2}\!\left(- i H_+\right)^{-1}\!\ket{m_{1}\!+\!2} \\
\sin_{m_{2}}^2\!(\phi) \bra{m_{2}}\!\left(- i H_+\right)^{-1}\!\ket{m_{2}} \\
\sin_{m_{1}}(\phi)\sin_{m_{2}}(\phi)\bra{m_{2}}\!\left(- i H_+\right)^{-1}\!\ket{m_{1}\!+\!2} \\
\sin_{m_{1}}(\phi)\sin_{m_{2}}(\phi) \bra{m_{1}\!+\!2}\!\left(- i H_+\right)^{-1}\!\ket{m_{2}}  \end{array}\right]
&=& 
\left[\begin{array}{c} - |c_{\rm e}|^2 \,\sin_{m_{1}}^2(\phi) |c_{m_{1}}^{(0)}|^2 |\beta_0|^2\\ - \left[ |c_{\rm e}|^2\,\sin_{m_{1}}^2(\phi)\,|c_{m_{1}}^{(0)}|^2 +|c_{\rm g}|^2 \,\sin_{m_{2}}^2(\phi)\, |c_{m_{2}+2}^{(2)}|^2\right]\beta_0\beta_2^*/2\\ |c_{\rm e}|^2 \,\sin_{m_{1}}^2(\phi) |c_{m_{1}}^{(0)}|^2 |\beta_0|^2 +|c_{\rm g}|^2 \,\sin_{m_{2}}^2(\phi)\,   |c_{m_{2}+2}^{(2)}|^2 |\beta_2|^2 \\  - \left[ |c_{\rm e}|^2\,\sin_{m_{1}}^2(\phi)\,|c_{m_{1}}^{(0)}|^2 +|c_{\rm g}|^2 \,\sin_{m_{2}}^2(\phi)\, |c_{m_{2}+2}^{(2)}|^2\right]\beta_2\beta_0^*/2\\ -|c_{\rm g}|^2 \,\sin_{m_{2}}^2(\phi)\,   |c_{m_{2}+2}^{(2)}|^2 |\beta_2|^2\end{array}\right]\!\!,\qquad
\end{eqnarray}
where 
\begin{eqnarray}
%\small
Y=\left[\begin{array}{cccc}
2 |c_{\rm g}c_{\rm e}^*|^2 |c_{m_{1}+2}^{(1)}|^2       |\beta_0|^2 &0&-  (c_{\rm e}  c_{\rm g}^* )^2  c_{m_{1}}^{(0)} [c_{m_{2}+2}^{(2)} ]^*     \beta_0\beta_2^*&-  (c_{\rm e}^*  c_{\rm g} )^2   [c_{m_{1}}^{(0)} ]^*c_{m_{2}+2}^{(2)}    \beta_2\beta_0^*\\
|c_{\rm g}c_{\rm e}^*|^2|c_{m_{1}}^{(0)}|^2     \beta_0\beta_2^* &|c_{\rm g}c_{\rm e}^*|^2|c_{m_{2}+2}^{(2)}|^2      \beta_0\beta_2^*&0&- (c_{\rm e}^*  c_{\rm g})^2   [c_{m_{1}}^{(0)}  ]^*c_{m_{2}+2}^{(2)}    \\
-2 |c_{\rm g}c_{\rm e}^*|^2 |c_{m_{1}+2}^{(1)}|^2     |\beta_0|^2&-2  |c_{\rm g}c_{\rm e}^*|^2 |c_{m_{2}+2}^{(2)}|^2    |\beta_2|^2&2  (c_{\rm e}  c_{\rm g}^* )^2  c_{m_{1}}^{(0)}  [c_{m_{2}+2}^{(2)} ]^*     \beta_0\beta_2^*&2  (c_{\rm e} ^* c_{\rm g} )^2   [c_{m_{1}}^{(0)} ]^* c_{m_{2}+2}^{(2)}    \beta_2\beta_0^*\\
|c_{\rm g}c_{\rm e}^*|^2|c_{m_{1}}^{(0)}|^2       \beta_2\beta_0^* &|c_{\rm g}c_{\rm e}^*|^2|c_{m_{2}+2}^{(2)}|^2       \beta_2\beta_0^*&- (c_{\rm e}  c_{\rm g}^*)^2   c_{m_{1}}^{(0)}  [c_{m_{2}+2}^{(2)}]^*    &0\\
0&2  |c_{\rm g}c_{\rm e}^*|^2 |c_{m_{2}+2}^{(2)}|^2  |\beta_2|^2&-  (c_{\rm e}  c_{\rm g}^* )^2  c_{m_{1}}^{(0)} [c_{m_{2}+2}^{(2)} ]^*     \beta_0\beta_2^*&-  (c_{\rm e}^*  c_{\rm g} )^2   [c_{m_{1}}^{(0)} ]^*c_{m_{2}+2}^{(2)}    \beta_2\beta_0^*\\ \end{array}\right]\!\!,\qquad
\end{eqnarray}
so that we can find analytically the columns of the dynamics generator that correspond to the support of the stationary state~\eqref{eq:rhoss_soft_wall} [cf.~Eqs.~\eqref{eq:Leff_soft_wall2d_0},~\eqref{eq:Leff_soft_wall2d_2}, and~\eqref{eq:Leff_soft_wall2d_02}].

	\section{Review of metastability theory}
	\label{app:meta_theory}
	Here we summarize the properties of Markovian dynamics of open quantum systems which lead to metastability~\cite{macieszczak2016towards,Rose2016}.

	\subsection{Markovian dynamics}
	 We consider an open quantum system dynamics described by a Markovian master equation~\cite{Lindblad1976,Gorini1976}, 
	\begin{equation} \label{eq:master_general}
	\frac{ d }{ d t}\rho(t) = \mathcal{L} [\rho(t)] = - i [H,\rho(t)] +\frac{1}{2}\sum_j \left [ 2\, J_j \,\rho(t)\, J_j^\dagger - J_j^\dagger J_j\, \rho(t) - \rho(t) \,  J_j^\dagger J_j \right],
	\end{equation} 
	where $H$ is the system Hamiltonian and $J_j$ denote so called jump operators which describe the interaction of the system with the environment. In the case of the dynamics of micromasers, Eq.~\eqref{eq:master}, the system is the cavity which interacts with the environment constituted by passing atoms. The Hamiltonian $H=0$ (dynamics is considered in the rotating frame with the Hamiltonian as explained in~Appendix~\ref{app:micromaser}), while the jump operators are given by the Kraus operators, Eq.~\eqref{eq:KrausFull}. 
	
	Timescales of the dynamics in~\eqref{eq:master_general} are given by the spectrum of the  master operator $\mathcal{L}$. 
	Although in general $\mathcal{L}$ is not Hermitian, and thus not necessarily diagonalizable, in all studied cases it could be diagonalized. We label the corresponding eigenvalues as $\{\lambda_{k}\}_{k\geq 1}$, ordered in the decreasing order of their real part, ${\rm Re}\, \lambda_1 \geq {\rm Re}\, \lambda_2 \geq ...$, and the corresponding left- and right-eigenmodes $L_k$ and $R_k$, $\mathcal{L} R_k = \lambda_k R_k$, $ L_k\mathcal{L}=\lambda_k L_k $ [normalized as $\mathrm{Tr}(L_j R_k)=\delta_{jk}$]. For an initial state $\rho$ we have that the system state at time $t$ is given by
	\begin{equation}\label{eq:Lseries}
	\rho(t) = e^{t {\mathcal{L}}}(\rho) = \rho_{\rm ss} + \sum_{k \geq 2}e^{t\lambda_{k}} \,\text{Tr}\left(L_k\, \rho\right) R_{k} ,
	\end{equation}
	where we used the fact that $\lambda_1=0$, which corresponds to a stationary state $R_1 =\rho_\text{ss}$, and $L_1=\mathds{1}$ due to trace preservation. When the stationary state is unique, $\rho(t)$ relaxes to $\rho_{\rm ss}$ at the timescale given by the inverse of the \emph{gap} to the second eigenvalue, $\tau=(-\mathrm{Re}\lambda_2)^{-1}$.
	
	\subsection{Metastability}
		 When there exists a separation between real parts of the eigenvalues,  $-\mathrm{Re}\lambda_m \ll -\mathrm{Re}\lambda_{m+1}$, there exists a time regime $(-\mathrm{Re}\lambda_{m+1})^{-1} \ll t\ll (-\mathrm{Re}\lambda_{m})^{-1}$, where after the initial fast relaxation of modes $k>m$, the system state appears steady, i.e., is \emph{metastable}, and can be approximated as [cf.~Eq.~\eqref{eq:Lseries}]
	\begin{equation}\label{eq:ms_general}
	\rho(t) \approx \rho_{\rm ss} + \sum_{k = 2}^m \text{Tr}\left(L_k\, \rho\right) R_{k} \equiv \Pi\ken (\rho),
	\end{equation}
	where we denoted by $\Pi$ the projection on the low-lying eigenmodes of the dynamics. The manifold of metastable states is  described by the coefficients $\{ \mathrm{Tr}\left(L_k\, \rho\right) \}_{k=2}^m$ that depend on the initial state $\rho$, and thus this manifold is $(m-1)$-dimensional.  Beyond the metastable regime, $t \gtrsim(-\text{Re}\lambda_m)^{-1}$, the decay of low-lying eigenmodes can no longer be neglected, and the system undergoes final relaxation inside the metastable manifold [cf.~Eq.~\eqref{eq:Lseries}]
	\begin{equation}\label{eq:ms_Leff}
	\rho(t) \approx \rho_{\rm ss} + \sum_{k = 2}^m e^{t\lambda_{k}} \,\text{Tr}\left(L_k\, \rho\right) R_{k}=e^{t \,	\mathcal{L}_\text{eff}}\,\Pi\, (\rho), 
	\end{equation}	
	which is governed by the low-lying modes as 
	\begin{equation}\label{eq:PLP}
	\mathcal{L}_\text{eff}=\Pi \,\mathcal{L}\,\Pi.
	\end{equation}    We note that several metastable regimes can exist if there are multiple separations in the spectrum of $\mathcal{L}$, which leads to hierarchy of the corresponding metastable manifolds. 	In Appendix~\ref{app:Leff}, we consider the case in which metastability is a consequence of perturbing dynamics which features multiple stationary states.

	%===================================================================================================================================================
	
	\section{Derivations of metastable dynamics}
	\label{app:Leff}

	Here we consider metastability and effective long-time dynamics in the case of perturbing the dynamics which features multiple stationary states. We derive the effective dynamics due to parity-conserving and parity-swapping perturbations, which leads to Eqs.~\eqref{eq:Leff_corr} and~\eqref{eq:Leff_loss}. We also discuss the corresponding dynamics in the presence of hard walls.

		\subsection{Metastability due to perturbations of multiple stationary states}
		 One class of open quantum dynamics where metastability arises is the case when the dynamics $\mathcal{L}_0$, which features multiple stationary states, is perturbed by  $\delta \mathcal{L}$, i.e., $\mathcal{L}=\mathcal{L}_0+\delta \mathcal{L}$.  By means of non-Hermitian perturbation theory, it can be shown~\cite{Kato1995}, that the slow (low-lying) eigenmodes which contribute to the metastable states, Eq.~\eqref{eq:ms_general}, correspond to the stationary states of $\mathcal{L}_0$,
	\begin{equation} \label{eq:P}
	\Pi=\Pi_0 +... %\qquad \text{so\,that}\qquad \rho_\text{ms}=\Pi_0\ken \rho+...\,
	\end{equation}
	where $\Pi_0$ is the projection on the stationary states of $\mathcal{L}_0$. Furthermore, the effective long-time dynamics, Eq.~\eqref{eq:PLP}, is  well approximated by 
	\begin{equation} 
	\Pi\,\mathcal{L}\,\Pi =\Pi_0\,\delta \mathcal{L}\,\Pi_0  +... ,\label{eq:1storder}
	\end{equation}
	which corresponds to completely positive and trace-preserving dynamics of the metastable states~\cite{zanardi_geometry_2015,zanardi_dissipative_2016,macieszczak2016towards,azouit2016adiabatic}.

	 In this work, we consider dynamics of the cavity, $\mathcal{L}_0$ in~\eqref{eq:L0}, which conserves the parity $P=(-1)^{a^\dagger a}$, Eq.~\eqref{eq:parity}, and features a stationary DFS spanned by states $|\Psi_{+}\rangle$ and $|\Psi_{-}\rangle$ of the opposite parity. In this case, the projection on the stationary subspace also conserves the parity, and is given by
	\begin{equation}
	\Pi_0 (\rho)=\ketbra{\Psi_+}{\Psi_+}\mathrm{Tr}(\mathds{1}_+ \rho)+\ketbra{\Psi_-}{\Psi_-}\mathrm{Tr}(\mathds{1}_- \rho)+\ketbra{\Psi_+}{\Psi_-}\mathrm{Tr}(L_{+-} \rho)+\ketbra{\Psi_-}{\Psi_+}\mathrm{Tr}(L_{-+} \rho). \label{eq:P0}
	\end{equation}
	where $\mathds{1}_-$ and $\mathds{1}_+$  are identity operators on the odd and even subspace, while $L_{+-}=L_{-+}^\dagger$ is a conserved quantity supported in odd-even coherences;  see~Sec.~\ref{sec:DFS}. For discussion of metastability in the case with hard wall in the dynamics, see Appendix~\ref{app:Leff:hard_walls}.
	
	%===================================================================================================================================================
%===================================================================================================================================================

%===================================================================================================================================================
%===================================================================================================================================================

	\subsection{Metastable dynamics with weak parity symmetry}
	\label{app:Leff:weak}
	
	Here we derive Eqs.~\eqref{eq:Leff_corr} and~\eqref{eq:Leff_loss}.

	\subsubsection{General results}
	\noindent
	\emph{Dynamics with weak parity symmetry}. We consider a perturbation by the purely dissipative dynamics with jumps $J$ [cf.~Eq.~\eqref{eq:master_general}]
	\begin{equation} \label{eq:Lprime}
	 \delta\mathcal{L}\,(\rho)=J\rho J^\dagger-\frac{1}{2} J^\dagger J\, \rho-\frac{1}{2}\rho\,J^\dagger J.
	 \end{equation}
	 We furthermore assume that the action of a jump $J$ flips (swaps) the cavity parity $P=(-1)^{a^\dagger a}$, 
	 \begin{equation} \label{eq:flip}
	 J\,P + P\, J=0,
	 \end{equation}
	 as is the case for a single-photon loss $J=\sqrt{\kappa_{\text{1ph}}} a$ in Eq.~\eqref{eq:L1ph}. Therefore, $\mathcal{L}=\mathcal{L}_0+ \delta\mathcal{L}$ features the weak-parity symmetry [cf.~Eqs.~\eqref{eq:weak_symmetry} and~\eqref{eq:weak_symmetry_loss}]. 
	 
	 ~\\
	 \emph{Effective dynamics}. Below we prove that the first-order dynamics due to~\eqref{eq:Lprime} is given by (in the basis $\{ \ket{\Psi_+}\bra{\Psi_+}, \ket{\Psi_-}\bra{\Psi_-}, \ket{\Psi_+}\bra{\Psi_-}, \ket{\Psi_-}\bra{\Psi_+} \}$) 
	 \begin{equation}
	 \frac{ d }{ d t}\rho(t)=
	 \left[\begin{array}{rrrr}
	 -\langle J^\dagger J\rangle_{+}& \langle J^\dagger J\rangle_{-}&0&0\\
	 \langle J^\dagger J\rangle_{+}& -\langle J^\dagger J\rangle_{-}&0&0\\
	 0&0&-\frac{1}{2}(\langle J^\dagger J\rangle_{+}+\langle J^\dagger J\rangle_{-})&\eta(\langle J^\dagger J\rangle_{+}\langle J^\dagger J\rangle_{-})^{1/2}\\ 
	 0&0&\eta^*(\langle J^\dagger J\rangle_{+}\langle J^\dagger J\rangle_{-})^{1/2}&-\frac{1}{2}(\langle J^\dagger J\rangle_{+}+\langle J^\dagger J\rangle_{-})\\
	 \end{array}\right] \rho(t), \label{eq:Leff_flip}
	 \end{equation}
	 where
	 \begin{equation}\label{eq:etaJ}
	 \eta=\frac{\mathrm{Tr}(L_{+-}  J\ketbra{\Psi_-}{\Psi_+}J^\dagger)}{(\langle J^\dagger J\rangle_{+}\langle J^\dagger J\rangle_{-})^{1/2}}\qquad \text{and}\qquad|\eta|\leq 1.
 \end{equation} 
 This gives Eq.~\eqref{eq:Leff_loss} and the dissipative contribution in Eq.~\eqref{eq:Leff_corr}. Although $L_{+-}$ is not known in general (i.e., beyond the weak-coupling limit~\cite{mirrahimi2014dynamically}), $\eta$ can be determined \emph{numerically} for a given coupling strength $\phi$ as [cf.~Eq.~\eqref{eq:P0}]
\begin{equation}
\mathrm{Tr}(L_{+-}  J\ketbra{\Psi_-}{\Psi_+}J^\dagger)=   \langle\Psi_+| \left[\Pi_0 \left(   J\ketbra{\Psi_-}{\Psi_+}J^\dagger \right) \right] |\Psi_-\rangle =
\langle\Psi_+| \left( \lim_{t\rightarrow\infty}e^{t\mathcal{L}_0}   J\ketbra{\Psi_-}{\Psi_+}J^\dagger \right) |\Psi_-\rangle.
\end{equation}

	 ~\\
	 \emph{Effective master equation}. Equation~\eqref{eq:Leff_flip} corresponds to \emph{biased bit flip noise} in the DFS,
	 \begin{eqnarray}
	 &&\frac{ d }{ d t}\rho(t)=  \sum_{j=1,2} \gamma_j \left[ s_j\,\rho(t)\, s_j^\dagger -\frac{1}{2}\left(s_j^\dagger  s_j\, \rho(t)+ \rho(t)\,s_j^\dagger  s_j\right) \right], \label{eq:master_flip}\\
	 \label{spin_flip}
	 &&s_{1,2}= \frac{e^{ i\varphi}(\epsilon+2\gamma \pm\sqrt{\epsilon ^2+4|\gamma|^2 })\ketbra{\Psi_+}{\Psi_-}+e^{- i\varphi}(-\epsilon+2\gamma \pm\sqrt{\epsilon ^2+4|\gamma|^2 })\ketbra{\Psi_-}{\Psi_+} }{N_{1,2}}.
	 \end{eqnarray}
	 Here $\gamma_{1,2}=(2\kappa\pm\sqrt{\epsilon ^2+4|\gamma|^2 }  )/4$ are the individual spin-flip rates, $N_{1,2}^2=\epsilon^2+[2\gamma \pm\sqrt{\epsilon ^2+4|\gamma|^2 })]^2$ are the normalization factors, and we have introduced: $\epsilon= \langle J^\dagger J\rangle_+-\langle J^\dagger J\rangle_-$, $\gamma= \eta (\langle J^\dagger J\rangle_+ \langle J^\dagger J\rangle_+)^{1/2}$, and the phase $e^{2 i\varphi}|\eta|=\eta$. Note that  the total dissipation rate $\kappa=(\langle J^\dagger J\rangle_++\langle J^\dagger J\rangle_-)/2$. 	
	 When $|\eta|= 1$, there is only a single jump, $s_1$. This corresponds to the case when the jump $J$ leaves the cavity state within the DFS [cf.~Eq.~\eqref{eq:etaJ}].  This takes place for single-photon losses and the cavity dynamics in the weak-coupling limit (see Sec.~\ref{sec:losses} and Refs.~\cite{minganti2016exact,Bartolo_2016,azouit2015convergence}).

	 ~\\
	 \emph{Steady state}. The effective dynamics in Eq.~\eqref{eq:Leff_flip} features a \emph{unique stationary state}, 
	 \begin{equation}
	 \rho_\text{ss}=\frac{\langle J^\dagger J\rangle_-\ketbra{\Psi_+}{\Psi_+}+\langle J^\dagger J\rangle_+\ketbra{\Psi_-}{\Psi_-}}{\langle J^\dagger J\rangle_++\langle J^\dagger J\rangle_-}, 
	 \label{eq:rho_ss_J}
	 \end{equation}
	 which approximates, in the zeroth order of the perturbation by $J$, the stationary state of the dynamics $\mathcal{L}=\mathcal{L}_0+\delta\mathcal{L}$.
	 
	 ~\\
	 \emph{Derivation of Eq.~\eqref{eq:Leff_flip}}. As $\Pi_0$ conserves the parity, the first-order corrections~\eqref{eq:1storder} must also feature the weak-parity symmetry.   Indeed, in the basis $\{\ketbra{\Psi_+}{\Psi_+}$, $\ketbra{\Psi_-}{\Psi_-}$, $\ketbra{\Psi_+}{\Psi_-}$, $\ketbra{\Psi_-}{\Psi_+} \}$, the effective dynamics  is \emph{block-diagonal}, $\frac{ d }{ d t}\rho(t)=$
	\begin{equation}\label{eq:Leff_J} 
	\scriptsize
	\left[\begin{array}{cccc}
	-\langle J^\dagger J\rangle_{+}& \mathrm{Tr}(\mathds{1}_{+}  J\ketbra{\Psi_-}{\Psi_-} J^\dagger) &0&0\\
	\mathrm{Tr}(\mathds{1}_{-}  J\ketbra{\Psi_+}{\Psi_+} J^\dagger) & -\langle J^\dagger J\rangle_{-}&0&0\\
	0&0&-\frac{1}{2}\mathrm{Tr}[L_{+-} (J^\dagger J\ketbra{\Psi_+}{\Psi_-}-\ketbra{\Psi_+}{\Psi_-}J^\dagger J)]&\mathrm{Tr}(L_{+-}  J\ketbra{\Psi_+}{\Psi_-}J^\dagger)\\ 
	0&0& \mathrm{Tr}(L_{-+}  J\ketbra{\Psi_-}{\Psi_+}J^\dagger)&-\frac{1}{2}\mathrm{Tr}[L_{-+} (J^\dagger J\ketbra{\Psi_-}{\Psi_+}-\ketbra{\Psi_-}{\Psi_+}J^\dagger J)]\\
	\end{array}\right] \rho(t). \nonumber
	\end{equation} %This matrix does not contain much information...
	The diagonal terms stem from the parity-conserving terms in~\eqref{eq:Lprime}, i.e., $(J^\dagger J\,\rho+\rho\,J^\dagger J)/2$, while the off-diagonal terms originate from the parity swap $J\rho J^\dagger$. Here, we denoted the averages  as $\langle J^\dagger J\rangle_{\pm}\equiv\mathrm{Tr}(\mathds{1}_{\pm} J^\dagger J\ketbra{\Psi_\pm}{\Psi_\pm} )=\langle\Psi_\pm|J^\dagger J|\Psi_\pm\rangle $. 
	
	We can further simplify the effective dynamics. First, from the trace preservation of Eq.~(\ref{eq:PLP}), we have that $\mathrm{Tr}(\mathds{1}_{\mp}  J\ketbra{\Psi_\pm}{\Psi_\pm} J^\dagger)=\langle J^\dagger J\rangle_{\pm}$. Second, we note that  $|\Psi_{+}\rangle$ and $|\Psi_{-}\rangle$ are the dark states of the dynamics~\eqref{eq:Kraus_shifted} and~\eqref{eq:master_noloss2}, i.e., $\widetilde{M}_g |\Psi_{\pm}\rangle =\widetilde{M}_e |\Psi_{\pm}\rangle =0$. Therefore, as the dynamics of coherences to a dark state is governed by the effective Hamiltonian of~\eqref{eq:master_noloss2},  $\frac{i}{2}(\widetilde{M}_g^\dagger \widetilde{M}_g+\widetilde{M}_e^\dagger \widetilde{M}_e)$, the projection $\Pi_0$ reduces to the orthogonal projection onto the dark states $|\Psi_{+}\rangle$, $|\Psi_{-}\rangle$~\cite{albert2016geometry} 
	\begin{equation}\label{eq:Pdark}
	\Pi_0(J^\dagger J\ketbra{\Psi_+}{\Psi_-})=\lim_{t\rightarrow\infty}e^{t\mathcal{L}_0}(J^\dagger J\ketbra{\Psi_+}{\Psi_-})=\lim_{t\rightarrow\infty}[e^{-\frac{1}{2} (\widetilde{M}_g^\dagger \widetilde{M}_g+\widetilde{M}_e^\dagger \widetilde{M}_{\rm e})t} J^\dagger J|\Psi_+\rangle ]\langle\Psi_-|=\langle J^\dagger J\rangle_{+}\ketbra{\Psi_+}{\Psi_-}.
	\end{equation}
	Finally, as the effective dynamics is completely-positive~\cite{zanardi_geometry_2015,zanardi_dissipative_2016, macieszczak2016towards,azouit2016adiabatic}, we have that [cf.~Eq.~\eqref{eq:etaJ}]
     \begin{equation}\label{eq:etaJ2}
      \mathrm{Tr}(L_{+-}  J\ketbra{\Psi_-}{\Psi_+}J^\dagger)=\eta\,(\langle J^\dagger J\rangle_{+}\langle J^\dagger J\rangle_{-})^{1/2}\qquad \text{where}\qquad|\eta|\leq 1.
      \end{equation} 
      Moreover, when $\mathcal{L}_0+\delta\mathcal{L}$ corresponds to the real dynamics (see ~Sec.~\ref{sec:dynamics}), $\eta$ is also real.

	%===================================================================================================================================================
%===================================================================================================================================================

	\subsection{Metastable dynamics with parity conservation}
	\label{app:Leff:conserved}
	
	\subsubsection{General results}
	\noindent		 
	 \emph{Effective dynamics with parity conservation}. We now consider a perturbation $\delta \mathcal{L}$ of the cavity dynamics $\mathcal{L}_0$  and assume that $\delta \mathcal{L}$  conserves the photon-number parity (see~Sec.~\ref{sec:dynamics}). As we derive below, the effective first-order dynamics in the DFS basis $\ketbra{\Psi_+}{\Psi_+}$, $\ketbra{\Psi_-}{\Psi_-}$, $\ketbra{\Psi_+}{\Psi_-}$, $\ketbra{\Psi_-}{\Psi_+}$ is
	 \emph{diagonal},
	 \begin{equation}
	 \frac{ d }{ d t}\rho(t)= 
	 \left[\begin{array}{cccc}
	 0&0&0&0\\
	 0& 0&0&0\\
	 0&0&- i\Omega-\gamma_\text{deph}&0\\ 
	 0&0&0& i\Omega-\gamma_\text{deph}\\
	 \end{array}\right] \rho(t), \label{eq:Leff_deph}
	 \end{equation}
	 where $- i\Omega-\gamma_\text{deph}=\mathrm{Tr}(L_{+-}\delta\mathcal{L}\ketbra{\Psi_+}{\Psi_-})$, which corresponds to  effective \emph{dephasing} at the rate  $\gamma_\text{deph}$ and \emph{unitary rotation} at frequency $\Omega$, along the direction of the DFS parity, 
	 \begin{eqnarray}
	 &&\frac{ d }{ d t}\rho(t)= - i\Omega\,[s_z,\,\rho(t)] + \gamma_\text{deph}\bigg[ s_z\,\rho(t)\, s_z^\dagger -\frac{1}{2}\left(s_z^\dagger  s_z\, \rho(t)+ \rho(t)\,s_z^\dagger  s_z\right)\bigg], 
	 \\
	 &&s_z=\frac{1}{\sqrt{2}}(\ketbra{\Psi_+}{\Psi_+}-\ketbra{\Psi_-}{\Psi_-}). 
	 \end{eqnarray} 
	 For discussion of metastability in the case with a hard wall in the dynamics, see Appendix~\ref{app:Leff:hard_walls}.

	 ~\\
	 \emph{Steady states}. Any dynamics conserving the parity features at least two stationary states~\cite{albert2014symmetries}, corresponding to the conserved quantities $\mathds{1}_+$ and $\mathds{1}_-$ (cf.~Sec.~\ref{sec:dynamics}). Indeed, in~\eqref{eq:Leff_deph} the even-odd coherences dephase to $0$ whenever  $\gamma_{\rm deph}> 0$ (cf.~Fig.~\ref{fig:metastability}) and asymptotic states are mixtures of the odd and even stationary states 
	 \begin{equation}
	 \rho_\text{ss}=p\,\ketbra{\Psi_{+}}{\Psi_{+}}+(1-p)\,\ketbra{\Psi_{-}}{\Psi_{-}},
	 \end{equation}
	 where $p$ is determined by the initial support in the even parity subspace.  $\rho_\text{ss}$ approximates (in the zeroth order of $\delta \mathcal{L}$) the asymptotic state of $\mathcal{L}=\mathcal{L}_0+\delta \mathcal{L}$.  
	 
	 ~\\
	 \emph{Derivation of Eq.~\eqref{eq:Leff_deph}}. As the projection on the stationary subspace $\Pi_0$ also conserves the parity, Eq.~\eqref{eq:P0}, so does the first-order effective dynamics, Eq.~\eqref{eq:1storder}. Therefore, in the basis $\ketbra{\Psi_+}{\Psi_+}$, $\ketbra{\Psi_-}{\Psi_-}$, $\ketbra{\Psi_+}{\Psi_-}$, $\ketbra{\Psi_-}{\Psi_+}$, the effective dynamics must be diagonal. The first two terms on the diagonal are $0$ from the trace preservation of the effective dynamics~\cite{zanardi_geometry_2015,zanardi_dissipative_2016, macieszczak2016towards,azouit2016adiabatic}. Furthermore, from $\mathcal{L}_0+\mathcal{L}\delta$ being Hermiticity-preserving we have $[\mathrm{Tr}(L_{-+}\delta\mathcal{L}\ketbra{\Psi_-}{\Psi_+})]^*=\mathrm{Tr}(L_{+-}\delta\mathcal{L}\ketbra{\Psi_+}{\Psi_-})$, which is in general complex so that we set $\mathrm{Tr}(L_{+-}\delta\mathcal{L}\ketbra{\Psi_+}{\Psi_-}\equiv - i\Omega-\gamma_\text{deph}$.

%===================================================================================================================================================

	\subsubsection{Metastable dynamics due to parity conserving higher order corrections in far-detuned regime}
	\label{app:Leff:conserved:higher_order}
	
	Here we derive the Hamiltonian contribution to Eq.~\eqref{eq:Leff_corr}.

	~\\
	\emph{Unitary first-order dynamics of dark states}. We now consider the case of  $\delta\mathcal{L}$ corresponding to the perturbation of the Hamiltonian $H$ by $\delta H$ and a jump $J$ by $\delta J$ in the master equation~\eqref{eq:master_general}, 
		\begin{eqnarray} \label{eq:Lprime'}
	\mathcal{L}\,(\rho)&=&	(\mathcal{L}_0+\delta\mathcal{L})(\rho)=- i\left[H+\delta H,\, \rho\right]+ (J+\delta J)\rho (J+\delta J)^\dagger-\frac{1}{2}\left \{(J+\delta J)^\dagger (J+\delta J),\,\rho\right\}\\\nonumber
		 &=&- i\left[H,\, \rho\right]+ J\rho \,J^\dagger-\frac{1}{2}\left \{J^\dagger J,\, \rho\right\}\\\nonumber
		 &&- i\left[\delta H,\, \rho\right]+ \delta J\rho \,J^\dagger +J\rho \,\delta J^\dagger -\frac{1}{2}\left \{\delta J^\dagger J+J^\dagger \delta J,\, \rho\right\}+\delta J\rho \,\delta J^\dagger-\frac{1}{2}\left \{\delta J^\dagger \delta J,\, \rho\right\},
		\end{eqnarray}
		where $\{X,Y\}=XY+YX$ denotes the anti-commutator, which corresponds to the first-order correction $\delta \mathcal{L}_1$ and second-order correction $\delta \mathcal{L}_2$  in $\delta H$ and $\delta J$. In the case when stationary states of $\mathcal{L}_0$ are pure, $|\Psi_+\rangle$ and $|\Psi_-\rangle$, and dark with respect to the jump operator $J$, i.e., $J|\Psi_\pm\rangle=0$, so that they form a DFS, the first-order corrections to the dynamics in the DFS are unitary~\cite{zanardi_coherent_2014,zanardi_geometry_2015} and only due to the Hamiltonian $\delta H$,
			\begin{eqnarray}
			\Pi_0  \,\delta \mathcal{L}_1 \left(\ketbra{\Psi_+}{\Psi_-}\right)&&=\Pi_0  \left(- i \left[\delta H,\, \ketbra{\Psi_+}{\Psi_-} \right] +  \delta J\ketbra{\Psi_+}{\Psi_-}  \,J^\dagger +J\ketbra{\Psi_+}{\Psi_-} \, \delta J^\dagger -\frac{1}{2}\left \{\delta J^\dagger J+J^\dagger \delta J,\, \ketbra{\Psi_+}{\Psi_-} \right\} \right)\qquad\\\nonumber
			&&=\Pi_0\left(- i \left[\delta H,\,\ketbra{\Psi_+}{\Psi_-} \right]  -\frac{1}{2} J^\dagger \delta J  \ketbra{\Psi_+}{\Psi_-} -\frac{1}{2}  \ketbra{\Psi_+}{\Psi_-} \delta J^\dagger J  \right)\\\nonumber
			&&
			= - i\left(\left\langle \delta H \right \rangle_+  -\left\langle \delta H \right \rangle_-\right) \ketbra{\Psi_+}{\Psi_-} -i \left\langle \delta H \right \rangle_{-+} \left(\ketbra{\Psi_-}{\Psi_-}-\ketbra{\Psi_+}{\Psi_+} \right),	
			\end{eqnarray}
			where $\left\langle \delta H \right \rangle_{-+}=\langle\Psi_-|\delta H |\Psi_+ \rangle$, and in the last line we used the fact that coherences to dark states are orthogonally projected on the dark states (cf.~Eq.~\eqref{eq:Pdark} and see Ref.~\cite{albert2016geometry}). When both $\mathcal{L}_0$ and $\mathcal{L}$ conserve the parity, the parity is necessarily conserved by $H$,  $J$, and $\delta H$, $\delta J$~\cite{albert2014symmetries}, and thus the first-order correction is given by [cf.~Eq.~\eqref{eq:Leff_deph}]
			\begin{eqnarray} \label{eq:Omega_proof}
			&&\Pi_0  \,\delta \mathcal{L}_1\left(\ketbra{\Psi_+}{\Psi_-}\right)
			= - i\left(\left\langle \delta H \right \rangle_+  -\left\langle \delta H \right \rangle_-\right) \ketbra{\Psi_+}{\Psi_-}  =- i\Omega\, \ketbra{\Psi_+}{\Psi_-}.	
			\end{eqnarray}
			
			~\\
			\emph{Higher order corrections in far-detuned regime}. 
The result in Eq.~\eqref{eq:Omega_proof} is directly used in Eq.~\eqref{eq:Omega}, which corresponds to the higher order corrections in the parity conserving Kraus operators due to finite detuning, Eqs.~\eqref{eq:KrausM1time} and~\eqref{eq:KrausM3time}. The parity-conserving operators can be shifted so that $|\Psi_{+}\rangle$ and $|\Psi_{-}\rangle$ are the dark states of the adiabatic dynamics [see Eq.~\eqref{eq:Kraus_shifted} and~\eqref{eq:master_noloss2}]. In this case, we can identify $H=0$ and 
\begin{equation}\label{eq:deltaH}
\delta H=\frac{i}{2}\left(c_{\rm g}^* e^{ i\tau \frac{|g_2|^2}{\Delta}}  M_1-c_{\rm g} e^{- i\tau \frac{|g_2|^2}{\Delta}} M_1^\dagger-c_{\rm e}^* e^{ i\tau \frac{|g_2|^2}{\Delta}} M_3+c_{\rm e} e^{- i\tau \frac{|g_2|^2}{\Delta}} M_e^\dagger\right),
\end{equation}
while the changes in the shifted Kraus operators
\begin{subequations}
	\begin{align}
 \widetilde{M}_1&=M_1-c_{\rm g} e^{- i\tau \frac{|g_2|^2}{\Delta}} \mathds{1}, \\
 \widetilde{M}_3&=M_3+c_{\rm e} e^{- i\tau \frac{|g_2|^2}{\Delta}}\mathds{1},
 \end{align}
\end{subequations}
that play the role of jump operators, do not contribute. Note that here we use the definitions of the Kraus operators $M_1$ and $M_3$ from Eqs.~\eqref{eq:KrausM1time} and~\eqref{eq:KrausM3time}, which differ from the Kraus operators defined in Eq.~\eqref{eq:Kraus} by the global phase $e^{ i\tau \frac{|g_2|^2}{\Delta} }$ due to constant terms neglected in~\eqref{eq:Heff}.

				%===================================================================================================================================================
\subsubsection{Metastable dynamics due to relaxing conditions for Stark-shift cancellation}
\label{app:Leff:conserved:cond}	
We now consider relaxing the conditions in Eq.~\eqref{eq:conditions}. Because of parity conservation, this leads to dephasing of odd-even coherences, Eq.~\eqref{eq:Leff_deph}, but only in a higher than the first order.

~\\			
\emph{Corrections to two photon interaction}. Relaxing the conditions in Eq.~\eqref{eq:conditions}, which cancel the Stark shifts from the atom-cavity interactions in Eq.~\eqref{eq:Heff}, leads to the higher order corrections to this Hamiltonian, as given by~Eq.~\eqref{eq:Heff0}. Therefore, the dynamics remain parity conserving, but with modified Kraus operators $M_1$ and $M_3$ [cf. Appendix~\ref{app:2photon}]. This is analogous to the case of the fourth-order corrections to the atom-cavity interactions, Eq.~\eqref{eq:Hdiag4}, contributing to cavity dynamics, which is discussed in Appendix~\ref{app:Leff:conserved:higher_order}.

~\\
\emph{Resulting cavity dynamics}. In the lowest order, the perturbation away from Eq.~\eqref{eq:conditions}, contributes to the unitary dynamics via Eq.~\eqref{eq:deltaH}, while in the higher order it can also lead to dephasing of coherences [cf.~Eq.~\eqref{eq:Leff_deph}]. Dephasing manifests mixedness of the odd and even stationary states, and thus only takes place when the perturbed interaction in Eq.~\eqref{eq:Heff0} does not lead to different set of pure stationary states [for the general dynamics leading to pure stationary states, see  Appendix~\ref{app:2photon}].

\subsubsection{Metastable dynamics due to mixed atom states}
\label{app:Leff:conserved:atom_mixed}

In the main text, we discussed the properties of two-photon micromaser dynamics under the assumption that all atoms entering the cavity are prepared in an identical pure state, Eq.~(\ref{eq:psi}). Here we investigate how the imperfections of the atom preparation influence the resulting cavity dynamics.
	
~\\
\emph{Micromaser dynamics with mixed atom state}. The most general state of the atom invariant to the Hamiltonian~\eqref{eq:H0} (as required by Assumption 3. in Appendix~\ref{app:micromaser}) is
\begin{equation}
\rho_{\rm at}= p_a \ket{\psi_a}\!\!\bra{\psi_a} + p_b\ket{\psi_b}\!\!\bra{\psi_b}+\sum_{j=0,2,4,\mathrm{a}} p_j\ket{j}\!\!\bra{j},\\
\label{eq:rho_at_mixed}
\end{equation}
where $p_a+p_b+\sum_{j=0,2,4,\mathrm{a}} p_j=1$ and coherent superpositions
\begin{equation}\label{eq:psi12}
|\psi_{a}\rangle=c_{\rm g}|1\rangle+c_{\rm e}|3\rangle, \qquad |\psi_{b}\rangle=c_{\rm e}^*|1\rangle-c_{\rm g}^*|3\rangle
\end{equation}
 are allowed due to the two-photon resonance in Eq.~\eqref{eq:resonance} [cf.~Eq.~\eqref{eq:psi}]. Note that the states $|\psi_{a}\rangle$ and $|\psi_{b}\rangle$ are orthonormal. 

The cavity dynamics due to a passage of a single atom in the mixed state~\eqref{eq:rho_at_mixed} is given by [cf.~Eq.~\eqref{eq:discrete}] 
\begin{equation}
\label{eq:discreteTduplicate}
\rho^{(k)}= \sum_{\substack{j= g,e\\l=a,b}} p_l \,M_{jl}\ken \rho^{(k-1)} \ken M_{j l}^{\dagger}+\sum_{\substack{j=0,2,4,\mathrm{a}}} p_j \,M_{j} \rho^{(k-1)} \ken M_{j}^{\dagger} \equiv\mathcal{M}\left[\rho^{(k-1)}\right],
\end{equation}
where for the initial states $|\psi_a\rangle$ and $|\psi_b\rangle$ we have two pairs of Kraus operators [cf.~Eqs.~\eqref{eq:KrausFULL} and~\eqref{eq:Kraus}],
\begin{subequations}\label{eq:Kraus_mixed}
\begin{align}
	M_{ga}=\bra{1} U_{\rm eff}(\tau) \ket{\psi_a}, &\qquad M_{ea}= \,\bra{3} U_{\rm eff}(\tau) \ket{\psi_a}, \qquad\text{and}\\
	M_{gb}= \bra{1} U_{\rm eff}(\tau) \ket{\psi_b}, &\qquad M_{eb}= \, \bra{3} U_{\rm eff}(\tau) \ket{\psi_b}
\end{align}
\end{subequations}
with the effective Hamiltonian $H_{\rm eff}$ coupling the resonant levels given by (\ref{eq:Heff}), while for $|0\rangle$, $|2\rangle$, $|4\rangle$, $|\text{a}\rangle$ 
\begin{equation}\label{eq:Kraus_mixed2}
M_{0}= e^{i \tau a^\dagger a \frac{|g_2|^2}{\Delta}}, \quad M_{2}= e^{-i \tau a\, a^\dagger  \frac{|g_2|^2+|g_3|^2}{\Delta }},\quad M_4 =e^{i \tau a\,a^\dagger\frac{|g_3|^2}{\Delta}},\quad\text{and}\quad M_{\rm a}=\mathds{1},
\end{equation}
up to a global phase [see Eqs.~\eqref{eq:KrausFULL},~\eqref{eq:conditions}, and~\eqref{eq:Hdiag}]. The continuous dynamics is then given by  Eq.~\eqref{eq:master}. We note that Eq.~\eqref{eq:Kraus_mixed2} depends on the specific (5+1)-model implementing the effective Hamiltonian in Eq.~\eqref{eq:Heff}, but below we discuss the effects from Eq.~\eqref{eq:Kraus_mixed} and~Eqs.~\eqref{eq:Kraus_mixed2} separately, and thus our results will be applicable to other realisations of two-photon dynamics without Stark shifts.

We note that, exactly as in the case of a pure atom state, the cavity dynamics is \emph{parity-conserving}, which is due to the far-detuned limit, Eq.~\eqref{eq:Heff}. Furthermore, it also corresponds to real-valued dynamics when $p_0=p_2=p_4=0$, as in this case the relative phase between coefficients of both atom states $|\psi_a\rangle$ and $|\psi_b\rangle$ is the same (see Sec.~\ref{sec:dynamics}).

~\\
\emph{Mixed stationary states of the dynamics}.
As discussed in Sec.~\ref{sec:recurrence}, a pair of Kraus operators in Eq.~\eqref{eq:Kraus} corresponding to a pure atom state in Eq.~\eqref{eq:psi} features two even and odd pure eigenstates,  which are determined by the recurrence relation in Eq.~(\ref{eq:pure_cond}). In order for stationary states of the cavity to be pure in the dynamics with the mixed atom state~\eqref{eq:rho_at_mixed} it is necessary for it to be an eigenstate of all Kraus operators in Eqs.~\eqref{eq:Kraus_mixed} and~\eqref{eq:Kraus_mixed2}. However, for the orthogonal states $\ket{\psi_a}$ and $\ket{\psi_b}$, Eq.~\eqref{eq:psi12}, the corresponding recurrence relations features  the factors $c_{\rm e}/c_{\rm g}$ and $-c_{\rm g}^*/c_{\rm e}^*$, respectively, which are always different, as $|c_{\rm g}|^2\neq-|c_{\rm e}|^2$. Therefore, \emph{no pure stationary states exist} if the atom state is mixed between levels $|j\rangle$ with $j=0,...,4$ (i.e., except $|\mathrm{a}\rangle$). Nevertheless, the cavity features at least two, odd and even, mixed stationary states, since the photon-number parity is conserved~\cite{albert2014symmetries}. We note that, even in the case when $p_a=0$ (or $p_b$=0), the Kraus operators $M_0$, $M_2$, and $M_4$ in Eq.~\eqref{eq:Kraus_mixed2} cannot feature pure cavity states as eigenstates unless the cavity state is a fixed photon number state or the interaction time $\tau$ is such that $\frac{|g_2|^2}{\Delta } \tau=\frac{|g_3|^2}{\Delta }\tau=2\pi$, so that $M_0=M_2=M_4=\mathds{1}$. This is because those Kraus operators imprint a nontrivial phase on the cavity state and thus lead to its dephasing [but if the state of outgoing atoms is measured, the conditional cavity state can become asymptotically pure for $p_a=0$ (or $p_b$=0), however, only the probability of the photon number will be stationary with the phases changing due to $M_0$, $M_2$, and $M_4$].  \\

~\\
\emph{Coherent stationary state of cavity from coherent states of atoms}. In Appendix~\ref{app:classical} we show that the dynamics with the atom state diagonal in atom levels leads to effective classical detailed balance dynamics of the cavity with odd and even thermal steady states diagonal in the photon-number basis, with the temperature determined as $\exp[-2\omega/(k_B T)]=p_a|c_{\rm e}|^2+p_b|c_{\rm g}|^2/(p_a|c_{\rm g}|^2+p_b|c_{\rm e}|^2)$. We now prove that whenever the atom state, Eq.~\eqref{eq:rho_at_mixed}, is not diagonal in the atom levels, i.e., coherent ($p_a\neq p_b$ and $|c_{\rm e}|\neq 1,0$), the even and odd stationary states of the cavity are coherent in the photon-number basis. 

Consider a diagonal even state $\rho^+=\sum_{n=0}^\infty p_{2n}|2n\rangle\!\langle2n|$. We have [cf.~Eqs.~\eqref{eq:discrete} and~\eqref{eq:Kraus}]
\begin{equation}\label{eq:discrete_mixed_atom}
\mathcal{M}(\rho_+)=\mathcal{M}_\text{diag}(\rho_+)+ \sum_{n=0}^\infty\Big\{ -i c_{\rm e} c_{\rm g}^* \sin_{2n}(\phi)\big(\sqrt{p_a}-\sqrt{p_b}\big) \left[\cos_{2n-1}(\phi) p_{2n} - \cos_{2n+2}(\phi) p_{2n+2}\right] |2n+2\rangle\!\langle2n| +\text{H.c.}\Big\},
\end{equation} 
where $\mathcal{M}_\text{diag}$ is the dynamics with a diagonal atom state, $\sum_{\substack{j=0,2,4,\mathrm{a}}} p_j |j\rangle\langle j| + [p_a|c_{\rm g}|^2+p_b|c_{\rm e}|^2] |1\rangle\!\langle 1| +[p_a|c_{\rm e}|^2+p_b|c_{\rm g}|^2]|3\rangle\!\langle 3|$, which leaves diagonal states diagonal. Therefore, for $\rho^+$ to be a stationary state, no coherences can appear in Eq.~\eqref{eq:discrete_mixed_atom}, and thus $c_{\rm e} c_{\rm g}^*=0$, or $p=1-p$, or $\cos_{2n-1}(\phi) p_{2n} - \cos_{2n+2}(\phi) p_{2n+2}=0$. The first two conditions correspond to incoherent states of the atom, while the last condition cannot be fulfilled for a stationary state of the diagonal dynamics $\mathcal{M}_\text{diag}$, as it is effectively thermal (see Appendix~\ref{app:classical}) and thus independent from the interaction strength. The proof for the odd stationary state is analogous.

~\\
\emph{Metastable dephasing dynamics for almost pure states}. When atom state in Eq.~\eqref{eq:rho_at_mixed} is almost pure, $p_a\approx 1$ (or $p_b\approx 1$) so that $\rho_\text{at}\approx |\psi_a\rangle\!\langle\psi_a|$ (or $|\psi_b\rangle\!\langle\psi_b|$), the Kraus operators $M_0$, $M_2$, $M_4$, and $M_{gb}$ and $M_{eb}$ (or $M_{ga}$ and $M_{ea}$) can be treated as the perturbation of the dynamics with the pure states $|\Psi_{+}\rangle$, $|\Psi_{-}\rangle$, that takes place at the \emph{reduced} rate $\nu p_a$ (or $\nu p_b$). 

From parity conservation, this perturbation necessarily leads to \emph{dephasing} of the even-odd coherences $|\Psi_{+}\rangle\!\langle \Psi_{-}|$ and $|\Psi_{-}\rangle\!\langle \Psi_{+}|$ [cf.~Eq.~\eqref{eq:Leff_deph}]. The dephasing manifests the fact that the even and odd stationary states of the dynamics are mixed (although in the zeroth order they are approximated by the pure state $|\Psi_{+}\rangle$ and $|\Psi_{-}\rangle$), and coherences between them are not stationary. The rate of dephasing and the frequency of unitary dynamics are \emph{bounded} as [cf.~Eq.~\eqref{eq:rho_at_mixed}]
\begin{eqnarray}
\label{eq:deph_bound1}
 \gamma_{\rm deph}&\leq& 2\nu(1-p_a-p_\text{a}), \\
\label{eq:deph_bound2}
  |\Omega|&\leq& \nu(p_0+p_2+p_4).
\end{eqnarray} 
This is follows from the fact that for the mixed atoms we have [cf. Eqs.~\eqref{eq:1storder} and~\eqref{eq:Leff_deph}] 
\begin{equation}
-i\Omega-\gamma_{\rm deph}= \nu \sum_{j=b,0,2,4} p_j \left\{\mathrm{Tr} \left[ L_{+-} \mathcal{M}_j' (|\Psi_+\rangle\!\langle \Psi_-|)\right] - 1\right\}
\end{equation}
where $\mathcal{M}'_b(\rho)=  M_{gb} \rho M_{gb}^\dagger +M_{eb} \rho M_{eb}^\dagger$ and $\mathcal{M}'_j(\rho)= M_{j} \rho M_{j}^\dagger$ for $j=0,2,4$, are all quantum channels conserving the parity [cf. Eqs.~\eqref{eq:Kraus_mixed} and~\eqref{eq:Kraus_mixed2}]. For any quantum channel $\mathcal{M}'$, we have that  $\Pi_0[\mathcal{M}' (|\Psi\rangle\!\langle \Psi|) ] $ is a quantum state, and thus its fidelity with any other state is between $0$ and $1$. Therefore, for $|\Psi\rangle=(|\Psi_+\rangle+|\Psi_-\rangle)/\sqrt{2}$, any parity-conserving $\mathcal{M}'$ we have that $0 \leq 2\langle \Psi|\Pi_0[\mathcal{M}' (|\Psi\rangle\!\langle \Psi|) ] |\Psi\rangle= 1+ \mathrm{Re} \{\mathrm{Tr} \left[ L_{+-}\mathcal{M}' (|\Psi_+\rangle\!\langle \Psi_-|)\right] )$ and, similarly, for $|\Psi'\rangle=(|\Psi_+\rangle\pm  i|\Psi_-\rangle)/\sqrt{2}$, we have $0 \leq 2\langle \Psi'|\Pi_0[ \mathcal{M}' (|\Psi\rangle\!\langle \Psi|) ] |\Psi'\rangle= 1\mp \mathrm{Im} \{\mathrm{Tr} \left[ L_{+-} \mathcal{M}' (|\Psi_+\rangle\!\langle \Psi_-|)\right] ) \leq 2$. Noting that $1-p_a-p_\text{a}= \sum_{j=b,0,2,4\text{a}} p_j$, and that both $\mathcal{M}'_a$ and $\mathcal{M}'_b$ can be considered real valued, so $\mathcal{M}'_b$ does not contribute to the unitary dynamics, we finally arrive at Eqs.~\eqref{eq:deph_bound1} and~\eqref{eq:deph_bound2}, respectively. We also note that the rate of coherence decay can be simply bounded by the mixedness of the atom state (defined as $1$ minus the purity), as from Eq.~\eqref{eq:deph_bound1} we have $\gamma_{\rm deph}\leq 2\nu(1-p_a)\approx 1-\mathrm{Tr}(\rho_\text{at}^2)$ for $p_a\approx 1$.

~\\
\emph{Dynamics in weak-coupling limit}. In the limit of small integrated coupling, $|\phi|\ll 1$,  the stationary states of the dynamics with a pure atom state are given by Schr\"{o}dinger cat states [cf.~Eqs.~\eqref{eq:psi_ss_lin} and~\eqref{eq:master_lin}]. For the mixed state, dynamics in Eq.~\eqref{eq:master_lin} in general additionally features \emph{two-photon injections}, \emph{photon-number Hamiltonian} and \emph{dephasing in photon number} (see also Refs.~\cite{mirrahimi2014dynamically, roy2015continuous, minganti2016exact})
\begin{eqnarray}
\frac{ d }{ d t} \rho&=& -i [g_\text{2ph}^* a^2+g_\text{2ph} a^{\dagger 2},\rho]+ \kappa_\text{2ph}\left( a^2 \rho a^{\dagger 2} -\frac{1}{2}\left\{a^{2\dagger }a^2,\rho\right\}\right)+ \gamma_\text{2ph}\left( a^{\dagger 2}  \rho a^2 -\frac{1}{2}\left\{a^2a^{2\dagger },\rho\right\}\right)\\\nonumber&&-i [\omega_\text{0}\,n ,\rho]+\gamma_\text{0}\left( n  \rho n -\frac{1}{2}\left\{ n^2,\rho\right\}\right),\label{eq:master_lin_mixed}
\end{eqnarray}
Here $n=a^\dagger a$ and the parameters $g_\text{2ph}=\nu (p_a-p_b) c_{\rm g}^* c_{\rm e}\, \phi$,  $\kappa_\text{2ph}=\nu p_a\, |c_{\rm g}|^2 \phi^2$, $\gamma_\text{2ph}=\nu p_b|c_{\rm g}|^2 \phi^2$,  $\omega_\text{0}=\nu [\phi_2(p_0-p_2)  +\phi_3(p_4-p_2)]$, and $\gamma_0= \nu[p_0 \phi_2^2 +p_2(\phi_2+\phi_3)^2+p_4 \phi_3^2]$. The terms in the first line of Eq.~\eqref{eq:master_lin_mixed} arise from the expansion of  $M_{ga}$, $M_{ea}$ [see Eqs.~\eqref{eq:Kraus_lin} and~\eqref{eq:Kraus_mixed}], and 
\begin{subequations}
	\begin{align}
	M_{gb}&\approx c_{\rm e}^*\mathds{1}+ i c_{\rm g}^* \phi\, a^{\dagger  2} %- c_{\rm e}^* \frac{\phi^2}{2}\,a^{\dagger  2}a^2
	,\\
	M_{eb}&\approx -c_{\rm g}^*\mathds{1} -i c_{\rm e}^* \phi\, a^2+ c_{\rm g}^*\frac{\phi^2}{2}\,a^{2}a^{\dagger 2},
	\end{align}
\end{subequations}
where we kept terms contributing to up to second order in $\phi$ and $c_{\rm e}$ to the master equation [cf. Eq.~\eqref{eq:master_lin}]. The terms in the second line of Eq.~\eqref{eq:master_lin_mixed} originate from the first and the second orders in the expansion of $M_0$, $M_2$, and $M_4$ of Eq.~\eqref{eq:Kraus_mixed2} in $\phi_j=|g_j|^2/\Delta$, which we assumed small, $|\phi_j|\ll 1$, $j=2,3$. 

For $p_a\approx 1$, we arrive at the metastable dynamics in with  
\begin{eqnarray} 
\Omega &=& \omega_0\left(\langle n \rangle_+ - \langle n \rangle_-\right)  ,\label{eq:deph_lin}\\
\gamma_\text{deph}&=&\gamma_\text{0} \left[\langle n^2 \rangle_+ + \langle n^2 \rangle_- - 2{\rm Tr}\left(L_{+-} n \ketbra{\Psi_{+}}{\Psi_{-}} n \right)\right] \label{eq:deph_lin2}\\\nonumber&&+\gamma_\text{2ph} \left[\langle n^2 \rangle_+ + \langle n^2 \rangle_-+3\left(\langle n\rangle_+ + \langle n^2 \rangle_-\right) +6- 2{\rm Tr}\left(L_{+-} a^{\dagger 2} \ketbra{\Psi_{+}}{\Psi_{-}} a^2 \right)\right] ,\qquad
\end{eqnarray}
where the stationary states are Schr\"{o}dinger cat states in Eq.~\eqref{eq:psi_ss_lin} with a modified parameter $\alpha=(1-p_b/2)e^{- i\pi/4} \sqrt{2 c_e/(c_g \phi)} $ in Eq.~\eqref{eq:psi_ss_lin}. Since $|\alpha|^2 |\phi| = 2(1-p_b)|c_e/c_g| \ll 1$, the bounds in Eqs.~\eqref{eq:deph_bound1} and~\eqref{eq:deph_bound2} indeed hold true even for large $|\alpha|$, where $\langle n\rangle_\pm\approx|\alpha|^2$ and $\langle n^2\rangle_\pm\approx|\alpha|^4$. Furthermore, since $L_{+-}$ is known, Eqs.~\eqref{eq:deph_lin} and~\eqref{eq:deph_lin2}  can be computed exactly~\cite{mirrahimi2014dynamically}. Finally, we note that for the mixed state supported only on $|1\rangle$ and $|3\rangle$ levels, which corresponds to dissipative dynamics with two-photon injections only ($\omega_0=0=\gamma_0$), we indeed observe that $\Omega=0$, as argued above.

\subsubsection{Metastable dynamics due to decaying atom levels}
\label{app:Leff:conserved:atom_decay}

\noindent
\emph{Finite lifetime of atom levels}. In this work so far, we have assumed that all atoms are prepared identically in an initial state $\rho_{\rm at}$ which only changes due to the interaction with the cavity (see Appendix~\ref{app:micromaser}). In general, however, atoms interact also with the external environment of continuum modes, which leads to decay of the atomic levels. Such decay may include transitions between the states  $|0\rangle,..,|4\rangle,|{\rm a}\rangle$ as well to other atom levels which are not coupled to the cavity field, and is described in the frame of the free Hamiltonian [Eq.~\eqref{eq:H0prime}] as   
\begin{equation}
\frac{\rm d}{{\rm d}t} \rho_\text{at}(t)= \sum_{k=-1,0,1,2,3,4,\text{a}} \sum_{j<k} \gamma_{jk} \left[\sigma_{jk}\rho_\text{at}(t)\sigma_{kj} -\frac{1}{2}\sigma_{kk}\rho_\text{at}(t)-\frac{1}{2} \rho_\text{at}(t)\sigma_{kk} \right] 
\label{eq:master_decay},
\end{equation}
where the state $|-1\rangle$, without loss of generality, describes all the other atom levels not coupled to the cavity and we consider only the transitions corresponding to the loss of atom energy.

~\\
\emph{Mixed atom states}.  For the initial pure state in Eq.~\eqref{eq:psi}, the dynamics in Eq.~\eqref{eq:master_decay} gives a mixed atomic state $\rho_\text{at}(t)$  [cf.~Eq.~\eqref{eq:rho_at_mixed}], where
\begin{eqnarray}\label{eq:rho_decay}
&&\rho_\text{at}(t) -[1-\overline{p}(t)]\,|-1\rangle\! \langle -1|\\\nonumber
&&=   \left[e^{- \Gamma_1 t}|c_g|^2  +\gamma_{13} \frac{e^{- \Gamma_1 t}-e^{- \Gamma_3 t}}{\Gamma_3-\Gamma_1}|c_e|^2 + \frac{\gamma_{12} \gamma_{23}}{\Gamma_3-\Gamma_2} \left(\frac{e^{- \Gamma_1 t}-e^{- \Gamma_2 t}}{\Gamma_2-\Gamma_1} - \frac{e^{- \Gamma_1 t}-e^{- \Gamma_3t}}{\Gamma_3-\Gamma_1} \right)|c_e|^2 \right]\,|1\rangle\! \langle 1| \\\nonumber
&&\quad +\, e^{- \frac{\Gamma_1+\Gamma_3}{2} t}\, c_g c_e^*\,
|1\rangle\! \langle 3|+e^{- \frac{\Gamma_1+\Gamma_3}{2} t}\,c_g^* c_e \,|3\rangle\! \langle 1|+e^{- \Gamma_3 t} |c_e|^2 \,|3\rangle\! \langle 3|+\, \gamma_{23} \frac{e^{- \Gamma_2 t}-e^{- \Gamma_3 t}}{\Gamma_3-\Gamma_2}  |c_e|^2 \,|2\rangle\! \langle 2|  \\\nonumber
&&\quad+\bigg\{\!\gamma_{01} \frac{e^{- \Gamma_0 t}-e^{- \Gamma_1 t}}{\Gamma_1-\Gamma_0} |c_g|^2+ \gamma_{03} \frac{e^{- \Gamma_0 t}-e^{- \Gamma_3 t}}{\Gamma_3-\Gamma_0} |c_e|^2\\\nonumber
&&\quad\quad
+\frac{ \gamma_{01}\gamma_{13}}{\Gamma_3-\Gamma_1} \!\left(\!\frac{e^{- \Gamma_0 t}-e^{- \Gamma_1 t}}{\Gamma_1-\Gamma_0} - \frac{e^{- \Gamma_0 t}-e^{- \Gamma_3t}}{\Gamma_3-\Gamma_0} \!\right) \!+ \frac{ \gamma_{02}\gamma_{23}}{\Gamma_3-\Gamma_2} \!\left(\!\frac{e^{- \Gamma_0 t}-e^{- \Gamma_2 t}}{\Gamma_2-\Gamma_0} - \frac{e^{- \Gamma_0 t}-e^{- \Gamma_3t}}{\Gamma_3-\Gamma_0} \!\right)\\\nonumber
&&\quad\quad+ \frac{ \gamma_{01}\gamma_{12}\gamma_{23}}{\Gamma_3-\Gamma_2} \!\left[\frac{e^{- \Gamma_0 t}-e^{- \Gamma_1 t}}{(\Gamma_2-\Gamma_1)(\Gamma_1-\Gamma_0)}-\frac{e^{- \Gamma_0 t}-e^{- \Gamma_2 t}}{(\Gamma_2-\Gamma_1)(\Gamma_2-\Gamma_0)}-\frac{e^{- \Gamma_0 t}-e^{- \Gamma_1 t}}{(\Gamma_3-\Gamma_1)(\Gamma_1-\Gamma_0)} +\frac{e^{- \Gamma_0 t}-e^{- \Gamma_3 t}}{(\Gamma_3-\Gamma_1)(\Gamma_3-\Gamma_0)} \right]
\!\bigg\}|0\rangle\! \langle 0| \\\nonumber
&&\equiv \overline{p}(t)\, \overline{\rho}_\text{at}(t),
\end{eqnarray}
and we defined $\Gamma_k=\sum_{j<k} \gamma_{jk}$, $\overline{p}(t) = \sum_{j=0,1,2,3,4,\text{a}} \langle j|\rho_\text{at}(t)|j\rangle $.

Below we show that the possible decay of atoms during the time $T$ between the preparation of the initial atomic state [Eq.~\eqref{eq:psi}] and entering the cavity leads to an effective micromaser with the \emph{reduced rate} $\overline{\nu}=\overline{p}(T)\nu$ and the \emph{mixed atom state} given by $\rho_\text{at}\equiv\overline{\rho}_\text{at}(T)$  (cf.~\emph{Assumption 2.} in Appendix~\ref{app:micromaser}). In particular, for transitions only toward levels not coupled to the cavity ($\gamma_{jk}=0$ unless $j=-1$), the cavity interacts with the effective \emph{pure} atom state 
\begin{equation} 
	\overline{\rho}_\text{at}(T)= \frac{e^{- \Gamma_1 T}|c_g|^2\,|1\rangle\! \langle 1|+e^{- \frac{\Gamma_1+\Gamma_3}{2} T}\, c_g c_e^*\,
	|1\rangle\! \langle 3|+e^{- \frac{\Gamma_1+\Gamma_3}{2} T}\,c_g^* c_e \,|3\rangle\! \langle 1|+e^{- \Gamma_3 T} |c_e|^2 \,|3\rangle\! \langle 3|}{e^{- \Gamma_1 T}|c_g|^2 +e^{- \Gamma_3 T}|c_e|^2} =|\overline\psi_\text{at}(T)\rangle\!\langle\overline \psi_\text{at}(T)|
\end{equation} 
where [cf.~Eq.~\eqref{eq:psi}]
\begin{equation} 
\label{eq:psi_decay}
|\overline\psi_\text{at}(T)\rangle= \overline c_g (T)\,|1\rangle+\overline c_e (T)\,|3\rangle=\frac{e^{- \frac{\Gamma_1}{2} T}  c_g \,|1\rangle + e^{- \frac{\Gamma_3}{2} T}  c_e \,|3\rangle}{\sqrt{e^{- \Gamma_1 T}|c_g|^2 +e^{- \Gamma_3 T}|c_e|^2}}
\end{equation} 
arriving at the reduced rate $\overline{\nu}=(e^{- \Gamma_1 T}|c_g|^2 +e^{- \Gamma_3 T}|c_e|^2)\nu$. We note that for the uniform decay, $\Gamma_1=\Gamma_3=\Gamma$, the atom state remains the same, $|\overline\psi_\text{at}(T)\rangle=|\psi_\text{at}\rangle$, but the rate is exponentially reduced, $\overline{\nu}=e^{- \Gamma T} \nu$.

~\\
\emph{Modified cavity dynamics}. In the first order, the interaction with the external environment and the cavity are independent, leading to the change in the cavity given by [cf.~Eq.~\eqref{eq:discrete0} and Eq.~\eqref{eq:master_decay}]
\begin{eqnarray}\label{eq:discrete_decay}
\rho^{(k)}&=&\text{Tr}_{\mathrm{at}}\left\{\Lambda(\tau) \ken \left[ 	\rho_\text{at}(T)\otimes \rho^{(k-1)} \right] \right\}\\\nonumber
&=&\overline{p}(T)\,\text{Tr}_{\mathrm{at}}\left\{\Lambda(\tau) \ken \left[ 	\overline{\rho}_\text{at}(T)\otimes \rho^{(k-1)} \right] \right\} +[1-\overline{p}(T)]\,\rho^{(k-1)}\\\nonumber
&\equiv&  \overline{p}(T)\,\overline{\mathcal{M}}[\rho^{(k-1)}] + [1-\overline{p}(T)]\,\rho^{(k-1)} , 
\end{eqnarray}
where we introduced [cf.~Eq.~\eqref{eq:U}]
\begin{equation}\label{eq:U_decay}
\Lambda(\tau)  = \mathcal{T} e^{- i \int_{0}^\tau  d t \left[\mathcal{H}(t)+\mathcal{L}_\text{at}\right] } ,
\end{equation}
with $\mathcal{H}(t)(\cdot)= -i [H_\text{int}(t)+ H_0, (\cdot)]$, and we again consider the frame rotating with the free Hamiltonian [Eq.~\eqref{eq:H0prime}]. The continuous dynamics takes place at the reduced rate $\overline{\nu}=\overline{p}(T)\nu$ [cf.~Eq.~\eqref{eq:master}]
\begin{eqnarray}
\frac{ d }{ d t} \rho(t)&=&\bar{\nu}\,\overline{\mathcal{M}}\left[\rho(t) \right]-\bar{\nu} \,\rho(t)\equiv\overline{\mathcal{L}}\left[ \rho(t)\right].
\label{eq:continuous_decay}
\end{eqnarray}

The cavity dynamics will be modified for two reasons. First, the mixed atom state $\overline{\rho}(T)$ [cf.~Eq.~\eqref{eq:rho_decay}] will lead to the cavity dynamics being a mixture of dynamics for different pure states, as discussed in Appendix~\ref{app:Leff:conserved:atom_mixed}. Namely, dynamics for the eigenstates of $\overline{\rho}(T)$: $|\overline{\psi}_k(T)\rangle$ [of the general form $|\overline{\psi}_a(T)\rangle$ and $|\overline{\psi}_b(T)\rangle$ supported on $|1\rangle$ and $|3\rangle$, and $|j\rangle$ with $j=0,2$; cf.~Eq.~\eqref{eq:rho_at_mixed}], chosen with the probability $\overline{p}_k(T)$ given by the corresponding eigenvalues, $k=a,b,0,2$. Second, possible atom decays during the interaction with cavity will lead to the evolution with a non-Hermitian Hamiltonian $\overline H(t)=H_0+H_\text{int}(t)+\frac{i}{2}\sum_{jk} \gamma_{jk}\sigma_{jj}$ intercepted by the updates of the atom state according to the occurring decay events. This will lead to a generally different set of Kraus operators for any sequence of decay events and the average dynamics given by the integral over all events [cf.~Eq.~\eqref{eq:Kraus_mixed}],
\begin{eqnarray}\label{eq:discrete_decay2}
\overline{\mathcal{M}}&=& \!\!\!\!\!\!\!\!\!\!\sum_{\substack{j=-1,0,1,2,3,4,\text{a}\\ k=a,b,0,2}}\!\!\!\!\!\!\!\!\!\! \overline{p}_k(T) \Bigg[\overline{\mathcal{M}}_{jk}(\tau) +\sum_{n=1}^\infty \int_{0}^\tau \!\!\!\! d t_n \!\! \int_{0}^{t_n} \!\!\!\! d t_{n-1} \cdots \!\!\int_{0}^{t_2}\!\!\!\!  d t_{1}\!\!\!\!\!\!\!\!\!\!\! \sum_{\substack{j_1,...,j_n=-1,0,1,2,3,4,\text{a}\\k_1,...,k_n=0,1,2,3,4,\text{a}}}\!\!\!\!\!\!\!\!\!\!\!\!\!\!\!\!\! \gamma_{j_nk_n}\cdots\gamma_{j_1k_1} \,\,\overline{\mathcal{M}}_{j; j_nk_n\cdots j_1k_1;k}(\tau;t_1,...,t_n)]\Bigg],\qquad\quad
\end{eqnarray}
where
\begin{eqnarray}\label{eq:M_decay}
&&\overline{\mathcal{M}}_{jk}(\tau)=\overline{M}_{jk}(\tau) \rho\, \overline{M}_{jk}(\tau) ^\dagger, \\
&&\overline{\mathcal{M}}_{j; j_nk_n\cdots j_1k_1;k}(\tau;t_1,...,t_n)=\overline{M}_{j; j_nk_n\cdots j_1k_1;k}(\tau;t_1,...,t_n) \rho\, \overline{M}_{j; j_nk_n\cdots j_1k_1;k}(\tau;t_1,...,t_n) ^\dagger
\label{eq:M_decay2}
\end{eqnarray}
with
\begin{equation}\label{eq:Kraus_decay1}
	\overline{M}_{jk}(\tau) =\langle j|\mathcal{T} e^{-i \int_{0}^t  d t' \overline H(t') }|\overline\psi_k(T)\rangle
\end{equation}
and
\begin{eqnarray}\label{eq:Kraus_decay2}
&&\overline{M}_{jk; j_nk_n\cdots j_1k_1}(\tau;t_1,...,t_n) \\\nonumber
&&=\langle j|\mathcal{T} e^{-i \int_{t_n}^{\tau}  d t' \overline H(t') }|j_n\rangle\!\langle k_{n}|\mathcal{T} e^{-i \int_{t_{n-1}}^{t_n}  d t' \overline H(t') }|j_{n-1}\rangle \cdots\langle k_{2}|\mathcal{T} e^{-i \int_{t_{1}}^{t_2}  d t' \overline H(t') }|j_{1}\rangle \! \langle k_1| e^{-i \int_{0}^{t_1}  d t' \overline H(t') }|\psi_k(T)\rangle.
\end{eqnarray}
In the far detuned limit, the levels $4$ and $\text{a}$ are not coupled to the dynamics.

In particular,  for decay transitions only toward levels not coupled to the cavity and the far detuned limit, we obtain [cf.~Eq.~\eqref{eq:psi_decay}]
\begin{equation}\label{eq:discrete_0_decay}
\overline{\mathcal{M}}_0= \overline{\mathcal{M}}_g(\tau)+\overline{\mathcal{M}}_e(\tau) +\int_{0}^\tau  d t\, \left[\Gamma_1\, \overline{\mathcal{M}}_g(t)+\Gamma_3\,\overline{\mathcal{M}}_e(t)\right],
\end{equation}
where 
\begin{subequations}
	\label{eq:M_0_decay}
	\begin{align}
	\overline{\mathcal{M}}_g(t)\, (\rho)   &=\overline{M}_{g}(t) \rho\, \overline{M}_g(t) ^\dagger,\\
	\overline{\mathcal{M}}_e(t)\, (\rho)   &=\overline{M}_{e}(t) \rho\, \overline{M}_e(t) ^\dagger,
	\end{align}
\end{subequations}
with [cf.~Eq.~\eqref{eq:Kraus} and see Appendix~\ref{app:2photon}]
\begin{subequations}
\label{eq:Kraus_0_decay}
\begin{align}
&\overline{M}_{\rm g}(t) =\langle 1|\mathcal{T} e^{-i \int_{0}^t  d t' \left[H_\text{eff}(t')-\frac{i}{2} (\Gamma_1 \sigma_{11}+\Gamma_3\sigma_{33}) \right] }|\overline\psi_\text{at}(T)\rangle\nonumber\\
&= e^{- \frac{\Gamma_1+\Gamma_3}{4} t}\left[\overline c_{\mathrm{g}}(T)  \ken \cos\left(t  \sqrt{|\lambda|^2 \,a^{\dagger  2}\ken a^2- \frac{(\Gamma_1-\Gamma_3)^2}{16}}\right)- \left(\! \overline  c_{\mathrm{g}}(T)\,\frac{\Gamma_1-\Gamma_3}{4}+  i\overline c_{\mathrm{e}}(T)\ken \lambda^* a^{\dagger  2}\right)\frac{\sin\left(t  \sqrt{|\lambda|^2 \,a^2\ken a^{\dagger  2} - \frac{(\Gamma_1-\Gamma_3)^2}{16}}\right)}{\sqrt{|\lambda|^2 \,a^2\ken a^{\dagger  2} - \frac{(\Gamma_1-\Gamma_3)^2}{16}}}\right]\!\! ,\qquad\quad\label{eq:Kraus1_decay}\\
&\overline{M}_{\rm e}(t) =\langle 3|\mathcal{T} e^{-i \int_{0}^t  d t' \left[H_\text{eff}(t')-\frac{i}{2} (\Gamma_1 \sigma_{11}+\Gamma_3\sigma_{33}) \right] }|\overline\psi_\text{at}(T)\rangle\nonumber\\
&= e^{- \frac{\Gamma_1+\Gamma_3}{4}t}\left[\left(\!-  i\overline c_{\mathrm{g}}(T)\ken \lambda a^{  2}+  \overline c_{\mathrm{e}}(T)\,\frac{\Gamma_1-\Gamma_3}{4}\right)\frac{\sin\left(t  \sqrt{|\lambda|^2 \,a^{\dagger  2}\ken a^2  - \frac{(\Gamma_1-\Gamma_3)^2}{16}}\right)}{ \sqrt{|\lambda|^2 \,a^{\dagger  2}\ken a^2  - \frac{(\Gamma_1-\Gamma_3)^2}{16}}}+\overline c_{\mathrm{e}}(T)  \ken \cos\left(t  \sqrt{|\lambda|^2 \,a^2\ken a^{\dagger 2}- \frac{(\Gamma_1-\Gamma_3)^2}{16}}\right)\right]\!\! ,\qquad\quad \label{eq:Kraus2_decay}
\end{align}
\end{subequations}
where we neglected a global phase $e^{-i\tau (\Delta_1+\frac{|g_2|^2}{\Delta})}$ (cf. Appendix~\ref{app:Kraus}). 
In Eq.~\eqref{eq:discrete_0_decay}, the terms $\overline{\mathcal{M}}_g(\tau)+\overline{\mathcal{M}}_e(\tau)$ describe the situation when no decay events occur during the interaction time $\tau$, while the decay from $|1\rangle$ or from $|3\rangle$,  happening at time $t$ is described by  $\overline{\mathcal{M}}_g(t)$ or $\overline{\mathcal{M}}_e(t)$, respectively, as in those cases the atom interacts with the cavity only for time $t$. 
For the case of uniform decay, $\Gamma_1=\Gamma_3=\Gamma$, we simply have $\overline{M}_{ j}(t)=e^{- \frac{\Gamma}{2} t} M_j(t)$, $j=g,e$, where $M_e(t)$ and $M_g(t)$ are Kraus operators in Eq.~\eqref{eq:Kraus} for the interaction time $t$. Therefore, the dynamics in Eq.~\eqref{eq:discrete_0_decay} simplifies to 
\begin{equation}\label{eq:discrete_0_decay_uniform}
\overline{\mathcal{M}}_0= e^{- \Gamma \tau} \mathcal{M}_0(\tau) +\Gamma \int_{0}^\tau  d t\, e^{- \Gamma t} \mathcal{M}_0(t).
\end{equation}

~\\
\emph{Even and odd stationary states of cavity}. We note that in the far-detuned limit the modified cavity dynamic in Eq.~\eqref{eq:discrete_decay2} conserved the photon number parity $P$ in Eq.~\eqref{eq:parity} as the Kraus operators in Eqs.~\eqref{eq:Kraus_decay1} and~\eqref{eq:Kraus_decay2} commute with $P$. Therefore, there exist both odd and even stationary states. Those stationary states, however, are \emph{mixed}, as already due to the mixed atom state $\overline\rho_\text{at}(T)$, the Kraus operators in Eq.~\eqref{eq:Kraus_decay1} corresponding to the initial pure states $|\overline\psi_a(T)\rangle$ and $|\overline\psi_b(T)\rangle$ could only imply contradictory recurrence relations for pure stationary states (cf. Appendixes~\ref{app:2photon} and~\ref{app:Leff:conserved:atom_mixed}).

Even in the case of decay only toward levels not coupled to the cavity, Eq.~\eqref{eq:discrete_0_decay}, when the effective atom state is pure, Eq.~\eqref{eq:psi_decay}, the Kraus operators in Eq.~\eqref{eq:Kraus_0_decay} do not feature a pure steady state when $\Gamma_1\neq \Gamma_3$ [cf. case B in Appendix~\ref{app:2photon}]. Furthermore, even in the case of the uniform decay,  since pure stationary states vary with the interaction time [cf.~Eq.~\eqref{eq:psi_ss}], random interaction times caused by decay events will lead to mixed stationary states of the overall dynamics.

~\\
\emph{Dynamics in the limit of weak atom decay}. We now consider a limit of weak decay with respect to two timescales  $T$ and $\tau$, which determine the effective atom state and the atom-cavity interaction, respectively. In the first order, due to the parity conservation, atom decay will lead to dephasing dynamics of odd-even coherences, as given by Eq.~\eqref{eq:Leff_deph}. \\

  \emph{(1) Contribution from $T$}. For $\Gamma_1,\Gamma_3 \ll T^{-1}$, the effective atom state entering the cavity is approximately pure. Assuming further weak decay from all relevant levels, i.e., $\Gamma_j\ll  T^{-1}$ also for $j=0,2$, from Eq.~\eqref{eq:rho_decay} we simply have [cf. Eq.~\eqref{eq:rho_at_mixed}]
\begin{eqnarray}\label{eq:rho_decay_lin}
 &&\overline p_a(T)=1-\left(\Gamma_3-\gamma_{-13} \right) T\, |c_e|^2-\left(\Gamma_1 -\gamma_{13}-\gamma_{-11} |c_e|^2\right)T\, |c_g|^2,\qquad\overline p_b(T)=\gamma_{13} T |c_e|^4,\\\nonumber 
 &&\overline p_0(T)=\gamma_{01} T\,|c_g|^2+\gamma_{03}T\,|c_e|^2 ,\qquad \overline p_2(T)=\gamma_{23}T\,|c_e|^2 , \qquad p_4=p_\text{a}=0,
\end{eqnarray}
with [cf. Eq.~\eqref{eq:psi} and~\eqref{eq:psi_decay}]
\begin{equation}\label{eq:psi_decay2}
|\overline \psi_a(T)\rangle= c_g [1-(\Gamma_1-\Gamma_3 )T \,|c_e|^2/2 +\gamma_{13 } T\, |c_e|^4 ]|1 \rangle + c_e [1+(\Gamma_1-\Gamma_3 )T \,|c_g|^2/2 -\gamma_{13 } T\, |c_g|^2 |c_e|^2 ]|3 \rangle,
\end{equation}
 which is normalized up to the first order, and $|\overline \psi_b(T)\rangle$ supported on $|1\rangle$  and $|3\rangle$ and orthogonal to $|\overline \psi_a(T)\rangle$ [cf. Eq.~\eqref{eq:psi12}]. The bound in Eq.~\eqref{eq:deph_bound1} gives the resulting contribution to the dephasing as
\begin{eqnarray}\label{eq:deph_bound1_T}
\gamma_\text{deph}(T)\leq 2 \overline \nu \,\left[1-\overline p_a(T)\right]% &\approx& 2 \nu \left[\overline p(T)-1+\Gamma_1 T |c_g|^2+\left(\Gamma_3-\gamma_{13}|c_g|^2 \right) T |c_e|^2 \right]\\\nonumber
&\approx & 2 \nu \left[\left(\Gamma_1-\gamma_{-11}\right) |c_g|^2+\left(\Gamma_3-\gamma_{-13} -\gamma_{13}|c_g|^2 \right)  |c_e|^2 \right] T
\end{eqnarray}
(since we consider first-order effects, we assumed a unitary atom-cavity dynamics). Here dephasing takes place between pure stationary states of the cavity obtained with the atom state $|\overline\psi_{a}(T)\rangle$ instead of Eq.~\eqref{eq:psi}. 
%Here the dephasing  won't take place if I change the state, for example the first-order corrections would be 0 due to real dynamics. But actually all corrections in the dynamics would be 0, with perturbation theory adding corrections only to the state (eigenmatrices). 
In particular,  for the decay only to the uncoupled levels, the bound in Eq.~\eqref{eq:deph_bound1_T} indicates no dephasing, which is indeed due to the effective state being pure [cf.~Eq.~\eqref{eq:psi_decay}; this  observation actually holds for any $T$]. An analogous bound holds for the frequency of unitary dynamics [cf.~Eqs.~\eqref{eq:deph_bound2} and~\eqref{eq:rho_decay}]
\begin{eqnarray}\label{eq:deph_bound2_T}
|\Omega(T)|\leq  \overline \nu\left[\overline p_0(T)+\overline p_2 (T)\right] &\approx&  \nu \left[\gamma_{01} |c_g|^2+\left(\gamma_{03}+\gamma_{23}\right) |c_e|^2\right] T.
\end{eqnarray}
The inequalities in Eqs.~\eqref{eq:deph_bound1_T} and~\eqref{eq:deph_bound2_T} hold also in the case of weak decay only for $|1\rangle$ and $|3\rangle$ levels~\footnote{In this case, we can estimate the leading probability $\overline{p}_a(T)$ from below by considering the fidelity of $\overline\rho_\text{at}(T)$ to any state ${\overline{p}_a(T)\geq \langle\psi|\overline\rho_\text{at}(T)|\psi\rangle}$. For example, for ${|\psi\rangle=|\overline\psi_\text{at}(T)\rangle}$ we have  $\overline{p}_a(T) \geq[ 1-\Gamma_1 T |c_g|^2-\Gamma_3 T |c_e|^2 + \gamma_{13} T |c_e|^2 |c_g|^2 ]/\overline{p}(T)$ in the limit $\Gamma_1,\Gamma_3 \ll T^{-1}$ [by considering the contribution to  $\rho_\text{at}(t)$ from nondecaying events and decay ${|3\rangle}$ to ${|1\rangle}$]. The bound in Eq.~\eqref{eq:deph_bound1} gives the resulting contribution to the dephasing as $\gamma_\text{deph}(T)\lesssim 2 \nu \left[\overline p(T)-1+\Gamma_1 T |c_g|^2+\left(\Gamma_3-\gamma_{13}|c_g|^2 \right) T |c_e|^2 \right]$.}.\\

\emph{(2) Contribution from $\tau$}. Second, we discuss the dephasing due to atom decay during its interaction with the cavity for $\Gamma_1,\Gamma_3\ll \tau^{-1}$. 

For the decay only to uncoupled levels we have
\begin{eqnarray}\label{eq:deph_atom_decay_1}
\frac{\gamma_\text{deph}(\tau)}{\nu}&=&  \frac{\Gamma_1+\Gamma_3}{2}\tau - \int_{0}^\tau  d t\,  \sum_{j=g,e}\Gamma_j \mathrm{Tr}\left[L_{+-} \,\mathcal{M}_{j}(\lambda t) \left(|\Psi_+\rangle\!\langle \Psi_-| \right)\right] -\frac{\Gamma_1-\Gamma_3}{4}\tau \left(\langle Y\rangle_+ +\langle Y\rangle_- \right),\\\label{eq:deph_atom_decay_2}
\Omega&=&0,
\end{eqnarray}
where
\begin{equation}\label{eq:Y}
Y= |c_g|^2 \frac{\sin(\phi\sqrt{a^2\ken a^{\dagger  2}})}{\phi\sqrt{a^2\ken a^{\dagger  2}}}+|c_e|^2 \frac{\sin(\phi\sqrt{a^{\dagger  2}\ken a^2 })}{\phi\sqrt{a^{\dagger  2}\ken a^2 }},
\end{equation}
and no unitary dynamics follows from the fact that since $\mathcal{M}_g(t)$ and $\mathcal{M}_e(t)$ can be considered real valued for all $t$ [cf.~Sec.~\ref{sec:dynamics}]. Furthermore, in the case of the \emph{uniform} decay [cf.~Eq.~\eqref{eq:discrete_0_decay_uniform}], the modified dynamics $\overline{\mathcal{M}}_0$ can be seen as a perturbation of the dynamics without decay $\mathcal{M}_0(\tau)$ at a rate $\overline{\nu}$ further reduced by $e^{-\Gamma \tau}$, by a quantum channel $\Gamma/(1-e^{-\Gamma \tau}) \int_{0}^\tau  d t\, e^{- \Gamma t} \mathcal{M}_0(t)$ at the rate  $\overline{\nu} (1-e^{-\Gamma \tau})$. Therefore, in the limit $\Gamma\ll \tau^{-1}$, the resulting dephasing rate in Eq.~\eqref{eq:deph_atom_decay_1} is bounded as [cf. the derivation of Eq.~\eqref{eq:deph_bound1}]
\begin{equation}
\label{eq:deph_bound1_tau_uniform}
\gamma_{\rm deph}(\tau)\leq 2\overline{\nu} \,\Gamma \tau, 
\end{equation} 
while there is no unitary dynamics, $\Omega=0$, since $\mathcal{M}_0(t)$ can be considered real valued for all $t$ [cf.~Sec.~\ref{sec:dynamics}]. Similarly, for the \emph{nonuniform decay} to the uncoupled levels only [cf.~Eq.~\eqref{eq:discrete_0_decay}], we have $\Omega=0$, since $\overline{\mathcal{M}}_{g,e}(t)$ can be considered real valued for all $t$ and 
\begin{equation}
\label{eq:deph_bound1_tau}
\gamma_{\rm deph}(\tau)\leq 2\overline{\nu} \,\max\left(\Gamma_1,\Gamma_3\right) \tau. 
\end{equation} 
Equation~\eqref{eq:deph_bound1_tau} follows from Eq.~\eqref{eq:deph_bound1_tau_uniform} and the fact that we can consider first-order effects. Indeed, for $\Gamma_1>\Gamma_3$ ($\Gamma_3>\Gamma_1$), increasing the decay rate from $|3\rangle$ ($|1\rangle$) by $\Gamma_1-\Gamma_3$ ($\Gamma_3-\Gamma_1$), while leading to the uniform decay at the rate $\max\left(\Gamma_1,\Gamma_3\right)$,  can only increase the effective dephasing rate $\gamma_{\rm deph}(\tau)$. Note that here we consider the unperturbed dynamics with respect to the pure atom state in Eq.~\eqref{eq:psi_decay} rather than Eq.~\eqref{eq:psi}, which is modified due to atom decay before entering the cavity.

Finally, for general case, we note that Eq.~\eqref{eq:deph_bound1_tau} also holds true provided that we also assume $\Gamma_1,\Gamma_3 \ll T^{-1}$  [cf.~Eq.~\eqref{eq:discrete_decay2}], in which limit $\overline{\nu}$ can be further replaced by $\nu$,
\begin{equation}
\label{eq:deph_bound1_tau2}
\gamma_{\rm deph}(\tau)\leq 2\nu \,\max\left(\Gamma_1,\Gamma_3\right) \tau. 
\end{equation}
Indeed, in that case, since we consider first-order contributions, we can assume pure atom state $|\overline\psi_a(T)\rangle$ in Eq.~\eqref{eq:psi_decay2} entering the cavity, so that Eq.~\eqref{eq:Kraus_decay1} is given by Eq.~\eqref{eq:Kraus_0_decay} with $|\overline\psi_a(T)\rangle$ [instead of $|\overline\psi_\text{at}(T)\rangle$ in Eq.~\eqref{eq:psi_decay}]. Thus, increasing the decay rate from $|1\rangle$ or $|3\rangle$ to achieve the uniform decay rate equal $\max\left(\Gamma_1,\Gamma_3\right)$  gives again a perturbation of $\mathcal{M}_0(\tau)$ by a quantum channel multiplied by $\max(\Gamma_1,\Gamma_3)\tau$. In this general case, the unitary dynamics is possible with the frequency bounded as 
\begin{equation}
\label{eq:deph_bound2_tau}
|\Omega(\tau)|\leq \nu \,\left(\gamma_{01}+\gamma_{03}+\gamma_{23}\right) \tau, 
\end{equation} 
where we further assumed weak decay from all relevant levels, i.e., $\Gamma_j\ll  T^{-1}$ also for $j=0,2$, so that the contribution to the frequency comes from $\int_{0}^\tau  d t\, [\gamma_{01}\, \overline{\mathcal{M}}_g(t)+(\gamma_{03}+\gamma_{23})\,\overline{\mathcal{M}}_e(t)]$ [cf.~Eq.~\eqref{eq:M_0_decay}], and we used the fact that $\mathcal{M}_g(t)$ and $\mathcal{M}_e(t)$ are completely positive and do not increase trace [cf. the derivation of Eq.~\eqref{eq:deph_bound2}].\\ % I do not know whether by incerasing noise I would incerase of decrease the frequency as its sign is not fixed...

We conclude that in the limit of weak decay with respect to both $T$ and $\tau$, we obtain dephasing dynamics of odd-even coherences at the rate $\gamma_{\rm deph}(\tau)+\gamma_{\rm deph}(\tau)$ in Eqs.~\eqref{eq:deph_bound1_T} and~\eqref{eq:deph_bound1_tau2}, with unitary rotation at the frequency $\Omega(T)+\Omega(\tau)$  in Eqs.~\eqref{eq:deph_bound2_T} and~\eqref{eq:deph_bound2_tau} [see Eqs.~\eqref{eq:deph_bound1_ad} and~\eqref{eq:deph_bound2_ad}]. We also note that even in the case of nonuniform decay, the dephasing rate is bounded by the change in the purity of the atom state during the total time $T+\tau$. Finally, for the decay only to levels not coupled to the cavity, the timescale $T$ does not contribute to noise, but it modifies the rate of the unperturbed dynamics to $\overline\nu$, so that its relaxation timescales are rescaled by $ (e^{-\Gamma_1 T} |c_g|^2+e^{-\Gamma_3 T} |c_e|^2)$.

~\\
\emph{Dynamics in the limits of weak coupling and weak atom decay}. In the weak-coupling limit, we can expand Eqs.~\eqref{eq:Kraus_decay1} and~\eqref{eq:Kraus_decay2} up to quadratic order in $\phi$ and $c_e$ [see Sec.~\ref{sec:schrodinger_cats}] and linear order in decay. %Here we cannot do taht in two steps, as in the case of mixed atoms, as the decay goes into the argument of trigonometic functions.

For the case of decay only toward uncoupled levels with $\Gamma_1,\Gamma_3\ll \tau^{-1}$, [Eqs.~(\ref{eq:discrete_0_decay})-(\ref{eq:Kraus_0_decay})], we obtain 
\begin{eqnarray}
\frac{ d }{ d t} \rho&=& -i [\overline g_\text{2ph}^* a^2+ \overline g_\text{2ph} a^{\dagger 2},\rho]+ \overline \kappa_\text{2ph}\left( a^2 \rho a^{\dagger 2} -\frac{1}{2}\left\{a^{2\dagger }a^2,\rho\right\}\right),
\end{eqnarray}
where 
\begin{equation}
\overline g_\text{2ph}=\nu e^{-\frac{\Gamma_1+\Gamma_3}{2}T} c_{\rm g}^* c_{\rm e}\, \phi\left(1-\frac{\Gamma_1+\Gamma_3}{4} \tau\right),\qquad 
\overline\kappa_\text{2ph}=\nu e^{-\Gamma_1 T} |c_{\rm g}|^2  \phi^2\left(1-\frac{\Gamma_1 }{2}\tau-\frac{\Gamma_3 }{6}\tau\right).
\end{equation}
This result follows from the expansion
$\overline{M}_g(t)\approx \overline c_g(T)\, (1-\Gamma_1 t/2)-\overline c_e (T)\phi {a^\dagger}^2  -\overline c_g(T)\,\phi^2 {a^\dagger}^2  a^2 [1- (\Gamma_1-\Gamma_3) t/12 -(\Gamma_1+\Gamma_3) t/4  ]/2$ and $\overline{M}_e(t)\approx \overline c_e(T)\, (1-\Gamma_3 t/2)  -i \overline c_g(T)\,\phi a^2 [1-(\Gamma_1+\Gamma_3) t/4  ]$, which holds for $|\overline c_e(T)|/|c_g(T)|=e^{-(\Gamma_3-\Gamma_1) T }|c_e|/|c_g| \ll1$  (which in general does \emph{not} require $\Gamma_1,\Gamma_3\ll T^{-1}$), and considering the contributions to Eq.~\eqref{eq:continuous_decay} up to quadratic order in $\phi$ and $\overline c_e(T)$. Therefore, in the presence of the decay to only uncoupled levels, there is no effective dephasing, and the DFS of pure Schr\"{o}dinger cat states with a modified parameter $\overline\alpha$,
\begin{equation}
\frac{|\overline \alpha\rangle\pm|-\overline\alpha\rangle}{\sqrt{2\pm 2 e^{-2|\overline\alpha|^2}}},\qquad \overline\alpha= e^{- i\frac{\pi}{4}}\sqrt{\frac{2 \overline g_\text{2ph} }{\overline\kappa_\text{2ph} }}=e^{\frac{\Gamma_1-\Gamma_3}{4} T }\left(1+ \frac{3\Gamma_1-\Gamma_3}{24}\tau \right) e^{- i\frac{\pi}{4}} \sqrt{\frac{2 c_e  }{c_g\phi}}, \label{eq:psi_ss_lin_decay}
\end{equation} 
remains stationary [cf.~Eqs.~\eqref{eq:psi_ss_lin} and~\eqref{eq:master_lin}].

For the general case of the atom decay in Eq.~(\ref{eq:master_decay}), we will have two contributions to the metastable dephasing dynamics in Eq.~\eqref{eq:Leff_deph}, from the mixedness of atom state $\overline\rho_\text{at}(T)$ [Eq.~\eqref{eq:rho_decay}], and via the decay during interaction with the cavity [Eqs.~\eqref{eq:Kraus_decay1} and~\eqref{eq:Kraus_decay2}].  In the first order of the limit of weak decay from all relevant levels $\Gamma_j\ll \tau^{-1},T^{-1}$ for $j=0,1,2,3$, the mixedness of atom state $\overline\rho_\text{at}(T)$ will contribute to the dephasing dynamics as given in Eqs.~\eqref{eq:deph_lin} and~\eqref{eq:deph_lin2} with the probabilities from Eq.~\eqref{eq:rho_decay_lin} [see also Eq.~\eqref{eq:master_lin_mixed}]. Second, since we can neglect the contribution from the mixedness of atoms when considering decay during the interaction with the cavity, the Kraus operators in Eq.~\eqref{eq:Kraus_decay1} can be expanded as in the case of decay only to uncoupled levels but for $|\overline\psi_a(T)\rangle$ in Eq.~\eqref{eq:psi_decay2} instead of $|\overline\psi_\text{at}(T)\rangle$ in Eq.~\eqref{eq:psi_decay}, while Kraus operators in Eq.~\eqref{eq:Kraus_decay2} will feature only a single decay event with $\overline{H}(t)$ replaced by  $\overline{H}_\text{eff}(t)$, and can be further expanded to consider only the contributions up to quadratic order in $\phi$ and $c_e$ for Eq.~\eqref{eq:continuous_decay}.

~\\
\emph{Stationary states in the limit of weak atom decay}. The atom decay will not only change the long-time dynamics, but also introduce corrections to the steady states, rendering them no longer pure, but mixed [cf.~Sec.~\ref{sec:QFI_loss}]. In order to generate approximately pure steady states, it is thus important to achieve $\Gamma_1,\Gamma_3 \ll \tau^{-1}, T^{-1}$. Interestingly, for the decay only toward levels not coupled to the cavity, the timescale $T$ does not play a role, but the only approximately pure stationary states are still altered by the decay, since in general the effective atom state in Eq.~\eqref{eq:psi_decay} changes with $T$ [see also Eq.~\eqref{eq:psi_ss_lin_decay}].

	%===================================================================================================================================================

				\subsubsection{Metastable dynamics due to nonmonochromatic atom beam}
				\label{app:Leff:conserved:atom_beam}

		In Secs.~\ref{sec:model}-\ref{sec:metastability}, we assumed that the atomic beam is monochromatic, i.e., the velocity $v$ of all atoms passing through the cavity is the same, leading to identical time $\tau$ spent in the cavity, and thus the uniform value of the integrated coupling strength $\phi$ [see Eqs.~\eqref{eq:KrausFULL} and~\eqref{eq:master}, and cf.~Appendix~\ref{app:micromaser}]. Here, we discuss how the micromaser dynamics is changed for a nonmonochromatic atomic beam.
			
		~\\
		\emph{Micromaser dynamics}.	
		We consider atom velocities drawn from a probability distribution $p(v)$, which can be, for example, a Maxwell-Boltzmann distribution, i.e., a Gaussian distribution with thermal width $\sqrt{k_B T/ m}$ and the corresponding average velocity of the atoms $\overline v = \int  d v\,p(v) v$. The velocity distribution determines the probability distribution of the integrated coupling given by $g(\phi)d\phi= p(l/\phi)l/\phi^2 d\phi$, where $l$ is the length of the cavity (note that in general $\overline \phi =l \overline{v^{-1}} \neq l /\overline v$). The dynamics of the cavity due to a single atom passage is now described by the average  [cf.~Eqs.~\eqref{eq:discrete} and~\eqref{eq:discrete_2}]
\begin{equation}\label{eq:Kraus_average}
\overline{\mathcal{M}} = \int d \phi\, g(\phi)\mathcal{M}(\phi),
\end{equation}			
where $\mathcal{M}(\phi)$ denotes the dynamics with the integrated coupling strength $\phi$ [see Eq.~\eqref{eq:Kraus}]. 

~\\
\emph{Mixed stationary states of even and odd parities}.	 As the recurrence relation in Eq.~(\ref{eq:psi_ss}) obeyed by pure stationary states depends on $\phi$, it can no longer be fulfilled for all velocities so that the stationary state becomes in general mixed [cf.~Appendix~\ref{app:Leff:conserved:atom_mixed}]. Nevertheless, due to the far-detuned limit in Eq.~\eqref{eq:Heff}, the parity is conserved by the dynamics, and thus there exists two even and odd stationary states~\cite{albert2014symmetries}, which are mixed [cf. Fig.~\ref{fig:phi_spread}(a)]. 

%Well this conditions have to be really only fulfilled by the orthogonal Kraus operators for the average channel, the representation by $M(\phi)$ is in general redundant. Yes, but if pure stationary state, it has to be  REMAIN pure for all representations, I.E. 	on all trajectories, so OK!!!

%what about cat states? They are invariant under Heff (this is averaged over all \phi!), and each jump corresponds really to double loss, so the states are left invariant! But each Kraus operator has a different stationary state really... Actually, in the approximation we have they are all just a sum of 1 and double loss, so they do not have unique stationary state, and this is what allows for the final state of the averaged dynamics to be pure! Beyond weak-coupling limit though, Kraus operators have unique stationary state??? Yes, in the waek coupling limit, we effectively have only one equation (22b)!

~\\
\emph{Metastable dephasing dynamics}. In the case in which the distribution of the integrated coupling is sufficiently peaked around  its average, we expect $\delta \mathcal{M}\equiv\overline{ \mathcal{M}} -\mathcal{M}(\overline{ \phi}) $ can be treated as a perturbation of $\mathcal{M}(\overline{ \phi})$. In such a case, it induces the dephasing dynamics  within the DFS of the pure stationary states of $\mathcal{M}(\overline{ \phi}) $, as the parity is conserved [see Eq.~\eqref{eq:Leff_deph}]. Furthermore, as the dynamics of $\mathcal{M}( \phi)$ corresponds to real-valued dynamics for all $\phi$ (cf.~Sec.~\ref{sec:dynamics}), there is no associated Hamiltonian contribution and $\Omega=0$ in  Eq.~\eqref{eq:Leff_deph}. In particular, by expanding Kraus operator in Eq.~\eqref{eq:Kraus} in $\phi=\overline\phi+\delta \phi$, we arrive at
\begin{eqnarray}\label{eq:deph_atom_decay}
\gamma_\text{deph} &=& \nu \,\overline {\delta \phi^2} \, \bigg[\frac{|c_g|^2}{2}  \left(\langle a^{\dagger 2}\,a^2 \rangle_++\langle  a^{\dagger 2}\,a^2\rangle_- \right) +\frac{|c_e|^2}{2} \left(\langle a^2 \,a^{\dagger 2}\rangle_++\langle a^2 \,a^{\dagger 2}\rangle_- \right)  \\\nonumber
&& \qquad\quad - |c_g|^2 \mathrm{Tr} \left(L_{+-} a^2  |\Psi_+\rangle\!\langle \Psi_-|a^{\dagger 2}\right) - |c_e|^2  \mathrm{Tr} \left(L_{+-} a^{\dagger 2}|\Psi_+\rangle\!\langle \Psi_-| a^2   \right)  \bigg],
\end{eqnarray}
with $ \overline{\delta \phi^2}$ being the variance of the distribution of $\phi$. 
\begin{comment}
The second term corresponds to $\sum_{j=e,g}\mathrm{Tr} \left(L_{+-} M_{j}^{(1)}|\Psi_+\rangle\!\langle \Psi_-|  M_{j}^{(1)\dagger}  \right)$ with the first-order corrections to Kraus operators
\begin{subequations}
	\label{eq:Kraus_atom_beam}
	\begin{align}
	M_{\rm g}^{(1)} &=-c_{\mathrm{g}}  \ken \sin\left(\overline \phi \sqrt{a^{\dagger  2}\ken a^2}\right)\sqrt{a^{\dagger  2}\ken a^2}- ic_{\mathrm{e}}\ken a^{\dagger  2}\ken\cos\left(\overline\phi \sqrt{a^2\ken a^{\dagger  2}}\right),\label{eq:Kraus1_atom_beam}\\
	M_{\rm e}^{(1)} &=- ic_{\mathrm{g}} \ken a^2\ken\cos\left(\overline\phi \sqrt{a^{\dagger 2}\ken a^2}\right)-c_{\mathrm{e}}\sin\left(\overline\phi\sqrt{a^{2}\ken a^{\dagger  2}}\right)\sqrt{a^{2}\ken a^{\dagger  2}}.\label{eq:Kraus2_atom_beam}
	\end{align}
\end{subequations}
We obtain flipped values...
\end{comment}
To arrive at Eq.~\eqref{eq:deph_bound1_atom_decay}, we assumed that $\delta \phi \, n \ll 1$ within the support of the pure stationary states, which corresponds to the condition
\begin{equation}\label{eq:deph_atom_decay_cond}
 \left(\langle n\rangle_\pm + \sqrt{ \langle n^2\rangle_\pm- \langle n\rangle^2_\pm} \right)\overline {\delta \phi^2} \ll 1.
\end{equation}
The effective dephasing dynamics  manifests the fact that the even and odd stationary states of the dynamics with $\overline{ \mathcal{M}} $ are mixed (and only in zero order are they approximated by the pure states $|\Psi_+\rangle$ and $|\Psi_-\rangle$), and coherences between them are not  stationary.

We note that Eq.~\eqref{eq:deph_bound1_atom_decay} can be interpreted as originating from dynamics with two photon losses   and and two-photon injections at the respective rates $\nu \,\overline {\delta \phi^2} |c_g|^2$ and $\nu \,\overline {\delta \phi^2} |c_e|^2$. Therefore, as  $a^2 \rho a^{\dagger 2}$ is a positive matrix, so is its projection on the DFS, $\Pi_0(a^2 \rho a^{\dagger 2})$ [cf. Eq.~\eqref{eq:P00}], and thus we have $\mathrm{Tr} (L_{+-}a^2 |\Psi_+\rangle\!\langle \Psi_-|a^{\dagger 2}   )|\leq  \sqrt{\langle a^{\dagger 2}\,a^2 \rangle_+\langle  a^{\dagger 2}\,a^2\rangle_- }$ [see also Eq.~\eqref{eq:etaJ}]. Analogously, $|\mathrm{Tr} (L_{+-} a^{\dagger 2}|\Psi_+\rangle\!\langle \Psi_-| a^2   )|\leq  \sqrt{\langle a^2 \,a^{\dagger 2}\rangle_+\langle a^2 \,a^{\dagger 2}\rangle_-}$ and we arrive at the lower bound   
\begin{eqnarray}\label{eq:deph_bound0_atom_decay}
 \gamma_\text{deph} \geq  \nu \,\overline {\delta \phi^2} \,\left[\frac{|c_g|^2}{2}  \left(\sqrt{\langle a^{\dagger 2}\,a^2 \rangle_+}-\sqrt{\langle  a^{\dagger 2}\,a^2\rangle_-} \right)^2 +\frac{|c_e|^2}{2} \left(\sqrt{\langle a^2 \,a^{\dagger 2}\rangle_+}-\sqrt{\langle a^2 \,a^{\dagger 2}\rangle_-} \right)^2\right],
\end{eqnarray}
and the upper bound [cf. Eq.~\eqref{eq:deph_bound1_ab}]
\begin{eqnarray}\label{eq:deph_bound1_atom_decay}
\gamma_\text{deph} \leq  \nu \,\overline {\delta \phi^2} \,\left[\frac{|c_g|^2}{2} \left(\sqrt{\langle a^{\dagger 2}\,a^2 \rangle_+}+\sqrt{\langle  a^{\dagger 2}\,a^2\rangle_-} \right)^2  +\frac{|c_e|^2}{2} \left(\sqrt{\langle a^2 \,a^{\dagger 2}\rangle_+}+\sqrt{\langle a^2 \,a^{\dagger 2}\rangle_-} \right)^2\right].
\end{eqnarray}
%I get an upper and lower bound cause the action of jump is not the same in odd and even subspace - the effective qubit is unbalanced. Indeed for the same averages the lower bound is simply 0 [so if I have a quantum channel the averages are simply 1].

%Below we discuss that for linear states this does not matter --- a good platform to implement Schoredinger cat states. 

~\\
\emph{No metastable dephasing in weak-coupling limit}. In the weak-coupling limit, however, from Eq.~\eqref{eq:Kraus_average}, we obtain the dynamics described by Eq.~\eqref{eq:master_lin} with the averaged coefficients $\braket{g_\text{2ph}}=\nu   c_{\rm g}^* c_{\rm e}\braket{\phi}$,  $\braket{\kappa_\text{2ph}}=\nu  |c_{\rm g}|^2 \braket{\phi^2}$. Therefore, in the weak-coupling limit, the stationary states are pure Schr\"{o}dinger cat states of Eq.~\eqref{eq:psi_ss_lin} with $\alpha= e^{- i\pi/4}\sqrt{2 \braket{ g_\text{2ph}}/\braket{\kappa_\text{2ph}} }=e^{- i\pi/4}\sqrt{2  c_e\braket{ \phi}/(c_g\braket{\phi^2}) }$, and their coherences are stationary as well. Indeed, in Eq.~\eqref{eq:deph_atom_decay}, we only have contribution from two-photon losses [cf. Eq.~\eqref{eq:cond_lin}], which preserve the DFS of cat states and give $\gamma_\text{deph}=0$. We emphasize that this approximation requires the weak-coupling limit to be valid for all values of $\phi$ attainable in the distribution $g(\phi)$ [cf. Eq.~\eqref{eq:deph_atom_decay_cond}].

%HOW DOES IT HAPPEN THAT I HAVE A PURE STATE ALTHOUGH MANY DIFFERENT KRAUS OPERATORS? The two-photon loss, no matter at what (random) rate does not change a cat state, only the effective Hamiltonian [averaged over all phi] can change it with the decay toward the correct alpha. But effective Hamiltonian will be different in each realisation of phi (time)... yes, but this never interacts for too long, so they add up translating randomness of phi into randomness of time between jumps.... the effective Hamiltonian is -i g^* a^2 + a^dagger^2[g-k/2 a^2], and this is only ever applied in the linear order, and again a^2 acts as multiplying by a number, so the first term is just a phase while the second needs to go to 0 by this still varies by phi... 

~\\
\emph{Phase estimation precision}.  In the lowest-order in $\delta \mathcal{M}$, the nonmonochromaticity of the atom beam leads to the dephasing of the odd-even coherences, so that the QFI of the states of fixed parity is not affected. However, those stationary states are only approximately pure [cf.~Fig.~\ref{fig:phi_spread}(a)] with corrections proportional to $\delta \mathcal{M}$ and  the relaxation time of $\mathcal{M}(\overline{\phi})$ (cf.~Sec.~\ref{sec:QFI_loss}). This mixedness introduced by the nonmonochromaticity of atom beam affects the QFI in phase estimation,~\eqref{eq:qfi_formula} [cf.~Fig.~\ref{fig:phi_spread}(b)].

This can be understood as follows. The enhancement in estimation precision and the long relaxation time is due to the presence of soft walls (cf.~Sec.~\ref{sec:metrology}). The height and position of soft walls, $\sin_m(\phi)\approx 0$, however, depends on $\phi$, leading to strong variations of the structure of the stationary states of $\mathcal{M}(\phi)$ (see Fig.~\ref{fig:wig_gall}) and thus also the QFI (cf.~Fig.~\ref{fig:qfi}). Therefore, for a broad enough distribution $g(\phi)$, the individual stationary states of $\mathcal{M}(\phi)$ differ significantly from the stationary state of $\mathcal{M}(\overline{\phi})$, and the state of the averaged dynamics, Eq.~\eqref{eq:Kraus_average}, is mixed. But, importantly, even when the purity of the final state is significantly reduced, it can still yield an enhancement over the standard quantum limit [cf.~Fig.~\ref{fig:phi_spread}(b)]. See also Appendix~\ref{app:Leff:hard_walls_atom_beam}.

%===================================================================================================================================================

\subsection{Metastable dynamics  in the presence of hard walls}
\label{app:Leff:hard_walls}

We now discuss the effective dynamics due to: higher order corrections in the far-detuned limit [cf.~Eq.~\eqref{eq:Leff_corr}], single-photon losses [cf.~Eq.~\eqref{eq:Leff_corr}] and decay of atom levels or nonmonochromaticity of the atom beam [cf.~Eq.~\eqref{eq:Leff_deph_0}] in the case when the unperturbed dynamics features hard walls and thus multiple stationary states of the same parity (see Sec.~\ref{sec:walls_hard}). In the presence of weak noise or small imperfections these states become metastable and undergo long-time dynamics with local transitions between the hard walls [see Fig.~\ref{fig:phi_wall}].  As a consequence, there exist no trapping states in a cavity pumped by excited atoms. These results also inform Sec.~\ref{sec:beyond}, in which we discuss dynamics of a realistic micromaser with noise and corrections faster than the timescales of relaxation across soft walls, as in such case soft walls can be replaced by hard by means of approximately degenerate perturbation theory.

\subsubsection{Summary of results on effective dynamics}
\begin{figure}[t]
	\centering
	\subfloat{\includegraphics[width=0.8\columnwidth]{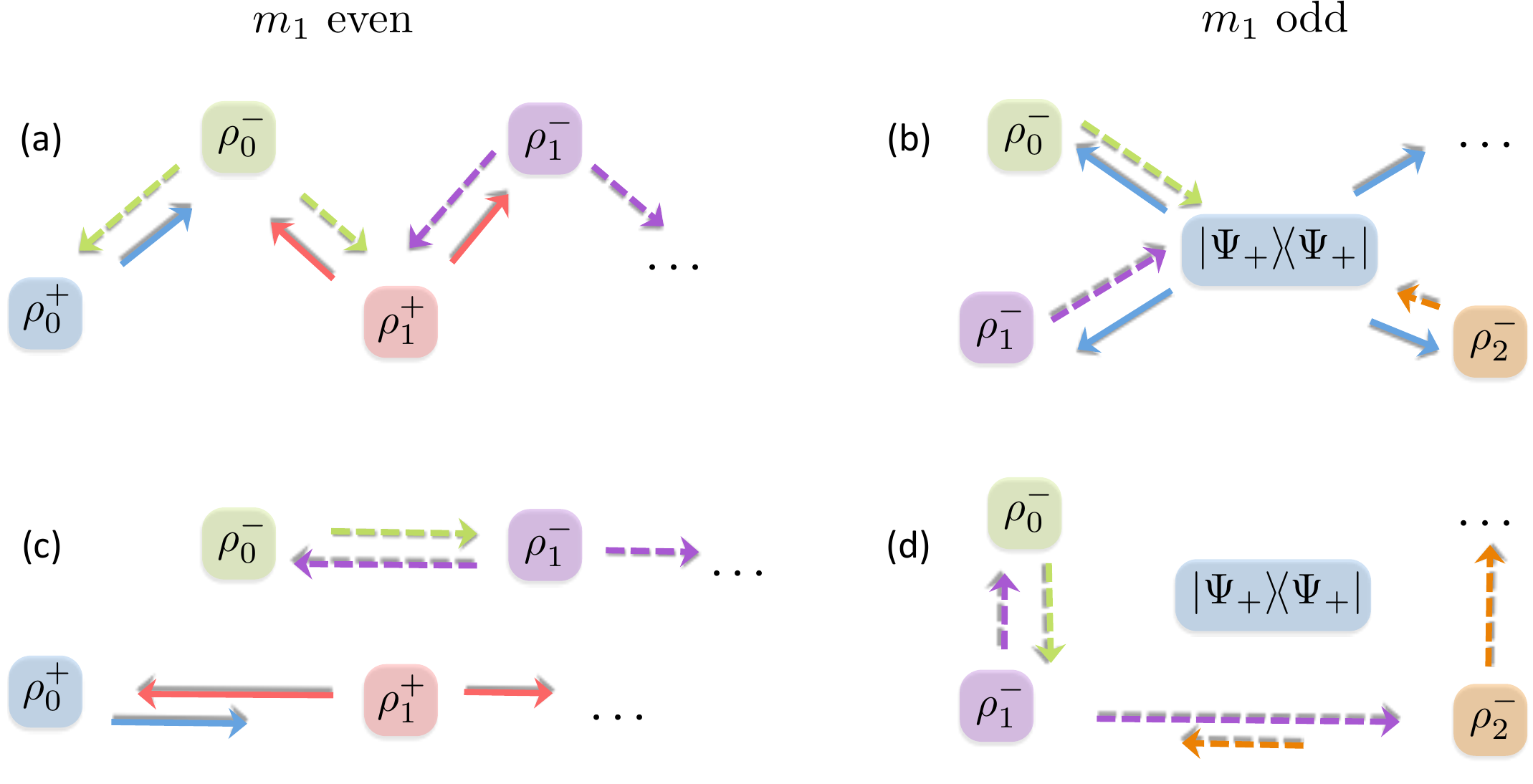}}
	\vspace*{-1mm}
	\caption{\textbf{Effective dynamics of realistic micromaser with hard walls}. %\textbf{(a)} Photon number distribution of the stationary states between hard walls: even $\rho_0^{+}$ (blue) and $\rho_1^{+}$ (red), odd $\rho_0^{-}$ (green), and the stationary state $\rho_{\rm ss}$ of the dynamics with losses (black), Eq. (\ref{eq:master_loss}). Here the first wall is even (vertical dashed line at $m_1=30$) leading to hard walls of both parities [cf. Table~\ref{tab:mK}] and multiple stationary states of both parities. The stationary state $\rho_{\rm ss}$ is supported mostly after the first hard wall, which is due to large overlap of the supports of $\rho_0^{-}$ and $\rho_1^{+}$ [see Eq.~\eqref{eq:rhoss_wall_even}]. Parameters: $c_{\rm e} = 0.58$ and $K=31$ in Eq.~\eqref{eq:phi_wall} leading to $\phi\approx 3.09$. 
		 First wall at even $m_1$ leads to hard walls of both parities (cf. Table~\ref{tab:mK}) and multiple even and odd stationary states [(a),(c)], while first wall at odd $m_1$ leads to only odd hard walls and multiple odd stationary states [(b),(d)]. 
		 \textbf{(a),(b)}  Single-photon losses and higher order-corrections to far-detuned limit induce \emph{local} transitions between states of opposite parity: from $\rho_k^+$ only to $\rho_{k-1}^-$ and  $\rho_{k}^-$ (solid arrows), and from $\rho_k^-$  only to $\rho_{k}^+$ or $\rho_{k+1}^+$ (dashed arrows) for $m_1$ even, and from the unique even state $|\Psi_+\rangle\!\langle\Psi_+|$ to  odd states $\rho_{k}^-$  (solid arrows), and from the odd states to the even state (dashed arrows) for $m_2$ odd, $k=0,1,...$. 
		\textbf{(c),(d)} Atom decay and nonmonochromaticity of atom beam lead to \emph{local} transitions only between the states of the same parity: from $\rho_{k}^\pm$ only to $\rho_{k-1}^\pm$ and $\rho_{k+1}^\pm$.  All effective dynamics feature \emph{detailed balance}.
	}	 
	\label{fig:phi_wall}
	\vspace*{-2mm}
\end{figure}

\noindent
\emph{No hard walls beyond far-detuned limit or in the presence of single-photon losses}. A hard wall at $m$ refers to the case of the zero amplitude of connecting states $|m\rangle$ and $|m+2\rangle$ [cf.~Eqs.~\eqref{eq:Kraus} and~\eqref{eq:phi_wall}]. As the wall affects only the states of the same parity [the subsequent walls are exponentially separated; see~Eq.~\eqref{eq:m}], any perturbations in the dynamics that swap the parity allow for circumventing hard walls and lead to a unique stationary state (see Fig.~\ref{fig:phi_wall}). As we discuss below, this is indeed the case for higher order corrections in the far-detuned limit and single-photon losses from the cavity.

For the first wall being even, there exist infinitely many even and odd stationary states between hard walls, which we denote $\rho_k^+$ and $\rho_k^-$, $k=0,1,...$ (cf. Table~\ref{tab:mK}). In the presence of weak single-photon losses or small higher order corrections, these states become metastable and at long times undergo transitions: from $\rho_k^+$ to $\rho_{k-1}^-$ or to $\rho_{k}^-$ at the respective rates $\gamma_{k-1,k}^{-+}$ and $\gamma_{k,k}^{-+}$, and from $\rho_k^-$  to $\rho_{k}^+$ or to $\rho_{k+1}^+$ at the respective rates $\gamma_{k,k}^{+-}$ and $\gamma_{k+1,k}^{+-}$,  where
\begin{align}\label{eq:gamma_wall_even}
\gamma_{k,k'}^{-+}&= \kappa \langle n\rangle_{k,k'}^+ + \nu \langle X\rangle_{k,k'}^+,\qquad
\gamma_{k,k'}^{+-}= \kappa \langle n\rangle_{k,k'}^- + \nu \langle X\rangle_{k,k'}^-,
\end{align}
and $\langle n\rangle_{k,k'}^\pm=\mathrm{Tr}(\mathds{1}_{k}^\mp \,a \rho_{k'}^\pm a^\dagger)$ and $\langle X\rangle_{k,k'}^\pm=\sum_{j=0,2,4}\mathrm{Tr}(\mathds{1}_{k}^\mp  \,M_j \rho_{k'}^\pm M_j^\dagger)$, while $\mathds{1}_{k}^\pm$  is the projection on the support of $\rho_k^\pm$ [cf.~Eqs.~\eqref{eq:Leff_corr} and~\eqref{eq:Leff_loss}]. The  rates in Eq.~\eqref{eq:gamma_wall_even} simply depend on the overlap of the perturbed state, i.e.,  the state after a photon loss, with the support of a state of the opposite parity.  Note that the \emph{ladder structure} of the transitions obeys detailed balance [see Fig.~\ref{fig:phi_wall}(a)]. Thus, the stationary state is approximated as  [cf.~Eq.~\eqref{eq:rhoss_combined}]
\begin{equation}\label{eq:rhoss_wall_even}
\rho_\text{ss}\approx  \sum_{k=0}^\infty  p_k^+ \,\rho_k^+ +  \sum_{k=0}^\infty  p_{k}^-\, \rho_k^-, 
\end{equation}
which is determined by the rates, in the recurrence relation
\begin{equation}\label{eq:pss_wall_even}
\frac{p_{k}^+}{p_{k-1}^-}= \frac{\gamma_{k,k-1}^{+-}}{\gamma_{k-1,k}^{-+}},\qquad \frac{p_{k}^-}{p_{k}^+}= \frac{\gamma_{k,k}^{-+}}{\gamma_{k,k}^{+-}}, 
\end{equation}
where $p_0^+$ is determined by the normalization $\sum_{k} (p_k^++p_k^-)=1$. %The distribution of the stationary state in Eq.~\eqref{eq:pss_wall_even} is then dominated by the fastest transitions, and thus the stationary state is supported on the closest states of the opposite parity  [see Fig.~\ref{fig:phi_wall}(a)].

When the first hard wall is odd, there are no hard walls of even parity. As the effective dynamics features only the transitions between the states of opposite parity, we only have transitions from $|\Psi_+\rangle\!\langle \Psi_+|$ into $\rho_k^-$, and from $\rho_k^-$ to $|\Psi_+\rangle\!\langle \Psi_+|$, for $k=0,1,...$, with the respective rates  $\gamma_{k}^{-+}$ and $\gamma_{k}^{+-}$,
\begin{align}\label{eq:gamma_wall_odd}
\gamma_{k}^{-+}&= \kappa \langle n\rangle_{k}^+ + \nu \langle X\rangle_{k}^+,\qquad \gamma_{k}^{+-}= \kappa \langle n\rangle_{k}^- + \nu \langle X\rangle_{k}^-,
\end{align}
where $ \langle n\rangle_k^-=\mathrm{Tr}(n\, \rho_k^-)$ and $\langle X\rangle_k^-=\mathrm{Tr}(X\, \rho_k^-)$, while $\langle n\rangle_k^+=\mathrm{Tr}(\mathds{1}_{k}^- \,a |\Psi_+\rangle\!\langle \Psi_+| a^\dagger)$ and $\langle X\rangle_k^+=\sum_{j=0,2,4}\mathrm{Tr}(\mathds{1}_{k}^- \,M_j |\Psi_+\rangle\!\langle \Psi_+| M_j^\dagger)$ with the projection $\mathds{1}_{k}^-$ on the support of $\rho_k^-$. Note that the \emph{star structure} also obeys detailed balance [see Fig.~\ref{fig:phi_wall}(b)]. Thus, the stationary state for $\cos_{m_1}(\phi)=-1$ is approximated by  
\begin{eqnarray}\label{eq:rhoss_wall_odd}
\rho_\text{ss}&\approx&  p^+ \,|\Psi_+\rangle\!\langle \Psi_+| +  \sum_{k=0}^\infty  p_{k}^- \rho_k^- \qquad\text{with}\qquad
\frac{p_{k}^-}{p^+}= \frac{\gamma_{k}^{-+}}{\gamma_{k}^{+-}}.\qquad 
\end{eqnarray} 
For  $\cos_{m_1}(\phi)=1$, the dynamics can additionally create odd coherences $|\Psi_k^-\rangle\!\langle \Psi_{k'}^-|$ from the even state at the rate $\kappa \mathrm{Tr}(L^-_{k,k'} \,a|\Psi_+\rangle\!\langle \Psi_+|a^\dagger)+\nu \sum_{j=0,2,4}\mathrm{Tr}(L^-_{k,k'} \,M_j|\Psi_+\rangle\!\langle \Psi_+|M_j^\dagger) $. These coherences decay at the rate  $\kappa(\langle n\rangle_k^- + \langle n\rangle_{k'}^-)/2+\nu(\langle X\rangle_k^- + \langle X\rangle_{k'}^-)/2$, leading to the stationary state 
\begin{equation}\label{eq:rhoss_wall_odd_2}
\rho_\text{ss}=  p^+ \, |\Psi_+\rangle\!\langle \Psi_+|+ \sum_{k=0}^\infty  p_{k}^-\, |\Psi_{k}^-\rangle\!\langle \Psi_{k}^-| +\sum_{k=0}^{\infty}\sum_{\substack{k'> k:\\(k'-k)|2 }}  \left[ c_{k,k'}^{--} |\Psi_k^-\rangle\!\langle \Psi_{k'}^-| +(c_{k,k'}^{--})^* |\Psi_{k'}^-\rangle\!\langle \Psi_{k}^-| \right] 
\end{equation}
with the probabilities as in Eq.~\eqref{eq:rhoss_wall_odd} and
\begin{equation}
c_{k,k'}^{--}= 2\frac{\,\kappa\,\mathrm{Tr}\left(L^-_{k,k'} \,a|\Psi_+\rangle\!\langle \Psi_+|a^\dagger\right)+\nu\sum_{j=0,2,4} \mathrm{Tr}\left(L^-_{k,k'} \,M_j|\Psi_+\rangle\!\langle \Psi_+|M_j^\dagger\right)}{\kappa\langle n\rangle_k^- + \kappa\langle n\rangle_{k'}^-+\nu \langle X\rangle_{k}^-+\nu \langle X\rangle_{k'}^-} p_\text{ss}^+.
\end{equation}

~\\
\emph{No hard walls for finite-life time of atom levels}. Similarly, the atom decay or nonmonochromatic atom beam lead to a random distribution of interaction times between atoms and the cavity, and thus the value of the integrated coupling $\phi$ fluctuates, and so does the presence of hard walls. Therefore, in the limit of weak noise, formerly stationary states between hard walls become metastable and effectively connected to the preceding and following states of the same parity  [see Figs.~\ref{fig:phi_wall}(c) and~\ref{fig:phi_wall}(d)].  The dynamics takes place independently in the odd- and even-parity subspace as a consequence of the parity conservation, and leads to two, rather than one, odd and even stationary states. 

For the first wall being even, $\rho_k^\pm$ can transition to $\rho_{k-1}^\pm$ or to $\rho_{k}^\pm$ at the respective rates $\gamma_{k-1,k}^\pm$ and $\gamma_{k+1,k}^\pm$,  where
\begin{subequations}\label{eq:gamma_wall_even2}
\begin{align}
\gamma_{k+1,k}^+&= \nu \,|c_e|^2\left[ \frac{\Gamma_1}{2} \tau + \overline{\sin^2_{m_{2k+1}}\!(\phi)} \right] \!\langle m_{2k+1}|\rho_{k}^+|m_{2k+1} \rangle,\\
\gamma_{k-1,k}^+&=\nu \,|c_g|^2\left[ \frac{4\Gamma_3-3\gamma_{13}}{8}\tau + \overline{\sin^2_{m_{2k-1}}\!(\phi)} \right]\! \langle m_{2k-1}\!+\!2|\rho_{k}^+|m_{2k-1}\!+\!2 \rangle,\\
\gamma_{k+1,k}^-&=  \nu \,|c_e|^2\left[ \frac{\Gamma_1 }{2} \tau+ \overline{\sin^2_{m_{2k+2}}\!(\phi)} \right]\! \langle m_{2k+2}|\rho_{k}^-|m_{2k+2} \rangle,\\
\gamma_{k-1,k}^-&=\nu \,|c_g|^2\left[ \frac{4\Gamma_3-3\gamma_{13}}{8} \tau+ \overline{\sin^2_{m_{2k}}\!(\phi)} \right]\! \langle m_{2k}\!+\!2|\rho_{k}^+|m_{2k}\!+\!2 \rangle,
\end{align}
\end{subequations}
with $m_k$ being the position of the $k$th wall and $\overline{\sin^2_{m}\!(\phi)}$ denoting the average of $\sin^2_{m}\!(\phi)$ with respect to the distribution of integrated coupling from a nonmonochromatic atom beam. The rates simply depend on the local density of the state at the wall it is transformed across. This \emph{ladder structure} within each of the parity subspaces again obeys detailed balance [see Fig.~\ref{fig:phi_wall}(c)] and the asymptotic state is a probabilistic mixture of two odd and even stationary states with probability $p=\mathrm{Tr}(\mathds{1}_+\rho)$ approximated by [cf. Eq.~\eqref{eq:rhoss_deph}]
\begin{equation}\label{eq:rhoss_wall_even2}
\rho_\text{ss}\approx  p \sum_{k=0}^\infty  p_k^+ \,\rho_k^+ +  (1-p)\sum_{k=0}^\infty  p_{k}^-\, \rho_k^-, 
\end{equation}
where the stationary states  are determined by the rates in the recurrence relations
\begin{equation}\label{eq:pss_wall_even2}
\frac{p_{k}^\pm}{p_{k-1}^\pm}= \frac{\gamma_{k,k-1}^\pm}{\gamma_{k-1,k}^\pm}
\end{equation}
with $p_0^\pm$ determined by the normalization $\sum_{k} p_k^\pm=1$.

For the first hard wall being odd, due to parity conservation, the effective dynamics features only the transitions between the states of odd parity, from $\rho_k^-$ to $\rho_{k-1}^-$ and $\rho_{k-1}^+$, with the respective rates
\begin{subequations}\label{eq:gamma_wall_odd2}
\begin{align}
\gamma_{k+1,k}^-&=  \nu \,|c_e|^2\left[ \frac{\Gamma_1 }{2} \tau+ \overline{\sin^2_{m_{k+1}}\!(\phi)} \right]\! \langle m_{k+1}|\rho_{k}^-|m_{k+1} \rangle,\\
\gamma_{k-1,k}^-&=\nu \,|c_g|^2\left[ \frac{4\Gamma_3-3\gamma_{13}}{8} \tau+ \overline{\sin^2_{m_{k}}\!(\phi)} \right]\! \langle m_{k}\!+\!2|\rho_{k}^+|m_{k}\!+\!2 \rangle,
\end{align}
\end{subequations}
This dynamics structure also obeys detailed balance [see Fig.~\ref{fig:phi_wall}(d)], and the asymptotic state is a probabilistic mixture of two odd and even stationary states with probability $p=\mathrm{Tr}(\mathds{1}_+\rho)$ approximated by [cf. Eq.~\eqref{eq:pss_wall_even2}]
\begin{equation}\label{eq:rhoss_wall_odd2}
\rho_\text{ss}\approx  p \,|\Psi_+\rangle\!\langle \Psi_+| +  (1-p)\sum_{k=0}^\infty  p_{k}^-\, \rho_k^-, \qquad \text{with} \qquad \frac{p_{k}^-}{p_{k-1}^-}= \frac{\gamma_{k,k-1}^-}{\gamma_{k-1,k}^-}. 
\end{equation}
and $p=\mathrm{Tr}(\mathds{1}_+\rho)$.

Finally, we note that there is no contribution to the dynamics from the atom decay during time of preparation and entering cavity, which leads to mixed rather than pure atom state  (cf. Appendixes~\ref{app:Leff:conserved:atom_mixed} and~\ref{app:Leff:conserved:atom_decay}). Indeed, dynamics with mixed atom state conserves the support of the states between hard walls, leading only to dephasing of coherences between pure states (see below).

~\\
\emph{Dynamics of realistic micromaser}. In a realistic micromaser, the dynamics features both the transitions between the states of the opposite parity in Eq.~\eqref{eq:gamma_wall_even} [Eq.~\eqref{eq:gamma_wall_odd}] and the states of the same parity in Eqs.~\eqref{eq:gamma_wall_even2} and~\eqref{eq:gamma_wall_odd2}.  Such dynamics does not feature detailed balance, unless $\gamma_{k,k-1}^{+-} \gamma_{k,k}^{-+}/( \gamma_{k-1,k}^{-+}\gamma_{k,k}^{+-}) = \gamma_{k,k-1}^-/\gamma_{k-1,k}^-$ and $\gamma_{k,k-1}^{+-} \gamma_{k-1,k-1}^{-+}/( \gamma_{k-1,k}^{-+}\gamma_{k-1,k-1}^{+-}) = \gamma_{k,k-1}^+/\gamma_{k-1,k}^+$ [ $\gamma_{k}^{-+}\gamma_{k-1}^{+-}/(\gamma_{k}^{+-}\gamma_{k-1}^{-+})= \gamma_{k,k-1}^-/\gamma_{k-1,k}^-$], in which case the steady states in Eqs.~\eqref{eq:rhoss_wall_even} and~\eqref{eq:rhoss_wall_even2} [Eqs.~\eqref{eq:rhoss_wall_odd} and~\eqref{eq:rhoss_wall_odd2}] coincide.

~\\
\emph{No trapping states}. In the cavity with the first wall even and pumped by excited atoms, $|c_{\rm e}|=1$, the long-time dynamics, Eqs.~\eqref{eq:gamma_wall_even} and~\eqref{eq:gamma_wall_even2}, features only the transitions that increase the photon number: $|m_{k}\rangle$ is transformed into $|m_{k+1}\rangle$  at the rate $\kappa\, m_{k} +\nu\,\langle m_{k} | X| m_{k} \rangle$ and into $|m_{k+2}\rangle$ at the rate $\nu \,[ \Gamma_1 \tau/2+ \overline{\sin^2_{m_{k}}\!(\phi)}]$. Similarly, in the cavity with the first wall odd and pumped by excited atoms, no even stationary state exists and the odd trapping states are connected to this subspace at the rate $\kappa\, m_{k} +\nu\,\langle m_{k} | X| m_{k} \rangle$,  while $|m_{k}\rangle$ is transformed into $|m_{k+1}\rangle$ at the rate $\nu \,[ \Gamma_1 \tau/2+ \overline{\sin^2_{m_{k}}\!(\phi)}]$. We thus conclude there exists no trapping states in a realistic micromaser.

~\\
Below we derive Eqs.~(\ref{eq:gamma_wall_even})-(\ref{eq:rhoss_wall_odd2}) and the corresponding dynamics of coherences.

\subsubsection{Multiple stationary states for hard walls without single-photon losses}

Hard walls in the far-detuned dynamics of Eq.~\eqref{eq:Kraus} lead to presence of multiple stationary states (see Sec.~\ref{sec:walls_hard}). If the first wall appears at even $m_1$, $\sin_{m_1}(\phi)=0$, there are infinitely many stationary states of both parities, as the parity of subsequent walls alternates. If the first wall appears at odd $m_1$, however, there are only odd walls, leading to multiple odd stationary states (cf. Table~\ref{tab:mK}). Furthermore, pure stationary states exist only when the first wall is odd with the integrated coupling strength such that $\cos_{m_1}(\phi)=1$. In this case, also the coherences between the pure stationary states with the same boundary conditions  are stationary.

In derivations below, we \emph{assume} there is a unique stationary state between each two walls. In such a case, for the first hard wall at even $m_1$, the asymptotic state is given by
\begin{equation}\label{eq:rhoss_m_even}
\lim_{t\rightarrow\infty}\rho(t)= \sum_{k=0}^\infty p_k^+ \rho_k^++\sum_{k=0}^\infty p_k^- \rho_k^-,
\end{equation}
where $\rho_k^+$ [$\rho_k^-$] denotes $k$th even (odd) stationary states, i.e., the stationary state supported between walls at $m_{2k-1}$ and $m_{2k+1}$ (at $m_{2k}$ and $m_{2k+2}$), and we formally expressed the boundary conditions (of non-negative photon number) as $m_{-1}=-2$ and $m_{0}=-1$. The probabilities are given by the initial support between the hard walls, $p_k^\pm=\mathrm{Tr}(\mathds{1}_k^\pm \rho) $ with $\mathds{1}_k^+=\sum_{m=m_{2k-1}+2}^{m_{2k+1}}|m\rangle\!\langle m|$ and $\mathds{1}_k^-=\sum_{m=m_{2k}+2}^{m_{2k+2}}|m\rangle\!\langle m|$. Similarly, for the first wall being odd, 
\begin{eqnarray}\label{eq:rhoss_m_odd}
\lim_{t\rightarrow\infty}\rho(t)&=&  p^+ |\Psi_+\rangle\!\langle \Psi_+| +\sum_{k=0}^\infty p_k^- \rho_k^- 
\\\nonumber 
&&+ \sum_{k=0}^\infty \left(c_{2k}^{+-} |\Psi_+\rangle \langle \Psi_{2k}^-| + \text{H.c.}\right)+\sum_{k=0}^\infty \sum_{\substack{ k'>k: \\ (k'-k) |2}}\left(c_{k ,k'}^{--}  |\Psi_{k}^-\rangle \langle \Psi_{k'}^-| + \text{H.c.}\right) ,
\end{eqnarray} 
and $p^+=\mathrm{Tr}(\mathds{1}^+ \rho) $ with $\mathds{1}^+=\sum_{m=0}^\infty|2m\rangle\!\langle 2m|$. The second line in Eq.~\eqref{eq:rhoss_m_odd} is present only when the first wall corresponds to $\cos_{m_1}(\phi)=1$, i.e., the odd stationary states are pure, $\rho_k^-=|\Psi_k^-\rangle\!\langle\Psi_k^-|$ allowing for stationary coherences with $c_{2k}^{+-}=\mathrm{Tr}(L_{2k}^{+-}\rho)$ and $c_{k,k'}^{--}=\mathrm{Tr}(L_{k,k'}^{-}\rho)$, where $L_{2k}^{+-}$ is a conserved quantity in odd-even coherences with the odd part within the support of $\rho_{2k}^-$, while $L_{k,k'}^{-}$ is the conserved quantity between the supports of $\rho_{k}^-$ and $\rho_{k'}^-$ (where $k'>k$ such that the difference $k'-k$ is divisible by $2$).  

\subsubsection{Effective dynamics due to single-photon losses}

As a single-photon loss changes the parity of a state, consequently only the states of opposite parity in Eqs.~\eqref{eq:rhoss_m_even} and~\eqref{eq:rhoss_m_odd} get connected. Furthermore, a single-photon loss reduces photon number by $1$ in each state. Therefore, for the states to get connected, their supports need to overlap after the loss.  

~\\
\emph{Case of the even first wall}. For the probability $p^\pm_k$ of being in the state $\rho^\pm_k$ [cf.~Eq.~\eqref{eq:rhoss_m_even}] single-photon losses  induce the following dynamics [see~Eq.~\eqref{eq:Leff_flip}] 
\begin{equation}\label{eq:Leff_loss:m_even} 
\frac{ d }{ d t}\left(\begin{array}{c}  p_0^+\\p_0^-\\p_1^+\\p_1^-\\\vdots \end{array}\right) = \kappa\left(\begin{array}{ccccc} 
-\langle n\rangle_0^+ & \phantom{+}\langle n\rangle_{0,0}^- &&& \\
 \phantom{+}\langle n\rangle_0^+ &-\langle n\rangle_0^-& \phantom{+}\langle n\rangle_{1,0}^+ &&\\
 &\phantom{+}\langle n\rangle_{0,1}^-&-\langle n\rangle_1^+&\phantom{+}\langle n\rangle_{1,1}^- &\\
 &&\phantom{+}\langle n\rangle_{1,1}^+ &-\langle n\rangle_1^-&\ddots\\
 &&&\ddots& \ddots\end{array}\right)
 \left(\begin{array}{c}  p_0^+\\p_0^-\\p_1^+\\p_1^-\\\vdots \end{array}\right) ,
\end{equation}
where  $\langle n\rangle_k^\pm= \mathrm{Tr}(n\, \rho_k^\pm)$, $\langle n\rangle_{k,k'}^\pm= \mathrm{Tr}(\mathds{1}_{k'}^\mp \,a \rho_k^\pm a^\dagger)$, and empty entries correspond to $0$. 
Since the parity of the subsequent walls alternates, the support of a given state between two walls  shifted by $1$ overlaps only with two states of opposite parity, so that $\langle n\rangle_{k,k}^\pm+\langle n\rangle_{k,k\mp1}^\pm=\langle n\rangle_k^\pm$ (except the case of $\rho_0^+$). 

The dynamics in Eq.~\eqref{eq:Leff_loss:m_even} obeys \emph{detailed balance}, leading to the \emph{unique stationary state} given by 
\begin{equation}\label{eq:rhoss_loss:m_even} 
\rho_\text{ss}=  \sum_{k=0}^\infty \left (p_{\text{ss},k}^+ \,\rho_k^+ + p_{\text{ss},k}^-\, \rho_k^- \right), \qquad\text{where} \qquad  \frac{p_{\text{ss} ,k}^+}{p_{\text{ss},k-1}^-}= \frac{\langle n\rangle_{k-1,k}^-  }{\langle n\rangle_{k,k-1}^+} \qquad\text{and} \qquad \frac{p_{\text{ss} ,k}^-}{p_{\text{ss},k}^+}= \frac{\langle n\rangle_{k,k}^+  }{\langle n\rangle_{k,k}^-},
\end{equation}
and $p_{\text{ss},0}^+$ is determined by the normalization $\sum_{k=0}^\infty (p_{\text{ss},k}^++p_{\text{ss},k}^-)=1$, and $\langle n\rangle_{0,0}^+\equiv\langle n\rangle_{0}^+$. Equation~\eqref{eq:rhoss_loss:m_even} follows from Eq.~\eqref{eq:Leff_loss:m_even}  corresponding to the classical \emph{birth-death process}.

~\\
\emph{Trapping states}. In the case when the cavity is being pumped by the atoms in the excited state ($|c_{\rm e}|=1$), the stationary states of the cavity are pure and correspond to the position of hard walls $\rho_k^+=|m_{2k+1}\rangle\!\langle m_{2k+1}|$ and $\rho_k^-=| m_{2k+2}\rangle\!\langle m_{2k+2}|$. In this case, a single-photon loss transforms the states into  $|m_{2k+1}-1\rangle\!\langle m_{2k+1}-1|$ and $|m_{2k+2}-1\rangle\!\langle m_{2k+2}-1|$, which evolve into $\rho_k^-$ and $\rho_{k+1}^+$, respectively. Therefore, the effective dynamics due to single-photon losses leads to the stochastic increase of the photon number of the cavity [cf.~Eq.~\eqref{eq:Leff_loss:m_even}]  
\begin{equation}
\frac{ d }{ d t}\left(\begin{array}{c}  p_0^+\\p_0^-\\p_1^+\\p_1^-\\\vdots \end{array}\right) = \kappa\left(\begin{array}{ccccc} 
-m_1 & &&& \\
\phantom{+}m_1 &-m_2& &&\\
&\phantom{+}m_2&-m_3& &\\
&&\phantom{+}m_3 &-m_4&\\
&&&\ddots& \ddots\end{array}\right)
\left(\begin{array}{c}  p_0^+\\p_0^-\\p_1^+\\p_1^-\\\vdots \end{array}\right) ,
\end{equation}
%\begin{equation}
%\frac{ d }{ d t}\left(\begin{array}{c}  p_0^+\\p_0^-\\p_1^+\\p_1^-\\\vdots \end{array}\right) = \kappa\left(\begin{array}{ccccc} 
%-\langle n\rangle_0^+ & &&& \\
%\phantom{+}\langle n\rangle_0^+ &-\langle n\rangle_0^-& &&\\
%&\phantom{+}\langle n\rangle_{1}^-&-\langle n\rangle_1^+& &\\
%&&\phantom{+}\langle n\rangle_{1}^+ &-\langle n\rangle_1^-&\\
%&&&\ddots& \ddots\end{array}\right)
%\left(\begin{array}{c}  p_0^+\\p_0^-\\p_1^+\\p_1^-\\\vdots \end{array}\right) ,
%\end{equation}
and no stationary state exists. This is due to the assumption, that $\kappa\ll \nu$, so that cavity is pumped at a much higher rate than it loses photons. Furthermore, the formerly stationary coherences between trapping states of the same parity decay as
\begin{equation} \label{eq:Leff_loss:m_even_trap} 
\frac{ d }{ d t}\left(\begin{array}{c} 
\vdots \\  c_{k,k'}^{++} \\ c_{k,k'}^{--}\\ c_{k+1,k'+1}^{++} \\\vdots
\end{array}\right)
=
\kappa\left(\begin{array}{ccccc} 
\ddots& && &\\
\ddots &-\frac{m_{2k+1} + m_{2k'+1}}{2}& &&\\
& \bar{\eta}_{k,k'}^{++} &-\frac{m_{2k+2} + m_{2k'+2}}{2}&  &\\  && \bar{\eta}_{k,k'}^{--}& -\frac{m_{2k+3} + m_{2k'+3}}{2}&\\
 & &&\ddots&\ddots
\end{array}\right)
\left(\begin{array}{c} 
\vdots \\  c_{k,k'}^{++} \\ c_{k,k'}^{--}\\ c_{k+1,k'+1}^{++} \\\vdots
\end{array}\right)
\end{equation}
where $c_{k,k'}^{++}$ is the coefficient corresponding to the even-even coherence $|m_{2k+1}\rangle\!\langle m_{2k'+1}|$ and $c_{k,k'}^{--}$ is the coefficient for the odd-odd coherence $|m_{2k+2}\rangle\!\langle m_{2k'+2}|$. We have defined  $\bar{\eta}_{k,k'}^{++}=\sqrt{m_{2k+1} m_{2k'+1}} \langle m_{2k'+1}\!-\!1|L_{k,k'}^- |m_{2k+1}\!-\!1\rangle $ and $\bar{\eta}_{k,k'}^{--}=\sqrt{m_{2k+2} m_{2k'+2}} \langle m_{2k'+2}\!-\!1|L_{k+1,k'+1}^+ |m_{2k+2}\!-\!1\rangle $, where $L_{k,k'}^-$ and $L_{k,k'}^+$  are the conserved quantities corresponding to $|m_{2k+2}\rangle\!\langle m_{2k'+2}|$ and  $|m_{2k+1}\rangle\!\langle m_{2k'+1}|$, respectively. Furthermore, when $\cos_{m_1}(\phi)=1$,  the formerly stationary even-odd  and odd-even coherences similarly decay as
 \begin{equation} \label{eq:Leff_loss:m_even_trap2} 
 \frac{ d }{ d t}\left(\begin{array}{c} 
 \vdots \\  c_{k,k'}^{+-} \\ c_{k,k'+1}^{-+}\\ c_{k+1,k'+1}^{+-} \\\vdots
 \end{array}\right)
 =
 \kappa\left(\begin{array}{ccccc} 
 \ddots& && &\\
 \ddots &-\frac{m_{2k+1} + m_{2k'+2}}{2}p& &&\\
 & \bar{\eta}_{k,k'}^{+-} &-\frac{m_{2k+2} + m_{2k'+3}}{2}&  &\\  && \bar{\eta}_{k,k'+1}^{-+}& -\frac{m_{2k+3} + m_{2k'+4}}{2}&\\
 & &&\ddots&\ddots
 \end{array}\right)
 \left(\begin{array}{c} 
 \vdots \\  c_{k,k'}^{+-} \\ c_{k,k'}^{-+}\\ c_{k+1,k'+1}^{+-} \\\vdots
 \end{array}\right),
 \end{equation}
 where $c_{k,k'}^{+-}$ is the coefficient corresponding to the even-odd coherence $|m_{2k+1}\rangle\!\langle m_{2k'+2}|$, $c_{k,k'}^{-+}$ is the coefficient corresponding to the odd-even coherence $|m_{2k+2}\rangle\!\langle m_{2k'+1}|$, and we have defined
$\bar{\eta}_{k,k'}^{+-}=\sqrt{m_{2k+1} m_{2k'+2}} \langle m_{2k'+2}\!-\!1|L_{k,k'+1}^{-+} |m_{2k+1}\!-\!1\rangle $ and $\bar{\eta}_{k,k'}^{-+}=\sqrt{m_{2k+2} m_{2k'+1}} \langle m_{2k'+1}\!-\!1|L_{k+1,k'}^{+-} |m_{2k+2}\!-\!1\rangle $ with $L_{k,k'}^{-+}$ and $L_{k,k'}^{+-}$  being the conserved quantities corresponding to $|m_{2k+2}\rangle\!\langle m_{2k'+1}|$ and  $|m_{2k+1}\rangle\!\langle m_{2k'+2}|$, respectively.

~\\
\emph{Case of the odd first wall}. For the case of the first wall with $\cos_{m_1}(\phi)=-1$, there exist a single even pure stationary state and multiple odd mixed stationary states between odd hard walls [cf. Eq.~\eqref{eq:mK}]. In the presence of single-photon losses the corresponding probabilities [cf.~Eq.~\eqref{eq:rhoss_m_odd}] undergo the following dynamics  [see~Eq.~\eqref{eq:Leff_flip}]
\begin{equation}\label{eq:Leff_loss:m_odd} 
\frac{ d }{ d t}\left(\begin{array}{c} 
p^+\\p_0^-\\p_1^-\\\vdots
\end{array}\right)=\kappa \left(\begin{array}{cccc} 
-\langle n\rangle_+ & \phantom{+}\langle n\rangle_{0}^- &\phantom{+}\langle n\rangle_{1}^-&\cdots\\
\phantom{+}\langle n\rangle_0^+ &-\langle n\rangle_0^-\\
\phantom{+}\langle n\rangle_1^+&&-\langle n\rangle_1^-\\
\vdots& & &\ddots\\
\end{array}\right)
\left(\begin{array}{c} 
p^+\\p_0^-\\p_1^-\\\vdots
\end{array}\right),
\end{equation}
where $\langle n\rangle_+=\langle \Psi_+| n|\Psi_+\rangle$, $\langle n\rangle_k^-=\mathrm{Tr}( n\,\rho_k^-)$, and $\langle n\rangle_k^+=\mathrm{Tr}(\mathds{1}_{k}^- \,a |\Psi_+\rangle\!\langle \Psi_+| a^\dagger)$. For the first wall with $\cos_{m_1}(\phi)=-1$, the dynamics in Eq.~\eqref{eq:Leff_loss:m_odd}  leads to the stationary state
\begin{equation}\label{eq:rhoss_loss:m_odd} 
\rho_\text{ss}=  p_{\text{ss}}^+ \, |\Psi_+\rangle\!\langle \Psi_+|+ \sum_{k=0}^\infty  p_{\text{ss},k}^-\, \rho_k^- , \qquad\text{where} \qquad  \frac{p_{\text{ss},k}^-}{p_\text{ss}^+}= \frac{\langle n\rangle_{k}^+}{\langle n\rangle_{k}^-  }, 
\end{equation}
in which the structure is due to the dynamics obeying the \emph{detailed balance}, as the odd states are only coupled to the unique even state. In Eq.~\eqref{eq:rhoss_loss:m_odd},  $p_{\text{ss}}^+ $ is determined by the normalization $p_{\text{ss}}^++\sum_{k=0}^\infty p_{\text{ss},k}^-=1$.

For the first wall with $\cos_{m_1}(\phi)=1$, coherences can also be stationary in the absence of single-photon losses [cf.~Eq.~\eqref{eq:rhoss_m_odd}], but the single-photon losses lead to their partial decay, as follows. For the coherences between the even state and odd states, we have
\begin{equation}\label{eq:Leff_loss:m_odd_coh} 
\frac{ d }{ d t}\left(\begin{array}{c} 
c_0^{+-} \\ c_0^{-+}\\c_2^{+-}\\ c_2^{-+}\\\vdots  
\end{array}\right)
=
\kappa\left(\begin{array}{ccccc} 
-\frac{\langle n\rangle_+ + \langle n\rangle_0^-}{2}& \bar{\eta}_{0,0}  && \bar{\eta}_{2,0}&\cdots\\
 \bar{\eta}_{0,0}  &-\frac{\langle n\rangle_+ + \langle n\rangle_0^-}{2}&\bar{\eta}_{2,0} &&\ddots\\
  & \bar{\eta}_{0,2} &-\frac{\langle n\rangle_+ + \langle n\rangle_2^-}{2}& \bar{\eta}_{2,2} &\\
  \bar{\eta}_{0,2}  && \bar{\eta}_{2,2}& -\frac{\langle n\rangle_+ + \langle n\rangle_2^-}{2}&\ddots\\
   \vdots & \ddots &&\ddots&\ddots
\end{array}\right)
\left(\begin{array}{c} 
c_0^{+-} \\ c_0^{-+}\\c_2^{+-}\\ c_2^{-+}\\\vdots  
\end{array}\right) ,
\end{equation}
where $c_{2k}^{+-}$, $c_{2k}^{-+}$  are  the coefficients for the coherences  $|\Psi_+\rangle\!\langle \Psi_{2k}^-|$ and  $|\Psi_{2k}^-\rangle\!\langle\Psi_+ |$, respectively, and we have defined $\bar{\eta}_{2k,2k'}=\mathrm{Tr}[(L_{2k'}^{+})^\dagger \,a |\Psi_+\rangle\!\langle \Psi_{2k}^-| a^\dagger] $, $k,k'=0,1,...$. Furthermore, the coherences between odd states  decay as %we should check complete positivity here, what are the conditions for eta!
\begin{equation}\label{eq:Leff_loss:m_odd_coh2} 
\frac{ d }{ d t}\,
 c_{k,k'}^{--} 
=-\kappa\frac{\langle n\rangle_k^- + \langle n\rangle_{k'}^-}{2} \, c_{k,k'}^{--} ,
\end{equation}
where $c_{k,k'}^{--}$ is the coefficient for the coherences $|\Psi_{k}^-\rangle\!\langle\Psi_{k'}^-|$ and $(k'-k)$ is divisible by $2$ (then they correspond to states with the same boundary conditions). Finally, coherences between the odd states can be created by the single-photon loss from the even state [cf.~Eq.~\eqref{eq:Leff_loss:m_odd}]
\begin{equation}
\kappa^{-1} \frac{ d }{ d t} |\Psi_+\rangle\!\langle \Psi_+| =- \langle n\rangle^+    |\Psi_+\rangle\!\langle \Psi_+|  +\sum_{k=0}^{\infty} \langle n\rangle_{k}^+  |\Psi_k^-\rangle\!\langle \Psi_k^-|  +\sum_{k=0}^{\infty}\!\!\sum_{\substack{k'> k:\\(k'-k)|2 }} \!\!\!\! \left[\mathrm{Tr}\left(L^-_{k,k'} \,a|\Psi_+\rangle\!\langle \Psi_+|a^\dagger\right) |\Psi_k^-\rangle\!\langle \Psi_{k'}^-|  +\text{H.c.}\right]\!\!.
 \end{equation}
 Therefore, the coherences between the even state and odd states decay at long times, while the coherences between odd states can be featured in the stationary state [cf.~Eq.~\eqref{eq:rhoss_loss:m_odd}]
 \begin{equation}\label{eq:rhoss_loss:m_odd2} 
 \rho_\text{ss}=  p_{\text{ss}}^+ \, |\Psi_+\rangle\!\langle \Psi_+|+ \sum_{k=0}^\infty  p_{\text{ss},k}^-\, |\Psi_{k}^-\rangle\!\langle \Psi_{k}^-|  +\sum_{k=0}^{\infty}\sum_{\substack{k'> k:\\(k'-k)|2 }} \left[ c_{\text{ss},k,k'}^{--} |\Psi_k^-\rangle\!\langle \Psi_{k'}^-| +(c_{\text{ss},k,k'}^{--})^* |\Psi_{k'}^-\rangle\!\langle \Psi_{k}^-| \right], 
 \end{equation}
 where $p_{\text{ss},k}^-/p_\text{ss}^+= \langle n\rangle_{k}^+/\langle n\rangle_{k}^-$ as before, while
 \begin{equation}
  c_{\text{ss},k,k'}^{--}= \frac{2\,\mathrm{Tr}\left(L^-_{k,k'} \,a|\Psi_+\rangle\!\langle \Psi_+|a^\dagger\right)}{\langle n\rangle_k^- + \langle n\rangle_{k'}^-} p_\text{ss}^+.
 \end{equation}

%What is the symmetry here? Occupations can be transformed into coherences, yes cause they have the same parity symmetry!!!. Is there any stronger symmetry possible here? Well that symmetry would need to leave the pure even state invariant... so $e^{i\phi n}$ is not a symmetry. There is weak symmetry with respect to n of losses, and for Karusoperators for strong symmetry with respect to states supports (but not n). This gives a weak symmetry with respect to supports.

\subsubsection{Effective dynamics due to corrections to the far-detuned limit}

The corrections to the far-detuned limit lead to the introduction of the parity-swapping Kraus operators $M_0$, $M_2$, and $M_4$, and modification of the parity-conserving Kraus operators $M_1$, $M_3$ (as well as the introduction of $M_a$) [cf.~Eq.~\eqref{eq:KrausFull} and~Appendix~\ref{app:Kraus}]. 

~\\
\emph{Dissipative dynamics}. The parity-swapping Kraus operators $M_0$, $M_2$, and $M_4$ can change the support of a state between hard walls only by a single-photon number (analogously to adding or removing a single photon)  in the first order of the ratio between couplings and detunings (see~Appendix~\ref{app:Kraus}). Therefore, repeating  the arguments for the dynamics with single-photon losses, we conclude that the parity-swapping Kraus operators lead to the second-order dynamics as in Eqs.~(\ref{eq:Leff_loss:m_even})-(\ref{eq:Leff_loss:m_odd_coh2}), but with  $\sqrt{\kappa}a$ replaced by $\sqrt{\nu}M_0$, $\sqrt{\nu}M_2$ or $\sqrt{\nu}M_4$, and then summed [compare~Eqs.~\eqref{eq:Leff_corr} and~\eqref{eq:eta_corr} and Eqs.~\eqref{eq:Leff_loss} and~\eqref{eq:eta_loss}]. 

~\\
~\emph{Unitary dynamics}.
The parity-conserving Kraus operators $M_1$, $M_3$ change the support of a state between hard walls by two photons in the second order  of the ratio between couplings and detunings (see~Appendix~\ref{app:Kraus}). Therefore, these corrections contribute unitarily to the dynamics of coherences as follows [cf.~Eqs.~\eqref{eq:Leff_corr}  and~\eqref{eq:Omega}]: For the first wall being even and trapping states [cf.~Eqs.~\eqref{eq:Leff_loss:m_even_trap} and~\eqref{eq:Leff_loss:m_even_trap2}]
\begin{eqnarray}\label{eq:Leff_corr:m_even_coh} 
\frac{ d }{ d t}c_{k,k'}^{++} &=& -i \left[\langle \delta H\rangle_k^+-\langle \delta H\rangle_{k'}^+\right] c_{2k}^{++},\\
\frac{ d }{ d t}c_{k,k'}^{--} &=& -i \left[\langle \delta H\rangle_k^--\langle \delta H\rangle_{k'}^-\right]   c_{2k}^{--} \times\cos_{m_1}(\phi),\\
\frac{ d }{ d t}c_{k,k'}^{+-}&=& -i \left[\langle \delta H \rangle_k^+-\langle \delta H\rangle_{k'}^-\right] c_{k,k'}^{+-},
\end{eqnarray}
where $\delta H$ is given by Eq.~\eqref{eq:deltaH}. For the first wall being odd [cf.~Eqs.~\eqref{eq:Leff_loss:m_odd_coh} and~\eqref{eq:Leff_loss:m_odd_coh2}]
\begin{eqnarray}\label{eq:Leff_corr:m_odd_coh} 
\frac{ d }{ d t}
c_{2k}^{+-}&=& - i \left[\langle \delta H\rangle_+-\langle \delta H\rangle_k^-\right] c_{2k}^{+-},\\
\frac{ d }{ d t}c_{k,k'}^{--} &=& -i \left[\langle \delta H \rangle_k^--\langle \delta H\rangle_{k'}^-\right] c_{k,k'}^{--} \cos_{m_k}(\phi),
\end{eqnarray}
 and we further have $\cos_{m_k}(\phi)=(-1)^k$.

~\\
\emph{Steady states}. From the above considerations, the stationary state for the first wall being even is, cf.
 Eq.~\eqref{eq:rhoss_loss:m_even},
\begin{equation}\label{eq:rhoss_corr:m_even} 
\rho_\text{ss}=  \sum_{k=0}^\infty \left (p_{\text{ss},k}^+ \,\rho_k^+ + p_{\text{ss},k}^-\, \rho_k^- \right), \qquad\text{where} \qquad  \frac{p_{\text{ss} ,k}^+}{p_{\text{ss},k-1}^-}= \frac{\langle X\rangle_{k-1,k}^-  }{\langle X\rangle_{k,k-1}^+} \qquad\text{and} \qquad \frac{p_{\text{ss} ,k}^-}{p_{\text{ss},k}^+}= \frac{\langle X\rangle_{k,k}^+  }{\langle X\rangle_{k,k}^-},
\end{equation}
where $\langle X\rangle_{k,k'}^\pm= \sum_{j=0,2,4}\mathrm{Tr}(\mathds{1}_{k'}^\mp \,M_j \rho_k^\pm M_j^\dagger)$. For the first wall being odd with $\cos_{m_1}(\phi)=-1$ [cf.~Eq.~\eqref{eq:rhoss_loss:m_odd}]
\begin{equation}\label{eq:rhoss_corr:m_odd} 
\rho_\text{ss}=  p_{\text{ss}}^+ \, |\Psi_+\rangle\!\langle \Psi_+|+ \sum_{k=0}^\infty  p_{\text{ss},k}^-\, \rho_k^- , \qquad\text{where} \qquad  \frac{p_{\text{ss},k}^-}{p_\text{ss}^+}= \frac{\langle X\rangle_{k}^+}{\langle X\rangle_{k}^-  }, 
\end{equation}
where $X=\sum_{j=0,2,4} M_j^\dagger M_j$, $\langle X\rangle_k^-=\mathrm{Tr}(X\, \rho_k^-)$ and $\langle X\rangle_k^+=\sum_{j=0,2,4}\mathrm{Tr}(\mathds{1}_{k}^- \,M_j |\Psi_+\rangle\!\langle \Psi_+| M_j^\dagger)$, while for $\cos_{m_1}(\phi)=1$ the stationary state features coherence between odd states [cf.~Eq.~\eqref{eq:rhoss_loss:m_odd2}]
\begin{equation}\label{eq:rhoss_corr:m_odd2} 
\rho_\text{ss}=  p_{\text{ss}}^+ \, |\Psi_+\rangle\!\langle \Psi_+|+ \sum_{k=0}^\infty  p_{\text{ss},k}^-\, |\Psi_{k}^-\rangle\!\langle \Psi_{k}^-|  +\sum_{k=0}^{\infty}\sum_{\substack{k'> k:\\(k'-k)|2 }} \left[ c_{\text{ss},k,k'}^{--} |\Psi_k^-\rangle\!\langle \Psi_{k'}^-| +(c_{\text{ss},k,k'}^{--})^* |\Psi_{k'}^-\rangle\!\langle \Psi_{k}^-|\right], 
\end{equation}
where $p_{\text{ss},k}^-/p_\text{ss}^+= \langle X\rangle_{k}^+/\langle X\rangle_{k}^- $ as before and
\begin{equation}
c_{\text{ss},k,k'}^{--}=  \frac{2\sum_{j=0,2,4} \mathrm{Tr}\left(L^-_{k,k'} \,M_j |\Psi_+\rangle\!\langle \Psi_+|M_j^\dagger\right)}{\langle X\rangle_k^- + \langle X\rangle_{k'}^-} p_\text{ss}^+.
\end{equation}

\subsubsection{Effective dynamics due to mixed atom state}
\label{app:Leff:hard_walls_atom_mixed}

We now show that a mixed, rather than pure, atom state can only lead to
to dephasing of coherences between pure states that were stationary in the unperturbed dynamics (the case of the odd first wall with $cos_{m_1}(\phi)=1$ (cf. Appendix~\ref{app:Leff:conserved:atom_mixed}). We can further bound the dephasing rates and frequencies analogously to Eqs.~\eqref{eq:deph_bound1_ad} and~\eqref{eq:deph_bound2_ad}. These results are due to the fact that the modified dynamics, preserves not only the parity, but also the support of the states between the hard walls.

~\\
\emph{Case of the even first wall}. We now argue that the probability $p^\pm_k$ of being in the state $\rho^\pm_k$ [cf.~Eq.~\eqref{eq:rhoss_m_even}] is stationary:
\begin{equation}\label{eq:Leff_atom_mixed:m_even} 
\frac{ d }{ d t}\left(\begin{array}{c}  p_0^+\\p_0^-\\p_1^+\\p_1^-\\\vdots \end{array}\right) = 
\left(\begin{array}{c}  0\\0\\0\\0\\\vdots \end{array}\right).
\end{equation}
Indeed, from Eq.~\eqref{eq:Kraus_mixed2}, we have that $M_0$, $M_2$, $M_4$, and $M_\text{a}$ as function of the photon number $n$ conserve the support of $\rho^\pm_k$, i.e.,  $[M_j,\mathds{1}_{k}^\pm]=0$ [cf. Eq.~\eqref{eq:parity_cons}], $j=0,2,4,\text{a}$. Similarly, Kraus operators $M_{gb}$ and $M_{eb}$ in Eq.~\eqref{eq:Kraus_mixed} are defined for the same integrated coupling as $M_{ga}$ and $M_{ea}$, and thus feature the same hard walls leading to $[M_{ja},\mathds{1}_{k}^\pm]=0=[M_{jb},\mathds{1}_{k}^\pm]$, $j=g,e$. We conclude that Eq.~\eqref{eq:Leff_atom_mixed:m_even}  holds true to all orders, while the states $\rho^\pm_k$ are modified by higher order corrections.

~\\
\emph{Case of the odd first wall}. For the case of the first wall with $\cos_{m_1}(\phi)=-1$, there exist a single even pure stationary state and multiple odd mixed stationary states between odd hard walls [cf. Eq.~\eqref{eq:mK}]. Analogously to Eq.~\eqref{eq:Leff_atom_mixed:m_even} we have 
\begin{equation}\label{eq:Leff_atom_mixed:m_odd} 
\frac{ d }{ d t}\left(\begin{array}{c} 
p^+\\p_0^-\\p_1^-\\\vdots
\end{array}\right)=
\left(\begin{array}{c} 
0\\0\\0\\\vdots
\end{array}\right).
\end{equation}
In the limit of almost pure atom state, coherences, which are stationary for the unperturbed dynamics, again undergo dephasing [cf. Eq.~\eqref{eq:Leff_deph_0}]. For the first wall with $\cos_{m_1}(\phi)=1$ [cf.~Eq.~\eqref{eq:rhoss_m_odd}], 
\begin{equation}\label{eq:Leff_atom_mixed:m_odd_coh} 
\frac{ d }{ d t}\, c_{2k}^{+-}
=-\left(\gamma_{2k} + i\Omega_{2k}\right) \, c_{2k}^{+-} ,\qquad c_{2k}^{-+}
=-\left(\gamma_{2k} - i\Omega_{2k}\right) \, c_{2k}^{-+} 
\end{equation}
where $c_{2k}^{+-}$, $c_{2k}^{-+}$  are  the coefficients for the coherences $|\Psi_+\rangle\!\langle \Psi_{2k}^-|$ and  $|\Psi_{2k}^-\rangle\!\langle\Psi_+ |$, respectively, and we have defined $-\gamma_{k} - i\Omega_{k}=p_b \sum_{j=g,e}\mathrm{Tr}(L_{2k}^{+} \,M_{jb} |\Psi_+\rangle\!\langle \Psi_{2k}^-| M_{jb}^\dagger) +\sum_{j=0,2,4} p_j\mathrm{Tr}(L_{2k}^{+} \,M_j |\Psi_+\rangle\!\langle \Psi_{2k}^-| M_j^\dagger)$ for $k=0,1,...$. Analogously,
the coherence between odd states decay as 
\begin{equation}\label{eq:Leff_atom_mixed:m_odd_coh2} 
\frac{ d }{ d t}\,
c_{k,k'}^{--} 
=-\left(\gamma_{k,k'} + i\Omega_{k,k'}\right) \, c_{k,k'}^{--} ,
\end{equation}
where $c_{k,k'}^{--}$ is the coefficient for the coherences $|\Psi_{k}^-\rangle\!\langle\Psi_{k'}^-|$, $(k'-k)$ is divisible by $2$ (then they correspond to states with the same boundary conditions) and we defined $-\gamma_{k} - i\Omega_{k}=p_b \sum_{j=g,e}\mathrm{Tr}(L_{k,k'}^{-} \,M_{jb} |\Psi_{k}^-\rangle\!\langle\Psi_{k'}^-| M_{jb}^\dagger) +\sum_{j=0,2,4} p_j\mathrm{Tr}(L_{k,k'}^{+} \,M_j |\Psi_{k}^-\rangle\!\langle\Psi_{k'}^-| M_j^\dagger)$ (note that $\gamma_{k,k'}=\gamma_{k',k}$ and $\Omega_{k,k'}=-\Omega_{k',k}$).

The dephasing rates can be further bounded as [cf. Eq.~\eqref{eq:deph_bound1}]
\begin{equation}
\label{eq:deph_bound1:hard_wall}
\gamma_{2k}, \gamma_{k,k'}\leq 2\nu(1-p_a-p_\text{a}),
\end{equation}
and frequencies as [cf. Eq.~\eqref{eq:deph_bound2}]
\begin{equation} 
\label{eq:deph_bound2:hard_wall}
|\Omega_{2k}|, |\Omega_{k,k'}|\leq \nu(p_0+p_2+p_4).
\end{equation} 
These results follow from the derivation in  Appendix~\ref{app:Leff:conserved:atom_mixed} by considering dynamics restricted to the sum of the even subspace and support of $|\Psi_{2k}^-\rangle$, or the sum of supports of $|\Psi_{2k}^-\rangle$ and $|\Psi_{2k'}^-\rangle$, respectively.

~\\
\emph{Steady states}. The mixed atom state leads to stationary states  of the cavity being probabilistic mixtures of states between hard walls, i.e., given by Eqs.~\eqref{eq:rhoss_m_even} and~\eqref{eq:rhoss_m_odd} (without the second line), where the probabilities are determined by the support of the initial cavity state between the walls.

~\\
\emph{Dynamics for trapping states}. Finally, we note that in the case of the atoms prepared in the excited state $|c_e|=1$, the probabilities are again conserved [cf. Eqs.~\eqref{eq:Leff_atom_mixed:m_even}  and~\eqref{eq:Leff_atom_mixed:m_odd}], while the coherences simply undergo dephasing with bounds analogous to Eqs.~\eqref{eq:deph_bound1:hard_wall} and~\eqref{eq:deph_bound2:hard_wall}.

\subsubsection{Effective dynamics due to atom decay}
\label{app:Leff:hard_walls_atom_decay}

We now discuss how decay of atoms, leads to the mixing dynamics of states between hard walls with the same parity. This leads to two mixed stationary states of even and odd parities, which is due to the fact that the parity remains conserved. 

There are two contributions arising from the finite lifetime of atom levels that modify the dynamics of micromaser (cf.~Appendix~\ref{app:Leff:conserved:atom_decay}). First, atoms arrive at the cavity in the mixed rather than pure state with probabilities given by Eq.~\eqref{eq:rho_decay_lin}. In the limit of weak decay, this only leads to the dephasing of coherences between pure stationary states between the walls [see Eqs.~\eqref{eq:Leff_atom_mixed:m_odd_coh} and~\eqref{eq:Leff_atom_mixed:m_odd_coh}] with the rates bounded as in Eqs.~\eqref{eq:deph_bound1_ad} and~\eqref{eq:deph_bound2_ad} [cf. Eqs.~\eqref{eq:deph_bound1:hard_wall} and~\eqref{eq:deph_bound2:hard_wall}]. Second, the possible atom decay during the interaction with the cavity modifies Kraus operators in Eq.~\eqref{eq:Kraus}. We now discuss this contribution in the limit of weak decay and only toward the levels uncoupled to the cavity [see Eqs.~(\ref{eq:discrete_0_decay})-(\ref{eq:Kraus_0_decay})]. We comment on the general case at the end of this appendix.

~\\
\emph{Case of the even first wall}. For the probability $p^\pm_k$ of being in the state $\rho^\pm_k$ [cf.~Eq.~\eqref{eq:rhoss_m_even}], atom decay induces the following dynamics,
\begin{equation}\label{eq:Leff_atom_decay:m_even} 
\frac{ d }{ d t}\left(\begin{array}{c}  p_0^+\\p_1^+\\p_2^+\\\vdots \end{array}\right) = \frac{\nu}{2}\left(\begin{array}{ccccc} 
-|c_e|^2 \,\Gamma_1 \tau\,\rho_{0;m_1}^{+} & \phantom{+}|c_g|^2 \,\Gamma_3 \tau\, \rho_{1;m_1+2}^{+} &&& \\
\phantom{+}|c_e|^2 \,\Gamma_1 \tau \, \rho_{0;m_1}^{+} &-|c_g|^2 \,\Gamma_3 \tau\, \rho_{1;m_1+2}^{+}-|c_e|^2 \,\Gamma_1 \tau\,  \rho_{1;m_3}^{+} & \phantom{+}|c_g|^2 \,\Gamma_3 \tau \rho_{2;m_3+2}^{+}  &&\\
&\phantom{+}|c_e|^2 \,\Gamma_1 \tau\, \rho_{1;m_3}^{+}&\ddots&\ddots &\\
&&\ddots&&\end{array}\right)
\left(\begin{array}{c}  p_0^+\\p_1^+\\p_2^+\\\vdots \end{array}\right) ,
\end{equation}
and, analogously,
\begin{equation}\label{eq:Leff_atom_decay:m_even2} 
\frac{ d }{ d t}\left(\begin{array}{c}  p_0^-\\p_1^-\\p_2^-\\\vdots \end{array}\right) = \frac{\nu}{2}\left(\begin{array}{ccccc} 
-|c_e|^2\,\Gamma_1 \tau\,\rho_{0;m_2}^{-} & \phantom{+}|c_g|^2 \,\Gamma_3 \tau\, \rho_{1;m_2+2}^{-} &&& \\
\phantom{+}|c_e|^2\,\Gamma_1 \tau\,\rho_{0;m_2}^{-} &-|c_g|^2 \,\Gamma_3 \tau\,\rho_{1;m_2+2}^{-}-|c_e|^2 \,\Gamma_1 \tau\, \rho_{1;m_4}^{-} & \phantom{+}|c_g|^2 \,\Gamma_3 \tau\, \rho_{2;m_4+2}^{-}  &&\\
&\phantom{+}|c_e|^2 \,\Gamma_1 \tau\, \rho_{1;m_4}^{-}&\ddots&\ddots &\\
&&\ddots&&\end{array}\right)
\left(\begin{array}{c}  p_0^-\\p_1^-\\p_2^-\\\vdots \end{array}\right) ,
\end{equation}
where we introduced $\rho_{k;m}^{\pm}=\langle m|\rho_k^\pm|m\rangle$ as the local density of the state $\rho_k^\pm$. Here we considered contribution from Eqs.~(\ref{eq:discrete_0_decay})-(\ref{eq:Kraus_0_decay}). Since the nontrivial dynamics is only induced when the support of the state is changed beyond the hard wall, only the contribution decay events, described by the integral in Eq.~(\ref{eq:discrete_0_decay}), lead to the long-time dynamics of the state between hard walls. In the limit of the weak decay, keeping terms up to linear order allows us to replace $\overline{\mathcal{M}}_g(t)$ and  $\overline{\mathcal{M}}_e(t)$ in Eq.~(\ref{eq:discrete_0_decay}) by $\mathcal{M}_g(t)$ and  $\mathcal{M}_e(t)$, which can allow a transition of $\rho_{k}^\pm$ to $\rho_{k+1}^\pm$ and $\rho_{k-1}^\pm$, respectively. The integral, however,  effectively gives a random interaction time $t$ described by a uniform distribution within the interval $[0,\tau)$. In that case, the rates of the long-time dynamics are proportional to the averaged probability of crossing a hard wall at $m$ (see Appendix~\ref{app:Leff:hard_walls_atom_beam}), and we simply have $\overline{\sin^2_{m}(\phi)} =\tau^{-1}\int_0^\tau  d t \sin_{m}^2\!(\lambda t)= K \pi /(2 \phi \sqrt{(m+1)(m+2)} )=1/2$ [cf. Eq.~\eqref{eq:phi_wall} and see  Eqs.~\eqref{eq:Leff_atom_beam:m_even} and~\eqref{eq:Leff_atom_beam:m_even2}]. 

The dynamics within the even and odd subspaces, Eqs.~\eqref{eq:Leff_atom_decay:m_even} and~\eqref{eq:Leff_atom_decay:m_even2}, respectively,  obeys \emph{detailed balance}, leading to an asymptotic state being a general mixture of odd and even stationary states given by 
\begin{equation}\label{eq:rhoss_atom_decay:m_even} 
\rho_\text{ss}=  p \sum_{k=0}^\infty p_{\text{ss},k}^+ \,\rho_k^+ +(1-p)\sum_{k=0}^\infty p_{\text{ss},k}^-\, \rho_k^- ,
\end{equation}
where
\begin{equation}\label{eq:rhoss_atom_decay:m_even2} 
\frac{p_{\text{ss} ,k}^+}{p_{\text{ss},k-1}^+}= \frac{|c_e|^2}{|c_g|^2} \frac{\Gamma_1}{\Gamma_3} \frac{  \rho_{k-1;m_{2k-2}}^{+} }{\rho_{k;m_{2k-1}+2}^{+} } \qquad\text{and} \qquad \frac{p_{\text{ss} ,k}^-}{p_{\text{ss},k-1}^-}= \frac{|c_e|^2}{|c_g|^2} \frac{\Gamma_1}{\Gamma_3} \frac{  \rho_{k-1;m_{2k}}^{-} }{\rho_{k;m_{2k}+2}^{-} } ,
\end{equation}
and $p_{\text{ss},0}^\pm$ are determined by the normalization $\sum_{k=0}^\infty p_{\text{ss},k}^\pm=1$, but $p=\mathrm{Tr}(\mathds{1}_+\rho)$ is determined by the support of the initial state. In the case of the uniform decay, the stationary state is the same as the stationary state of the micromaser with a nonmonochromatic atom beam [see Eq.~\eqref{eq:rhoss_atom_beam:m_even2} below], as in that case atom decay leads exactly to the random interaction time described by the uniform distribution [cf. Eqs.~\eqref{eq:discrete_0_decay} and~\eqref{eq:Kraus_average}].

~\\
\emph{Case of the odd first wall}.  Similarly, for the case of the first wall with $\cos_{m_1}(\phi)=-1$, we have [cf.~Eq.~\eqref{eq:rhoss_m_odd}] 
\begin{equation}\label{eq:Leff_atom_decay:m_odd} 
\frac{ d }{ d t}\,
p^+
=0,
\end{equation}
and
\begin{equation}\label{eq:Leff_atom_decay:m_odd2} 
\frac{ d }{ d t}\left(\begin{array}{c}  p_0^-\\p_1^-\\p_2^-\\\vdots \end{array}\right) = \frac{\nu}{2}\left(\begin{array}{ccccc} 
-|c_e|^2\, \Gamma_1 \tau \,\rho_{0;m_1}^{-} & \phantom{+}|c_g|^2\, \Gamma_3 \tau \, \rho_{1;m_1+2}^{-} &&& \\
\phantom{+}|c_e|^2\, \Gamma_1 \tau \, \rho_{0;m_1}^{-} &-|c_g|^2\, \Gamma_3 \tau \, \rho_{1;m_1+2}^{-}-|c_e|^2\, \Gamma_1 \tau \, \rho_{1;m_2}^{-} & \phantom{+}|c_g|^2\, \Gamma_3 \tau \, \rho_{2;m_2+2}^{-}  &&\\
&\phantom{+}|c_e|^2\, \Gamma_1 \tau \,\rho_{1;m_2}^{-}&\ddots&\ddots &\\
&&\ddots&&\end{array}\right)
\left(\begin{array}{c}  p_0^-\\p_1^-\\p_2^-\\\vdots \end{array}\right) ,
\end{equation}
where the first equation is a direct consequence of the parity conservation and uniqueness of the even steady state. Therefore, the dynamics in  in Eqs.~\eqref{eq:Leff_atom_decay:m_odd} and~\eqref{eq:Leff_atom_decay:m_odd2}   leads to the stationary state
\begin{equation}\label{eq:rhoss_atom_decay:m_odd} 
\rho_\text{ss}=  p \, |\Psi_+\rangle\!\langle \Psi_+|+ (1-p)\sum_{k=0}^\infty  p_{\text{ss},k}^-\, \rho_k^- , \qquad\text{where} \qquad  \frac{p_{\text{ss} ,k}^-}{p_{\text{ss},k-1}^-}= \frac{|c_e|^2}{|c_g|^2} \frac{\Gamma_1}{\Gamma_3}\frac{  \rho_{k-1;m_{k}}^{-} }{\rho_{k;m_{k}+2}^{-} }  
\end{equation}
and $p_{\text{ss},0}^-$ is determined by the normalization $\sum_{k=0}^\infty p_{\text{ss},k}^-=1$, but $p=\mathrm{Tr}(\mathds{1}_+\rho) $ is a free parameter.

%EDIT HERE!!!

For the first wall with $\cos_{m_1}(\phi)=1$, atom decay also leads to dynamics of coherences in Eq.~\eqref{eq:rhoss_m_odd}. We have that coherences between even and every second odd state decay as 
\begin{equation}\label{eq:Leff_atom_decay:m_odd_coh} 
\frac{ d }{ d t}\,
c_k^{+-}
=-\left(\overline\gamma_{k} + i\overline\Omega_{k}\right) \, c_k^{+-} ,\qquad c_k^{-+}
=-\left(\overline\gamma_{k} - i\overline\Omega_{k}\right) \, c_k^{-+} 
\end{equation}
where $c_{2k}^{+-}$, $c_{2k}^{-+}$  are  the coefficients for the coherences  $|\Psi_+\rangle\!\langle \Psi_{2k}^-|$ and  $|\Psi_{2k}^-\rangle\!\langle\Psi_+ |$, 
while coherences between odd states decay as
\begin{equation}\label{eq:Leff_atom_decay:m_odd_coh2} 
\frac{ d }{ d t}\left(\begin{array}{c} 
\vdots \\  c_{k-1,k'-1} \\ c_{k,k'}\\ c_{k+1,k'+1} \\\vdots
\end{array}\right)
=
\left(\begin{array}{ccccc} 
\ddots& \ddots&& &\\
\ddots &-\overline\gamma_{k-1,k'-1} - i\overline\Omega_{k-1,k'-1}& \nu \,\bar{\eta}_{k,k'}^{--} &&\\
& \nu \,\bar{\eta}_{k-1,k'-1}^{++} &-\overline\gamma_{k,k'} - i\overline\Omega_{k,k'}& \nu \,\bar{\eta}_{k+1,k'+1}^{--} &\\  && \nu \,\bar{\eta}_{k,k'}^{++}& -\overline\gamma_{k+1,k'+1} - i\overline\Omega_{k+1,k'+1}&\ddots\\
& &&\ddots&\ddots
\end{array}\right)
\left(\begin{array}{c} 
\vdots \\  c_{k-1,k'-1} \\ c_{k,k'}\\ c_{k+1,k'+1} \\\vdots
\end{array}\right),
\end{equation}
where $c_{k,k'}^{--}$ is the coefficient for the coherences $|\Psi_{k}^-\rangle\!\langle\Psi_{k'}^-|$ and $(k'-k)$ is divisible by $2$.
We have introduced [cf. Eq.~\eqref{eq:deph_atom_decay_1}]
\begin{subequations}\label{eq:Leff_atom_decay:odd_coh_par}
	\begin{align}\label{eq:Leff_atom_decay:odd_coh_par1}
\frac{\overline\gamma_{2k}}{\nu}&=  \frac{\Gamma_1+\Gamma_3}{2}\tau - \int_{0}^\tau  d t\,  \sum_{j=g,e}\Gamma_j \mathrm{Tr}\left[L_{2k}^{+} \,\mathcal{M}_{j}(\lambda t) \left(|\Psi_+\rangle\!\langle \Psi_{2k}^-| \right)\right] -\frac{\Gamma_1+\Gamma_3}{4}\tau \left(\langle Y\rangle_+ +\langle Y\rangle_{k}^- \right),\\ \label{eq:Leff_atom_decay:odd_coh_par2}
\frac{\overline\gamma_{k,k'}}{\nu}&=  \frac{\Gamma_1+\Gamma_3}{2}\tau - \int_{0}^\tau  d t\,  \sum_{j=g,e}\Gamma_j \mathrm{Tr}\left[L_{k,k'}^{-} \,\mathcal{M}_{j}(\lambda t) \left(|\Psi_k^-\rangle\!\langle \Psi_{k'}^-| \right)\right] +\frac{\Gamma_1+\Gamma_3}{4}\tau \left(\langle Y\rangle_k^- +\langle Y\rangle_{k'}^- \right) ,\\\label{eq:Leff_atom_decay:odd_coh_par3}
\bar{\eta}_{k,k'}^{++}&=|c_e|^2\, \Gamma_1 \tau\, c_{m_{k+1}}^{(k)} [c_{m_{k'+1}}^{(k')}]^* \langle m_{k'+1}\!+\!2| L_{k+1,k'+1}^{-}  |m_{k+1}\!+\!2\rangle \,\tau^{-1}\int_{0}^\tau  d t\, \sin_{m_{k+1}}\!(\lambda t)\sin_{m_{k'+1}}\! (\lambda t)=0,\\\label{eq:Leff_atom_decay:odd_coh_par4}
\bar{\eta}_{k,k'}^{--}&=|c_g|^2 \,\Gamma_3 \tau c_{m_{k}+2}^{(k)} [c_{m_{k'}+2}^{(k')}]^* \langle m_{k'}| L_{k-1,k'-1}^{-}  |m_{k}\rangle \,\tau^{-1}\int_{0}^\tau  d t\, \, \sin_{m_{k}}\!(\lambda t)\sin_{m_{k'}}\! (\lambda t)=0,
	\end{align}
\end{subequations}
where $Y$ is defined in Eq.~\eqref{eq:Y} and we used $\tau^{-1}\int_{0}^\tau  d t\, \, \sin_{m}\!(\lambda t)\sin_{m'}\! (\lambda t)=\tau^{-1}\int_{0}^\tau  d t\, \, \{\cos[\lambda t\sqrt{(m+1)(m+2)}-\lambda t\sqrt{(m+1)(m+2)}]-\cos[\lambda t\sqrt{(m+1)(m+2)}+\lambda t\sqrt{(m+1)(m+2)}]\}/2=0$ (see also Appendix~\ref{app:Leff:hard_walls_atom_beam}), so that all coherences simply undergo \emph{dephasing} (this, however, will not be the case for general decay; see below). Moreover, there is no contribution to the unitary dynamics, $\overline\Omega_{2k}=0$ and $\overline\Omega_{k,k'}=0$ due to real-valued dynamics [cf. Eq.~\eqref{eq:deph_atom_decay_2}].  As a result of the dephasing of all coherences, the stationary state for $\cos_{m_1}(\phi)=1$ is again given by Eq.~\eqref{eq:rhoss_atom_decay:m_odd}.

~\\
\emph{General decay}. For the case of general atom decay [see Eq.~\eqref{eq:master_decay} and Eqs.~(\ref{eq:discrete_decay2})-(\ref{eq:Kraus_decay2})], decay toward levels $|0\rangle$, $|2\rangle$, or $|1\rangle$, leads to the nontrivial cavity dynamics also after the decay event [cf. Eqs.~\eqref{eq:discrete_0_decay} and~\eqref{eq:Kraus_average}]. In particular, the dynamics of probabilities in Eqs.~\eqref{eq:Leff_atom_decay:m_even},~\eqref{eq:Leff_atom_decay:m_even2}, and~\eqref{eq:Leff_atom_decay:m_odd2} is modified by replacing $\Gamma_3$ by $(\Gamma_3-\gamma_{13}) +\gamma_{13}/4=\Gamma_3-3\gamma_{13}/4$. Here, the new term corresponds to decay from $|3\rangle$ to $|1\rangle$ followed by atom leaving the cavity in a state $|1\rangle$, which contributes with the average probability as $\tau^{-1}\int_0^\tau  d t \cos_m^2[\lambda (\tau-t)]\sin_{m}^2\!(\lambda t)=\tau^{-1}\int_0^\tau  d t \cos_m^2(\lambda t)]\sin_{m}^2\!(\lambda t)= K \pi /(8\phi \sqrt{(m+1)(m+2)} )=1/8$ for a hard wall at $m$ described by Eq.~\eqref{eq:phi_wall}. Similarly, the dynamics of coherences will have the structure of Eqs.~\eqref{eq:Leff_atom_decay:m_odd_coh} and~\eqref{eq:Leff_atom_decay:m_odd_coh2}, but with modified parameters due to a more complex single decay contribution in Eq.~(\ref{eq:discrete_decay2}). In particular, $\tau^{-1}\int_{0}^\tau  d t\, \, \sin_{m}\!(\lambda t)\sin_{m'}\! (\lambda t)$ will be replaced by $\tau^{-1}\int_{0}^\tau  d t\, \, \cos_m\!(\lambda t)\cos_{m'}\!(\lambda t)\sin_{m}\!(\lambda t)\sin_{m'}\! (\lambda t)$ in Eqs.~\eqref{eq:Leff_atom_decay:odd_coh_par3} and~\eqref{eq:Leff_atom_decay:odd_coh_par4} due to possible decay from $|3\rangle$ to $|1\rangle$.

%can we say that the structure is the same?

\subsubsection{Effective dynamics due to nonmonochromatic atom beam}
\label{app:Leff:hard_walls_atom_beam}

Finally, we consider a nonmonochromatic atom beam, which leads to the fluctuating integrated coupling $\tau$ described by a probability distribution (see Appendix~\ref{app:Leff:conserved:atom_beam}). Since the existence and positions of the hard wall depend on $\phi$ (see Sec.~\ref{sec:walls_hard} and Appendix~\ref{app:hard_walls}), the supports of the states between the hard walls are not conserved, leading to mixing dynamics between the states of the same parity.    
 
~\\
\emph{Case of the even first wall}. For the probability $p^\pm_k$ of being in the state $\rho^\pm_k$ [cf.~Eq.~\eqref{eq:rhoss_m_even}], nonmonochromatic beam induced the following dynamics [cf.~Eq.~\eqref{eq:Leff_loss:m_even}]:
\begin{equation}\label{eq:Leff_atom_beam:m_even} 
\frac{ d }{ d t}\left(\begin{array}{c}  p_0^+\\p_1^+\\p_2^+\\\vdots \end{array}\right) = \nu\left(\begin{array}{ccccc} 
-|c_e|^2 \overline{\sin_{m_1}^2\!(\phi)} \rho_{0;m_1}^{+} & \phantom{+}|c_g|^2 \overline{\sin_{m_1}^2\!(\phi)} \rho_{1;m_1+2}^{+} &&& \\
\phantom{+}|c_e|^2 \overline{\sin_{m_1}^2\!(\phi)} \rho_{0;m_1}^{+} &-|c_g|^2 \overline{\sin_{m_1}^2\!(\phi)} \rho_{1;m_1+2}^{+}-|c_e|^2 \overline{\sin_{m_3}^2\!(\phi)} \rho_{1;m_3}^{+} & \phantom{+}|c_g|^2 \overline{\sin_{m_3}^2\!(\phi)} \rho_{2;m_3+2}^{+}  &&\\
&\phantom{+}|c_e|^2 \overline{\sin_{m_3}^2\!(\phi)} \rho_{1;m_3}^{+}&\ddots&\ddots &\\
&&\ddots&&\end{array}\right)
\left(\begin{array}{c}  p_0^+\\p_1^+\\p_2^+\\\vdots \end{array}\right) ,
\end{equation}
and, analogously,
\begin{equation}\label{eq:Leff_atom_beam:m_even2} 
\frac{ d }{ d t}\left(\begin{array}{c}  p_0^-\\p_1^-\\p_2^-\\\vdots \end{array}\right) = \nu\left(\begin{array}{ccccc} 
-|c_e|^2 \overline{\sin_{m_2}^2\!(\phi)} \rho_{0;m_2}^{-} & \phantom{+}|c_g|^2 \overline{\sin_{m_2}^2\!(\phi)} \rho_{1;m_2+2}^{-} &&& \\
\phantom{+}|c_e|^2 \overline{\sin_{m_2}^2\!(\phi)} \rho_{0;m_2}^{-} &-|c_g|^2 \overline{\sin_{m_2}^2\!(\phi)} \rho_{1;m_2+2}^{-}-|c_e|^2 \overline{\sin_{m_4}^2\!(\phi)} \rho_{1;m_4}^{-} & \phantom{+}|c_g|^2 \overline{\sin_{m_4}^2\!(\phi)} \rho_{2;m_4+2}^{-}  &&\\
&\phantom{+}|c_e|^2 \overline{\sin_{m_4}^2\!(\phi)} \rho_{1;m_4}^{-}&\ddots&\ddots &\\
&&\ddots&&\end{array}\right)
\left(\begin{array}{c}  p_0^-\\p_1^-\\p_2^-\\\vdots \end{array}\right) ,
\end{equation}
where $\overline{\sin_{m}^2\!(\phi)}=\int d \phi\,g(\phi) \sin_{m}^2\!(\phi)$ is the average with respect to the distribution $g(\phi)$ of the integrated coupling strength, and $\rho_{k;m}^{\pm}=\langle m|\rho_k^\pm|m\rangle$ is the local density of the state $\rho_k^\pm$. Note that we expect $\overline{\sin_{m}^2\!(\phi)}\ll 1$ for a narrow enough distribution $g(\phi)$ with $\sin_{m}^2\!(\langle\phi\rangle)=0$, that is, when  $m\,(\overline{\phi^2}-\overline\phi^2)\,\ll 1$.

The dynamics within the even and odd subspaces, Eqs.~\eqref{eq:Leff_atom_beam:m_even} and~\eqref{eq:Leff_atom_beam:m_even2}, respectively,  again obeys \emph{detailed balance}, leading to the mixture of odd and even stationary states 
\begin{equation}\label{eq:rhoss_atom_beam:m_even} 
\rho_\text{ss}=  p \sum_{k=0}^\infty p_{\text{ss},k}^+ \,\rho_k^+ +(1-p)\sum_{k=0}^\infty p_{\text{ss},k}^-\, \rho_k^- ,
\end{equation}
where
\begin{equation}\label{eq:rhoss_atom_beam:m_even2} 
 \frac{p_{\text{ss} ,k}^+}{p_{\text{ss},k-1}^+}= \frac{|c_e|^2}{|c_g|^2} \frac{  \rho_{k-1;m_{2k-2}}^{+} }{\rho_{k;m_{2k-1}+2}^{+} } \qquad\text{and} \qquad \frac{p_{\text{ss} ,k}^-}{p_{\text{ss},k-1}^-}= \frac{|c_e|^2}{|c_g|^2} \frac{  \rho_{k-1;m_{2k}}^{-} }{\rho_{k;m_{2k}+2}^{-} } ,
\end{equation}
and $p_{\text{ss},0}^\pm$ are determined by the normalization $\sum_{k=0}^\infty p_{\text{ss},k}^\pm=1$, but $p$ is a free parameter and depends on the support of the initial cavity space $\rho$ in the even subspace, $p=\mathrm{Tr}(\mathds{1}_+\rho)$, which is a consequence of the parity conservation [cf. Eq.~\eqref{eq:parity_cons}]. We note that the structure of the stationary states is independent from the distribution of the integrated coupling.

~\\
\emph{Case of the odd first wall}. For the case of the first wall with $\cos_{m_1}(\phi)=-1$, there exist a single even pure stationary state and multiple odd mixed stationary states between odd hard walls. For the nonmonochromatic atom beam, the corresponding probabilities [cf.~Eq.~\eqref{eq:rhoss_m_odd}] undergo the following dynamics  [cf.~Eqs.~\eqref{eq:Leff_atom_beam:m_even} and~\eqref{eq:Leff_atom_beam:m_even2}]
\begin{equation}\label{eq:Leff_atom_beam:m_odd} 
\frac{ d }{ d t}\,
p^+
=0,
\end{equation}
and
\begin{equation}\label{eq:Leff_atom_beam:m_odd2} 
\frac{ d }{ d t}\left(\begin{array}{c}  p_0^-\\p_1^-\\p_2^-\\\vdots \end{array}\right) = \nu\left(\begin{array}{ccccc} 
-|c_e|^2 \overline{\sin_{m_1}^2\!(\phi)} \rho_{0;m_1}^{-} & \phantom{+}|c_g|^2 \overline{\sin_{m_1}^2\!(\phi)} \rho_{1;m_1+2}^{-} &&& \\
\phantom{+}|c_e|^2 \overline{\sin_{m_1}^2\!(\phi)} \rho_{0;m_1}^{-} &-|c_g|^2 \overline{\sin_{m_1}^2\!(\phi)} \rho_{1;m_1+2}^{-}-|c_e|^2 \overline{\sin_{m_2}^2\!(\phi)} \rho_{1;m_2}^{-} & \phantom{+}|c_g|^2 \overline{\sin_{m_2}^2\!(\phi)} \rho_{2;m_2+2}^{-}  &&\\
&\phantom{+}|c_e|^2 \overline{\sin_{m_2}^2\!(\phi)} \rho_{1;m_2}^{-}&\ddots&\ddots &\\
&&\ddots&&\end{array}\right)
\left(\begin{array}{c}  p_0^-\\p_1^-\\p_2^-\\\vdots \end{array}\right) .
\end{equation}
Therefore, for the first wall with $\cos_{m_1}(\phi)=-1$, the dynamics in Eqs.~\eqref{eq:Leff_atom_beam:m_odd} and~\eqref{eq:Leff_atom_beam:m_odd2}   leads to the stationary state [cf. Eqs.~\eqref{eq:rhoss_atom_beam:m_even}  and~\eqref{eq:rhoss_atom_beam:m_even2}]
\begin{equation}\label{eq:rhoss_atom_beam:m_odd} 
\rho_\text{ss}=  p \, |\Psi_+\rangle\!\langle \Psi_+|+ (1-p)\sum_{k=0}^\infty  p_{\text{ss},k}^-\, \rho_k^- , \qquad\text{where} \qquad  \frac{p_{\text{ss} ,k}^-}{p_{\text{ss},k-1}^-}= \frac{|c_e|^2}{|c_g|^2} \frac{  \rho_{k-1;m_{k}}^{-} }{\rho_{k;m_{k}+2}^{-} }  
\end{equation}
and $p_{\text{ss},0}^-$ is determined by the normalization $\sum_{k=0}^\infty p_{\text{ss},k}^-=1$, but $p=\mathrm{Tr}(\mathds{1}_+\rho) $ is a free parameter.

For the first wall with $\cos_{m_1}(\phi)=1$, nonmonochromatic beam also leads to decay of formerly stationary coherences, as given by Eqs.~\eqref{eq:Leff_atom_decay:m_odd_coh} and~\eqref{eq:Leff_atom_decay:m_odd_coh2}, but with parameters defined as
\begin{subequations}\label{eq:Leff_atom_beam:odd_coh_par}
	\begin{align}
	\frac{\overline\gamma_{2k}}{\nu} &= -\mathrm{Tr}\left[L_{2k}^{+} \,\overline{\mathcal{M}(\phi)} \left(|\Psi_+\rangle\!\langle \Psi_{2k}^-|\right)\right]+1,\\ 
	\frac{\overline\gamma_{k,k'}}{\nu} &= -\mathrm{Tr}\left[L_{k,k'}^{-} \,\overline{\mathcal{M}(\phi)} \left(|\Psi_{k}^-\rangle\!\langle \Psi_{k'}^-|\right)\right]+1,\\
	\bar{\eta}_{k,k'}^{++}&=|c_e|^2 c_{m_{k+1}}^{(k)} [c_{m_{k'+1}}^{(k')}]^* \langle m_{k'+1}\!+\!2| L_{k+1,k'+1}^{-}  |m_{k+1}\!+\!2\rangle \,\overline{\sin_{m_{k+1}}\!(\phi)\sin_{m_{k'+1}}\! (\phi)},\\
	\bar{\eta}_{k,k'}^{--}&=|c_g|^2 c_{m_{k}+2}^{(k)} [c_{m_{k'}+2}^{(k')}]^* \langle m_{k'}| L_{k-1,k'-1}^{-}  |m_{k}\rangle\, \overline{ \sin_{m_{k}}\!(\phi)\sin_{m_{k'}}\! (\phi)}
	\end{align}
\end{subequations}
with $c_{m}^{(k)}=\langle m|\Psi_{k}^-\rangle$.  Here $k\neq k'=0,1,2,....$ and  $(k'-k)$ is divisible by $2$. There is no unitary dynamics, $\overline\Omega_{2k}=0=\overline\Omega_{k,k'}$, as the dynamics is real valued (see Secs.~\ref{sec:dynamics} and~\ref{sec:meta_mixed}). The coherences decay at long times, leading to the same structure of the stationary state as in Eq.~\eqref{eq:rhoss_atom_beam:m_odd}.

%===================================================================================================================================================

%===================================================================================================================================================

	%===================================================================================================================================================
\section{Classical micromaser dynamics for thermal atoms}
\label{app:classical}
	
%WHAT BEYOND THE FAR-DETUNED LIMIT?		

%WHAT ABOUT WEAK COUPLING LIMIT---> two-photon losses at $p_T \phi$ and two-photon gain at $(1-p_T)\phi$ plus  4 photon Hamiltonian $-\phi^2/2 Im[p_T (a^\dagger )^2 a^2+(1-p_T)a^2(a^\dagger )^2] $. Does the Hamiltonian part simplify somehow from the approximation of linear birth and death rates? i.e., the temperature of the state must be low enough for this to be true.
	
Here we consider the micromaser dynamics, Eq.~\eqref{eq:master}, in the case of thermal atoms. The dynamics in the far-detuned limit is \emph{classical}  and obeys \emph{detailed balance}, resulting in thermal stationary states of even and odd parities, which are independent from the integrated coupling.

\subsection{Classical detailed-balance dynamics}
 Consider an atom in a thermal state
	\begin{equation}\label{eq:rho_thermal}
	\rho_{\rm at}=\sum_{j=0,...,4,{\rm a}} p_j \ketbra{j}{j},\qquad p_j\propto e^{-\frac{E_j}{k_B T}}
	\end{equation}
	where  $T$ denotes the atom temperature and $E_j$ is the energy of the atomic level (see Sec.~\ref{sec:model}).
	
	There are eight Kraus operators [cf.~Eqs.~\eqref{eq:KrausFULL},~\eqref{eq:Kraus}, and~\eqref{eq:Kraus_mixed2}]:
	\begin{subequations}
		\label{eq:KrausT}
		\begin{align}
		M_{\rm g g} =\cos\left(\phi \sqrt{a^{\dagger 2} a^2}\right), &\qquad M_{\rm e g} = - i \, a^2\ken\frac{\sin\left(\phi \sqrt{a^{\dagger 2}a^2}\right)}{\sqrt{a^{\dagger 2}a^2}} ,\\
		M_{\rm g e} =- i\,a^{\dagger 2}\ken\frac{\sin\left(\phi \sqrt{a^2 a^{\dagger 2}}\right)}{\sqrt{a^2 a^{\dagger 2}}}, &\qquad
		M_{\rm e e} =\,\cos\left(\phi\sqrt{a^{2} a^{\dagger 2}}\right),\\
		M_{0}= e^{i \tau a^\dagger a \frac{|g_2|^2}{\Delta}}, \quad M_{2}= e^{-i \tau a\, a^\dagger  \frac{|g_2|^2+|g_3|^2}{\Delta }},&\quad M_4 =e^{i \tau a\,a^\dagger\frac{|g_3|^2}{\Delta}},\quad\text{and}\quad M_{\rm a}=\mathds{1},
		\end{align}
	\end{subequations}
    which describe the change in the cavity state due to a passage of the atom as [cf.~Eq.~\eqref{eq:discrete}]
    \begin{equation}
    \label{eq:discreteT}
    \rho^{(k)}= \sum_{j,l = g,e} p_l\, M_{jl}\ken \rho^{(k-1)} \ken M_{j l}^{\dagger}+ \sum_{j=0,2,4,{\rm a}} p_j \,M_{j}\ken \rho^{(k-1)} \ken M_{j}^{\dagger}\equiv\mathcal{M}\left[\rho^{(k-1)}\right].
    \end{equation}
    
	The resulting continuous cavity dynamics in Eq.~\eqref{eq:master} conserves the parity, Eq.~\eqref{eq:parity_cons}, due to the approximation of far-detuned limit [cf.~Eq.~\eqref{eq:Heff}]. Furthermore, the dynamics  is  classical, with diagonal states in the photon-number basis remaining diagonal and thus evolving independently from the coherences. In particular, for diagonal states, Eqs.~\eqref{eq:KrausT} describe a detailed balance process between the photon number states of fixed parity, which corresponds to the so-called \emph{birth-death process} with the birth referring to the change from $\ketbra{n}{n}$ to $\ketbra{n+2}{n+2}$ due to the Kraus operator $M_{ge}$, and the death  referring to that from $\ketbra{n}{n}$ to $\ketbra{n-2}{n-2}$ due to the Kraus operator $M_{eg}$, while the other Kraus operators do not contribute. The respective rates are given by  
	\begin{equation}
	b_n= \nu\,p_3 \sin_n^2 (\phi), \quad d_n= \nu  \,p_1 \sin_{n-2}^2 (\phi).
	\label{eq:dynT}
	\end{equation}

	\subsection{Thermal stationary states}
	 From the detailed balance, it follows that two stationary states $\rho^+ = \sum_{n=0}^\infty p_{2n} \ket{2n}\bra{2n}$ and $\rho^- = \sum_{n=0}^\infty h_{2n+1} \ket{2n+1}\bra{2n+1}$  are thermal with the probabilities determined by the recurrence relation
	\begin{equation}\label{eq:dynT2}
	\frac{h_{n+2}}{h_n}= \frac{b_n}{d_{n+2}}=\frac{p_3}{p_1}=e^{\frac{-2\omega}{k_B T}},
	\end{equation}
	 where $2\omega=E_1-E_3$ due to the two-photon resonance in Eq.~\eqref{eq:resonance}. Furthermore, the detailed balance dynamics is present for any diagonal, not necessarily thermal, state of the atom. In this case, Eq.~\eqref{eq:dynT2} defines the effective temperature $T$. 	
	 
	The sequence of probabilities $h_n$ is convergent if $e^{-2\omega/k_BT} < 1$, which takes place for positive temperatures $T>0$ (or for a diagonal state when $p_1> p_3$). In the case of an initial state of the cavity $\rho$ with support on both even and odd subspaces, the asymptotic state is a probabilistic mixture of the even and odd stationary states
	\begin{align}
	\rho_{\rm ss} &=p\,\rho^+ + (1-p)\,\rho^-=  \frac{1}{1+e^{\frac{-2\omega}{k_B T}}} \sum_{n=0}^\infty e^{\frac{-2n\omega}{k_B T}}\Big[p\,\ketbra{2n}{2n} +   (1-p)\,\ketbra{2n+1}{2n+1} \Big], 
	\label{eq:rhossT}
	\end{align}
	where the probability $p=\mathrm{Tr}(\mathds{1}_{+}\rho)$ is determined  by the initial support on the even subspace.

	\subsection{Interaction-dependent timescales of dynamics}
	Because the initial atomic state is thermal, Eq.~\eqref{eq:rho_thermal}, the stationary states of the cavity are independent from the integrated coupling strength $\phi$. However, the dynamics of relaxation toward the stationary state depends crucially on the value of $\phi$. This follows from the birth and death rates, Eq.~(\ref{eq:dynT}) being dependent on $\sin_n^2(\phi)$. Therefore, the presence of a soft wall at $n=m$, $\sin_m(\phi)\approx 0$, leads to a slowdown of the dynamics, similarly to the case for the quantum micromaser dynamics discussed in~Sec.~\ref{sec:walls_soft}. In particular, the relaxation timescales to the stationary state are dominated by the slowest pairs of the birth and death rates, i.e., such $m$ within the support of the stationary state for which $b_m, d_{m+2} \propto \sin_m^2 (\phi)\approx 0$. Treating $b_m, d_{m+2}$ as a perturbation of the dynamics with $b_m^{(0)}=0$, $d_{m+2}^{(0)}=0$,    from~Eq.~\eqref{eq:1storder} we obtain the long-time dynamics between thermal states supported before and after a wall as~\cite{Rose2016}
	 \begin{align}\label{eq:Leff_wall_thermal}
	 \frac{d}{dt}\, p_{k}(t)=&-\left[ p_1\,\sin_{m_k}^2(\phi)\, p_{m_k+2}^{(k)} +p_3\,\sin_{m_{k+1}}^2(\phi)\, p_{m_{k+1}}^{(k)}\right] p_k(t) \\\nonumber&+p_1\,\sin_{m_k}^2(\phi)\, p_{m_k+2}^{(k)}\, p_{k-1}(t) +p_3\,\sin_{m_{k+1}}^2(\phi)\, p_{m_{k+1}}^{(k)}\,p_{k+1}(t),
	 \end{align}
	 where $p_{k}(t)$ denotes the probability of being in the $k$th state supported after $k$th wall, while  $h_{n}^{(k)}$ denotes the probability of finding $n$ photons in the $k$th state  (for simplicity we dropped the indices denoting the parity, but only the states of the same parity are coupled) (see also Appendix~\ref{app:soft_walls}). Note that the final stationary state is again given by Eq.~\eqref{eq:rhossT}.

	%Here the stationary states are not be dependent on the rates! Much simpler perturbation theory!

%===================================================================================================================================================

%===================================================================================================================================================

	 	\section{Continuous vs discrete cavity dynamics}
	\label{app:numerics}

In this appendix, we discuss similarities and differences between continuous dynamics, Eqs.~\eqref{eq:master} and~\eqref{eq:L0}, and the discrete dynamics, Eqs.~\eqref{eq:discrete} and~\eqref{eq:discrete_2}, where the number of atoms that has passed is known explicitly. In particular, the numerical simulations in Figs. \ref{fig:model}-\ref{fig:hard_walls}, \ref{fig:comparison}-\ref{fig:phi_wall}, \ref{fig:qfi}, and \ref{fig:phi_spread} utilize the discrete dynamics.

	\subsection{Discrete dynamics} 
	The master equations~\eqref{eq:master} and~\eqref{eq:L0} represent \emph{continuous dynamics} of the density matrix, which describes the cavity state averaged both over the possible measurement outcomes of the outgoing atomic states --- i.e., when the atoms are traced out --- and over the exponentially distributed arrival times of atoms into the cavity (see Appendix~\ref{app:micromaser}). The former average procedure results precisely in Kraus operators in Eqs.~\eqref{eq:discrete} and~\eqref{eq:KrausFull}, while the latter average yields the master equation~\eqref{eq:master} governing continuous evolution of the cavity in time. Note that by counting the number of atoms that have passed through the cavity, its state after the passage of $k$ atoms is simply given by [cf.~Eqs.~\eqref{eq:discrete} and~\eqref{eq:discrete_2}]  
	\begin{equation}
	\rho^{(k)}=\mathcal{M}^k (\rho), \label{eq:discrete_3}
	\end{equation}
    where $\rho\equiv\rho^{(0)}$ denotes the initial state of the cavity. Note that the conditional \emph{discrete dynamics} in~\eqref{eq:discrete_3} is independent from the atom rate $\nu$, but the probability of the passage of $k$ atoms up to time $t$ is given by $e^{-\nu t}\,(\nu t)^k/k! $, which depends solely on $\nu t$, as described by the Poisson point process (see also Appendix~\ref{app:micromaser}). 
	
	\subsection{Timescales of dynamics}
	 We first note that, in the far-detuned limit, the stationary states of the discrete dynamics~\eqref{eq:discrete_2} corresponding to the eigenvalue $1$ of $\mathcal{M}_0$ are also the stationary states of the continuous dynamics $\mathcal{L}_0$,~\eqref{eq:L0}, which is also the case beyond the adiabatic approximation for $\mathcal{M}$ and $\mathcal{L}$, Eqs.~\eqref{eq:discrete} and~\eqref{eq:master}. Actually, all eigenmodes of the discrete dynamics are also eigenmodes of continuous dynamics, with eigenvalues $\lambda_m^\text{discrete}$ of $\mathcal{M}$ rescaled to the eigenvalues $\lambda_m$ of $\mathcal{L}$ as~\footnote{The eigenvalues of $\mathcal{M}$ can be found within a unit circle in the complex plane, which implies that the eigenvalues of $\mathcal{L}$ are found within the circle of radius $\nu$ centered at $-\nu$.} 
	\begin{equation}
	\lambda_m=\nu(\lambda_m^\text{discrete}-1), \label{eq:val_DvsC}
	\end{equation}
	since $\mathcal{L}=\nu(\mathcal{M}-\mathcal{I})$. The relation~\eqref{eq:val_DvsC} plays an important role in the presence of a hard wall (see Sec.~\ref{sec:walls_hard}). For the discrete dynamics, all eigenmodes of $\mathcal{M}$ with eigenvalue of absolute value $1$ are nondecaying, while for the continuous dynamics only the modes corresponding to the eigenvalue $1$ are stationary. In particular, for a hard wall leading to different boundary conditions before and after the wall, the coherence between the pure stationary states after and before the wall is nondecaying in the discrete dynamics, but the coherence phase is flipped, i.e., shifted by $\pi$, with each passing atom, which in the continuous case leads to its dephasing (see Sec.~\ref{sec:walls_hard}). 
    
    %Furthermore, for time expressed in the unit of $\nu^{-1}$ replaced by the number of atoms that have passed through the cavity, the corresponding eigenvalues of the evolution to that moment are $e^{t \lambda_m}=  e^{\nu t(\lambda_m^\text{discrete}-1)} $  and $(\lambda_m^\text{discrete})^k$. Therefore, for positive real-valued eigenvalues $\lambda_m^\text{discrete}$ we have $1-|\lambda_m^\text{discrete}|\leq -\log(|\lambda_m^\text{discrete}|)$ and thus for the corresponding modes the discrete dynamics is faster. \\
	
	\subsection{Metastability in discrete dynamics}
	
	\subsubsection{Discrete dynamics in the presence of losses}
	 In Sec.~\ref{sec:losses}, we consider cavity dynamics in the presence of single-photon losses at rate $\kappa$. In the derivation of the dynamics governed by the master equation~\eqref{eq:master_loss}, it is assumed that photon loss takes place when there is no atom within the cavity, i.e.,  $\kappa\ken \tau \ll 1$ for  the atom passage time $\tau$, so that the single-photon losses can be considered independent of the atom-cavity dynamics~\cite{englert2002elements,davidovich1987two}. For the discrete dynamics, this assumption leads to the state of the cavity after the passage of $k$ atoms given by
		\begin{equation}
	\rho^{(k)}= \left(\mathcal{M}_\text{1ph}\right)^k \left( \mathcal{I}-\mathcal{L}_\text{1ph}/\nu\right)^{-1}(\rho), \qquad \text{where}\qquad\mathcal{M}_\text{1ph}\equiv \left(\mathcal{I}-\mathcal{L}_\text{1ph}/\nu\right)^{-1}\mathcal{M}_0 . \label{eq:discrete_loss}
	\end{equation}
	%Here we integrate things which decay may be bigger than I. How does it take place? Well, all the eignavlues are negative so OK! 
	Note that $\mathcal{M}_\text{1ph}$ describes the joint effect of the passage of an atom in the far-detuned limit given by $\mathcal{M}_0$, and the losses that can occur afterward, but before the passage of the next atom, $\int_0^\infty dt\, \nu e^{-\nu t} \,e^{t \mathcal{L}_\text{1ph}}=\left[ \mathcal{I}-\mathcal{L}_\text{1ph}/\nu\right]^{-1}$. Equation~\eqref{eq:discrete_loss} can be used to derive the master dynamics~\eqref{eq:master_loss} in the limit $\kappa\ll \nu$ (cf.~Appendix A in Ref.~\cite{davidovich1987two}).
	%Note that for initial state $|0\rangle$, the first loss term in (E.3) can be dropped, so Fig. 10 ok.
	Therefore, from Eq.~\eqref{eq:discrete_loss}, the stationary state of continuous dynamics in the presence of losses~\eqref{eq:master_loss}  corresponds to the stationary state of the discrete dynamics,
	\begin{equation}
	\rho_\text{ss}^\text{discrete}=\left( \mathcal{I}-\mathcal{L}_\text{1ph}/\nu\right)\rho_\text{ss},\label{eq:discrete_loss_ss}
	\end{equation}
	since $\mathcal{L} \rho_\text{ss}=0$, where $\mathcal{L} \equiv\nu(\mathcal{M}_0-\mathcal{I})+\mathcal{L}_\text{1ph}$, so that $\mathcal{M}_0\rho_\text{ss}=(\mathcal{I}-\mathcal{L}_\text{1ph}/\nu)\rho_\text{ss}$ and, thus,  $\mathcal{M}_\text{1ph}\rho_\text{ss}=\rho_\text{ss}$.
	
	\subsubsection{Metastability} 
	 In the metastable limit of a small rate of the single-photon losses, $\kappa\ll \nu$, we recover $\rho_\text{ss}^\text{discrete}\approx\rho_\text{ss}$ from Eq.~\eqref{eq:discrete_loss_ss}. Furthermore, the continuous dynamics of all the metastable modes discussed in Sec.~\ref{sec:losses} will be approximately the same in the discrete case, as follows. Recall from above that, without the losses, the DFS of pure stationary states $|\Psi_+\rangle$ and  $|\Psi_-\rangle$, Eq.~\eqref{eq:even_odd_ss}, is stationary both in the continuous case of $\mathcal{L}_0$ and discrete case of $\mathcal{M}_0$. Expanding $\mathcal{M}_{\rm 1ph}$ in (\ref{eq:discrete_loss}), we have
    \begin{equation}
\mathcal{M}_\text{1ph} = \mathcal{M}_0+\mathcal{L}_\text{1ph} \mathcal{M}_0 /\nu+O(\kappa^2/\nu^2).
	\end{equation}
    Therefore, within the DFS, the eigenvalues and eigenmodes of $\mathcal{M}_\text{1ph}$ in the lowest order of the expansion in $\kappa/\nu$ correspond to the eigenmodes of the continuous effective first-order dynamics in Eq.~\eqref{eq:Leff_loss}, as
       \begin{equation}
  \Pi_0 \mathcal{M}_\text{1ph} \Pi_0 = \Pi_0+\Pi_0\mathcal{L}_\text{1ph}\Pi_0 /\nu+O(\kappa^2/\nu^2), \label{eq:discrete_loss2}
    \end{equation}
    where $\Pi_0$ denotes the projection on the formerly stationary DFS (cf.~Sec.~\ref{sec:DFS} and Appendix~\ref{app:Leff:weak}), while the initial term $[ \mathcal{I}-\mathcal{L}_\text{1ph}/\nu]^{-1}$ in~\eqref{eq:discrete_loss} contributes only as the higher order corrections to the eigenmodes of the discrete dynamics. 
    
    Similarly, in the case of the metastability due the higher order corrections to the two-photon cavity dynamics (see Sec.~\ref{sec:higher-order}), the long-time discrete dynamics beyond adiabatic limit $\mathcal{M}$ can be approximated within the metastable DFS exactly as in Eq.~\eqref{eq:discrete_loss2}, but with $\Pi_0\mathcal{L}_\text{1ph}\Pi_0 $ replaced by the master operator of Eq.~\eqref{eq:Leff_corr}, which corresponds to $\nu(\Pi_0 \mathcal{M} \Pi_0 -\Pi_0)$.

%===================================================================================================================================================

%===================================================================================================================================================

	 	\section{Identifying possible (5+1)-level scheme in Rydberg atoms}
	\label{app:experimental}
	
	Here we provide discussion of the results on Rydberg atoms from Sec.~\ref{sec:Implementations}.

	\subsection{Methods}
	We have used the ARC package~\cite{ARC_package,Sibalic_CompPhysComm_2017} (see also Refs.~\cite{Weber_PairInteractions_package, Weber_JPhysB_2017} for related software) in order to evaluate the energies of levels $\ket{j}$, $j=0,..,4$, as well as the corresponding dipole moments 
	\begin{equation}
	d_{j-1,j}=\braket{j-1|e \hat{r} | j},
	\end{equation}
    $j>0$, where $e$ is the electron charge and $\hat{r}$ the position operator.  The dipole moments determine the single photon Rabi frequencies $g_j$ as 
    \begin{equation}
    g_j=d_{j-1,j} \sqrt{\frac{ \omega}{2 \hbar\varepsilon_0 V }},
    \end{equation}
    where $\omega$, $\varepsilon_0$ and $V$ are the cavity frequency, vacuum permittivity, and the volume of the cavity mode, respectively. 
    We take $V=70 \; {\rm mm}^3$ as a benchmark from Ref.~\cite{Brune_1987}. The number of possible transitions grows rapidly with the number of basis states considered. Considering Ref. \cite{Brune_1987}, which used a ladder configuration $39{\rm S}_{\frac{1}{2}} \leftrightarrow 39{\rm P}_{\frac{3}{2}} \leftrightarrow 40{\rm S}_{\frac{1}{2}}$, we limit the search to a set of 30 basis states $\ket{n,l,j}$ with $n=35,..,45$ and $l=0,1$, where $j=l \pm s$, with $s=1/2$ being the value of the electronic spin. For $\pi$ polarization, we identify 444 600 dipole allowed transitions. In order to satisfy the resonance condition in Eq.~\eqref{eq:resonance}, we further define the cavity frequency as 
    \begin{equation}
    \omega=(E_3-E_1)/2\hbar
    \end{equation}
    and the corresponding detunings $\Delta_j$ according to Eq. (\ref{eq:energygap}). Post-selecting on cases where the levels $\ket{j}$, $j=1,2,3$ form a ladder, i.e., $E_1 > E_2 > E_3$ or $E_1 < E_2 < E_3$ [cf. Fig. \ref{fig:model}(a)], and requiring the rotating-wave approximation, ${\rm max}(|\Delta_j/\omega|<0.1)$, and the far-detuned limit, ${\rm max}(|g_j/\Delta_j|)<0.1$), to hold, we are left with 104 transitions \footnote{The chosen upper bounds in the conditions for rotating wave approximation and far detuned limit can be of course made more stringent. We have chosen 0.1 which is the first order satisfying the (order of magnitude) relation $0.1 \ll 1$. It turns out that for the set of basis states used, the number of post-selected levels reduces to 16 (0) if we set the bound to 0.05 (0.03) instead.}. Having identified the possible candidates, we asses the conditions (\ref{eq:conditions_a}) and~(\ref{eq:conditions_b}) according to the following criterion.  We define factors $f_{a,b}$ as $|g_1|^2/\Delta_1 = f_a |g_2|^2/\Delta$, $|g_4|^2/\Delta_4 = -f_b |g_3|^2/\Delta$, so that the conditions are satisfied for $f_{a}=f_b=1$, %. We also note, that a value of, say, $f_a \neq 1$ requires to adjust the detuning $\Delta_1$ by a factor of $f_a$ or, equivalently, the detuning $\Delta$ by a factor of $1/f_a$, assuming the couplings $g$ remain constant. For this reason we seek a figure of merit which assigns to $f$ and $1/f$ the same distance from the ideal point $f=1$. 
    and minimize ${\rm max}(|1-f_a|,|1-1/f_a|)+{\rm max}(|1-f_b|,|1-1/f_b|)$. This leads us to the transitions $37{\rm S}_{\frac{1}{2}} \leftrightarrow 37{\rm P}_{\frac{3}{2}} \leftrightarrow 38{\rm S}_{\frac{1}{2}} \leftrightarrow 38{\rm P}_{\frac{3}{2}} \leftrightarrow 39{\rm S}_{\frac{1}{2}}$, as described in Sec. \ref{sec:Implementations}.

%THIS REALLY SHOULD LOOK AT WHEN THE CORRECTIONS FROM METASTABILITY ARE MINIMISED.... so in the rough approximation only $|1-f_a|+|1-f_b|$. But for small values of 1-f_a it is the same... it is though diffuclt to argue why such a figure of merit, as it is the consequence that counts. 

\subsection{Possible improvements}
  In order to increase the effective coupling strength $|\lambda|$, the search strategy could consider a larger set of basis states, and, in particular, the level manipulations with external electric field $\mathcal{E}$ which would allow for further modification of $\Delta_j$ through the static Stark effect. Here, in order to evaluate (\ref{eq:conditions}), one needs to compute not only the energies of the atomic levels but also the dipole elements of the allowed transitions. For $l \leq 3$ and small values of $\mathcal{E}$ one might attempt a perturbative approach with level energies given by
\begin{equation}
E_{nlj} = - \frac{E_{\rm Ry}}{n^{*2}} -\frac{1}{2} \alpha_0 \mathcal{E}^2,  
\end{equation}
where $E_{\rm Ry}$ is the Rydberg energy, $n^* = n - \delta_{nlj}$ with the quantum defect $\delta_{nlj}$~\cite{Gallagher_1994,Li_PRA_2003,Saffman_RMP_2010,Pritchard_thesis_2012}, while the static polarizability $\alpha_0=\beta_1  n^{*6}  + \beta_2  n^{*7}$. Here, $\beta_1, \beta_2$ are coefficients which can be obtained, e.g., by the ARC package~\cite{ARC_package} and have been found to be in good agreement with experimental values; see Refs.~\cite{VanWijngaarden_JSpec_1997,OSullivan_PRA_1986} for the case of rubidium. For higher $l$ and values of $\mathcal{E}$, a numerical approach requiring exact diagonalization of the Hamiltonian with the external electric field is necessary. A systematic exploration of the coupling strengths in this generalized scenario,  however, goes beyond the scope of this work.

%\bibliography{bibfile}

\begin{thebibliography}{127}%
\makeatletter
\providecommand \@ifxundefined [1]{%
 \@ifx{#1\undefined}
}%
\providecommand \@ifnum [1]{%
 \ifnum #1\expandafter \@firstoftwo
 \else \expandafter \@secondoftwo
 \fi
}%
\providecommand \@ifx [1]{%
 \ifx #1\expandafter \@firstoftwo
 \else \expandafter \@secondoftwo
 \fi
}%
\providecommand \natexlab [1]{#1}%
\providecommand \enquote  [1]{``#1''}%
\providecommand \bibnamefont  [1]{#1}%
\providecommand \bibfnamefont [1]{#1}%
\providecommand \citenamefont [1]{#1}%
\providecommand \href@noop [0]{\@secondoftwo}%
\providecommand \href [0]{\begingroup \@sanitize@url \@href}%
\providecommand \@href[1]{\@@startlink{#1}\@@href}%
\providecommand \@@href[1]{\endgroup#1\@@endlink}%
\providecommand \@sanitize@url [0]{\catcode `\\12\catcode `\$12\catcode
  `\&12\catcode `\#12\catcode `\^12\catcode `\_12\catcode `\%12\relax}%
\providecommand \@@startlink[1]{}%
\providecommand \@@endlink[0]{}%
\providecommand \url  [0]{\begingroup\@sanitize@url \@url }%
\providecommand \@url [1]{\endgroup\@href {#1}{\urlprefix }}%
\providecommand \urlprefix  [0]{URL }%
\providecommand \Eprint [0]{\href }%
\providecommand \doibase [0]{http://dx.doi.org/}%
\providecommand \selectlanguage [0]{\@gobble}%
\providecommand \bibinfo  [0]{\@secondoftwo}%
\providecommand \bibfield  [0]{\@secondoftwo}%
\providecommand \translation [1]{[#1]}%
\providecommand \BibitemOpen [0]{}%
\providecommand \bibitemStop [0]{}%
\providecommand \bibitemNoStop [0]{.\EOS\space}%
\providecommand \EOS [0]{\spacefactor3000\relax}%
\providecommand \BibitemShut  [1]{\csname bibitem#1\endcsname}%
\let\auto@bib@innerbib\@empty
%</preamble>
\bibitem [{\citenamefont {Zurek}(2001)}]{zurek2001sub}%
  \BibitemOpen
  \bibfield  {author} {\bibinfo {author} {\bibfnamefont {W.~H.}\ \bibnamefont
  {Zurek}},\ }\href@noop {} {\bibfield  {journal} {\bibinfo  {journal}
  {Nature}\ }\textbf {\bibinfo {volume} {412}},\ \bibinfo {pages} {712}
  (\bibinfo {year} {2001})}\BibitemShut {NoStop}%
\bibitem [{\citenamefont {Vlastakis}\ \emph {et~al.}(2013)\citenamefont
  {Vlastakis}, \citenamefont {Kirchmair}, \citenamefont {Leghtas},
  \citenamefont {Nigg}, \citenamefont {Frunzio}, \citenamefont {Girvin},
  \citenamefont {Mirrahimi}, \citenamefont {Devoret},\ and\ \citenamefont
  {Schoelkopf}}]{vlastakis2013deterministically}%
  \BibitemOpen
  \bibfield  {author} {\bibinfo {author} {\bibfnamefont {B.}~\bibnamefont
  {Vlastakis}}, \bibinfo {author} {\bibfnamefont {G.}~\bibnamefont
  {Kirchmair}}, \bibinfo {author} {\bibfnamefont {Z.}~\bibnamefont {Leghtas}},
  \bibinfo {author} {\bibfnamefont {S.~E.}\ \bibnamefont {Nigg}}, \bibinfo
  {author} {\bibfnamefont {L.}~\bibnamefont {Frunzio}}, \bibinfo {author}
  {\bibfnamefont {S.~M.}\ \bibnamefont {Girvin}}, \bibinfo {author}
  {\bibfnamefont {M.}~\bibnamefont {Mirrahimi}}, \bibinfo {author}
  {\bibfnamefont {M.~H.}\ \bibnamefont {Devoret}}, \ and\ \bibinfo {author}
  {\bibfnamefont {R.~J.}\ \bibnamefont {Schoelkopf}},\ }\href@noop {}
  {\bibfield  {journal} {\bibinfo  {journal} {Science}\ }\textbf {\bibinfo
  {volume} {342}},\ \bibinfo {pages} {607} (\bibinfo {year}
  {2013})}\BibitemShut {NoStop}%
\bibitem [{\citenamefont {Wollman}\ \emph {et~al.}(2015)\citenamefont
  {Wollman}, \citenamefont {Lei}, \citenamefont {Weinstein}, \citenamefont
  {Suh}, \citenamefont {Kronwald}, \citenamefont {Marquardt}, \citenamefont
  {Clerk},\ and\ \citenamefont {Schwab}}]{wollman2015quantum}%
  \BibitemOpen
  \bibfield  {author} {\bibinfo {author} {\bibfnamefont {E.~E.}\ \bibnamefont
  {Wollman}}, \bibinfo {author} {\bibfnamefont {C.}~\bibnamefont {Lei}},
  \bibinfo {author} {\bibfnamefont {A.}~\bibnamefont {Weinstein}}, \bibinfo
  {author} {\bibfnamefont {J.}~\bibnamefont {Suh}}, \bibinfo {author}
  {\bibfnamefont {A.}~\bibnamefont {Kronwald}}, \bibinfo {author}
  {\bibfnamefont {F.}~\bibnamefont {Marquardt}}, \bibinfo {author}
  {\bibfnamefont {A.}~\bibnamefont {Clerk}}, \ and\ \bibinfo {author}
  {\bibfnamefont {K.}~\bibnamefont {Schwab}},\ }\href@noop {} {\bibfield
  {journal} {\bibinfo  {journal} {Science}\ }\textbf {\bibinfo {volume}
  {349}},\ \bibinfo {pages} {952} (\bibinfo {year} {2015})}\BibitemShut
  {NoStop}%
\bibitem [{\citenamefont {Rashid}\ \emph {et~al.}(2016)\citenamefont {Rashid},
  \citenamefont {Tufarelli}, \citenamefont {Bateman}, \citenamefont {Vovrosh},
  \citenamefont {Hempston}, \citenamefont {Kim},\ and\ \citenamefont
  {Ulbricht}}]{Rashid_2016}%
  \BibitemOpen
  \bibfield  {author} {\bibinfo {author} {\bibfnamefont {M.}~\bibnamefont
  {Rashid}}, \bibinfo {author} {\bibfnamefont {T.}~\bibnamefont {Tufarelli}},
  \bibinfo {author} {\bibfnamefont {J.}~\bibnamefont {Bateman}}, \bibinfo
  {author} {\bibfnamefont {J.}~\bibnamefont {Vovrosh}}, \bibinfo {author}
  {\bibfnamefont {D.}~\bibnamefont {Hempston}}, \bibinfo {author}
  {\bibfnamefont {M.~S.}\ \bibnamefont {Kim}}, \ and\ \bibinfo {author}
  {\bibfnamefont {H.}~\bibnamefont {Ulbricht}},\ }\href {\doibase
  10.1103/PhysRevLett.117.273601} {\bibfield  {journal} {\bibinfo  {journal}
  {Phys. Rev. Lett.}\ }\textbf {\bibinfo {volume} {117}},\ \bibinfo {pages}
  {273601} (\bibinfo {year} {2016})}\BibitemShut {NoStop}%
\bibitem [{\citenamefont {Pirkkalainen}\ \emph {et~al.}(2015)\citenamefont
  {Pirkkalainen}, \citenamefont {Damsk{\"{a}}gg}, \citenamefont {Brandt},
  \citenamefont {Massel},\ and\ \citenamefont
  {Sillanp{\"{a}}{\"{a}}}}]{Pirkkalainen:15}%
  \BibitemOpen
  \bibfield  {author} {\bibinfo {author} {\bibfnamefont {J.~M.}\ \bibnamefont
  {Pirkkalainen}}, \bibinfo {author} {\bibfnamefont {E.}~\bibnamefont
  {Damsk{\"{a}}gg}}, \bibinfo {author} {\bibfnamefont {M.}~\bibnamefont
  {Brandt}}, \bibinfo {author} {\bibfnamefont {F.}~\bibnamefont {Massel}}, \
  and\ \bibinfo {author} {\bibfnamefont {M.~A.}\ \bibnamefont
  {Sillanp{\"{a}}{\"{a}}}},\ }\href {\doibase 10.1103/PhysRevLett.115.243601}
  {\bibfield  {journal} {\bibinfo  {journal} {Phys. Rev. Lett.}\ }\textbf
  {\bibinfo {volume} {115}},\ \bibinfo {pages} {243601} (\bibinfo {year}
  {2015})}\BibitemShut {NoStop}%
\bibitem [{\citenamefont {Lecocq}\ \emph {et~al.}(2015)\citenamefont {Lecocq},
  \citenamefont {Clark}, \citenamefont {Simmonds}, \citenamefont {Aumentado},\
  and\ \citenamefont {Teufel}}]{Lecocq:15}%
  \BibitemOpen
  \bibfield  {author} {\bibinfo {author} {\bibfnamefont {F.}~\bibnamefont
  {Lecocq}}, \bibinfo {author} {\bibfnamefont {J.~B.}\ \bibnamefont {Clark}},
  \bibinfo {author} {\bibfnamefont {R.~W.}\ \bibnamefont {Simmonds}}, \bibinfo
  {author} {\bibfnamefont {J.}~\bibnamefont {Aumentado}}, \ and\ \bibinfo
  {author} {\bibfnamefont {J.~D.}\ \bibnamefont {Teufel}},\ }\href {\doibase
  10.1103/PhysRevX.5.041037} {\bibfield  {journal} {\bibinfo  {journal} {Phys.
  Rev. X}\ }\textbf {\bibinfo {volume} {5}},\ \bibinfo {pages} {041037}
  (\bibinfo {year} {2015})}\BibitemShut {NoStop}%
\bibitem [{\citenamefont {Ourjoumtsev}\ \emph {et~al.}(2007)\citenamefont
  {Ourjoumtsev}, \citenamefont {Jeong}, \citenamefont {Tualle-Brouri},\ and\
  \citenamefont {Grangier}}]{ourjoumtsev2007generation}%
  \BibitemOpen
  \bibfield  {author} {\bibinfo {author} {\bibfnamefont {A.}~\bibnamefont
  {Ourjoumtsev}}, \bibinfo {author} {\bibfnamefont {H.}~\bibnamefont {Jeong}},
  \bibinfo {author} {\bibfnamefont {R.}~\bibnamefont {Tualle-Brouri}}, \ and\
  \bibinfo {author} {\bibfnamefont {P.}~\bibnamefont {Grangier}},\ }\href@noop
  {} {\bibfield  {journal} {\bibinfo  {journal} {Nature}\ }\textbf {\bibinfo
  {volume} {448}},\ \bibinfo {pages} {784} (\bibinfo {year}
  {2007})}\BibitemShut {NoStop}%
\bibitem [{\citenamefont {Etesse}\ \emph {et~al.}(2015)\citenamefont {Etesse},
  \citenamefont {Bouillard}, \citenamefont {Kanseri},\ and\ \citenamefont
  {Tualle-Brouri}}]{Etesse_2015}%
  \BibitemOpen
  \bibfield  {author} {\bibinfo {author} {\bibfnamefont {J.}~\bibnamefont
  {Etesse}}, \bibinfo {author} {\bibfnamefont {M.}~\bibnamefont {Bouillard}},
  \bibinfo {author} {\bibfnamefont {B.}~\bibnamefont {Kanseri}}, \ and\
  \bibinfo {author} {\bibfnamefont {R.}~\bibnamefont {Tualle-Brouri}},\ }\href
  {\doibase 10.1103/PhysRevLett.114.193602} {\bibfield  {journal} {\bibinfo
  {journal} {Phys. Rev. Lett.}\ }\textbf {\bibinfo {volume} {114}},\ \bibinfo
  {pages} {193602} (\bibinfo {year} {2015})}\BibitemShut {NoStop}%
\bibitem [{\citenamefont {Huang}\ \emph {et~al.}(2015)\citenamefont {Huang},
  \citenamefont {Le~Jeannic}, \citenamefont {Ruaudel}, \citenamefont {Verma},
  \citenamefont {Shaw}, \citenamefont {Marsili}, \citenamefont {Nam},
  \citenamefont {Wu}, \citenamefont {Zeng}, \citenamefont {Jeong},
  \citenamefont {Filip}, \citenamefont {Morin},\ and\ \citenamefont
  {Laurat}}]{Huang_2015}%
  \BibitemOpen
  \bibfield  {author} {\bibinfo {author} {\bibfnamefont {K.}~\bibnamefont
  {Huang}}, \bibinfo {author} {\bibfnamefont {H.}~\bibnamefont {Le~Jeannic}},
  \bibinfo {author} {\bibfnamefont {J.}~\bibnamefont {Ruaudel}}, \bibinfo
  {author} {\bibfnamefont {V.~B.}\ \bibnamefont {Verma}}, \bibinfo {author}
  {\bibfnamefont {M.~D.}\ \bibnamefont {Shaw}}, \bibinfo {author}
  {\bibfnamefont {F.}~\bibnamefont {Marsili}}, \bibinfo {author} {\bibfnamefont
  {S.~W.}\ \bibnamefont {Nam}}, \bibinfo {author} {\bibfnamefont
  {E.}~\bibnamefont {Wu}}, \bibinfo {author} {\bibfnamefont {H.}~\bibnamefont
  {Zeng}}, \bibinfo {author} {\bibfnamefont {Y.-C.}\ \bibnamefont {Jeong}},
  \bibinfo {author} {\bibfnamefont {R.}~\bibnamefont {Filip}}, \bibinfo
  {author} {\bibfnamefont {O.}~\bibnamefont {Morin}}, \ and\ \bibinfo {author}
  {\bibfnamefont {J.}~\bibnamefont {Laurat}},\ }\href {\doibase
  10.1103/PhysRevLett.115.023602} {\bibfield  {journal} {\bibinfo  {journal}
  {Phys. Rev. Lett.}\ }\textbf {\bibinfo {volume} {115}},\ \bibinfo {pages}
  {023602} (\bibinfo {year} {2015})}\BibitemShut {NoStop}%
\bibitem [{\citenamefont {Flurin}\ \emph {et~al.}(2012)\citenamefont {Flurin},
  \citenamefont {Roch}, \citenamefont {Mallet}, \citenamefont {Devoret},\ and\
  \citenamefont {Huard}}]{Flurin_2012}%
  \BibitemOpen
  \bibfield  {author} {\bibinfo {author} {\bibfnamefont {E.}~\bibnamefont
  {Flurin}}, \bibinfo {author} {\bibfnamefont {N.}~\bibnamefont {Roch}},
  \bibinfo {author} {\bibfnamefont {F.}~\bibnamefont {Mallet}}, \bibinfo
  {author} {\bibfnamefont {M.~H.}\ \bibnamefont {Devoret}}, \ and\ \bibinfo
  {author} {\bibfnamefont {B.}~\bibnamefont {Huard}},\ }\href {\doibase
  10.1103/PhysRevLett.109.183901} {\bibfield  {journal} {\bibinfo  {journal}
  {Phys. Rev. Lett.}\ }\textbf {\bibinfo {volume} {109}},\ \bibinfo {pages}
  {183901} (\bibinfo {year} {2012})}\BibitemShut {NoStop}%
\bibitem [{\citenamefont {Mallet}\ \emph {et~al.}(2011)\citenamefont {Mallet},
  \citenamefont {Castellanos-Beltran}, \citenamefont {Ku}, \citenamefont
  {Glancy}, \citenamefont {Knill}, \citenamefont {Irwin}, \citenamefont
  {Hilton}, \citenamefont {Vale},\ and\ \citenamefont {Lehnert}}]{Mallet_2011}%
  \BibitemOpen
  \bibfield  {author} {\bibinfo {author} {\bibfnamefont {F.}~\bibnamefont
  {Mallet}}, \bibinfo {author} {\bibfnamefont {M.~A.}\ \bibnamefont
  {Castellanos-Beltran}}, \bibinfo {author} {\bibfnamefont {H.~S.}\
  \bibnamefont {Ku}}, \bibinfo {author} {\bibfnamefont {S.}~\bibnamefont
  {Glancy}}, \bibinfo {author} {\bibfnamefont {E.}~\bibnamefont {Knill}},
  \bibinfo {author} {\bibfnamefont {K.~D.}\ \bibnamefont {Irwin}}, \bibinfo
  {author} {\bibfnamefont {G.~C.}\ \bibnamefont {Hilton}}, \bibinfo {author}
  {\bibfnamefont {L.~R.}\ \bibnamefont {Vale}}, \ and\ \bibinfo {author}
  {\bibfnamefont {K.~W.}\ \bibnamefont {Lehnert}},\ }\href {\doibase
  10.1103/PhysRevLett.106.220502} {\bibfield  {journal} {\bibinfo  {journal}
  {Phys. Rev. Lett.}\ }\textbf {\bibinfo {volume} {106}},\ \bibinfo {pages}
  {220502} (\bibinfo {year} {2011})}\BibitemShut {NoStop}%
\bibitem [{\citenamefont {Nakamura}\ and\ \citenamefont
  {Yamamoto}(2013)}]{nakamura2013breakthroughs}%
  \BibitemOpen
  \bibfield  {author} {\bibinfo {author} {\bibfnamefont {Y.}~\bibnamefont
  {Nakamura}}\ and\ \bibinfo {author} {\bibfnamefont {T.}~\bibnamefont
  {Yamamoto}},\ }\href@noop {} {\bibfield  {journal} {\bibinfo  {journal} {IEEE
  Photonics Journal}\ }\textbf {\bibinfo {volume} {5}},\ \bibinfo {pages}
  {0701406} (\bibinfo {year} {2013})}\BibitemShut {NoStop}%
\bibitem [{\citenamefont {Pezz\`e}\ \emph {et~al.}(2018)\citenamefont
  {Pezz\`e}, \citenamefont {Smerzi}, \citenamefont {Oberthaler}, \citenamefont
  {Schmied},\ and\ \citenamefont {Treutlein}}]{pezze2016non}%
  \BibitemOpen
  \bibfield  {author} {\bibinfo {author} {\bibfnamefont {L.}~\bibnamefont
  {Pezz\`e}}, \bibinfo {author} {\bibfnamefont {A.}~\bibnamefont {Smerzi}},
  \bibinfo {author} {\bibfnamefont {M.~K.}\ \bibnamefont {Oberthaler}},
  \bibinfo {author} {\bibfnamefont {R.}~\bibnamefont {Schmied}}, \ and\
  \bibinfo {author} {\bibfnamefont {P.}~\bibnamefont {Treutlein}},\ }\href
  {\doibase 10.1103/RevModPhys.90.035005} {\bibfield  {journal} {\bibinfo
  {journal} {Rev. Mod. Phys.}\ }\textbf {\bibinfo {volume} {90}},\ \bibinfo
  {pages} {035005} (\bibinfo {year} {2018})}\BibitemShut {NoStop}%
\bibitem [{\citenamefont {Stevenson}\ \emph {et~al.}(2016)\citenamefont
  {Stevenson}, \citenamefont {Min\'{a}\v{r}}, \citenamefont {Hofferberth},\
  and\ \citenamefont {Lesanovsky}}]{stevenson2016prospects}%
  \BibitemOpen
  \bibfield  {author} {\bibinfo {author} {\bibfnamefont {R.}~\bibnamefont
  {Stevenson}}, \bibinfo {author} {\bibfnamefont {J.}~\bibnamefont
  {Min\'{a}\v{r}}}, \bibinfo {author} {\bibfnamefont {S.}~\bibnamefont
  {Hofferberth}}, \ and\ \bibinfo {author} {\bibfnamefont {I.}~\bibnamefont
  {Lesanovsky}},\ }\href {\doibase 10.1103/PhysRevA.94.043813} {\bibfield
  {journal} {\bibinfo  {journal} {Phys. Rev. A}\ }\textbf {\bibinfo {volume}
  {94}},\ \bibinfo {pages} {043813} (\bibinfo {year} {2016})}\BibitemShut
  {NoStop}%
\bibitem [{\citenamefont {Kepesidis}\ \emph {et~al.}(2016)\citenamefont
  {Kepesidis}, \citenamefont {Lemonde}, \citenamefont {Norambuena},
  \citenamefont {Maze},\ and\ \citenamefont {Rabl}}]{Kepesidis_2016}%
  \BibitemOpen
  \bibfield  {author} {\bibinfo {author} {\bibfnamefont {K.~V.}\ \bibnamefont
  {Kepesidis}}, \bibinfo {author} {\bibfnamefont {M.-A.}\ \bibnamefont
  {Lemonde}}, \bibinfo {author} {\bibfnamefont {A.}~\bibnamefont {Norambuena}},
  \bibinfo {author} {\bibfnamefont {J.~R.}\ \bibnamefont {Maze}}, \ and\
  \bibinfo {author} {\bibfnamefont {P.}~\bibnamefont {Rabl}},\ }\href {\doibase
  10.1103/PhysRevB.94.214115} {\bibfield  {journal} {\bibinfo  {journal} {Phys.
  Rev. B}\ }\textbf {\bibinfo {volume} {94}},\ \bibinfo {pages} {214115}
  (\bibinfo {year} {2016})}\BibitemShut {NoStop}%
\bibitem [{\citenamefont {Diehl}\ \emph {et~al.}(2008)\citenamefont {Diehl},
  \citenamefont {Micheli}, \citenamefont {Kantian}, \citenamefont {Kraus},
  \citenamefont {B{\"{u}}chler},\ and\ \citenamefont
  {Zoller}}]{diehl_quantum_2008}%
  \BibitemOpen
  \bibfield  {author} {\bibinfo {author} {\bibfnamefont {S.}~\bibnamefont
  {Diehl}}, \bibinfo {author} {\bibfnamefont {A.}~\bibnamefont {Micheli}},
  \bibinfo {author} {\bibfnamefont {A.}~\bibnamefont {Kantian}}, \bibinfo
  {author} {\bibfnamefont {B.}~\bibnamefont {Kraus}}, \bibinfo {author}
  {\bibfnamefont {H.~P.}\ \bibnamefont {B{\"{u}}chler}}, \ and\ \bibinfo
  {author} {\bibfnamefont {P.}~\bibnamefont {Zoller}},\ }\href {\doibase
  10.1038/nphys1073} {\bibfield  {journal} {\bibinfo  {journal} {Nat. Phys.}\
  }\textbf {\bibinfo {volume} {4}},\ \bibinfo {pages} {878} (\bibinfo {year}
  {2008})}\BibitemShut {NoStop}%
\bibitem [{\citenamefont {Kraus}\ \emph {et~al.}(2008)\citenamefont {Kraus},
  \citenamefont {B\"uchler}, \citenamefont {Diehl}, \citenamefont {Kantian},
  \citenamefont {Micheli},\ and\ \citenamefont
  {Zoller}}]{kraus_preparation_2008}%
  \BibitemOpen
  \bibfield  {author} {\bibinfo {author} {\bibfnamefont {B.}~\bibnamefont
  {Kraus}}, \bibinfo {author} {\bibfnamefont {H.~P.}\ \bibnamefont
  {B\"uchler}}, \bibinfo {author} {\bibfnamefont {S.}~\bibnamefont {Diehl}},
  \bibinfo {author} {\bibfnamefont {A.}~\bibnamefont {Kantian}}, \bibinfo
  {author} {\bibfnamefont {A.}~\bibnamefont {Micheli}}, \ and\ \bibinfo
  {author} {\bibfnamefont {P.}~\bibnamefont {Zoller}},\ }\href {\doibase
  10.1103/PhysRevA.78.042307} {\bibfield  {journal} {\bibinfo  {journal} {Phys.
  Rev. A}\ }\textbf {\bibinfo {volume} {78}},\ \bibinfo {pages} {042307}
  (\bibinfo {year} {2008})}\BibitemShut {NoStop}%
\bibitem [{\citenamefont {Gottesman}\ \emph {et~al.}(2001)\citenamefont
  {Gottesman}, \citenamefont {Kitaev},\ and\ \citenamefont
  {Preskill}}]{Gottesman_2001}%
  \BibitemOpen
  \bibfield  {author} {\bibinfo {author} {\bibfnamefont {D.}~\bibnamefont
  {Gottesman}}, \bibinfo {author} {\bibfnamefont {A.}~\bibnamefont {Kitaev}}, \
  and\ \bibinfo {author} {\bibfnamefont {J.}~\bibnamefont {Preskill}},\ }\href
  {\doibase 10.1103/PhysRevA.64.012310} {\bibfield  {journal} {\bibinfo
  {journal} {Phys. Rev. A}\ }\textbf {\bibinfo {volume} {64}},\ \bibinfo
  {pages} {012310} (\bibinfo {year} {2001})}\BibitemShut {NoStop}%
\bibitem [{\citenamefont {Duivenvoorden}\ \emph {et~al.}(2017)\citenamefont
  {Duivenvoorden}, \citenamefont {Terhal},\ and\ \citenamefont
  {Weigand}}]{Duivenvoorden_2017}%
  \BibitemOpen
  \bibfield  {author} {\bibinfo {author} {\bibfnamefont {K.}~\bibnamefont
  {Duivenvoorden}}, \bibinfo {author} {\bibfnamefont {B.~M.}\ \bibnamefont
  {Terhal}}, \ and\ \bibinfo {author} {\bibfnamefont {D.}~\bibnamefont
  {Weigand}},\ }\href {\doibase 10.1103/PhysRevA.95.012305} {\bibfield
  {journal} {\bibinfo  {journal} {Phys. Rev. A}\ }\textbf {\bibinfo {volume}
  {95}},\ \bibinfo {pages} {012305} (\bibinfo {year} {2017})}\BibitemShut
  {NoStop}%
\bibitem [{\citenamefont {Terhal}\ and\ \citenamefont
  {Weigand}(2016)}]{Terhal_2016}%
  \BibitemOpen
  \bibfield  {author} {\bibinfo {author} {\bibfnamefont {B.~M.}\ \bibnamefont
  {Terhal}}\ and\ \bibinfo {author} {\bibfnamefont {D.}~\bibnamefont
  {Weigand}},\ }\href {\doibase 10.1103/PhysRevA.93.012315} {\bibfield
  {journal} {\bibinfo  {journal} {Phys. Rev. A}\ }\textbf {\bibinfo {volume}
  {93}},\ \bibinfo {pages} {012315} (\bibinfo {year} {2016})}\BibitemShut
  {NoStop}%
\bibitem [{\citenamefont {Brunelli}\ \emph {et~al.}(2018)\citenamefont
  {Brunelli}, \citenamefont {Houhou}, \citenamefont {Moore}, \citenamefont
  {Nunnenkamp}, \citenamefont {Paternostro},\ and\ \citenamefont
  {Ferraro}}]{Brunelli2018a}%
  \BibitemOpen
  \bibfield  {author} {\bibinfo {author} {\bibfnamefont {M.}~\bibnamefont
  {Brunelli}}, \bibinfo {author} {\bibfnamefont {O.}~\bibnamefont {Houhou}},
  \bibinfo {author} {\bibfnamefont {D.~W.}\ \bibnamefont {Moore}}, \bibinfo
  {author} {\bibfnamefont {A.}~\bibnamefont {Nunnenkamp}}, \bibinfo {author}
  {\bibfnamefont {M.}~\bibnamefont {Paternostro}}, \ and\ \bibinfo {author}
  {\bibfnamefont {A.}~\bibnamefont {Ferraro}},\ }\href@noop {} {\bibfield
  {journal} {\bibinfo  {journal} {arXiv preprint arXiv:1804.00014}\ } (\bibinfo
  {year} {2018})}\BibitemShut {NoStop}%
\bibitem [{\citenamefont {Brunelli}\ and\ \citenamefont
  {Houhou}(2018)}]{Brunelli2018b}%
  \BibitemOpen
  \bibfield  {author} {\bibinfo {author} {\bibfnamefont {M.}~\bibnamefont
  {Brunelli}}\ and\ \bibinfo {author} {\bibfnamefont {O.}~\bibnamefont
  {Houhou}},\ }\href@noop {} {\bibfield  {journal} {\bibinfo  {journal} {arXiv
  preprint arXiv:1809.05266}\ } (\bibinfo {year} {2018})}\BibitemShut {NoStop}%
\bibitem [{\citenamefont {Munro}\ \emph {et~al.}(2002)\citenamefont {Munro},
  \citenamefont {Nemoto}, \citenamefont {Milburn},\ and\ \citenamefont
  {Braunstein}}]{Munro_2002}%
  \BibitemOpen
  \bibfield  {author} {\bibinfo {author} {\bibfnamefont {W.~J.}\ \bibnamefont
  {Munro}}, \bibinfo {author} {\bibfnamefont {K.}~\bibnamefont {Nemoto}},
  \bibinfo {author} {\bibfnamefont {G.~J.}\ \bibnamefont {Milburn}}, \ and\
  \bibinfo {author} {\bibfnamefont {S.~L.}\ \bibnamefont {Braunstein}},\ }\href
  {\doibase 10.1103/PhysRevA.66.023819} {\bibfield  {journal} {\bibinfo
  {journal} {Phys. Rev. A}\ }\textbf {\bibinfo {volume} {66}},\ \bibinfo
  {pages} {023819} (\bibinfo {year} {2002})}\BibitemShut {NoStop}%
\bibitem [{\citenamefont {Gilchrist}\ \emph {et~al.}(2004)\citenamefont
  {Gilchrist}, \citenamefont {Nemoto}, \citenamefont {Munro}, \citenamefont
  {Ralph}, \citenamefont {Glancy}, \citenamefont {Braunstein},\ and\
  \citenamefont {Milburn}}]{gilchrist2004schrodinger}%
  \BibitemOpen
  \bibfield  {author} {\bibinfo {author} {\bibfnamefont {A.}~\bibnamefont
  {Gilchrist}}, \bibinfo {author} {\bibfnamefont {K.}~\bibnamefont {Nemoto}},
  \bibinfo {author} {\bibfnamefont {W.~J.}\ \bibnamefont {Munro}}, \bibinfo
  {author} {\bibfnamefont {T.}~\bibnamefont {Ralph}}, \bibinfo {author}
  {\bibfnamefont {S.}~\bibnamefont {Glancy}}, \bibinfo {author} {\bibfnamefont
  {S.~L.}\ \bibnamefont {Braunstein}}, \ and\ \bibinfo {author} {\bibfnamefont
  {G.}~\bibnamefont {Milburn}},\ }\href
  {http://iopscience.iop.org/article/10.1088/1464-4266/6/8/032/meta} {\bibfield
   {journal} {\bibinfo  {journal} {Journal of Optics B: Quantum and
  Semiclassical Optics}\ }\textbf {\bibinfo {volume} {6}},\ \bibinfo {pages}
  {S828} (\bibinfo {year} {2004})}\BibitemShut {NoStop}%
\bibitem [{\citenamefont {Knott}\ \emph {et~al.}(2016)\citenamefont {Knott},
  \citenamefont {Proctor}, \citenamefont {Hayes}, \citenamefont {Cooling},\
  and\ \citenamefont {Dunningham}}]{Knott_2016}%
  \BibitemOpen
  \bibfield  {author} {\bibinfo {author} {\bibfnamefont {P.~A.}\ \bibnamefont
  {Knott}}, \bibinfo {author} {\bibfnamefont {T.~J.}\ \bibnamefont {Proctor}},
  \bibinfo {author} {\bibfnamefont {A.~J.}\ \bibnamefont {Hayes}}, \bibinfo
  {author} {\bibfnamefont {J.~P.}\ \bibnamefont {Cooling}}, \ and\ \bibinfo
  {author} {\bibfnamefont {J.~A.}\ \bibnamefont {Dunningham}},\ }\href
  {\doibase 10.1103/PhysRevA.93.033859} {\bibfield  {journal} {\bibinfo
  {journal} {Phys. Rev. A}\ }\textbf {\bibinfo {volume} {93}},\ \bibinfo
  {pages} {033859} (\bibinfo {year} {2016})}\BibitemShut {NoStop}%
\bibitem [{\citenamefont {Anetsberger}\ \emph {et~al.}(2009)\citenamefont
  {Anetsberger}, \citenamefont {Arcizet}, \citenamefont {Unterreithmeier},
  \citenamefont {Rivi{\`e}re}, \citenamefont {Schliesser}, \citenamefont
  {Weig}, \citenamefont {Kotthaus},\ and\ \citenamefont
  {Kippenberg}}]{anetsberger2009near}%
  \BibitemOpen
  \bibfield  {author} {\bibinfo {author} {\bibfnamefont {G.}~\bibnamefont
  {Anetsberger}}, \bibinfo {author} {\bibfnamefont {O.}~\bibnamefont
  {Arcizet}}, \bibinfo {author} {\bibfnamefont {Q.~P.}\ \bibnamefont
  {Unterreithmeier}}, \bibinfo {author} {\bibfnamefont {R.}~\bibnamefont
  {Rivi{\`e}re}}, \bibinfo {author} {\bibfnamefont {A.}~\bibnamefont
  {Schliesser}}, \bibinfo {author} {\bibfnamefont {E.~M.}\ \bibnamefont
  {Weig}}, \bibinfo {author} {\bibfnamefont {J.~P.}\ \bibnamefont {Kotthaus}},
  \ and\ \bibinfo {author} {\bibfnamefont {T.~J.}\ \bibnamefont {Kippenberg}},\
  }\href@noop {} {\bibfield  {journal} {\bibinfo  {journal} {Nature Physics}\
  }\textbf {\bibinfo {volume} {5}},\ \bibinfo {pages} {909} (\bibinfo {year}
  {2009})}\BibitemShut {NoStop}%
\bibitem [{\citenamefont {Gavartin}\ \emph {et~al.}(2012)\citenamefont
  {Gavartin}, \citenamefont {Verlot},\ and\ \citenamefont
  {Kippenberg}}]{gavartin2012hybrid}%
  \BibitemOpen
  \bibfield  {author} {\bibinfo {author} {\bibfnamefont {E.}~\bibnamefont
  {Gavartin}}, \bibinfo {author} {\bibfnamefont {P.}~\bibnamefont {Verlot}}, \
  and\ \bibinfo {author} {\bibfnamefont {T.~J.}\ \bibnamefont {Kippenberg}},\
  }\href@noop {} {\bibfield  {journal} {\bibinfo  {journal} {Nature
  nanotechnology}\ }\textbf {\bibinfo {volume} {7}},\ \bibinfo {pages} {509}
  (\bibinfo {year} {2012})}\BibitemShut {NoStop}%
\bibitem [{\citenamefont {Bose}\ \emph {et~al.}(1999)\citenamefont {Bose},
  \citenamefont {Jacobs},\ and\ \citenamefont {Knight}}]{bose1999scheme}%
  \BibitemOpen
  \bibfield  {author} {\bibinfo {author} {\bibfnamefont {S.}~\bibnamefont
  {Bose}}, \bibinfo {author} {\bibfnamefont {K.}~\bibnamefont {Jacobs}}, \ and\
  \bibinfo {author} {\bibfnamefont {P.~L.}\ \bibnamefont {Knight}},\ }\href
  {\doibase 10.1103/PhysRevA.59.3204} {\bibfield  {journal} {\bibinfo
  {journal} {Phys. Rev. A}\ }\textbf {\bibinfo {volume} {59}},\ \bibinfo
  {pages} {3204} (\bibinfo {year} {1999})}\BibitemShut {NoStop}%
\bibitem [{\citenamefont {Arndt}\ and\ \citenamefont
  {Hornberger}(2014)}]{arndt2014testing}%
  \BibitemOpen
  \bibfield  {author} {\bibinfo {author} {\bibfnamefont {M.}~\bibnamefont
  {Arndt}}\ and\ \bibinfo {author} {\bibfnamefont {K.}~\bibnamefont
  {Hornberger}},\ }\href@noop {} {\bibfield  {journal} {\bibinfo  {journal}
  {Nature Physics}\ }\textbf {\bibinfo {volume} {10}},\ \bibinfo {pages} {271}
  (\bibinfo {year} {2014})}\BibitemShut {NoStop}%
\bibitem [{\citenamefont {Bateman}\ \emph {et~al.}(2015)\citenamefont
  {Bateman}, \citenamefont {McHardy}, \citenamefont {Merle}, \citenamefont
  {Morris},\ and\ \citenamefont {Ulbricht}}]{Bateman:15}%
  \BibitemOpen
  \bibfield  {author} {\bibinfo {author} {\bibfnamefont {J.}~\bibnamefont
  {Bateman}}, \bibinfo {author} {\bibfnamefont {I.}~\bibnamefont {McHardy}},
  \bibinfo {author} {\bibfnamefont {A.}~\bibnamefont {Merle}}, \bibinfo
  {author} {\bibfnamefont {T.~R.}\ \bibnamefont {Morris}}, \ and\ \bibinfo
  {author} {\bibfnamefont {H.}~\bibnamefont {Ulbricht}},\ }\href@noop {}
  {\bibfield  {journal} {\bibinfo  {journal} {Scientific reports}\ }\textbf
  {\bibinfo {volume} {5}},\ \bibinfo {pages} {8058} (\bibinfo {year}
  {2015})}\BibitemShut {NoStop}%
\bibitem [{\citenamefont {Davidovich}\ \emph {et~al.}(1987)\citenamefont
  {Davidovich}, \citenamefont {Raimond}, \citenamefont {Brune},\ and\
  \citenamefont {Haroche}}]{davidovich1987two}%
  \BibitemOpen
  \bibfield  {author} {\bibinfo {author} {\bibfnamefont {L.}~\bibnamefont
  {Davidovich}}, \bibinfo {author} {\bibfnamefont {J.~M.}\ \bibnamefont
  {Raimond}}, \bibinfo {author} {\bibfnamefont {M.}~\bibnamefont {Brune}}, \
  and\ \bibinfo {author} {\bibfnamefont {S.}~\bibnamefont {Haroche}},\ }\href
  {\doibase 10.1103/PhysRevA.36.3771} {\bibfield  {journal} {\bibinfo
  {journal} {Phys. Rev. A}\ }\textbf {\bibinfo {volume} {36}},\ \bibinfo
  {pages} {3771} (\bibinfo {year} {1987})}\BibitemShut {NoStop}%
\bibitem [{\citenamefont {Brune}\ \emph
  {et~al.}(1987{\natexlab{a}})\citenamefont {Brune}, \citenamefont {Raimond},
  \citenamefont {Goy}, \citenamefont {Davidovich},\ and\ \citenamefont
  {Haroche}}]{Brune_1987}%
  \BibitemOpen
  \bibfield  {author} {\bibinfo {author} {\bibfnamefont {M.}~\bibnamefont
  {Brune}}, \bibinfo {author} {\bibfnamefont {J.~M.}\ \bibnamefont {Raimond}},
  \bibinfo {author} {\bibfnamefont {P.}~\bibnamefont {Goy}}, \bibinfo {author}
  {\bibfnamefont {L.}~\bibnamefont {Davidovich}}, \ and\ \bibinfo {author}
  {\bibfnamefont {S.}~\bibnamefont {Haroche}},\ }\href {\doibase
  10.1103/PhysRevLett.59.1899} {\bibfield  {journal} {\bibinfo  {journal}
  {Phys. Rev. Lett.}\ }\textbf {\bibinfo {volume} {59}},\ \bibinfo {pages}
  {1899} (\bibinfo {year} {1987}{\natexlab{a}})}\BibitemShut {NoStop}%
\bibitem [{\citenamefont {Brune}\ \emph
  {et~al.}(1987{\natexlab{b}})\citenamefont {Brune}, \citenamefont {Raimond},\
  and\ \citenamefont {Haroche}}]{Brune_1987rydberg_two_phot}%
  \BibitemOpen
  \bibfield  {author} {\bibinfo {author} {\bibfnamefont {M.}~\bibnamefont
  {Brune}}, \bibinfo {author} {\bibfnamefont {J.~M.}\ \bibnamefont {Raimond}},
  \ and\ \bibinfo {author} {\bibfnamefont {S.}~\bibnamefont {Haroche}},\ }\href
  {\doibase 10.1103/PhysRevA.35.154} {\bibfield  {journal} {\bibinfo  {journal}
  {Phys. Rev. A}\ }\textbf {\bibinfo {volume} {35}},\ \bibinfo {pages} {154}
  (\bibinfo {year} {1987}{\natexlab{b}})}\BibitemShut {NoStop}%
\bibitem [{\citenamefont {Ashraf}\ \emph {et~al.}(1990)\citenamefont {Ashraf},
  \citenamefont {Gea-Banacloche},\ and\ \citenamefont
  {Zubairy}}]{Ashraf1990_2ph_micromaser_statistics}%
  \BibitemOpen
  \bibfield  {author} {\bibinfo {author} {\bibfnamefont {I.}~\bibnamefont
  {Ashraf}}, \bibinfo {author} {\bibfnamefont {J.}~\bibnamefont
  {Gea-Banacloche}}, \ and\ \bibinfo {author} {\bibfnamefont {M.~S.}\
  \bibnamefont {Zubairy}},\ }\href {\doibase 10.1103/PhysRevA.42.6704}
  {\bibfield  {journal} {\bibinfo  {journal} {Phys. Rev. A}\ }\textbf {\bibinfo
  {volume} {42}},\ \bibinfo {pages} {6704} (\bibinfo {year}
  {1990})}\BibitemShut {NoStop}%
\bibitem [{\citenamefont {Toor}\ \emph {et~al.}(1996)\citenamefont {Toor},
  \citenamefont {Zhu},\ and\ \citenamefont {Zubairy}}]{toor1996theory}%
  \BibitemOpen
  \bibfield  {author} {\bibinfo {author} {\bibfnamefont {A.~H.}\ \bibnamefont
  {Toor}}, \bibinfo {author} {\bibfnamefont {S.-Y.}\ \bibnamefont {Zhu}}, \
  and\ \bibinfo {author} {\bibfnamefont {M.~S.}\ \bibnamefont {Zubairy}},\
  }\href {\doibase 10.1103/PhysRevA.53.3529} {\bibfield  {journal} {\bibinfo
  {journal} {Phys. Rev. A}\ }\textbf {\bibinfo {volume} {53}},\ \bibinfo
  {pages} {3529} (\bibinfo {year} {1996})}\BibitemShut {NoStop}%
\bibitem [{\citenamefont {Sarlette}\ \emph {et~al.}(2011)\citenamefont
  {Sarlette}, \citenamefont {Raimond}, \citenamefont {Brune},\ and\
  \citenamefont {Rouchon}}]{Sarlette_2011}%
  \BibitemOpen
  \bibfield  {author} {\bibinfo {author} {\bibfnamefont {A.}~\bibnamefont
  {Sarlette}}, \bibinfo {author} {\bibfnamefont {J.~M.}\ \bibnamefont
  {Raimond}}, \bibinfo {author} {\bibfnamefont {M.}~\bibnamefont {Brune}}, \
  and\ \bibinfo {author} {\bibfnamefont {P.}~\bibnamefont {Rouchon}},\ }\href
  {\doibase 10.1103/PhysRevLett.107.010402} {\bibfield  {journal} {\bibinfo
  {journal} {Phys. Rev. Lett.}\ }\textbf {\bibinfo {volume} {107}},\ \bibinfo
  {pages} {010402} (\bibinfo {year} {2011})}\BibitemShut {NoStop}%
\bibitem [{\citenamefont {Facon}\ \emph {et~al.}(2016)\citenamefont {Facon},
  \citenamefont {Dietsche}, \citenamefont {Grosso}, \citenamefont {Haroche},
  \citenamefont {Raimond}, \citenamefont {Brune},\ and\ \citenamefont
  {Gleyzes}}]{facon2016sensitive}%
  \BibitemOpen
  \bibfield  {author} {\bibinfo {author} {\bibfnamefont {A.}~\bibnamefont
  {Facon}}, \bibinfo {author} {\bibfnamefont {E.-K.}\ \bibnamefont {Dietsche}},
  \bibinfo {author} {\bibfnamefont {D.}~\bibnamefont {Grosso}}, \bibinfo
  {author} {\bibfnamefont {S.}~\bibnamefont {Haroche}}, \bibinfo {author}
  {\bibfnamefont {J.-M.}\ \bibnamefont {Raimond}}, \bibinfo {author}
  {\bibfnamefont {M.}~\bibnamefont {Brune}}, \ and\ \bibinfo {author}
  {\bibfnamefont {S.}~\bibnamefont {Gleyzes}},\ }\href@noop {} {\bibfield
  {journal} {\bibinfo  {journal} {Nature}\ }\textbf {\bibinfo {volume} {535}},\
  \bibinfo {pages} {262} (\bibinfo {year} {2016})}\BibitemShut {NoStop}%
\bibitem [{\citenamefont {Neilinger}\ \emph {et~al.}(2015)\citenamefont
  {Neilinger}, \citenamefont {Reh\'ak}, \citenamefont {Grajcar}, \citenamefont
  {Oelsner}, \citenamefont {H\"ubner},\ and\ \citenamefont
  {Il'ichev}}]{Neilinger_2015}%
  \BibitemOpen
  \bibfield  {author} {\bibinfo {author} {\bibfnamefont {P.}~\bibnamefont
  {Neilinger}}, \bibinfo {author} {\bibfnamefont {M.}~\bibnamefont {Reh\'ak}},
  \bibinfo {author} {\bibfnamefont {M.}~\bibnamefont {Grajcar}}, \bibinfo
  {author} {\bibfnamefont {G.}~\bibnamefont {Oelsner}}, \bibinfo {author}
  {\bibfnamefont {U.}~\bibnamefont {H\"ubner}}, \ and\ \bibinfo {author}
  {\bibfnamefont {E.}~\bibnamefont {Il'ichev}},\ }\href {\doibase
  10.1103/PhysRevB.91.104516} {\bibfield  {journal} {\bibinfo  {journal} {Phys.
  Rev. B}\ }\textbf {\bibinfo {volume} {91}},\ \bibinfo {pages} {104516}
  (\bibinfo {year} {2015})}\BibitemShut {NoStop}%
\bibitem [{\citenamefont {Mirrahimi}\ \emph {et~al.}(2014)\citenamefont
  {Mirrahimi}, \citenamefont {Leghtas}, \citenamefont {Albert}, \citenamefont
  {Touzard}, \citenamefont {Schoelkopf}, \citenamefont {Jiang},\ and\
  \citenamefont {Devoret}}]{mirrahimi2014dynamically}%
  \BibitemOpen
  \bibfield  {author} {\bibinfo {author} {\bibfnamefont {M.}~\bibnamefont
  {Mirrahimi}}, \bibinfo {author} {\bibfnamefont {Z.}~\bibnamefont {Leghtas}},
  \bibinfo {author} {\bibfnamefont {V.~V.}\ \bibnamefont {Albert}}, \bibinfo
  {author} {\bibfnamefont {S.}~\bibnamefont {Touzard}}, \bibinfo {author}
  {\bibfnamefont {R.~J.}\ \bibnamefont {Schoelkopf}}, \bibinfo {author}
  {\bibfnamefont {L.}~\bibnamefont {Jiang}}, \ and\ \bibinfo {author}
  {\bibfnamefont {M.~H.}\ \bibnamefont {Devoret}},\ }\href
  {http://iopscience.iop.org/article/10.1088/1367-2630/16/4/045014} {\bibfield
  {journal} {\bibinfo  {journal} {New Journal of Physics}\ }\textbf {\bibinfo
  {volume} {16}},\ \bibinfo {pages} {045014} (\bibinfo {year}
  {2014})}\BibitemShut {NoStop}%
\bibitem [{\citenamefont {Roy}\ \emph {et~al.}(2015)\citenamefont {Roy},
  \citenamefont {Leghtas}, \citenamefont {Stone}, \citenamefont {Devoret},\
  and\ \citenamefont {Mirrahimi}}]{roy2015continuous}%
  \BibitemOpen
  \bibfield  {author} {\bibinfo {author} {\bibfnamefont {A.}~\bibnamefont
  {Roy}}, \bibinfo {author} {\bibfnamefont {Z.}~\bibnamefont {Leghtas}},
  \bibinfo {author} {\bibfnamefont {A.~D.}\ \bibnamefont {Stone}}, \bibinfo
  {author} {\bibfnamefont {M.}~\bibnamefont {Devoret}}, \ and\ \bibinfo
  {author} {\bibfnamefont {M.}~\bibnamefont {Mirrahimi}},\ }\href {\doibase
  10.1103/PhysRevA.91.013810} {\bibfield  {journal} {\bibinfo  {journal} {Phys.
  Rev. A}\ }\textbf {\bibinfo {volume} {91}},\ \bibinfo {pages} {013810}
  (\bibinfo {year} {2015})}\BibitemShut {NoStop}%
\bibitem [{\citenamefont {Leghtas}\ \emph {et~al.}(2015)\citenamefont
  {Leghtas}, \citenamefont {Touzard}, \citenamefont {Pop}, \citenamefont {Kou},
  \citenamefont {Vlastakis}, \citenamefont {Petrenko}, \citenamefont {Sliwa},
  \citenamefont {Narla}, \citenamefont {Shankar}, \citenamefont {Hatridge}
  \emph {et~al.}}]{leghtas2015confining}%
  \BibitemOpen
  \bibfield  {author} {\bibinfo {author} {\bibfnamefont {Z.}~\bibnamefont
  {Leghtas}}, \bibinfo {author} {\bibfnamefont {S.}~\bibnamefont {Touzard}},
  \bibinfo {author} {\bibfnamefont {I.~M.}\ \bibnamefont {Pop}}, \bibinfo
  {author} {\bibfnamefont {A.}~\bibnamefont {Kou}}, \bibinfo {author}
  {\bibfnamefont {B.}~\bibnamefont {Vlastakis}}, \bibinfo {author}
  {\bibfnamefont {A.}~\bibnamefont {Petrenko}}, \bibinfo {author}
  {\bibfnamefont {K.~M.}\ \bibnamefont {Sliwa}}, \bibinfo {author}
  {\bibfnamefont {A.}~\bibnamefont {Narla}}, \bibinfo {author} {\bibfnamefont
  {S.}~\bibnamefont {Shankar}}, \bibinfo {author} {\bibfnamefont {M.~J.}\
  \bibnamefont {Hatridge}},  \emph {et~al.},\ }\href@noop {} {\bibfield
  {journal} {\bibinfo  {journal} {Science}\ }\textbf {\bibinfo {volume}
  {347}},\ \bibinfo {pages} {853} (\bibinfo {year} {2015})}\BibitemShut
  {NoStop}%
\bibitem [{\citenamefont {Orszag}\ \emph {et~al.}(1993)\citenamefont {Orszag},
  \citenamefont {Roa},\ and\ \citenamefont
  {Ram\'{\i}rez}}]{orszag1993generation}%
  \BibitemOpen
  \bibfield  {author} {\bibinfo {author} {\bibfnamefont {M.}~\bibnamefont
  {Orszag}}, \bibinfo {author} {\bibfnamefont {L.}~\bibnamefont {Roa}}, \ and\
  \bibinfo {author} {\bibfnamefont {R.}~\bibnamefont {Ram\'{\i}rez}},\ }\href
  {\doibase 10.1103/PhysRevA.48.4648} {\bibfield  {journal} {\bibinfo
  {journal} {Phys. Rev. A}\ }\textbf {\bibinfo {volume} {48}},\ \bibinfo
  {pages} {4648} (\bibinfo {year} {1993})}\BibitemShut {NoStop}%
\bibitem [{\citenamefont {Gerry}(1988)}]{gerry1988two-photon}%
  \BibitemOpen
  \bibfield  {author} {\bibinfo {author} {\bibfnamefont {C.~C.}\ \bibnamefont
  {Gerry}},\ }\href {\doibase 10.1103/PhysRevA.37.2683} {\bibfield  {journal}
  {\bibinfo  {journal} {Phys. Rev. A}\ }\textbf {\bibinfo {volume} {37}},\
  \bibinfo {pages} {2683} (\bibinfo {year} {1988})}\BibitemShut {NoStop}%
\bibitem [{\citenamefont {Orszag}\ \emph {et~al.}(1992)\citenamefont {Orszag},
  \citenamefont {Ram\'{\i}rez}, \citenamefont {Retamal},\ and\ \citenamefont
  {Roa}}]{Orszag1992_squeezed}%
  \BibitemOpen
  \bibfield  {author} {\bibinfo {author} {\bibfnamefont {M.}~\bibnamefont
  {Orszag}}, \bibinfo {author} {\bibfnamefont {R.}~\bibnamefont
  {Ram\'{\i}rez}}, \bibinfo {author} {\bibfnamefont {J.~C.}\ \bibnamefont
  {Retamal}}, \ and\ \bibinfo {author} {\bibfnamefont {L.}~\bibnamefont
  {Roa}},\ }\href {\doibase 10.1103/PhysRevA.45.6717} {\bibfield  {journal}
  {\bibinfo  {journal} {Phys. Rev. A}\ }\textbf {\bibinfo {volume} {45}},\
  \bibinfo {pages} {6717} (\bibinfo {year} {1992})}\BibitemShut {NoStop}%
\bibitem [{\citenamefont {Roa}(1994)}]{roa1995phase}%
  \BibitemOpen
  \bibfield  {author} {\bibinfo {author} {\bibfnamefont {L.}~\bibnamefont
  {Roa}},\ }\href {\doibase 10.1103/PhysRevA.50.R1995} {\bibfield  {journal}
  {\bibinfo  {journal} {Phys. Rev. A}\ }\textbf {\bibinfo {volume} {50}},\
  \bibinfo {pages} {R1995} (\bibinfo {year} {1994})}\BibitemShut {NoStop}%
\bibitem [{\citenamefont {Helstrom}(1967)}]{Helstrom1967}%
  \BibitemOpen
  \bibfield  {author} {\bibinfo {author} {\bibfnamefont {C.}~\bibnamefont
  {Helstrom}},\ }\href {\doibase
  http://dx.doi.org/10.1016/0375-9601(67)90366-0} {\bibfield  {journal}
  {\bibinfo  {journal} {Phys. Lett. A}\ }\textbf {\bibinfo {volume} {25}},\
  \bibinfo {pages} {101 } (\bibinfo {year} {1967})}\BibitemShut {NoStop}%
\bibitem [{\citenamefont {Helstrom}(1968{\natexlab{a}})}]{Helstrom1968}%
  \BibitemOpen
  \bibfield  {author} {\bibinfo {author} {\bibfnamefont {C.}~\bibnamefont
  {Helstrom}},\ }\href {\doibase 10.1109/TIT.1968.1054108} {\bibfield
  {journal} {\bibinfo  {journal} {IEEE Trans. Inform. Theory}\ }\textbf
  {\bibinfo {volume} {14}},\ \bibinfo {pages} {234} (\bibinfo {year}
  {1968}{\natexlab{a}})}\BibitemShut {NoStop}%
\bibitem [{\citenamefont {Braunstein}\ and\ \citenamefont
  {Caves}(1994)}]{Braunstein1994}%
  \BibitemOpen
  \bibfield  {author} {\bibinfo {author} {\bibfnamefont {S.~L.}\ \bibnamefont
  {Braunstein}}\ and\ \bibinfo {author} {\bibfnamefont {C.~M.}\ \bibnamefont
  {Caves}},\ }\href {\doibase 10.1103/PhysRevLett.72.3439} {\bibfield
  {journal} {\bibinfo  {journal} {Phys. Rev. Lett.}\ }\textbf {\bibinfo
  {volume} {72}},\ \bibinfo {pages} {3439} (\bibinfo {year}
  {1994})}\BibitemShut {NoStop}%
\bibitem [{\citenamefont {Cahill}\ and\ \citenamefont
  {Glauber}(1969)}]{cahill1969density}%
  \BibitemOpen
  \bibfield  {author} {\bibinfo {author} {\bibfnamefont {K.~E.}\ \bibnamefont
  {Cahill}}\ and\ \bibinfo {author} {\bibfnamefont {R.~J.}\ \bibnamefont
  {Glauber}},\ }\href {\doibase 10.1103/PhysRev.177.1882} {\bibfield  {journal}
  {\bibinfo  {journal} {Phys. Rev.}\ }\textbf {\bibinfo {volume} {177}},\
  \bibinfo {pages} {1882} (\bibinfo {year} {1969})}\BibitemShut {NoStop}%
\bibitem [{\citenamefont {Macieszczak}\ \emph {et~al.}(2016)\citenamefont
  {Macieszczak}, \citenamefont {Gu\ifmmode \mbox{\c{t}}\else
  \c{t}\fi{}\ifmmode~\u{a}\else \u{a}\fi{}}, \citenamefont {Lesanovsky},\ and\
  \citenamefont {Garrahan}}]{macieszczak2016towards}%
  \BibitemOpen
  \bibfield  {author} {\bibinfo {author} {\bibfnamefont {K.}~\bibnamefont
  {Macieszczak}}, \bibinfo {author} {\bibfnamefont {M.}~\bibnamefont
  {Gu\ifmmode \mbox{\c{t}}\else \c{t}\fi{}\ifmmode~\u{a}\else \u{a}\fi{}}},
  \bibinfo {author} {\bibfnamefont {I.}~\bibnamefont {Lesanovsky}}, \ and\
  \bibinfo {author} {\bibfnamefont {J.~P.}\ \bibnamefont {Garrahan}},\ }\href
  {\doibase 10.1103/PhysRevLett.116.240404} {\bibfield  {journal} {\bibinfo
  {journal} {Phys. Rev. Lett.}\ }\textbf {\bibinfo {volume} {116}},\ \bibinfo
  {pages} {240404} (\bibinfo {year} {2016})}\BibitemShut {NoStop}%
\bibitem [{\citenamefont {{Azouit, R\'{e}mi}}\ \emph
  {et~al.}(2016)\citenamefont {{Azouit, R\'{e}mi}}, \citenamefont {{Sarlette,
  Alain}},\ and\ \citenamefont {{Rouchon, Pierre}}}]{azouit2016well}%
  \BibitemOpen
  \bibfield  {author} {\bibinfo {author} {\bibnamefont {{Azouit, R\'{e}mi}}},
  \bibinfo {author} {\bibnamefont {{Sarlette, Alain}}}, \ and\ \bibinfo
  {author} {\bibnamefont {{Rouchon, Pierre}}},\ }\href {\doibase
  10.1051/cocv/2016050} {\bibfield  {journal} {\bibinfo  {journal} {ESAIM:
  COCV}\ }\textbf {\bibinfo {volume} {22}},\ \bibinfo {pages} {1353} (\bibinfo
  {year} {2016})}\BibitemShut {NoStop}%
\bibitem [{\citenamefont {Minganti}\ \emph {et~al.}(2016)\citenamefont
  {Minganti}, \citenamefont {Bartolo}, \citenamefont {Lolli}, \citenamefont
  {Casteels},\ and\ \citenamefont {Ciuti}}]{minganti2016exact}%
  \BibitemOpen
  \bibfield  {author} {\bibinfo {author} {\bibfnamefont {F.}~\bibnamefont
  {Minganti}}, \bibinfo {author} {\bibfnamefont {N.}~\bibnamefont {Bartolo}},
  \bibinfo {author} {\bibfnamefont {J.}~\bibnamefont {Lolli}}, \bibinfo
  {author} {\bibfnamefont {W.}~\bibnamefont {Casteels}}, \ and\ \bibinfo
  {author} {\bibfnamefont {C.}~\bibnamefont {Ciuti}},\ }\href@noop {}
  {\bibfield  {journal} {\bibinfo  {journal} {Scientific reports}\ }\textbf
  {\bibinfo {volume} {6}},\ \bibinfo {pages} {26987} (\bibinfo {year}
  {2016})}\BibitemShut {NoStop}%
\bibitem [{\citenamefont {Bartolo}\ \emph {et~al.}(2016)\citenamefont
  {Bartolo}, \citenamefont {Minganti}, \citenamefont {Casteels},\ and\
  \citenamefont {Ciuti}}]{Bartolo_2016}%
  \BibitemOpen
  \bibfield  {author} {\bibinfo {author} {\bibfnamefont {N.}~\bibnamefont
  {Bartolo}}, \bibinfo {author} {\bibfnamefont {F.}~\bibnamefont {Minganti}},
  \bibinfo {author} {\bibfnamefont {W.}~\bibnamefont {Casteels}}, \ and\
  \bibinfo {author} {\bibfnamefont {C.}~\bibnamefont {Ciuti}},\ }\href
  {\doibase 10.1103/PhysRevA.94.033841} {\bibfield  {journal} {\bibinfo
  {journal} {Phys. Rev. A}\ }\textbf {\bibinfo {volume} {94}},\ \bibinfo
  {pages} {033841} (\bibinfo {year} {2016})}\BibitemShut {NoStop}%
\bibitem [{\citenamefont {Azouit}\ \emph {et~al.}(2015)\citenamefont {Azouit},
  \citenamefont {Sarlette},\ and\ \citenamefont
  {Rouchon}}]{azouit2015convergence}%
  \BibitemOpen
  \bibfield  {author} {\bibinfo {author} {\bibfnamefont {R.}~\bibnamefont
  {Azouit}}, \bibinfo {author} {\bibfnamefont {A.}~\bibnamefont {Sarlette}}, \
  and\ \bibinfo {author} {\bibfnamefont {P.}~\bibnamefont {Rouchon}},\ }in\
  \href {http://ieeexplore.ieee.org/document/7403235/} {\emph {\bibinfo
  {booktitle} {2015 IEEE 54th Annual Conference on Decision and Control
  (CDC)}}}\ (\bibinfo  {publisher} {IEEE},\ \bibinfo {year} {2015})\ pp.\
  \bibinfo {pages} {6447--6453}\BibitemShut {NoStop}%
\bibitem [{\citenamefont {Tannoudji}\ \emph {et~al.}(1998)\citenamefont
  {Tannoudji}, \citenamefont {Dupont-Roc},\ and\ \citenamefont
  {Grynberg}}]{Tannoudji_1998}%
  \BibitemOpen
  \bibinfo {editor} {\bibfnamefont {C.~C.}\ \bibnamefont {Tannoudji}}, \bibinfo
  {editor} {\bibfnamefont {J.}~\bibnamefont {Dupont-Roc}}, \ and\ \bibinfo
  {editor} {\bibfnamefont {G.}~\bibnamefont {Grynberg}},\ eds.,\ \href@noop {}
  {\emph {\bibinfo {title} {Atom-Photon Interactions}}}\ (\bibinfo  {publisher}
  {John Wiley \& Sons},\ \bibinfo {year} {1998})\BibitemShut {NoStop}%
\bibitem [{\citenamefont {Alexanian}\ and\ \citenamefont
  {Bose}(1995)}]{Alexanian1995unitary}%
  \BibitemOpen
  \bibfield  {author} {\bibinfo {author} {\bibfnamefont {M.}~\bibnamefont
  {Alexanian}}\ and\ \bibinfo {author} {\bibfnamefont {S.~K.}\ \bibnamefont
  {Bose}},\ }\href {\doibase 10.1103/PhysRevA.52.2218} {\bibfield  {journal}
  {\bibinfo  {journal} {Phys. Rev. A}\ }\textbf {\bibinfo {volume} {52}},\
  \bibinfo {pages} {2218} (\bibinfo {year} {1995})}\BibitemShut {NoStop}%
\bibitem [{\citenamefont {Klimov}\ \emph {et~al.}(2002)\citenamefont {Klimov},
  \citenamefont {S\'{a}nchez-Soto}, \citenamefont {Navarro},\ and\
  \citenamefont {Yustas}}]{Klimov2002}%
  \BibitemOpen
  \bibfield  {author} {\bibinfo {author} {\bibfnamefont {A.~B.}\ \bibnamefont
  {Klimov}}, \bibinfo {author} {\bibfnamefont {L.~L.}\ \bibnamefont
  {S\'{a}nchez-Soto}}, \bibinfo {author} {\bibfnamefont {A.}~\bibnamefont
  {Navarro}}, \ and\ \bibinfo {author} {\bibfnamefont {E.~C.}\ \bibnamefont
  {Yustas}},\ }\href {\doibase 10.1080/09500340210134675} {\bibfield  {journal}
  {\bibinfo  {journal} {Journal of Modern Optics}\ }\textbf {\bibinfo {volume}
  {49}},\ \bibinfo {pages} {2211} (\bibinfo {year} {2002})}\BibitemShut
  {NoStop}%
\bibitem [{\citenamefont {Toor}\ and\ \citenamefont
  {Zubairy}(1992)}]{Toor1992validity}%
  \BibitemOpen
  \bibfield  {author} {\bibinfo {author} {\bibfnamefont {A.~H.}\ \bibnamefont
  {Toor}}\ and\ \bibinfo {author} {\bibfnamefont {M.~S.}\ \bibnamefont
  {Zubairy}},\ }\href {\doibase 10.1103/PhysRevA.45.4951} {\bibfield  {journal}
  {\bibinfo  {journal} {Phys. Rev. A}\ }\textbf {\bibinfo {volume} {45}},\
  \bibinfo {pages} {4951} (\bibinfo {year} {1992})}\BibitemShut {NoStop}%
\bibitem [{\citenamefont {Boone}\ and\ \citenamefont
  {Swain}(1989)}]{Boone1989effective_hamiltonians}%
  \BibitemOpen
  \bibfield  {author} {\bibinfo {author} {\bibfnamefont {A.~W.}\ \bibnamefont
  {Boone}}\ and\ \bibinfo {author} {\bibfnamefont {S.}~\bibnamefont {Swain}},\
  }\href {http://stacks.iop.org/0954-8998/1/i=1/a=004} {\bibfield  {journal}
  {\bibinfo  {journal} {Quantum Optics: Journal of the European Optical Society
  Part B}\ }\textbf {\bibinfo {volume} {1}},\ \bibinfo {pages} {27} (\bibinfo
  {year} {1989})}\BibitemShut {NoStop}%
\bibitem [{\citenamefont {Dung}\ and\ \citenamefont {Huyen}(1994)}]{Dung1994}%
  \BibitemOpen
  \bibfield  {author} {\bibinfo {author} {\bibfnamefont {H.~T.}\ \bibnamefont
  {Dung}}\ and\ \bibinfo {author} {\bibfnamefont {N.~D.}\ \bibnamefont
  {Huyen}},\ }\href {\doibase 10.1103/PhysRevA.49.473} {\bibfield  {journal}
  {\bibinfo  {journal} {Phys. Rev. A}\ }\textbf {\bibinfo {volume} {49}},\
  \bibinfo {pages} {473} (\bibinfo {year} {1994})}\BibitemShut {NoStop}%
\bibitem [{\citenamefont {Alexanian}\ \emph {et~al.}(1998)\citenamefont
  {Alexanian}, \citenamefont {Bose},\ and\ \citenamefont
  {Chow}}]{Alexanian1998_trapped}%
  \BibitemOpen
  \bibfield  {author} {\bibinfo {author} {\bibfnamefont {M.}~\bibnamefont
  {Alexanian}}, \bibinfo {author} {\bibfnamefont {S.}~\bibnamefont {Bose}}, \
  and\ \bibinfo {author} {\bibfnamefont {L.}~\bibnamefont {Chow}},\ }\href
  {http://www.sciencedirect.com/science/article/pii/S0022231397002664}
  {\bibfield  {journal} {\bibinfo  {journal} {Journal of Luminescence}\
  }\textbf {\bibinfo {volume} {76}},\ \bibinfo {pages} {677 } (\bibinfo {year}
  {1998})}\BibitemShut {NoStop}%
\bibitem [{\citenamefont {Jaynes}\ and\ \citenamefont
  {Cummings}(1963)}]{jaynes1963comparison}%
  \BibitemOpen
  \bibfield  {author} {\bibinfo {author} {\bibfnamefont {E.~T.}\ \bibnamefont
  {Jaynes}}\ and\ \bibinfo {author} {\bibfnamefont {F.~W.}\ \bibnamefont
  {Cummings}},\ }\href@noop {} {\bibfield  {journal} {\bibinfo  {journal}
  {Proceedings of the IEEE}\ }\textbf {\bibinfo {volume} {51}},\ \bibinfo
  {pages} {89} (\bibinfo {year} {1963})}\BibitemShut {NoStop}%
\bibitem [{Note1()}]{Note1}%
  \BibitemOpen
  \bibinfo {note} {Here we assumed $\lambda (t)$ real. If the coupling $\lambda
  (t)$ is complex, its phase instead adds to the relative phase between the
  atom state amplitudes $c_{\protect \rm g}$ and $c_{\protect \rm
  e}$.}\BibitemShut {Stop}%
\bibitem [{\citenamefont {Englert}(2002)}]{englert2002elements}%
  \BibitemOpen
  \bibfield  {author} {\bibinfo {author} {\bibfnamefont {B.-G.}\ \bibnamefont
  {Englert}},\ }\href@noop {} {\bibfield  {journal} {\bibinfo  {journal} {arXiv
  preprint quant-ph/0203052}\ } (\bibinfo {year} {2002})}\BibitemShut {NoStop}%
\bibitem [{\citenamefont {Guerra}\ \emph {et~al.}(1991)\citenamefont {Guerra},
  \citenamefont {Khoury}, \citenamefont {Davidovich},\ and\ \citenamefont
  {Zagury}}]{guerra1991role}%
  \BibitemOpen
  \bibfield  {author} {\bibinfo {author} {\bibfnamefont {E.~S.}\ \bibnamefont
  {Guerra}}, \bibinfo {author} {\bibfnamefont {A.~Z.}\ \bibnamefont {Khoury}},
  \bibinfo {author} {\bibfnamefont {L.}~\bibnamefont {Davidovich}}, \ and\
  \bibinfo {author} {\bibfnamefont {N.}~\bibnamefont {Zagury}},\ }\href
  {\doibase 10.1103/PhysRevA.44.7785} {\bibfield  {journal} {\bibinfo
  {journal} {Phys. Rev. A}\ }\textbf {\bibinfo {volume} {44}},\ \bibinfo
  {pages} {7785} (\bibinfo {year} {1991})}\BibitemShut {NoStop}%
\bibitem [{\citenamefont {Lindblad}(1976)}]{Lindblad1976}%
  \BibitemOpen
  \bibfield  {author} {\bibinfo {author} {\bibfnamefont {G.}~\bibnamefont
  {Lindblad}},\ }\href@noop {} {\bibfield  {journal} {\bibinfo  {journal}
  {Comm. Math. Phys}\ }\textbf {\bibinfo {volume} {48}},\ \bibinfo {pages}
  {119} (\bibinfo {year} {1976})}\BibitemShut {NoStop}%
\bibitem [{\citenamefont {Gorini}\ \emph {et~al.}(1976)\citenamefont {Gorini},
  \citenamefont {Kossakowski},\ and\ \citenamefont {Sudarshan}}]{Gorini1976}%
  \BibitemOpen
  \bibfield  {author} {\bibinfo {author} {\bibfnamefont {V.}~\bibnamefont
  {Gorini}}, \bibinfo {author} {\bibfnamefont {A.}~\bibnamefont {Kossakowski}},
  \ and\ \bibinfo {author} {\bibfnamefont {E.~C.~G.}\ \bibnamefont
  {Sudarshan}},\ }\href {\doibase 10.1063/1.522979} {\bibfield  {journal}
  {\bibinfo  {journal} {Journal of Mathematical Physics}\ }\textbf {\bibinfo
  {volume} {17}},\ \bibinfo {pages} {821} (\bibinfo {year} {1976})}\BibitemShut
  {NoStop}%
\bibitem [{\citenamefont {Royer}(1977)}]{royer1977wigner}%
  \BibitemOpen
  \bibfield  {author} {\bibinfo {author} {\bibfnamefont {A.}~\bibnamefont
  {Royer}},\ }\href {\doibase 10.1103/PhysRevA.15.449} {\bibfield  {journal}
  {\bibinfo  {journal} {Phys. Rev. A}\ }\textbf {\bibinfo {volume} {15}},\
  \bibinfo {pages} {449} (\bibinfo {year} {1977})}\BibitemShut {NoStop}%
\bibitem [{\citenamefont {Zanardi}\ and\ \citenamefont
  {Rasetti}(1997)}]{Zanardi1997d}%
  \BibitemOpen
  \bibfield  {author} {\bibinfo {author} {\bibfnamefont {P.}~\bibnamefont
  {Zanardi}}\ and\ \bibinfo {author} {\bibfnamefont {M.}~\bibnamefont
  {Rasetti}},\ }\href {\doibase 10.1103/PhysRevLett.79.3306} {\bibfield
  {journal} {\bibinfo  {journal} {Phys. Rev. Lett.}\ }\textbf {\bibinfo
  {volume} {79}},\ \bibinfo {pages} {3306} (\bibinfo {year}
  {1997})}\BibitemShut {NoStop}%
\bibitem [{\citenamefont {Zanardi}(1997)}]{Zanardi1997e}%
  \BibitemOpen
  \bibfield  {author} {\bibinfo {author} {\bibfnamefont {P.}~\bibnamefont
  {Zanardi}},\ }\href {\doibase 10.1103/PhysRevA.56.4445} {\bibfield  {journal}
  {\bibinfo  {journal} {Phys. Rev. A}\ }\textbf {\bibinfo {volume} {56}},\
  \bibinfo {pages} {4445} (\bibinfo {year} {1997})}\BibitemShut {NoStop}%
\bibitem [{\citenamefont {Lidar}\ \emph {et~al.}(1998)\citenamefont {Lidar},
  \citenamefont {Chuang},\ and\ \citenamefont {Whaley}}]{Lidar1998c}%
  \BibitemOpen
  \bibfield  {author} {\bibinfo {author} {\bibfnamefont {D.~A.}\ \bibnamefont
  {Lidar}}, \bibinfo {author} {\bibfnamefont {I.~L.}\ \bibnamefont {Chuang}}, \
  and\ \bibinfo {author} {\bibfnamefont {K.~B.}\ \bibnamefont {Whaley}},\
  }\href@noop {} {\bibfield  {journal} {\bibinfo  {journal} {Phys. Rev. Lett.}\
  }\textbf {\bibinfo {volume} {81}},\ \bibinfo {pages} {2594} (\bibinfo {year}
  {1998})}\BibitemShut {NoStop}%
\bibitem [{\citenamefont {Dodonov}\ \emph {et~al.}(1974)\citenamefont
  {Dodonov}, \citenamefont {Malkin},\ and\ \citenamefont
  {Man'ko}}]{Dodonov1974}%
  \BibitemOpen
  \bibfield  {author} {\bibinfo {author} {\bibfnamefont {V.}~\bibnamefont
  {Dodonov}}, \bibinfo {author} {\bibfnamefont {I.}~\bibnamefont {Malkin}}, \
  and\ \bibinfo {author} {\bibfnamefont {V.}~\bibnamefont {Man'ko}},\ }\href
  {\doibase https://doi.org/10.1016/0031-8914(74)90215-8} {\bibfield  {journal}
  {\bibinfo  {journal} {Physica}\ }\textbf {\bibinfo {volume} {72}},\ \bibinfo
  {pages} {597 } (\bibinfo {year} {1974})}\BibitemShut {NoStop}%
\bibitem [{\citenamefont {Gerry}(1993)}]{gerry1993non}%
  \BibitemOpen
  \bibfield  {author} {\bibinfo {author} {\bibfnamefont {C.~C.}\ \bibnamefont
  {Gerry}},\ }\href
  {http://www.tandfonline.com/doi/pdf/10.1080/09500349314551131} {\bibfield
  {journal} {\bibinfo  {journal} {Journal of Modern Optics}\ }\textbf {\bibinfo
  {volume} {40}},\ \bibinfo {pages} {1053} (\bibinfo {year}
  {1993})}\BibitemShut {NoStop}%
\bibitem [{\citenamefont {Buca}\ and\ \citenamefont {Prosen}(2012)}]{Buca2012}%
  \BibitemOpen
  \bibfield  {author} {\bibinfo {author} {\bibfnamefont {B.}~\bibnamefont
  {Buca}}\ and\ \bibinfo {author} {\bibfnamefont {T.}~\bibnamefont {Prosen}},\
  }\href {http://stacks.iop.org/1367-2630/14/i=7/a=073007} {\bibfield
  {journal} {\bibinfo  {journal} {New Journal of Physics}\ }\textbf {\bibinfo
  {volume} {14}},\ \bibinfo {pages} {073007} (\bibinfo {year}
  {2012})}\BibitemShut {NoStop}%
\bibitem [{\citenamefont {Albert}\ and\ \citenamefont
  {Jiang}(2014)}]{albert2014symmetries}%
  \BibitemOpen
  \bibfield  {author} {\bibinfo {author} {\bibfnamefont {V.~V.}\ \bibnamefont
  {Albert}}\ and\ \bibinfo {author} {\bibfnamefont {L.}~\bibnamefont {Jiang}},\
  }\href {http://journals.aps.org/pra/abstract/10.1103/PhysRevA.89.022118}
  {\bibfield  {journal} {\bibinfo  {journal} {Physical Review A}\ }\textbf
  {\bibinfo {volume} {89}},\ \bibinfo {pages} {022118} (\bibinfo {year}
  {2014})}\BibitemShut {NoStop}%
\bibitem [{\citenamefont {Hach~III}\ and\ \citenamefont
  {Gerry}(1994)}]{hach1994generation}%
  \BibitemOpen
  \bibfield  {author} {\bibinfo {author} {\bibfnamefont {E.~E.}\ \bibnamefont
  {Hach~III}}\ and\ \bibinfo {author} {\bibfnamefont {C.~C.}\ \bibnamefont
  {Gerry}},\ }\href {\doibase 10.1103/PhysRevA.49.490} {\bibfield  {journal}
  {\bibinfo  {journal} {Phys. Rev. A}\ }\textbf {\bibinfo {volume} {49}},\
  \bibinfo {pages} {490} (\bibinfo {year} {1994})}\BibitemShut {NoStop}%
\bibitem [{\citenamefont {Gilles}\ \emph {et~al.}(1994)\citenamefont {Gilles},
  \citenamefont {Garraway},\ and\ \citenamefont
  {Knight}}]{gilles1994generation}%
  \BibitemOpen
  \bibfield  {author} {\bibinfo {author} {\bibfnamefont {L.}~\bibnamefont
  {Gilles}}, \bibinfo {author} {\bibfnamefont {B.~M.}\ \bibnamefont
  {Garraway}}, \ and\ \bibinfo {author} {\bibfnamefont {P.~L.}\ \bibnamefont
  {Knight}},\ }\href {\doibase 10.1103/PhysRevA.49.2785} {\bibfield  {journal}
  {\bibinfo  {journal} {Phys. Rev. A}\ }\textbf {\bibinfo {volume} {49}},\
  \bibinfo {pages} {2785} (\bibinfo {year} {1994})}\BibitemShut {NoStop}%
\bibitem [{\citenamefont {Guerra}\ \emph {et~al.}(1997)\citenamefont {Guerra},
  \citenamefont {Garraway},\ and\ \citenamefont {Knight}}]{Guerra_1997_PRA}%
  \BibitemOpen
  \bibfield  {author} {\bibinfo {author} {\bibfnamefont {E.~S.}\ \bibnamefont
  {Guerra}}, \bibinfo {author} {\bibfnamefont {B.~M.}\ \bibnamefont
  {Garraway}}, \ and\ \bibinfo {author} {\bibfnamefont {P.~L.}\ \bibnamefont
  {Knight}},\ }\href {\doibase 10.1103/PhysRevA.55.3842} {\bibfield  {journal}
  {\bibinfo  {journal} {Phys. Rev. A}\ }\textbf {\bibinfo {volume} {55}},\
  \bibinfo {pages} {3842} (\bibinfo {year} {1997})}\BibitemShut {NoStop}%
\bibitem [{\citenamefont {Xie}\ \emph {et~al.}(2013)\citenamefont {Xie},
  \citenamefont {Ciornea},\ and\ \citenamefont
  {Macovei}}]{Xie_2013_ProcRomAcademy}%
  \BibitemOpen
  \bibfield  {author} {\bibinfo {author} {\bibfnamefont {X.-T.}\ \bibnamefont
  {Xie}}, \bibinfo {author} {\bibfnamefont {V.}~\bibnamefont {Ciornea}}, \ and\
  \bibinfo {author} {\bibfnamefont {M.~A.}\ \bibnamefont {Macovei}},\
  }\href@noop {} {\bibfield  {journal} {\bibinfo  {journal} {Proceedings of The
  Romanian Academy, Series A}\ }\textbf {\bibinfo {volume} {14}},\ \bibinfo
  {pages} {48} (\bibinfo {year} {2013})}\BibitemShut {NoStop}%
\bibitem [{\citenamefont {Barbeau}(2003)}]{Barbeau}%
  \BibitemOpen
  \bibfield  {author} {\bibinfo {author} {\bibfnamefont {E.~J.}\ \bibnamefont
  {Barbeau}},\ }\href@noop {} {\emph {\bibinfo {title} {Pell's Equation}}}\
  (\bibinfo  {publisher} {Springer},\ \bibinfo {year} {2003})\BibitemShut
  {NoStop}%
\bibitem [{\citenamefont {Lenstra}(2008)}]{Lenstra08solvingthe}%
  \BibitemOpen
  \bibfield  {author} {\bibinfo {author} {\bibfnamefont {H.~W.}\ \bibnamefont
  {Lenstra}},\ }\href@noop {} {\enquote {\bibinfo {title} {Solving the pell
  equation},}\ } (\bibinfo {year} {2008})\BibitemShut {NoStop}%
\bibitem [{\citenamefont {Copley}(1959)}]{Copley1959}%
  \BibitemOpen
  \bibfield  {author} {\bibinfo {author} {\bibfnamefont {G.~N.}\ \bibnamefont
  {Copley}},\ }\href {http://www.jstor.org/stable/2309637} {\bibfield
  {journal} {\bibinfo  {journal} {The American Mathematical Monthly}\ }\textbf
  {\bibinfo {volume} {66}},\ \bibinfo {pages} {288} (\bibinfo {year}
  {1959})}\BibitemShut {NoStop}%
\bibitem [{Note2()}]{Note2}%
  \BibitemOpen
  \bibinfo {note} {This corresponds to the presence of a soft wall with
  {$\protect \qopname \relax o{cos}_{m'}(\phi )=-1$} between the hard walls
  (see Appendix~\ref {app:soft_walls}).}\BibitemShut {Stop}%
\bibitem [{Note3()}]{Note3}%
  \BibitemOpen
  \bibinfo {note} {Coherence to a pure stationary state {$|\Psi \delimiter
  "526930B $} would be of the form {$|\Psi \delimiter "526930B \protect
  \tmspace -\thinmuskip {.1667em}\delimiter "426830A \Phi |$}, where {$|\Phi
  \delimiter "526930B $} is within the disjoint support of a mixed stationary
  state $\rho $. Since {$|\Psi \delimiter "526930B $} can be considered a dark
  state [cf.~Eq.~\protect \textup {\hbox {\mathsurround \z@ \protect
  \normalfont (\ignorespaces \ref {eq:Kraus_shifted}\unskip \@@italiccorr )}}],
  any coherence decays with the corresponding effective Hamiltonian, unless it
  is a coherence to another dark state, i.e., pure state with the same boundary
  conditions as {$|\Psi \delimiter "526930B $}. {$|\Phi \delimiter "526930B $}
  however corresponds to the mixed boundary conditions, since $\rho $ is
  mixed.}\BibitemShut {Stop}%
\bibitem [{Note4()}]{Note4}%
  \BibitemOpen
  \bibinfo {note} {The coherence can be maintained using a feedback mechanism
  of counting the total number of atoms passing through the cavity, and then
  applying a phase flip, if necessary, to the final state (cf.~\cite
  {minganti2016exact}).}\BibitemShut {Stop}%
\bibitem [{Note5()}]{Note5}%
  \BibitemOpen
  \bibinfo {note} {The odd or even parity correspond to $\protect \qopname
  \relax o{sin}_{2n} (\phi )$ or $\protect \qopname \relax o{sin}_{2n+1} (\phi
  )$, respectively, which are approximated as $n$ rotations by $2\phi $ with
  the initial phase shift $3\phi /2$ or $5\phi /2$, respectively. Thus, as for
  the irrational $\phi /\pi $, $2\phi /\pi $ and $3\phi /2$ is also irrational,
  there exist infinitely many soft walls of both parities.}\BibitemShut {Stop}%
\bibitem [{\citenamefont {Rose}\ \emph {et~al.}(2016)\citenamefont {Rose},
  \citenamefont {Macieszczak}, \citenamefont {Lesanovsky},\ and\ \citenamefont
  {Garrahan}}]{Rose2016}%
  \BibitemOpen
  \bibfield  {author} {\bibinfo {author} {\bibfnamefont {D.~C.}\ \bibnamefont
  {Rose}}, \bibinfo {author} {\bibfnamefont {K.}~\bibnamefont {Macieszczak}},
  \bibinfo {author} {\bibfnamefont {I.}~\bibnamefont {Lesanovsky}}, \ and\
  \bibinfo {author} {\bibfnamefont {J.~P.}\ \bibnamefont {Garrahan}},\ }\href
  {\doibase 10.1103/PhysRevE.94.052132} {\bibfield  {journal} {\bibinfo
  {journal} {Phys. Rev. E}\ }\textbf {\bibinfo {volume} {94}},\ \bibinfo
  {pages} {052132} (\bibinfo {year} {2016})}\BibitemShut {NoStop}%
\bibitem [{\citenamefont {Kato}(1995)}]{Kato1995}%
  \BibitemOpen
  \bibfield  {author} {\bibinfo {author} {\bibfnamefont {T.}~\bibnamefont
  {Kato}},\ }\href@noop {} {\emph {\bibinfo {title} {{Perturbation Theory for
  Linear Operators}}}}\ (\bibinfo  {publisher} {Springer},\ \bibinfo {year}
  {1995})\BibitemShut {NoStop}%
\bibitem [{\citenamefont {Zanardi}\ and\ \citenamefont
  {Campos~Venuti}(2014)}]{zanardi_coherent_2014}%
  \BibitemOpen
  \bibfield  {author} {\bibinfo {author} {\bibfnamefont {P.}~\bibnamefont
  {Zanardi}}\ and\ \bibinfo {author} {\bibfnamefont {L.}~\bibnamefont
  {Campos~Venuti}},\ }\href {\doibase 10.1103/PhysRevLett.113.240406}
  {\bibfield  {journal} {\bibinfo  {journal} {Phys. Rev. Lett.}\ }\textbf
  {\bibinfo {volume} {113}},\ \bibinfo {pages} {240406} (\bibinfo {year}
  {2014})}\BibitemShut {NoStop}%
\bibitem [{\citenamefont {Zanardi}\ and\ \citenamefont
  {Campos~Venuti}(2015)}]{zanardi_geometry_2015}%
  \BibitemOpen
  \bibfield  {author} {\bibinfo {author} {\bibfnamefont {P.}~\bibnamefont
  {Zanardi}}\ and\ \bibinfo {author} {\bibfnamefont {L.}~\bibnamefont
  {Campos~Venuti}},\ }\href {\doibase 10.1103/PhysRevA.91.052324} {\bibfield
  {journal} {\bibinfo  {journal} {Phys. Rev. A}\ }\textbf {\bibinfo {volume}
  {91}},\ \bibinfo {pages} {052324} (\bibinfo {year} {2015})}\BibitemShut
  {NoStop}%
\bibitem [{\citenamefont {Zanardi}\ \emph {et~al.}(2016)\citenamefont
  {Zanardi}, \citenamefont {Marshall},\ and\ \citenamefont
  {Campos~Venuti}}]{zanardi_dissipative_2016}%
  \BibitemOpen
  \bibfield  {author} {\bibinfo {author} {\bibfnamefont {P.}~\bibnamefont
  {Zanardi}}, \bibinfo {author} {\bibfnamefont {J.}~\bibnamefont {Marshall}}, \
  and\ \bibinfo {author} {\bibfnamefont {L.}~\bibnamefont {Campos~Venuti}},\
  }\href {\doibase 10.1103/PhysRevA.93.022312} {\bibfield  {journal} {\bibinfo
  {journal} {Phys. Rev. A}\ }\textbf {\bibinfo {volume} {93}},\ \bibinfo
  {pages} {022312} (\bibinfo {year} {2016})}\BibitemShut {NoStop}%
\bibitem [{\citenamefont {Minganti}\ \emph {et~al.}(2018)\citenamefont
  {Minganti}, \citenamefont {Biella}, \citenamefont {Bartolo},\ and\
  \citenamefont {Ciuti}}]{Minganti2018}%
  \BibitemOpen
  \bibfield  {author} {\bibinfo {author} {\bibfnamefont {F.}~\bibnamefont
  {Minganti}}, \bibinfo {author} {\bibfnamefont {A.}~\bibnamefont {Biella}},
  \bibinfo {author} {\bibfnamefont {N.}~\bibnamefont {Bartolo}}, \ and\
  \bibinfo {author} {\bibfnamefont {C.}~\bibnamefont {Ciuti}},\ }\href
  {\doibase 10.1103/PhysRevA.98.042118} {\bibfield  {journal} {\bibinfo
  {journal} {Phys. Rev. A}\ }\textbf {\bibinfo {volume} {98}},\ \bibinfo
  {pages} {042118} (\bibinfo {year} {2018})}\BibitemShut {NoStop}%
\bibitem [{\citenamefont
  {Helstrom}(1968{\natexlab{b}})}]{Helstrom1968detection}%
  \BibitemOpen
  \bibfield  {author} {\bibinfo {author} {\bibfnamefont {C.~W.}\ \bibnamefont
  {Helstrom}},\ }\href
  {http://www.sciencedirect.com/science/article/pii/S0019995868907468}
  {\bibfield  {journal} {\bibinfo  {journal} {Information and Control}\
  }\textbf {\bibinfo {volume} {13}},\ \bibinfo {pages} {156 } (\bibinfo {year}
  {1968}{\natexlab{b}})}\BibitemShut {NoStop}%
\bibitem [{\citenamefont {Giovannetti}\ \emph {et~al.}(2004)\citenamefont
  {Giovannetti}, \citenamefont {Lloyd},\ and\ \citenamefont
  {Maccone}}]{Giovannetti2004}%
  \BibitemOpen
  \bibfield  {author} {\bibinfo {author} {\bibfnamefont {V.}~\bibnamefont
  {Giovannetti}}, \bibinfo {author} {\bibfnamefont {S.}~\bibnamefont {Lloyd}},
  \ and\ \bibinfo {author} {\bibfnamefont {L.}~\bibnamefont {Maccone}},\ }\href
  {\doibase 10.1126/science.1104149} {\bibfield  {journal} {\bibinfo  {journal}
  {Science}\ }\textbf {\bibinfo {volume} {306}},\ \bibinfo {pages} {1330}
  (\bibinfo {year} {2004})}\BibitemShut {NoStop}%
\bibitem [{\citenamefont {Giovannetti}\ \emph {et~al.}(2006)\citenamefont
  {Giovannetti}, \citenamefont {Lloyd},\ and\ \citenamefont
  {Maccone}}]{Giovannetti2006}%
  \BibitemOpen
  \bibfield  {author} {\bibinfo {author} {\bibfnamefont {V.}~\bibnamefont
  {Giovannetti}}, \bibinfo {author} {\bibfnamefont {S.}~\bibnamefont {Lloyd}},
  \ and\ \bibinfo {author} {\bibfnamefont {L.}~\bibnamefont {Maccone}},\ }\href
  {\doibase 10.1103/PhysRevLett.96.010401} {\bibfield  {journal} {\bibinfo
  {journal} {Phys. Rev. Lett.}\ }\textbf {\bibinfo {volume} {96}},\ \bibinfo
  {pages} {010401} (\bibinfo {year} {2006})}\BibitemShut {NoStop}%
\bibitem [{\citenamefont {Giovannetti}\ \emph {et~al.}(2011)\citenamefont
  {Giovannetti}, \citenamefont {Lloyd},\ and\ \citenamefont
  {Maccone}}]{Giovannetti2011}%
  \BibitemOpen
  \bibfield  {author} {\bibinfo {author} {\bibfnamefont {V.}~\bibnamefont
  {Giovannetti}}, \bibinfo {author} {\bibfnamefont {S.}~\bibnamefont {Lloyd}},
  \ and\ \bibinfo {author} {\bibfnamefont {L.}~\bibnamefont {Maccone}},\ }\href
  {\doibase 10.1038/nphoton.2011.35} {\bibfield  {journal} {\bibinfo  {journal}
  {Nat. Photon.}\ }\textbf {\bibinfo {volume} {5}},\ \bibinfo {pages} {222}
  (\bibinfo {year} {2011})}\BibitemShut {NoStop}%
\bibitem [{Note6()}]{Note6}%
  \BibitemOpen
  \bibinfo {note} {We note that in general steady states can vary very strongly
  with $\phi $. In particular, the support of the stationary state is limited
  by hard walls which exist at a given $m$ only for values of $\phi $ given by
  Eq.~\protect \textup {\hbox {\mathsurround \z@ \protect \normalfont
  (\ignorespaces \ref {eq:phi_wall}\unskip \@@italiccorr )}}. Nevertheless,
  when $\phi $ is varied, a given hard wall becomes a soft wall and the
  stationary state becomes metastable, so that it can be dissipatively
  generated in the corresponding metastable regime [cf.~Secs.~\ref
  {sec:walls_hard} and~\ref {sec:walls_soft}].}\BibitemShut {Stop}%
\bibitem [{Note7()}]{Note7}%
  \BibitemOpen
  \bibinfo {note} {Let {$\rho = p_1|E_1\delimiter "526930B \protect \tmspace
  -\thinmuskip {.1667em}\delimiter "426830A E_1|+p_2|E_2\delimiter "526930B
  \protect \tmspace -\thinmuskip {.1667em}\delimiter "426830A E_2|$}, with
  {$|E_1\delimiter "526930B =c_+|\Psi _+\delimiter "526930B +c_-|\Psi
  _-\delimiter "526930B $} and {$|E_2\delimiter "526930B =c_-^*|\Psi
  _+\delimiter "526930B -c_+^*|\Psi _-\delimiter "526930B $}. From~\protect
  \textup {\hbox {\mathsurround \z@ \protect \normalfont (\ignorespaces \ref
  {eq:qfi_formula}\unskip \@@italiccorr )}}, we have {$F_Q(\rho )= 4
  (p_1-p_2)^2 |\delimiter "426830A E_1| n|E_2\delimiter "526930B |^2+4 p_1
  [\protect \text {Var}(n,|E_1\delimiter "526930B )-|\delimiter "426830A E_1|
  n|E_2\delimiter "526930B |^2] +4 p_2 [\protect \text {Var}(n,|E_2\delimiter
  "526930B )-|\delimiter "426830A E_1| n|E_2\delimiter "526930B |^2] $}, where
  {$\protect \text {Var}(n,|E_{1,2}\delimiter "526930B )$} denotes the variance
  of $n$ in {$|E_{1,2}\delimiter "526930B $} [cf.~Eq.~\protect \textup {\hbox
  {\mathsurround \z@ \protect \normalfont (\ignorespaces \ref
  {eq:qfi_formula2}\unskip \@@italiccorr )}}]. Furthermore, {$\protect \text
  {Var}(n,|E_1\delimiter "526930B )= |c_+|^2 \protect \text {Var}(n,|\Psi
  _+\delimiter "526930B )+|c_-|^2 \protect \text {Var}(n,|\Psi _-\delimiter
  "526930B ) + |c_+ c_- |^2(\delimiter "426830A n\delimiter "526930B
  _+-\delimiter "426830A n\delimiter "526930B _-)^2 $} [cf.~Eq.~\protect
  \textup {\hbox {\mathsurround \z@ \protect \normalfont (\ignorespaces \ref
  {eq:qfi_modes}\unskip \@@italiccorr )}}], while from the parity conservation
  by {$n$}, {$|\delimiter "426830A E_1| n|E_2\delimiter "526930B |^2= |c_+
  c_-|^2(\delimiter "426830A n\delimiter "526930B _+-\delimiter "426830A
  n\delimiter "526930B _-)^2$}. Identifying {$p=p_1|c_+|^2+p_2|c_-|^2$} and
  {$c= (p_1-p_2) c_+ c_-^*$} in~\protect \textup {\hbox {\mathsurround \z@
  \protect \normalfont (\ignorespaces \ref {eq:rho}\unskip \@@italiccorr )}},
  we arrive at~\protect \textup {\hbox {\mathsurround \z@ \protect \normalfont
  (\ignorespaces \ref {eq:qfi_rho}\unskip \@@italiccorr )}}.}\BibitemShut
  {Stop}%
\bibitem [{Note8()}]{Note8}%
  \BibitemOpen
  \bibinfo {note} {In general the maximal QFI is achieved for: $p=0$ when
  {$F_Q(|\Psi _+\delimiter "526930B )\geq F_Q(|\Psi _-\delimiter "526930B
  )+4(\delimiter "426830A n\delimiter "526930B _+-\delimiter "426830A
  n\delimiter "526930B _-)^2$}, $p=1$ when {$F_Q(|\Psi _-\delimiter "526930B
  )\geq F_Q(|\Psi _+\delimiter "526930B )+4(\delimiter "426830A n\delimiter
  "526930B _+-\delimiter "426830A n\delimiter "526930B _-)^2$} and {$p=1/2+
  [F_Q(|\Psi _+\delimiter "526930B )-F_Q(|\Psi _-\delimiter "526930B
  )]/[8(\delimiter "426830A n\delimiter "526930B _+-\delimiter "426830A
  n\delimiter "526930B _-)^2]$} otherwise. The minimal QFI is {$\protect
  \qopname \relax m{min}[F_Q(|\Psi _+\delimiter "526930B ),F_Q(|\Psi
  _-\delimiter "526930B )]$}}\BibitemShut {NoStop}%
\bibitem [{Note9()}]{Note9}%
  \BibitemOpen
  \bibinfo {note} {The minimal enhancement in the DFS is {$\protect \qopname
  \relax m{min}[F_Q(|\Psi _+\delimiter "526930B )/\delimiter "426830A
  n\delimiter "526930B _+,F_Q(|\Psi _-\delimiter "526930B )/\delimiter "426830A
  n\delimiter "526930B _-]$}.}\BibitemShut {Stop}%
\bibitem [{\citenamefont {Ko\l{}ody\ifmmode~\acute{n}\else \'{n}\fi{}ski}\ and\
  \citenamefont {Demkowicz-Dobrza\ifmmode~\acute{n}\else
  \'{n}\fi{}ski}(2010)}]{Kolodynski2010}%
  \BibitemOpen
  \bibfield  {author} {\bibinfo {author} {\bibfnamefont {J.}~\bibnamefont
  {Ko\l{}ody\ifmmode~\acute{n}\else \'{n}\fi{}ski}}\ and\ \bibinfo {author}
  {\bibfnamefont {R.}~\bibnamefont {Demkowicz-Dobrza\ifmmode~\acute{n}\else
  \'{n}\fi{}ski}},\ }\href {\doibase 10.1103/PhysRevA.82.053804} {\bibfield
  {journal} {\bibinfo  {journal} {Phys. Rev. A}\ }\textbf {\bibinfo {volume}
  {82}},\ \bibinfo {pages} {053804} (\bibinfo {year} {2010})}\BibitemShut
  {NoStop}%
\bibitem [{\citenamefont {Escher}\ \emph {et~al.}(2011)\citenamefont {Escher},
  \citenamefont {de~Matos~Filho},\ and\ \citenamefont
  {Davidovich}}]{Escher2011}%
  \BibitemOpen
  \bibfield  {author} {\bibinfo {author} {\bibfnamefont {B.~M.}\ \bibnamefont
  {Escher}}, \bibinfo {author} {\bibfnamefont {R.~L.}\ \bibnamefont
  {de~Matos~Filho}}, \ and\ \bibinfo {author} {\bibfnamefont {L.}~\bibnamefont
  {Davidovich}},\ }\href {\doibase 10.1038/nphys1958} {\bibfield  {journal}
  {\bibinfo  {journal} {Nat. Phys.}\ }\textbf {\bibinfo {volume} {7}},\
  \bibinfo {pages} {406–411} (\bibinfo {year} {2011})}\BibitemShut {NoStop}%
\bibitem [{\citenamefont {Rafal
  Demkowicz-Dobrza{\'n}ski}(2012)}]{Demkowicz2012}%
  \BibitemOpen
  \bibfield  {author} {\bibinfo {author} {\bibfnamefont {M.~G. m. c. u.~u.}\
  \bibnamefont {Rafal Demkowicz-Dobrza{\'n}ski}, \bibfnamefont
  {Jan~Ko{\l}ody{\'n}ski}},\ }\href {\doibase 10.1038/ncomms2067} {\bibfield
  {journal} {\bibinfo  {journal} {Nat. Comm.}\ }\textbf {\bibinfo {volume} {3}}
  (\bibinfo {year} {2012}),\ 10.1038/ncomms2067}\BibitemShut {NoStop}%
\bibitem [{\citenamefont {Girolami}(2014)}]{Girolami2015}%
  \BibitemOpen
  \bibfield  {author} {\bibinfo {author} {\bibfnamefont {D.}~\bibnamefont
  {Girolami}},\ }\href {\doibase 10.1103/PhysRevLett.113.170401} {\bibfield
  {journal} {\bibinfo  {journal} {Phys. Rev. Lett.}\ }\textbf {\bibinfo
  {volume} {113}},\ \bibinfo {pages} {170401} (\bibinfo {year}
  {2014})}\BibitemShut {NoStop}%
\bibitem [{Note10()}]{Note10}%
  \BibitemOpen
  \bibinfo {note} {We have used $w \approx 2$ mm as compared with $w \approx
  10$ mm used in \cite {Sarlette_2011} to make it compatible with the mode
  volume $V=70$ mm${}^3$ taken from \cite {Brune_1987}}\BibitemShut {NoStop}%
\bibitem [{\citenamefont {\v{S}ibali\'{c}}\ \emph {et~al.}()\citenamefont
  {\v{S}ibali\'{c}}, \citenamefont {Pritchard}, \citenamefont {Adams},\ and\
  \citenamefont {Weatherill}}]{ARC_package}%
  \BibitemOpen
  \bibfield  {author} {\bibinfo {author} {\bibfnamefont {N.}~\bibnamefont
  {\v{S}ibali\'{c}}}, \bibinfo {author} {\bibfnamefont {J.~D.}\ \bibnamefont
  {Pritchard}}, \bibinfo {author} {\bibfnamefont {C.~S.}\ \bibnamefont
  {Adams}}, \ and\ \bibinfo {author} {\bibfnamefont {K.~J.}\ \bibnamefont
  {Weatherill}},\ }\href@noop {} {\enquote {\bibinfo {title} {Arc package},}\
  }\bibinfo {howpublished}
  {\url{https://arc-alkali-rydberg-calculator.readthedocs.io/en/latest/}}\BibitemShut
  {NoStop}%
\bibitem [{\citenamefont {{\v{S}}ibali{\'c}}\ \emph {et~al.}(2017)\citenamefont
  {{\v{S}}ibali{\'c}}, \citenamefont {Pritchard}, \citenamefont {Adams},\ and\
  \citenamefont {Weatherill}}]{Sibalic_CompPhysComm_2017}%
  \BibitemOpen
  \bibfield  {author} {\bibinfo {author} {\bibfnamefont {N.}~\bibnamefont
  {{\v{S}}ibali{\'c}}}, \bibinfo {author} {\bibfnamefont {J.~D.}\ \bibnamefont
  {Pritchard}}, \bibinfo {author} {\bibfnamefont {C.~S.}\ \bibnamefont
  {Adams}}, \ and\ \bibinfo {author} {\bibfnamefont {K.~J.}\ \bibnamefont
  {Weatherill}},\ }\href@noop {} {\bibfield  {journal} {\bibinfo  {journal}
  {Computer Physics Communications}\ }\textbf {\bibinfo {volume} {220}},\
  \bibinfo {pages} {319} (\bibinfo {year} {2017})}\BibitemShut {NoStop}%
\bibitem [{\citenamefont {Campagne-Ibarcq}\ \emph {et~al.}(2018)\citenamefont
  {Campagne-Ibarcq}, \citenamefont {Zalys-Geller}, \citenamefont {Narla},
  \citenamefont {Shankar}, \citenamefont {Reinhold}, \citenamefont {Burkhart},
  \citenamefont {Axline}, \citenamefont {Pfaff}, \citenamefont {Frunzio},
  \citenamefont {Schoelkopf},\ and\ \citenamefont
  {Devoret}}]{Campagne-Ibarcq_PRL_2018}%
  \BibitemOpen
  \bibfield  {author} {\bibinfo {author} {\bibfnamefont {P.}~\bibnamefont
  {Campagne-Ibarcq}}, \bibinfo {author} {\bibfnamefont {E.}~\bibnamefont
  {Zalys-Geller}}, \bibinfo {author} {\bibfnamefont {A.}~\bibnamefont {Narla}},
  \bibinfo {author} {\bibfnamefont {S.}~\bibnamefont {Shankar}}, \bibinfo
  {author} {\bibfnamefont {P.}~\bibnamefont {Reinhold}}, \bibinfo {author}
  {\bibfnamefont {L.}~\bibnamefont {Burkhart}}, \bibinfo {author}
  {\bibfnamefont {C.}~\bibnamefont {Axline}}, \bibinfo {author} {\bibfnamefont
  {W.}~\bibnamefont {Pfaff}}, \bibinfo {author} {\bibfnamefont
  {L.}~\bibnamefont {Frunzio}}, \bibinfo {author} {\bibfnamefont {R.~J.}\
  \bibnamefont {Schoelkopf}}, \ and\ \bibinfo {author} {\bibfnamefont {M.~H.}\
  \bibnamefont {Devoret}},\ }\href {\doibase 10.1103/PhysRevLett.120.200501}
  {\bibfield  {journal} {\bibinfo  {journal} {Phys. Rev. Lett.}\ }\textbf
  {\bibinfo {volume} {120}},\ \bibinfo {pages} {200501} (\bibinfo {year}
  {2018})}\BibitemShut {NoStop}%
\bibitem [{\citenamefont {Zeytinoglu}\ \emph {et~al.}(2015)\citenamefont
  {Zeytinoglu}, \citenamefont {Pechal}, \citenamefont {Berger}, \citenamefont
  {Abdumalikov}, \citenamefont {Wallraff},\ and\ \citenamefont
  {Filipp}}]{Zeytinoglu_2015}%
  \BibitemOpen
  \bibfield  {author} {\bibinfo {author} {\bibfnamefont {S.}~\bibnamefont
  {Zeytinoglu}}, \bibinfo {author} {\bibfnamefont {M.}~\bibnamefont {Pechal}},
  \bibinfo {author} {\bibfnamefont {S.}~\bibnamefont {Berger}}, \bibinfo
  {author} {\bibfnamefont {A.~A.}\ \bibnamefont {Abdumalikov}}, \bibinfo
  {author} {\bibfnamefont {A.}~\bibnamefont {Wallraff}}, \ and\ \bibinfo
  {author} {\bibfnamefont {S.}~\bibnamefont {Filipp}},\ }\href {\doibase
  10.1103/PhysRevA.91.043846} {\bibfield  {journal} {\bibinfo  {journal} {Phys.
  Rev. A}\ }\textbf {\bibinfo {volume} {91}},\ \bibinfo {pages} {043846}
  (\bibinfo {year} {2015})}\BibitemShut {NoStop}%
\bibitem [{\citenamefont {Beige}\ \emph {et~al.}(2000)\citenamefont {Beige},
  \citenamefont {Braun}, \citenamefont {Tregenna},\ and\ \citenamefont
  {Knight}}]{beige2000quantum}%
  \BibitemOpen
  \bibfield  {author} {\bibinfo {author} {\bibfnamefont {A.}~\bibnamefont
  {Beige}}, \bibinfo {author} {\bibfnamefont {D.}~\bibnamefont {Braun}},
  \bibinfo {author} {\bibfnamefont {B.}~\bibnamefont {Tregenna}}, \ and\
  \bibinfo {author} {\bibfnamefont {P.~L.}\ \bibnamefont {Knight}},\ }\href
  {\doibase 10.1103/PhysRevLett.85.1762} {\bibfield  {journal} {\bibinfo
  {journal} {Phys. Rev. Lett.}\ }\textbf {\bibinfo {volume} {85}},\ \bibinfo
  {pages} {1762} (\bibinfo {year} {2000})}\BibitemShut {NoStop}%
\bibitem [{\citenamefont {Kazakov}(2001)}]{Kazakov_JOptB_2001}%
  \BibitemOpen
  \bibfield  {author} {\bibinfo {author} {\bibfnamefont {A.~Y.}\ \bibnamefont
  {Kazakov}},\ }\href@noop {} {\bibfield  {journal} {\bibinfo  {journal} {J.
  Opt. B: Quantum and Semiclassical Optics}\ }\textbf {\bibinfo {volume} {3}},\
  \bibinfo {pages} {97} (\bibinfo {year} {2001})}\BibitemShut {NoStop}%
\bibitem [{Note11()}]{Note11}%
  \BibitemOpen
  \bibinfo {note} {Alternatively, the auxiliary level $\mathinner {|{\protect
  \rm a}\delimiter "526930B }$ can be coupled to the level $\mathinner
  {|{1}\delimiter "526930B }$, instead of $\mathinner {|{3}\delimiter "526930B
  }$, in order to compensate for the Stark shifts in the effective Hamiltonian.
  The coupling of three-level atom to a classical Rabi field was considered
  in~\cite {Kazakov_JOptB_2001}.}\BibitemShut {Stop}%
\bibitem [{\citenamefont {Scully}\ and\ \citenamefont
  {Zubairy}(2001)}]{AtomicOpticsBook}%
  \BibitemOpen
  \bibinfo {editor} {\bibfnamefont {M.~O.}\ \bibnamefont {Scully}}\ and\
  \bibinfo {editor} {\bibfnamefont {M.~S.}\ \bibnamefont {Zubairy}},\ eds.,\
  \href@noop {} {\emph {\bibinfo {title} {Quantum Optics}}}\ (\bibinfo
  {publisher} {Cambridge University Press},\ \bibinfo {address} {Cambridge,
  UK},\ \bibinfo {year} {2001})\BibitemShut {NoStop}%
\bibitem [{Note12()}]{Note12}%
  \BibitemOpen
  \bibinfo {note} {$S_k$ is only determined up to an anti-symmetric operator
  commuting with $H_0$ [cf.~Eq.~\protect \textup {\hbox {\mathsurround \z@
  \protect \normalfont (\ignorespaces \ref {eq:Scond}\unskip \@@italiccorr
  )}}]. This freedom, therefore, corresponds only to unitary transformations in
  degenerate eigenbasis of $H_0$, so that $H_\protect \text {diag}$ in~\protect
  \textup {\hbox {\mathsurround \z@ \protect \normalfont (\ignorespaces \ref
  {eq:expansion}\unskip \@@italiccorr )}} remains diagonal up to this
  degeneracy. Here we assume this transformation to be identity.}\BibitemShut
  {Stop}%
\bibitem [{\citenamefont {Albert}\ \emph {et~al.}(2016)\citenamefont {Albert},
  \citenamefont {Bradlyn}, \citenamefont {Fraas},\ and\ \citenamefont
  {Jiang}}]{albert2016geometry}%
  \BibitemOpen
  \bibfield  {author} {\bibinfo {author} {\bibfnamefont {V.~V.}\ \bibnamefont
  {Albert}}, \bibinfo {author} {\bibfnamefont {B.}~\bibnamefont {Bradlyn}},
  \bibinfo {author} {\bibfnamefont {M.}~\bibnamefont {Fraas}}, \ and\ \bibinfo
  {author} {\bibfnamefont {L.}~\bibnamefont {Jiang}},\ }\href {\doibase
  10.1103/PhysRevX.6.041031} {\bibfield  {journal} {\bibinfo  {journal} {Phys.
  Rev. X}\ }\textbf {\bibinfo {volume} {6}},\ \bibinfo {pages} {041031}
  (\bibinfo {year} {2016})}\BibitemShut {NoStop}%
\bibitem [{\citenamefont {Azouit}\ \emph {et~al.}(2016)\citenamefont {Azouit},
  \citenamefont {Sarlette},\ and\ \citenamefont
  {Rouchon}}]{azouit2016adiabatic}%
  \BibitemOpen
  \bibfield  {author} {\bibinfo {author} {\bibfnamefont {R.}~\bibnamefont
  {Azouit}}, \bibinfo {author} {\bibfnamefont {A.}~\bibnamefont {Sarlette}}, \
  and\ \bibinfo {author} {\bibfnamefont {P.}~\bibnamefont {Rouchon}},\ }in\
  \href {\doibase 10.1109/CDC.2016.7798963} {\emph {\bibinfo {booktitle} {2016
  IEEE 55th Conference on Decision and Control (CDC), Las Vegas}}}\ (\bibinfo
  {year} {2016})\ pp.\ \bibinfo {pages} {4559--4565}\BibitemShut {NoStop}%
\bibitem [{Note13()}]{Note13}%
  \BibitemOpen
  \bibinfo {note} {In this case, we can estimate the leading probability
  $\protect \overline {p}_a(T)$ from below by considering the fidelity of
  $\protect \overline \rho _\protect \text {at}(T)$ to any state ${\protect
  \overline {p}_a(T)\geq \delimiter "426830A \psi |\protect \overline \rho
  _\protect \text {at}(T)|\psi \delimiter "526930B }$. For example, for ${|\psi
  \delimiter "526930B =|\protect \overline \psi _\protect \text
  {at}(T)\delimiter "526930B }$ we have $\protect \overline {p}_a(T) \geq [
  1-\Gamma _1 T |c_g|^2-\Gamma _3 T |c_e|^2 + \gamma _{13} T |c_e|^2 |c_g|^2
  ]/\protect \overline {p}(T)$ in the limit $\Gamma _1,\Gamma _3 \ll T^{-1}$
  [by considering the contribution to $\rho _\protect \text {at}(t)$ from
  nondecaying events and decay ${|3\delimiter "526930B }$ to ${|1\delimiter
  "526930B }$]. The bound in Eq.~\protect \textup {\hbox {\mathsurround \z@
  \protect \normalfont (\ignorespaces \ref {eq:deph_bound1}\unskip
  \@@italiccorr )}} gives the resulting contribution to the dephasing as
  $\gamma _\protect \text {deph}(T)\lesssim 2 \nu \left [\protect \overline
  p(T)-1+\Gamma _1 T |c_g|^2+\left (\Gamma _3-\gamma _{13}|c_g|^2 \right ) T
  |c_e|^2 \right ]$.}\BibitemShut {Stop}%
\bibitem [{Note14()}]{Note14}%
  \BibitemOpen
  \bibinfo {note} {The eigenvalues of $\protect \mathcal {M}$ can be found
  within a unit circle in the complex plane, which implies that the eigenvalues
  of $\protect \mathcal {L}$ are found within the circle of radius $\nu $
  centered at $-\nu $.}\BibitemShut {Stop}%
\bibitem [{\citenamefont {Weber}\ \emph {et~al.}()\citenamefont {Weber},
  \citenamefont {Tresp}, \citenamefont {Menke}, \citenamefont {Urvoy},
  \citenamefont {Firstenberg}, \citenamefont {B{\"u}chler},\ and\ \citenamefont
  {Hofferberth}}]{Weber_PairInteractions_package}%
  \BibitemOpen
  \bibfield  {author} {\bibinfo {author} {\bibfnamefont {S.}~\bibnamefont
  {Weber}}, \bibinfo {author} {\bibfnamefont {C.}~\bibnamefont {Tresp}},
  \bibinfo {author} {\bibfnamefont {H.}~\bibnamefont {Menke}}, \bibinfo
  {author} {\bibfnamefont {A.}~\bibnamefont {Urvoy}}, \bibinfo {author}
  {\bibfnamefont {O.}~\bibnamefont {Firstenberg}}, \bibinfo {author}
  {\bibfnamefont {H.~P.}\ \bibnamefont {B{\"u}chler}}, \ and\ \bibinfo {author}
  {\bibfnamefont {S.}~\bibnamefont {Hofferberth}},\ }\href@noop {} {\enquote
  {\bibinfo {title} {Pairinteractions package},}\ }\bibinfo {howpublished}
  {\url{https://pairinteraction.github.io/pairinteraction/sphinx/html/index.html}}\BibitemShut
  {NoStop}%
\bibitem [{\citenamefont {Weber}\ \emph {et~al.}(2017)\citenamefont {Weber},
  \citenamefont {Tresp}, \citenamefont {Menke}, \citenamefont {Urvoy},
  \citenamefont {Firstenberg}, \citenamefont {B{\"u}chler},\ and\ \citenamefont
  {Hofferberth}}]{Weber_JPhysB_2017}%
  \BibitemOpen
  \bibfield  {author} {\bibinfo {author} {\bibfnamefont {S.}~\bibnamefont
  {Weber}}, \bibinfo {author} {\bibfnamefont {C.}~\bibnamefont {Tresp}},
  \bibinfo {author} {\bibfnamefont {H.}~\bibnamefont {Menke}}, \bibinfo
  {author} {\bibfnamefont {A.}~\bibnamefont {Urvoy}}, \bibinfo {author}
  {\bibfnamefont {O.}~\bibnamefont {Firstenberg}}, \bibinfo {author}
  {\bibfnamefont {H.~P.}\ \bibnamefont {B{\"u}chler}}, \ and\ \bibinfo {author}
  {\bibfnamefont {S.}~\bibnamefont {Hofferberth}},\ }\href@noop {} {\bibfield
  {journal} {\bibinfo  {journal} {J. Phys. B}\ }\textbf {\bibinfo {volume}
  {50}},\ \bibinfo {pages} {133001} (\bibinfo {year} {2017})}\BibitemShut
  {NoStop}%
\bibitem [{Note15()}]{Note15}%
  \BibitemOpen
  \bibinfo {note} {The chosen upper bounds in the conditions for rotating wave
  approximation and far detuned limit can be of course made more stringent. We
  have chosen 0.1 which is the first order satisfying the (order of magnitude)
  relation $0.1 \ll 1$. It turns out that for the set of basis states used, the
  number of post-selected levels reduces to 16 (0) if we set the bound to 0.05
  (0.03) instead.}\BibitemShut {Stop}%
\bibitem [{\citenamefont {Gallagher}(1994)}]{Gallagher_1994}%
  \BibitemOpen
  \bibinfo {editor} {\bibfnamefont {T.~F.}\ \bibnamefont {Gallagher}},\ ed.,\
  \href@noop {} {\emph {\bibinfo {title} {Rydberg atoms}}}\ (\bibinfo
  {publisher} {Cambridge University Press},\ \bibinfo {address} {Cambridge},\
  \bibinfo {year} {1994})\BibitemShut {NoStop}%
\bibitem [{\citenamefont {Li}\ \emph {et~al.}(2003)\citenamefont {Li},
  \citenamefont {Mourachko}, \citenamefont {Noel},\ and\ \citenamefont
  {Gallagher}}]{Li_PRA_2003}%
  \BibitemOpen
  \bibfield  {author} {\bibinfo {author} {\bibfnamefont {W.}~\bibnamefont
  {Li}}, \bibinfo {author} {\bibfnamefont {I.}~\bibnamefont {Mourachko}},
  \bibinfo {author} {\bibfnamefont {M.~W.}\ \bibnamefont {Noel}}, \ and\
  \bibinfo {author} {\bibfnamefont {T.~F.}\ \bibnamefont {Gallagher}},\ }\href
  {\doibase 10.1103/PhysRevA.67.052502} {\bibfield  {journal} {\bibinfo
  {journal} {Phys. Rev. A}\ }\textbf {\bibinfo {volume} {67}},\ \bibinfo
  {pages} {052502} (\bibinfo {year} {2003})}\BibitemShut {NoStop}%
\bibitem [{\citenamefont {Saffman}\ \emph {et~al.}(2010)\citenamefont
  {Saffman}, \citenamefont {Walker},\ and\ \citenamefont
  {M\o{}lmer}}]{Saffman_RMP_2010}%
  \BibitemOpen
  \bibfield  {author} {\bibinfo {author} {\bibfnamefont {M.}~\bibnamefont
  {Saffman}}, \bibinfo {author} {\bibfnamefont {T.~G.}\ \bibnamefont {Walker}},
  \ and\ \bibinfo {author} {\bibfnamefont {K.}~\bibnamefont {M\o{}lmer}},\
  }\href {\doibase 10.1103/RevModPhys.82.2313} {\bibfield  {journal} {\bibinfo
  {journal} {Rev. Mod. Phys.}\ }\textbf {\bibinfo {volume} {82}},\ \bibinfo
  {pages} {2313} (\bibinfo {year} {2010})}\BibitemShut {NoStop}%
\bibitem [{\citenamefont {Pritchard}(2012)}]{Pritchard_thesis_2012}%
  \BibitemOpen
  \bibfield  {author} {\bibinfo {author} {\bibfnamefont {J.~D.}\ \bibnamefont
  {Pritchard}},\ }\emph {\bibinfo {title} {{Cooperative Optical Non-linearity
  in a blockaded Rydberg ensemble}}},\ \href@noop {} {Ph.D. thesis},\ \bibinfo
  {school} {Durham University, United Kingdom} (\bibinfo {year}
  {2012})\BibitemShut {NoStop}%
\bibitem [{\citenamefont {Van~Wijngaarden}(1997)}]{VanWijngaarden_JSpec_1997}%
  \BibitemOpen
  \bibfield  {author} {\bibinfo {author} {\bibfnamefont {W.}~\bibnamefont
  {Van~Wijngaarden}},\ }\href@noop {} {\bibfield  {journal} {\bibinfo
  {journal} {Journal of Quantitative Spectroscopy and Radiative Transfer}\
  }\textbf {\bibinfo {volume} {57}},\ \bibinfo {pages} {275} (\bibinfo {year}
  {1997})}\BibitemShut {NoStop}%
\bibitem [{\citenamefont {O'Sullivan}\ and\ \citenamefont
  {Stoicheff}(1986)}]{OSullivan_PRA_1986}%
  \BibitemOpen
  \bibfield  {author} {\bibinfo {author} {\bibfnamefont {M.~S.}\ \bibnamefont
  {O'Sullivan}}\ and\ \bibinfo {author} {\bibfnamefont {B.~P.}\ \bibnamefont
  {Stoicheff}},\ }\href {\doibase 10.1103/PhysRevA.33.1640} {\bibfield
  {journal} {\bibinfo  {journal} {Phys. Rev. A}\ }\textbf {\bibinfo {volume}
  {33}},\ \bibinfo {pages} {1640} (\bibinfo {year} {1986})}\BibitemShut
  {NoStop}%
\end{thebibliography}
%\bibliographystyle{apsrev4-1}

%\input{Micromasers_PRX_v1_suppl.bbl}

	%*******************************************************************************************************************************************************

\end{document}